\documentclass{aa}  

\usepackage{natbib}
\usepackage{txfonts}
\usepackage{hyperref}
\usepackage{graphicx,xcolor,textpos}
\usepackage{rotating}
\usepackage{longtable,ltcaption}
\usepackage{lscape} 
\usepackage[utf8]{inputenc}
\usepackage{gensymb}
\usepackage{caption}
\usepackage{subcaption}
\usepackage{float}
\usepackage{placeins}
\usepackage[maxfloats=150]{morefloats}
\newcommand{\Msun}{M$_{\odot}$\,}

\begin{document} 
  \title{Barium and related stars, and their white-dwarf companions\\
  III. The masses of the white dwarfs}
  \author{A. Escorza \inst{1} \and
          R. J. De Rosa\inst{1}
         }

  \institute{European Southern Observatory, Alonso de C\'{o}rdova 3107, Vitacura, Santiago, Chile\\
              \email{ana.escorza@eso.org}
            }

  \date{\today}

  \abstract
   {Masses are one of the most difficult stellar properties to measure. In the case of the white-dwarf (WD) companions of Barium (Ba) stars, the situation is worse. These stars are dim, cool, and difficult to observe via direct methods. However, Ba stars were polluted by the Asymptotic Giant Branch (AGB) progenitors of these WDs with matter rich in heavy elements, and the properties of their WD companions contain key information about binary interaction processes involving AGB stars and about the slow neutron capture process (s-process) of nucleosynthesis.}
   {With this study, we aim to determine accurate and assumption-free masses for the WD companions of as many Ba stars as possible. We want to provide new observational constraints that can help us learn about the formation and evolution of these post-interaction binary systems and about the nucleosythesis processes that took place in the interiors of their AGB progenitors.}
   {We combined archival radial-velocity data with Hipparcos and Gaia astrometry using the software package \textsc{orvara}, a code designed to simultaneously fit a single Keplerian model to any combination of these types of data using a parallel-tempering Markov chain Monte Carlo method. We adopted Gaussian priors for the Ba star masses and for the parallaxes, and assumed uninformative priors for the orbital elements and the WD masses.}
   {We determined new orbital inclinations and companion masses for 60 Ba star systems. These results include a couple of new orbits and several improved orbits for the longest-period systems. Additionally, we unravelled a new triple system that was not known before and constrained the orbits and the masses of the two companions.}
   {The WD mass distribution presented in this work is compatible with that of field WDs and with the distributions published before for Ba star companions. A few WD companions have masses higher than 0.8~M$_{\odot}$, considering 1-$\sigma$ uncertainties. This indicates that they might come from AGB stars that are more massive than 3~M$_{\odot}$. These masses are higher than what the abundance ratios on Ba star atmospheres and theoretical models of the s-process of nucleosynthesis seem to expect, raising interesting questions about the formation of these systems.}


\maketitle


\section{Introduction}\label{sec:intro}

About half of the elements heavier than iron are synthesized by the slow neutron capture (s-) process of nucleosynthesis \citep[e.g.][]{Burbidge57, Clayton61, Kappeler11}. The main astrophysical site that meets the appropriate conditions for the s-process to operate is the helium-rich intershell in the interiors of thermally pulsing Asymptotic Giant Branch (tp-AGB) stars \citep[e.g.][]{Lugaro03long, Cristallo09, Karakas10, Kappeler11}. However, the overabundance of s-process elements on the surface of a star is not a unique feature of AGB stars. Barium (Ba) stars are an example of s-process enriched objects that have not reached the tp-AGB phase yet. They are known to form when an AGB companion pollutes them in a binary system \citep[e.g.][]{McClure80, McClure84, Udry98, Jorissen98}. The mass donors in these systems evolved off the AGB long ago and are now dim white dwarfs (WD), while the accretors -- the Ba stars -- are observed on the main sequence \citep[e.g.][]{NorthDuquennoy91, JorissenBoffin92, North94, North00, Pereira2005, Kong18, Escorza19}, the red-giant \citep[e.g.][]{BidelmanKeenan51, McClure83, Udry98II, Jorissen04, Escorza17, Jorissen19}, and the AGB \citep[as extrinsic S stars, e.g.][]{Jorissen98,Jorissen19,Shetye20} phases.

Although their exact formation channel and the mass-transfer mechanisms involved are not well understood \citep[e.g.][]{ToutEggleton88, Han95, Soker00, Pols03, BonacicMarinovic08, Izzard10, Dermine13, Abate18, Saladino19, Gao22}, our knowledge about the spectroscopic orbital parameters of Ba star systems and about the stellar properties of the Ba stars themselves is generally well established \citep[e.g.][and references therein]{Escorza19, Jorissen19}. Additionally, the evolutionary link between dwarf and giant Ba stars is well accepted \citep[e.g.][]{Escorza20}. However, not much is known about the WD companions. The mass-function distribution of Ba star systems is consistent with a narrow distribution of companion masses peaking at 0.6~\Msun \citep[e.g.][]{Webbink86, McClureWoodsworth90, Jorissen98, Merle16, Jorissen19, Escorza19EWASS}, but very few absolute masses have been determined, since there is normally no information about the orbital inclinations of these systems (a few exceptional cases were published by \citealt{PJ2000, Escorza19, Jorissen19}, among others, by combining the orbital parameters of Ba stars with Hipparcos astrometric data). These WDs are cool, dim, and directly undetectable in most cases; although, \cite{Bohm-Vitense84, Bohm-Vitense00, Gray11}, among others detected UV excess flux attributable to the WD in a few Ba star systems.

The masses of the WD companions of Ba stars contain important information about the AGB progenitors and the nucleosynthesis processes that took place in their interiors, and they are important input for binary interaction models. Even though mixing and dilution processes such as thermohaline mixing \citep[e.g.][]{Proffitt89, Charbonnel07, Stancliffe07, Stancliffe08, Aoki08}, rotationally induced mixing \citep[e.g.][]{Denissenkov00}, or atomic diffusion \citep[e.g.][]{Matrozis16, Matrozis17} might impact the final level of s-process abundance on Ba stars, correlations between these abundances and the WD mass can give us observational information about the efficiency of the s-process at different masses and metallicities and help us constrain AGB models \citep[e.g.][]{Cseh21} and mass-transfer and dilution models \citep[e.g.][]{Stancliffe21}. The ratio between the amount of heavy s-process elements (hs), such as Ba, La, or Ce, and light s-process elements (ls), such as Sr, Y, or Zr, on the surface of Ba stars suggests that the material accreted by these stars was synthesized by low-mass AGB stars (< 3 \Msun; \citealt{Lugaro03, Lugaro12, Lugaro16, Cseh18, Karinkuzhi18}), which still needs to be confirmed by measuring these WD masses. Additionally, \cite{Jorissen19} suggested that WD companions of strong Ba giants (based on the Ba index introduced by \citealt{Warner1965}) are more massive on average than the WD companions of mild Ba stars. However, most of their masses were determined under the assumption of a constant (or very narrow distribution of) $Q = M^{3}_{\rm WD}/(M_{\rm Ba} + M_{\rm WD})^{2}$ as proposed by \citet{Webbink1988} and \cite{McClureWoodsworth90}, so this trend still needs to be confirmed with assumption-free measurements of WD masses.

In the first two papers of this series, \cite{Jorissen19} and \cite{Escorza19} collected old and new radial-velocity (RV) data to study the orbits of giant and dwarf Ba stars, respectively. Additionally, we used spectroscopically-determined stellar parameters and Gaia DR2 distances \citep{Lindegren18, Bailer-Jones18} to locate these stars on the Hertzsprung–Russell diagram (HRD). By comparing their location on the HRD with STAREVOL evolutionary tracks \citep{Siess00, Siess2006, Siess08} and following the methodology described in \cite{Escorza17}, we also determined accurate masses for the primary stars of these systems, the Ba stars. In this third article, we focus on the faint WD companions. We used the \textsc{orvara} software \citep{Brandt21} to combine all the radial-velocity data available, the astrometric measurements from the Hipparcos mission \citep{HippCat97}, the Gaia positions and proper motions \citep{GaiaEDR3}, and the information in the Hipparcos-Gaia Catalogue of Accelerations (HGCA; \citealt{Brandt18,HGCA-EDR3}) to determine the astrometric orbital parameters of as many Ba star systems as possible (see Sect. \ref{sec:targets} for the description of the sample), and then derive the mass of the secondary stars. All these data sets are described in Sect. \ref{sec:data}. An important improvement with respect to what has been attempted before for these objects is that we use a joint astrometric-spectroscopic model (see Sect. \ref{sec:methods}) to find new best-fitting orbital parameters instead of relying only on RV data or imposing the spectroscopic solution on the astrometric data. Our results are presented in Sect. \ref{sec:results} and their implications are discussed in Sect. \ref{sec:discussion}. We also discuss the feasibility of the direct detection of the WD companion for a subset of the longest-period systems in Sect. \ref{sec:sphereplans}.


\section{Target selection}\label{sec:targets}
For our methodology (see Sect. \ref{sec:methods}) to be applicable, a target must fulfil three requirements: (i) it must be part of the  HGCA, (ii) we must have a good initial estimate of the mass of the primary star in the system, and (iii) the Hipparcos solution cannot not be more complex than the 5-parameter solution. As a starting point, we selected all the Ba stars from the samples studied by \cite{Jorissen19}, \cite{Escorza19} and \citealt{North20} that have Hipparcos identifiers. We excluded confirmed triple systems, stars whose Ba star nature was not certain or is under current investigation (e.g. \citealt{Escorza56UMa}), and a few systems that had an acceleration solution or an orbital solution in the Hipparcos data reduction (solution types, Sn, equal to 7, 9 and 75). We ended up with 60 systems.

Table \ref{table:targets} presents our target list. In addition to the most commonly used identifier, we include the Hipparcos identifier of each system and the Ba star type. We distinguish between \textit{preRGB}, which are all the stars classified as dwarfs or subgiants by \cite{Escorza19} and \cite{North20}, and mBag or sBag which are stars classified as mild or strong Ba giants by \cite{Jorissen19} based on their [La/Fe] and [Ce/Fe] values \citep[as measured by][]{Smith84, AllenBarbuy06I,AllenBarbuy06II, Pereira11, Karinkuzhi14, Karinkuzhi15, Luck14, deCastro16, Merle16, VanderSwaelmen17, Karinkuzhi18, Jorissen19} and on the Ba index introduced by \cite{Warner1965}. The table also lists the Ba star masses ($M_{\rm Ba}$) that we used as a prior in our MCMC model (see Sect. \ref{sec:methods}) and the metallicity of the system, both values collected from \cite{Jorissen19}, \cite{Escorza19} or \cite{North20} unless explicitly specified. For this work, we recomputed the primary masses for the ten systems that were part of the non-single-star (NSS) Gaia DR3 catalogues \citep{GaiaDR3binaries}. We followed the exact same procedure followed and described in the mentioned papers and used the same STAREVOL grids of models \citep{Siess00,SiessArnould08,Escorza17}, but we used the NSS Gaia DR3 parallaxes to recalculate their luminosities and masses. Finally, the last column of Table \ref{table:targets} includes the sources where we found the archival RV data used in our analysis.

\setlength{\tabcolsep}{5pt}
\renewcommand{\arraystretch}{1.2}
\begin{table*}
\caption{List of Ba star systems to which our methodology was applied. Column 1 lists the most commonly used identifiers, while column 2 lists the Hipparcos identifiers. Column 3 lists the Ba star type, which can be preRGB for stars classified as dwarfs or subgiants, or mBag or sBag for stars classified as mild or strong Ba giants, respectively. Column 4 lists the primary star masses and column 5, the metallicity of the system. These values were derived or collected by \cite{Escorza19} or \cite{Jorissen19} for preRGB and giant systems, respectively, unless otherwise indicated. Finally, the last column gives the sources of the archival RV data we used.}\label{table:targets}
{\centering
\begin{small}
\begin{tabular}{lccccc||lccccc}
\hline\hline       
\textbf{BD/HD} & \textbf{HIP} & \textbf{type} & \textbf{$M_{\rm Ba}$ [M$_{\odot}$]} & \textbf{[Fe/H]} & \textbf{RV ref}$^{*}$ & \textbf{HD} & \textbf{HIP} & \textbf{type} & \textbf{$M_{\rm Ba}$ [M$_{\odot}$]} & \textbf{[Fe/H]} & \textbf{RV ref}$^{*}$ \\
\hline                    
-10$^{o}$4311 & 80356 & preRGB & 0.8 $\pm$ 0.1 & $-0.65$ & E19 & 95241 & 53791 & preRGB & 1.3 $\pm$ 1.1 & $-0.37$ & E19,M04\\
-11$^{o}$3853 & 73444 & preRGB & 0.85 $\pm$ 0.04$^{(1)}$ & $-0.94^{(1)}$ & E19 & 98991 & 55598 & preRGB & 1.45 $\pm$ 0.08 & $-0.44$ & E19 \\
-14$^{o}$2678 & 43527 & mBag & 3.0 $\pm$ 0.2 & $0.01$ & U98a & 104979 & 58948 & mBag & 2.7 $\pm$ 0.2 & $-0.26$ & J19,M90 \\
2454 & 2235 & preRGB & 1.23 $\pm$ 0.07$^{(1)}$ & $-0.29^{(1)}$ & E19,M04 & 107541 & 60292 & sBag & 1.1 $\pm$ 0.2 & $-0.63$ & U98b\\
5424 & 4347 & sBag & 1.3 $\pm$ 0.4 & $-0.43$ & U98a & 107574 & 60299 & preRGB & 1.11 $\pm$ 0.05 & $-0.80$ & E19\\
16458 & 13055 & sBag & 1.9 $\pm$ 0.1 & $-0.64$ & M90 & 119185 & 66844 & mBag & 1.7 $\pm$ 0.2 & $-0.42$ & J19,U98a\\
18182 & 13596 & mBag & 1.8 $\pm$ 0.2 & $-0.17$ & J19 & 121447 & 68023 & sBag & 1.6 $\pm$ 0.1 & $-0.90$ & J95\\
20394 & 15264 & sBag & 2.0 $\pm$ 0.2 & $-0.27$ & G96 & 123585 & 69176 & preRGB & 1.0 $\pm$ 0.1$^{(0)}$ & $-0.50$ & E19\\
24035 & 17402 & sBag & 1.3 $\pm$ 0.3 & $-0.23$ & U98a & 123949 & 69290 & sBag & 1.3 $\pm$ 0.3 & $-0.23$ & J19,U98a\\
27271 & 20102 & mBag & 2.9 $\pm$ 0.2 & $-0.07$ & U98b,M04 & 127392 & 71058 & preRGB & 0.8 $\pm$ 0.3 & $-0.52$ & E19\\
31487 & 23168 & sBag & 2.5 $\pm$ 0.2$^{(2)}$ & $-0.04^{(2)}$ & M90 & 139195 & 76425 & mBag & 2.6 $\pm$ 0.1 & $-0.07$ & M90,G91\\
34654 & 25222 & preRGB & 1.19 $\pm$ 0.05$^{(0)}$ & $-0.09$ & E19 & 143899 & 78681 & mBag & 2.4 $\pm$ 0.1 & $-0.29$ & U98a\\
40430 & 28265 & mBag & 2.3 $\pm$ 0.2 & $-0.34$ & J19 & 178717 & 94103 & sBag & 1.6 $\pm$ 0.9 & $-0.52$ & M90\\
43389 & 29740 & sBag & 1.8 $\pm$ 0.4 & $-0.35$ & U98b & 180622 & 94785 & mBag & 1.8 $\pm$ 0.3 & $0.03$ & U98b\\
44896 & 30338 & sBag & 3.0 $\pm$ 1.2$^{(0)}$ & $-0.25$ & U98b & 182274 & 95293 & preRGB & 1.09 $\pm$ 0.05 & $-0.32$ & E19,M04\\
49641 & 32713 & sBag & 2.7 $\pm$ 1.2 & $-0.3$ & M90 & 183915 & 96024 & sBag & 1.8 $\pm$ 1.0 & $-0.59$ & J19,J98\\
49841 & 32831 & mBag & 2.85 $\pm$ 0.10$^{(0)}$ & $0.2$ & U98b & 199939 & 103546 & sBag & 2.7 $\pm$ 0.4$^{(0)}$ & $-0.22$ & M90,M04\\
50082 & 32960 & sBag & 1.6 $\pm$ 0.3 & $-0.32$ & U98b & 200063 & 103722 & mBag & 2.0 $\pm$ 1.3 & $-0.34$ & U98b\\
50264 & 32894 & preRGB & 0.9 $\pm$ 0.1$^{(0)}$ & $-0.34$ & E19 & 201657 & 104542 & sBag & 1.8 $\pm$ 0.5 & $-0.34$ & U98b\\
51959 & 33628 & mBag & 1.2 $\pm$ 0.1 & $-0.21$ & J19 & 201824 & 104684 & sBag & 1.7 $\pm$ 0.4 & $-0.40$ & G96\\
53199 & 34143 & mBag & 2.5 $\pm$ 0.1 & $-0.20$ & J19,M04 & 202400 & 105294 & preRGB & 0.98 $\pm$ 0.08$^{(3)}$ & $-0.7^{(3)}$ & N20\\
58121 & 35935 & mBag & 2.6 $\pm$ 0.5 & $-0.01$ & U98b & 204075 & 105881 & mBag & 4.5 $\pm$ 0.3 & $-0.09$ & M90\\
58368 & 36042 & mBag & 2.6 $\pm$ 0.2 & $0.04$ & M90 & 205011 & 106306 & mBag & 1.8 $\pm$ 0.3 & $-0.26$ & J98,M90\\
59852 & 36613 & mBag & 2.5 $\pm$ 0.3 & $-0.22$ & U98a & 207585 & 107818 & preRGB & 0.90 $\pm$ 0.10$^{(0)}$ & $-0.57$ & E19\\
77247 & 44464 & mBag & 3.9 $\pm$ 0.2 & $-0.13$ & M90 & 210946 & 109747 & mBag & 1.8 $\pm$ 0.5 & $-0.29$ & U98b,M04\\
87080 & 49166 & preRGB & 1.38 $\pm$ 0.15$^{(0)}$ & $-0.60$ & E19 & 211594 & 110108 & sBag & 2.0 $\pm$ 0.3 & $-0.29$ & U98b\\
88562 & 50006 & sBag & 1.0 $\pm$ 0.1 & $-0.53$ & U98a & 216219 & 112821 & preRGB & 1.45 $\pm$ 0.1 & $-0.17$ & E19\\
91208 & 51533 & mBag & 2.3 $\pm$ 0.2 & $-0.16$ & U98a & 218356 & 114155 & mBag & 4.3 $\pm$ 1.1 & $-0.06$ & G06,U98b\\
92626 & 52271 & sBag & 3.1 $\pm$ 0.6 & $-0.15$  & U98b & 221531 & 116233 & preRGB & 1.2 $\pm$ 0.1$^{(0)}$ & $-0.30$ & E19\\
95193 & 53717 & mBag & 2.7 $\pm$ 0.1 & $-0.04$ & U98a & 224621 & 118266 & preRGB &0.85 $\pm$ 0.06$^{(0)}$ & $-0.4$ & N20\\
\hline\hline
\end{tabular}
\end{small}\\}
\vspace{2mm}
$^{*}$ \textbf{RV reference abbreviations:} E19: \cite{Escorza19}, U98a: \cite{Udry98}, M04: \cite{Moultaka04}, M90: \cite{McClureWoodsworth90}, J19: \cite{Jorissen19}, G96: \cite{Griffin96}, U98b: \cite{Udry98II}, J98: \cite{Jorissen98}, J95: \cite{Jorissen95}, G91: \cite{Griffin91}, N20: \cite{North20}, G06: \cite{Griffin06}\\ 
\textbf{Mass \& metallicity references:} $^{(0)}$ This work; $^{(1)}$ \cite{Bensby18}; $^{(2)}$ \cite{Karinkuzhi18}; $^{(3)}$ \cite{North20}
\end{table*}


\section{Radial velocity and astrometric data}\label{sec:data}
\subsection{CORAVEL, HERMES and other radial-velocity data}\label{sec:RVdata}
The most important radial-velocity monitoring programs of Ba stars were carried out with the two CORAVEL spectrometers and with the HERMES high-resolution spectrograph. The CORAVEL spectrometers \citep{Baranne79} were installed on the 1-m Swiss telescope at the Haute-Provence Observatory and on the 1.54-m Danish telescope at ESO - La Silla, while HERMES \citep{Raskin11, Raskin14} is mounted on the 1.2-m Flemish Mercator telescope at the Observatory El Roque de Los Muchachos.

The main results of these radial-velocity programs were published by \cite{Jorissen88, Jorissen98, Udry98, Udry98II, North00, Gorlova13, Jorissen19, Escorza19} among others, and the strength of combining the two data sets, particularly for the longest-period systems, was discussed in the last two mentioned papers. \cite{Jorissen19} and \cite{Escorza19} also described the data reduction process for the two instruments and the existence of a non-zero radial-velocity offset between the data sets due to the use of a different system of standard stars. This zero-point offset depends on the stellar velocity and on the target's colour \textit{B-V}, and there is no real consensus about how to treat it. \cite{Jorissen19} derived it after fitting each orbit by minimizing the orbital residuals, while \cite{Escorza19} determined a relation between the offset and \textit{B-V} by comparing old and reprocessed CORAVEL data and calculated a fixed offset for each studied Ba star. For this work, we combined the two approaches. Where the RV data of a specific instrument spanned over a full orbit or more, we treated the offset as an additional free parameter that was optimized during the orbital fitting process. However, for systems with very few HERMES points or for some very long orbits, the offsets from \cite{Jorissen19} or \cite{Escorza19} were adopted and fixed. This will be clearly indicated in the captions of each RV fit shown in Appendix \ref{App}. Future monitoring with HERMES would remove the need for this assumption, allowing us to fit the offset term directly.

To complement the main CORAVEL and HERMES data, we collected additional radial-velocity measurements from other works and instruments, and the sources are listed in Table \ref{table:targets}. An optimizable RV offset, such as the one described above between CORAVEL and HERMES, was considered for each data set.

\subsection{Hipparcos astrometric data}\label{sec:hipp}
The Hipparcos satellite \cite{ESA1997}, launched in 1989, was the first space mission with precision astrometry as its main goal. Between 1989 and 1993, Hipparcos measured the location and motion on the sky of more than 100,000 stars many times, to figure out their astrometric path. For each target in Table \ref{table:targets}, we used the positions and the proper motions from the Hipparcos Catalogue \citep{HippCat97}. Additionally, we also queried the individual astrometric measurements from the re-reduction of the Hipparcos intermediate astrometric data (IAD; \citealt{vanLeeuwen07}). The coordinates are expressed in the International Celestial Reference Frame (ICRF) at the 1991.25 epoch.

Since the code we are using is not yet prepared to deal with Hipparcos solutions more complex than the 5-parameter solutions, we excluded a few targets with acceleration or orbital solutions from the initial sample. Some of our remaining targets have a stochastic Hipparcos solution (Sn = 1). These represent cases where the residuals are significantly larger than expected, but since the proper motions and the IAD were obtained using a 5-parameter solution, we included them and gave them no special treatment.

\subsection{Gaia astrometric data}\label{sec:gaia}

The Gaia mission \citep{GaiaMission, GaiaDR2, GaiaEDR3summary} was launched in 2013 as a successor of Hipparcos. For now, none of the Gaia Data Releases (DR) published individual astrometric measurements, so we queried the positions and proper motions published for our targets in the Early DR3 catalogue \citep{GaiaEDR3}. In contrast with the Hipparcos data, these are expressed in the ICRF at the 2016 epoch. The Gaia EDR3 parallaxes were also queried and used as prior in the fit (see Sect. \ref{sec:methods}). Finally, in order to use an equivalent to epoch astrometry, we also used the Gaia Observation Forecast Tool (GOST\footnote{https://gaia.esac.esa.int/gost/}). The GOST provides the predicted observations and scan angles for any Gaia source. We note that not all the planned observations will be used in the final astrometric solution, since some predicted scans might correspond to satellite dead times or might be unusable or rejected as outliers. For example, up to 20\% of the observations predicted by GOST were excluded from the analysis published in Gaia DR2 \citep{htof}. 

Ten of the 60 targets presented in this study had a non-single-star (NSS) solution in Gaia DR3 \citep{GaiaDR3binaries}. These targets are: HD\,50264, HD\,207585, HD\,221531, HD\,34654, HD\,49841, HD\,199939, HD\,224621 and HD\,87080, which had a non-single-star solution compatible with a combined astrometric and single lined spectroscopic model, and HD\,44896 and HD\,123585, which had a solution compatible with an astrometric binary. For these targets, we used the Gaia DR3 NSS parallax as priors, instead of the EDR3 value. Even though the Gaia DR3 NSS catalogue provided orbital inclinations for these 10 systems, we decided not to include an inclination prior in our calculations to first, treat all systems equally, and second, compare our independently determined inclinations with the new Gaia ones and validate our method.

\subsection{The Hipparcos-Gaia Catalogue of Accelerations.}\label{sec:HGCA}
As an additional astrometric constraint, we used the difference in Hipparcos and Gaia proper motions via the Hipparcos-Gaia Catalogue of Accelerations (HGCA; \citealt{Brandt18,HGCA-EDR3}). This catalogue puts the Hipparcos, Gaia, and Hipparcos-Gaia (H-G) proper motions into the same reference frame to make them suitable for orbital fitting. The Hipparcos-Gaia proper motion is derived from the right ascension and declination measurements from the two missions and is by far the most precise due to the long time elapsed between them (proper motion uncertainties scale inversely with the time baseline). This acceleration in the inertial frame can be used to improve the dynamical parameters of the companion and to measure its mass because it breaks the mass-inclination degeneracy that RV data suffers from. We used the EDR3 version of the HGCA \citep{HGCA-EDR3} for all our targets.

The EDR3 version of the HGCA also provides a $\chi^2$ value between the two most precise proper motion measurements (normally EDR3 and H-G). This value is meant to find accelerating candidates for follow-up and if it is higher than $\sim$11.8 \citep{HGCA-EDR3}, the measured acceleration is considered significant and statistically different, by 3$\sigma$, from constant proper motion. In our case, since all our targets are known binaries, we do not need this $\chi^2$ value to detect accelerators, but it can give us a hint about which systems are truly benefiting from the HGCA measurement. The queried HGCA $\chi^2$ values are included in the last column of our result table (Table \ref{table:results}).


\section{Orbital analysis with \textsc{orvara} }\label{sec:methods}

\textsc{Orvara}, developed by \cite{Brandt21}, is designed to simultaneously fit a single Keplerian model to any combination of radial velocity data and relative and absolute astrometry. The combination of these different data sets, using \textsc{Orvara} or not, has recently proven to be very powerful to improve the accuracy of orbits and to measure precise companion masses, even for very long period systems where the observations only cover part of the orbit (e.g. \citealt{DeRosa20, Kervella20, Brandt21, Venner21, Franson22, BrandtDupuy21, Leclerc22}).

\textsc{Orvara} integrates the Hipparcos and Gaia intermediate astrometry package (\textsc{htof}; \citealt{htof}) to fit the Hipparcos epoch astrometry and the times and scan angles of individual Gaia epochs. The code uses a parallel-tempering Markov chain Monte Carlo method (\textsc{ptmcmc}, \citealt{PTMCMC}) and first fits the RV data. \textsc{Orvara} allows RV points from each instrument to have a different RV zero point, which we need at least for the CORAVEL-HERMES combination as discussed in Sect. \ref{sec:RVdata}. Then the absolute astrometry is included and fit for the five astrometric parameters (positions, $\alpha$ and $\delta$, proper motions, $\mu_{\alpha}$ and $\mu_{\delta}$, and parallax, $\varpi$) using \textsc{htof} at each MCMC step. On top of the five astrometric parameters, we fitted the six Keplerian orbital elements (semimajor axis, $a$, eccentricity, $e$, time of periastron passage, T$_0$, argument of periastron, $\omega$, orbital inclination, $i$, and longitude of the ascending node, $\Omega$), the masses of the two components (M$_{\rm Ba}$ and M$_{\rm WD}$), and a radial-velocity jitter per instrument to be added to the uncertainties. Note that while the difference between the Hipparcos and Gaia reference frames is taken into account in the HGCA, this is not the case for the IAD. However, the rotation difference in the proper motions is $w = (-0.120, 0.173, 0.090)$~mas/yr \citep{Fabricius21}. These values are very small compared to the amplitudes of the proper motion curves that we are measuring (of the order from a few to a couple of tens mas/yr, see Appendix \ref{App}), and smaller than the residuals to these fits in most cases, so we did not take them into account.

For this work, we assumed uninformative priors for the orbital elements and for the WD mass, but we adopted Gaussian priors for the primary mass and for the parallax. For each target, we used the $M_{\rm Ba}$ value given in Table \ref{table:targets} but using three times the error bar as sigma to be conservative and take into account systematic errors not accounted for in the statistical uncertainty. Concerning the parallax, the Gaia EDR3 value was used as prior for most targets, and the Gaia DR3 NSS parallax was used for the 10 targets with a NSS solution. We used 15 temperatures and for each temperature we use 100 walkers with 100,000 steps per walker. In a few cases, we needed to run twice as long or repeat the calculations using an educated starting position based on our knowledge about the systems from the RV-only fits published by \cite{Jorissen19} or \cite{Escorza19}, however, in most cases, the MCMC chains converged quite quickly. We discarded the first 300 recorded steps (the first 15000 overall, as we saved every 50) as the burn-in phase to produce the results presented in Sect. \ref{sec:results}. 

For more details about the computational implementation in \textsc{orvara} and \textsc{htof} and for case studies showing the performance of the code, we refer to the mentioned publications.


\section{Results}\label{sec:results}

\setlength{\tabcolsep}{7.5pt}
\renewcommand{\arraystretch}{1.3}
\begin{sidewaystable*}
\caption{Orbital elements and WD masses derived following the method described in Sect. \ref{sec:methods} and listed in order of decreasing periods. The columns list, in order, the most commonly used identifier, the orbital period $P$, the eccentricity $e$, the time of periastron passage, the absolute semimajor axis of the orbit $a$, the argument of periastron of the visible star $\omega_*$, the longitudes of the ascending node $\Omega$, the orbital inclination $i$, and the WD companion mass $M_{\rm WD}$. To keep the table cleaner, we assumed symmetric error bars, and included only the largest one of the two. The last two columns include the $\chi^2$ of the best fitting model and the HGCA $\chi^2$ value discussed in Sect. \ref{sec:HGCA}.}\label{table:results}
{\centering
\begin{tabular}{lcccccccccc}
\hline\hline       
\textbf{Star ID} & \textbf{$P$ [days]} & \textbf{$e$} & \textbf{$T_0$ [HJD]} & \textbf{$a$ [AU]} & \textbf{$\omega_*$ [$^{\circ}$]} & \textbf{$\Omega$ [$^{\circ}$]} & \textbf{$i$ [$^{\circ}$]} & \textbf{$M_{\rm WD}$} & \textbf{Fit $\chi^2$} & \textbf{$\chi^2_{\rm HGCA}$}\\\hline 
HD\,2454 & 29220 $\pm$ 7670 & 0.59 $\pm$ 0.04 & 2458626 $\pm$ 173 & 22 $\pm$ 4 & 313 $\pm$ 19 & 11 $\pm$ 165 & 34 $\pm$ 6 & 0.50 $\pm$ 0.09 & 1.02 & 53537\\
HD\,119185 & 25385 $\pm$ 5114 & 0.61 $\pm$ 0.08 & 2477092 $\pm$ 5626 & 22.5 $\pm$ 3.0 & 105 $\pm$ 7 & 136 $\pm$ 6 & 98 $\pm$ 13 & 0.65 $\pm$ 0.08 & 2.82 & 869\\
BD-11$^{o}$3853 & 23376 $\pm$ 9862 & 0.46 $\pm$ 0.05 & 2472263 $\pm$ 10323 & 19 $\pm$ 6 & 199 $\pm$ 19 & 133 $\pm$ 10 & 102 $\pm$ 9 & 0.76 $\pm$ 0.14 & 0.46 & 2534\\
HD\,104979 & 18518 $\pm$ 1205 & 0.12 $\pm$ 0.04 & 2460663 $\pm$ 500 & 21.1 $\pm$ 1.7 & 180 $\pm$ 17 & 34 $\pm$ 2 & 147.8 $\pm$ 1.6 & 0.94 $\pm$ 0.14 & 0.087 & 5851\\
HD\,218356$^{*}$ & 15194 $\pm$ 2630 & 0.39 $\pm$ 0.13 & 2469014 $\pm$ 2835 & 22 $\pm$ 4 & 73 $\pm$ 24 & 153 $\pm$ 17 & 157 $\pm$ 5 & 0.85 $\pm$ 0.25 & 0.57 & 388\\
HD\,51959 & 11195 $\pm$ 475 & 0.30 $\pm$ 0.05 & 2459598 $\pm$ 329 & 11.8 $\pm$ 1.0 & 31 $\pm$ 11 & 81 $\pm$ 3 & 163.2 $\pm$ 0.8 & 0.51 $\pm$ 0.08 & 0.32 & 11044\\
HD\,123949 & 8544 $\pm$ 12 & 0.9167 $\pm$ 0.0007 & 2457772 $\pm$ 1 & 10.8 $\pm$ 0.9 & 97.1 $\pm$ 0.4 & 60 $\pm$ 30 & 122 $\pm$ 72 & 0.78 $\pm$ 0.15 & 0.52 & 418\\
HD\,182274 & 8393 $\pm$ 51 & 0.039 $\pm$ 0.013 & 2459883 $\pm$ 867 & 9.6 $\pm$ 0.5 & 67 $\pm$ 278 & 68.2 $\pm$ 0.8 & 50.5 $\pm$ 1.0 & 0.55 $\pm$ 0.06 & 0.85 & 65451\\
HD\,18182 & 8258 $\pm$ 300 & 0.35 $\pm$ 0.35 & 2461364 $\pm$ 3518 & 10.9 $\pm$ 1.5 & 194 $\pm$ 85 & 148 $\pm$ 46 & 33 $\pm$ 26 & 0.35 $\pm$ 0.35 & 1.46 & 179\\
HD\,53199 & 8233 $\pm$ 175 & 0.255 $\pm$ 0.010 & 2456826 $\pm$ 34 & 11.7 $\pm$ 0.5 & 66 $\pm$ 2 & 16 $\pm$ 3 & 103 $\pm$ 7 & 0.64 $\pm$ 0.05 & 1.51 & 1873\\
HD\,40430 & 6147 $\pm$ 278 & 0.26 $\pm$ 0.02 & 2457340 $\pm$ 54 & 9.4 $\pm$ 0.8 & 74 $\pm$ 5 & 177 $\pm$ 2 & 23.4 $\pm$ 0.6 & 0.70 $\pm$ 0.12 & 3.68 & 4774\\
HD\,95241 & 5344 $\pm$ 55 & 0.807 $\pm$ 0.004 & 2455739.7 $\pm$ 1.4 & 7.4 $\pm$ 1.5 & 107.8 $\pm$ 0.7 & 131.6 $\pm$ 1.4 & 108 $\pm$ 2 & 0.34 $\pm$ 0.12 & 4.61 & 7633\\
HD\,139195 & 5296 $\pm$ 14 & 0.32 $\pm$ 0.02 & 2460005 $\pm$ 78 & 8.8 $\pm$ 0.3 & 5.3 $\pm$ 3.5 & 29 $\pm$ 3 & 97.6 $\pm$ 1.1 & 0.66 $\pm$ 0.05 & 0.27 & 421\\
BD-10$^{o}$4311 & 4872 $\pm$ 14 & 0.047 $\pm$ 0.006 & 2456076 $\pm$ 105 & 6.2 $\pm$ 0.7 & 159 $\pm$ 8 & 53.5 $\pm$ 1.2 & 73 $\pm$ 3 & 0.52 $\pm$ 0.11 & 1.39 & 20039\\
HD\,183915 & 4344 $\pm$ 20 & 0.41 $\pm$ 0.04 & 2457609 $\pm$ 62 & 7.1 $\pm$ 1.2 & 118 $\pm$ 7 & 77 $\pm$ 3 & 174.3 $\pm$ 0.3 & 0.61 $\pm$ 0.19 & 0.18 & 5410\\
HD\,180622 & 4045 $\pm$ 30 & 0.08 $\pm$ 0.04 & 2458549 $\pm$ 3149 & 6.9 $\pm$ 1.2 & 293 $\pm$ 42 & 166 $\pm$ 10 & 100 $\pm$ 6 & 0.80 $\pm$ 0.25 & 1.03 & 1948\\
HD\,216219 & 3948 $\pm$ 23 & 0.085 $\pm$ 0.050 & 2456527 $\pm$ 275 & 6.2 $\pm$ 0.4 & 59 $\pm$ 26 & 52 $\pm$ 125 & 33.5 $\pm$ 1.8 & 0.63 $\pm$ 0.08 & 0.73 & 1059\\
HD\,107541 & 3583 $\pm$ 47 & 0.095 $\pm$ 0.035 & 2458478 $\pm$ 3180 & 5.5 $\pm$ 0.8 & 220 $\pm$ 16 & 9.2 $\pm$ 7.2 & 128 $\pm$ 3 & 0.55 $\pm$ 0.16 & 0.092 & 6173\\
BD-14$^{o}$2678 & 3481 $\pm$ 205 & 0.25 $\pm$ 0.05 & 2455846 $\pm$ 328 & 7.0 $\pm$ 0.5 & 283 $\pm$ 15 & 144 $\pm$ 11 & 93 $\pm$ 18 & 0.67 $\pm$ 0.10 & 0.28 & 399\\
HD\,59852 & 3477 $\pm$ 80 & 0.14 $\pm$ 0.08 & 2457280 $\pm$ 351 & 6.6 $\pm$ 0.8 & 95 $\pm$ 40 & 139 $\pm$ 129 & 26 $\pm$ 4 & 0.62 $\pm$ 0.15 & 3.52 & 1772\\
HD\,201824 & 2922 $\pm$ 23 & 0.30 $\pm$ 0.04 & 2456263 $\pm$ 102 & 5.5 $\pm$ 1.1 & 62 $\pm$ 9 & 143 $\pm$ 78 & 59 $\pm$ 66 & 0.78 $\pm$ 0.28 & 3.08 & 765\\
HD\,178717 & 2912 $\pm$ 14 & 0.46 $\pm$ 0.03 & 2455886 $\pm$ 57 & 5.2 $\pm$ 0.8 & 265 $\pm$ 5 & 20 $\pm$ 5 & 35 $\pm$ 4 & 0.53 $\pm$ 0.16 & 0.13 & 1357\\
HD\,50082 & 2883 $\pm$ 6 & 0.19 $\pm$ 0.02 & 2457496 $\pm$ 37 & 5.1 $\pm$ 0.9 & 207 $\pm$ 6 & 22 $\pm$ 7 & 63 $\pm$ 3 & 0.56 $\pm$ 0.18 & 1.11 & 291\\
HD\,98991 & 2849 $\pm$ 3 & 0.323 $\pm$ 0.004 & 2455898 $\pm$ 5 & 5.0 $\pm$ 0.3 & 27.8 $\pm$ 0.8 & 129.0 $\pm$ 1.3 & 124.9 $\pm$ 0.7 & 0.57 $\pm$ 0.08 & 14.3 & 5779\\
HD\,205011 & 2846 $\pm$ 5 & 0.23 $\pm$ 0.02 & 2455297 $\pm$ 36 & 5.3 $\pm$ 0.9 & 34 $\pm$ 5 & 56 $\pm$ 3 & 74 $\pm$ 5 & 0.61 $\pm$ 0.19 & 1.36 & 14392\\
HD\,204075 & 2367 $\pm$ 9 & 0.26 $\pm$ 0.05 & 2455474 $\pm$ 109 & 6.0 $\pm$ 0.4 & 255 $\pm$ 15 & 9 $\pm$ 165 & 133 $\pm$ 5 & 0.67 $\pm$ 0.10 & 5.62 & 41.1\\
HD\,20394 & 2248 $\pm$ 8 & 0.16 $\pm$ 0.06 & 2456928 $\pm$ 105 & 4.6 $\pm$ 0.5 & 145 $\pm$ 18 & 92 $\pm$ 6 & 31 $\pm$ 3 & 0.49 $\pm$ 0.10 & 0.57 & 489\\
\hline\hline
\end{tabular}}\\
\vspace{1mm}\\
$^{*}$For the triple system HD\,218356, we only include the outer orbit that hosts the WD. See Table \ref{paramsHD218356} for the remaining parameters.
\end{sidewaystable*}

\setlength{\tabcolsep}{7.5pt}
\renewcommand{\arraystretch}{1.3}
\begin{sidewaystable*}
\caption*{Table \ref{table:results} continues}
{\centering
\begin{tabular}{lcccccccccc}
\hline\hline       
\textbf{Star ID} & \textbf{$P$ [days]} & \textbf{$e$} & \textbf{$T_0$ [HJD]} & \textbf{$a$ [AU]} & \textbf{$\omega_*$ [$^{\circ}$]} & \textbf{$\Omega$ [$^{\circ}$]} & \textbf{$i$ [$^{\circ}$]} & \textbf{$M_{\rm WD}$} & \textbf{Fit $\chi^2$} & \textbf{$\chi^2_{\rm HGCA}$}\\
\hline 
HD\,16458 & 2017 $\pm$ 15 & 0.098 $\pm$ 0.025 & 2456430 $\pm$ 112 & 4.2 $\pm$ 0.5 & 119 $\pm$ 14 & 109 $\pm$ 10 & 61 $\pm$ 12 & 0.74 $\pm$ 0.15 & 2.53 & 3888\\
HD\,5424 & 1906 $\pm$ 17 & 0.19 $\pm$ 0.05 & 2455702 $\pm$ 145 & 3.7 $\pm$ 0.4 & 102 $\pm$ 14 & 40 $\pm$ 74 & 30 $\pm$ 3 & 0.52 $\pm$ 0.11 & 0.87 & 445\\
HD\,49641 & 1793 $\pm$ 21 & 0.06 $\pm$ 0.06 & 2456298 $\pm$ 352 & 4.9 $\pm$ 0.9 & 207 $\pm$ 63 & 138 $\pm$ 126 & 159.5 $\pm$ 1.6 & 1.2 $\pm$ 0.4 & 0.55 & 1384\\
HD\,91208 & 1770 $\pm$ 3 & 0.178 $\pm$ 0.018 & 2456240 $\pm$ 55 & 4.2 $\pm$ 0.4 & 83 $\pm$ 10 & 167 $\pm$ 3 & 133.6 $\pm$ 2.3 & 0.83 $\pm$ 0.14 & 6.19 & 486\\
HD\,200063 & 1743 $\pm$ 8 & 0.07 $\pm$ 0.04 & 2456510 $\pm$ 120 & 4.3 $\pm$ 0.5 & 228 $\pm$ 27 & 152 $\pm$ 17 & 115 $\pm$ 5 & 0.95 $\pm$ 0.26 & 0.96 & 4910\\
HD\,201657 & 1702 $\pm$ 4 & 0.16 $\pm$ 0.7 & 2456291 $\pm$ 280 & 3.9 $\pm$ 0.6 & 255 $\pm$ 61 & 82 $\pm$ 4 & 153 $\pm$ 3 & 0.66 $\pm$ 0.18 & 0.59 & 1282\\
HD\,43389 & 1688.4 $\pm$ 1.4 & 0.083 $\pm$ 0.017 & 2455663 $\pm$ 57 & 4.0 $\pm$ 0.7 & 190 $\pm$ 13 & 168 $\pm$ 150 & 111 $\pm$ 3 & 0.76 $\pm$ 0.25 & 1.03 & 279\\
HD\,27271 & 1681.1 $\pm$ 1.0 & 0.224 $\pm$ 0.007 & 2455519 $\pm$ 15 & 4.2 $\pm$ 0.3 & 210 $\pm$ 4 & 6 $\pm$ 2 & 103 $\pm$ 5 & 0.70 $\pm$ 0.09 & 3.93 & 1441\\
HD\,95193 & 1652 $\pm$ 5 & 0.13 $\pm$ 0.02 & 2456007 $\pm$ 52 & 4.12 $\pm$ 0.15 & 285 $\pm$ 10 & 88 $\pm$ 37 & 81 $\pm$ 25 & 0.71 $\pm$ 0.08 & 5.33 & 268\\
HD\,210946 & 1521.0 $\pm$ 0.9 & 0.109 $\pm$ 0.011 & 2455737 $\pm$ 35 & 3.9 $\pm$ 0.5 & 193 $\pm$ 8 & 15 $\pm$ 9 & 114 $\pm$ 8 & 0.86 $\pm$ 0.26 & 0.32 & 231\\
HD\,127392 & 1506.8 $\pm$ 1.8 & 0.088 $\pm$ 0.017 & 2456222 $\pm$ 91 & 3.1 $\pm$ 0.6 & 159 $\pm$ 20 & 4 $\pm$ 175 & 119 $\pm$ 15 & 0.73 $\pm$ 0.26 & 4.31 & 1531\\
HD\,143899 & 1461.8 $\pm$ 1.2 & 0.18 $\pm$ 0.04 & 2456481 $\pm$ 25 & 3.66 $\pm$ 0.15 & 279 $\pm$ 8 & 88 $\pm$ 3 & 125 $\pm$ 3 & 0.66 $\pm$ 0.06 & 2.00 & 640\\
HD\,88562 & 1451.0 $\pm$ 1.0 & 0.203 $\pm$ 0.013 & 2455931 $\pm$ 25 & 2.9 $\pm$ 0.3 & 352 $\pm$ 8 & 61 $\pm$ 97 & 87 $\pm$ 9 & 0.48 $\pm$ 0.09 & 0.84 & 11.0\\
HD\,221531 & 1402.3 $\pm$ 1.2 & 0.165 $\pm$ 0.005 & 2455563 $\pm$ 12 & 3.0 $\pm$ 0.2 & 189 $\pm$ 3 & 136 $\pm$ 3 & 59 $\pm$ 3 & 0.58 $\pm$ 0.09 & 0.49 & 794\\
HD\,202400 & 1391 $\pm$ 6 & 0.25 $\pm$ 0.09 & 2455555 $\pm$ 96 & 2.9 $\pm$ 0.2 & 37 $\pm$ 295 & 86 $\pm$ 86 & 61 $\pm$ 67 & 0.65 $\pm$ 0.13 & 1.48 & 564\\
HD\,107574 & 1384.8 $\pm$ 1.5 & 0.083 $\pm$ 0.005 & 2456521 $\pm$ 17 & 2.99 $\pm$ 0.12 & 216 $\pm$ 4 & 22 $\pm$ 2 & 166.3 $\pm$ 0.4 & 0.74 $\pm$ 0.06 & 0.24 & 3766\\
HD\,58121 & 1217 $\pm$ 5 & 0.135 $\pm$ 0.019 & 2455339 $\pm$ 50 & 3.4 $\pm$ 0.5 & 89 $\pm$ 10 & 58 $\pm$ 28 & 121 $\pm$ 4 & 0.67 $\pm$ 0.19 & 3.11 & 149\\
HD\,31487 & 1063.8 $\pm$ 0.4 & 0.037 $\pm$ 0.018 & 2455808 $\pm$ 98 & 3.3 $\pm$ 0.2 & 237 $\pm$ 33 & 48 $\pm$ 2 & 32.1 $\pm$ 1.2 & 1.59 $\pm$ 0.22 & 0.28 & 1064\\
HD\,211594 & 1018.5 $\pm$ 0.5 & 0.058 $\pm$ 0.013 & 2455675 $\pm$ 36 & 2.8 $\pm$ 0.4 & 76 $\pm$ 13 & 65 $\pm$ 17 & 123 $\pm$ 16 & 0.55 $\pm$ 0.16 & 1.85 & 5.02\\
HD\,34654 & 976.0 $\pm$ 0.3 & 0.1114 $\pm$ 0.0016 & 2455212 $\pm$ 2 & 2.36 $\pm$ 0.12 & 326.5 $\pm$ 0.7 & 91 $\pm$ 4 & 74 $\pm$ 5 & 0.64 $\pm$ 0.06 & 0.058 & 183\\
HD\,92626 & 921.7 $\pm$ 1.7 & 0.014 $\pm$ 0.014 & 2455685 $\pm$ 237 & 3.0 $\pm$ 0.5 & 116 $\pm$ 199 & 92 $\pm$ 62 & 85 $\pm$ 9 & 0.90 $\pm$ 0.27 & 0.61 & 2.47\\
HD\,50264 & 910.2 $\pm$ 0.8 & 0.077 $\pm$ 0.018 & 2455933 $\pm$ 43 & 2.1 $\pm$ 0.2 & 238 $\pm$ 17 & 63 $\pm$ 17 & 103 $\pm$ 40 & 0.63 $\pm$ 0.13 & 0.53 & 246\\
HD\,49841 & 897.5 $\pm$ 1.9 & 0.162 $\pm$ 0.015 & 2455518 $\pm$ 21 & 2.80 $\pm$ 0.10 & 350 $\pm$ 6 & 110 $\pm$ 51 & 109 $\pm$ 19 & 0.82 $\pm$ 0.16 & 0.51 & 14.5\\
HD\,58368 & 673.1 $\pm$ 1.5 & 0.22 $\pm$ 0.02 & 2455715 $\pm$ 28 & 2.23 $\pm$ 0.17 & 18 $\pm$ 6 & 99 $\pm$ 68 & 78 $\pm$ 27 & 0.66 $\pm$ 0.17 & 0.23 & 0.98\\
HD\,207585 & 671.7 $\pm$ 0.4 & 0.03 $\pm$ 0.03 & 2455613 $\pm$ 174 & 1.72 $\pm$ 0.16 & 109 $\pm$ 207 & 103 $\pm$ 17 & 93 $\pm$ 20 & 0.57 $\pm$ 0.12 & 1.06 & 20.6\\
HD\,44896 & 629.0 $\pm$ 1.1 & 0.019 $\pm$ 0.013 & 2455388 $\pm$ 76 & 2.29 $\pm$ 0.10 & 228 $\pm$ 35 & 101 $\pm$ 35 & 78 $\pm$ 35 & 1.0 $\pm$ 0.2 & 0.99 & 5.86\\
HD\,199939 & 585.39 $\pm$ 0.09 & 0.281 $\pm$ 0.012 & 2455207 $\pm$ 4 & 2.12 $\pm$ 0.18 & 48 $\pm$ 3 & 110 $\pm$ 84 & 82 $\pm$ 21 & 0.73 $\pm$ 0.14 & 2.24 & 4.61\\
HD\,123585 & 460.1 $\pm$ 1.4 & 0.03 $\pm$ 0.04 & 2455534 $\pm$ 152 & 1.41 $\pm$ 0.11 & 191 $\pm$ 73 & 27 $\pm$ 135 & 90 $\pm$ 19 & 0.65 $\pm$ 0.11 & 1.53 & 7.32\\
HD\,24035 & 378.0 $\pm$ 0.5 & 0.014 $\pm$ 0.016 & 2455307 $\pm$ 184 & 1.41 $\pm$ 0.15 & 204 $\pm$ 110 & 52 $\pm$ 106 & 71 $\pm$ 31 & 0.76 $\pm$ 0.25 & 0.74 & 6.69\\
HD\,224621 & 308.20 $\pm$ 0.11 & 0.020 $\pm$ 0.012 & 2455264 $\pm$ 32 & 1.03 $\pm$ 0.06 & 309 $\pm$ 57 & 64 $\pm$ 27 & 31 $\pm$ 7 & 0.66 $\pm$ 0.26 & 3.05 & 59.7\\
HD\,87080 & 274.31 $\pm$ 0.05 & 0.162 $\pm$ 0.016 & 2455230 $\pm$ 4 & 1.05 $\pm$ 0.10 & 128 $\pm$ 5 & 149 $\pm$ 112 & 60 $\pm$ 12 & 0.70 $\pm$ 0.18 & 3.58 & 15.6\\
HD\,121447 & 185.65 $\pm$ 0.05 & 0.012 $\pm$ 0.010 & 2455243 $\pm$ 51 & 0.85 $\pm$ 0.05 & 267 $\pm$ 104 & 90 $\pm$ 50 & 59 $\pm$ 76 & 0.72 $\pm$ 0.36 & 4.05 & 8.71\\
HD\,77247 & 80.5371 $\pm$ 0.0013 & 0.108 $\pm$ 0.005 & 2455236.5 $\pm$ 0.7 & 0.63 $\pm$ 0.03 & 40 $\pm$ 3 & 60 $\pm$ 35 & 98 $\pm$ 25 & 0.54 $\pm$ 0.11 & 3.11 & 2.31\\
\hline\hline\
\end{tabular}}
\end{sidewaystable*}

\begin{figure*}
\centering
\includegraphics[width=0.7\textwidth]{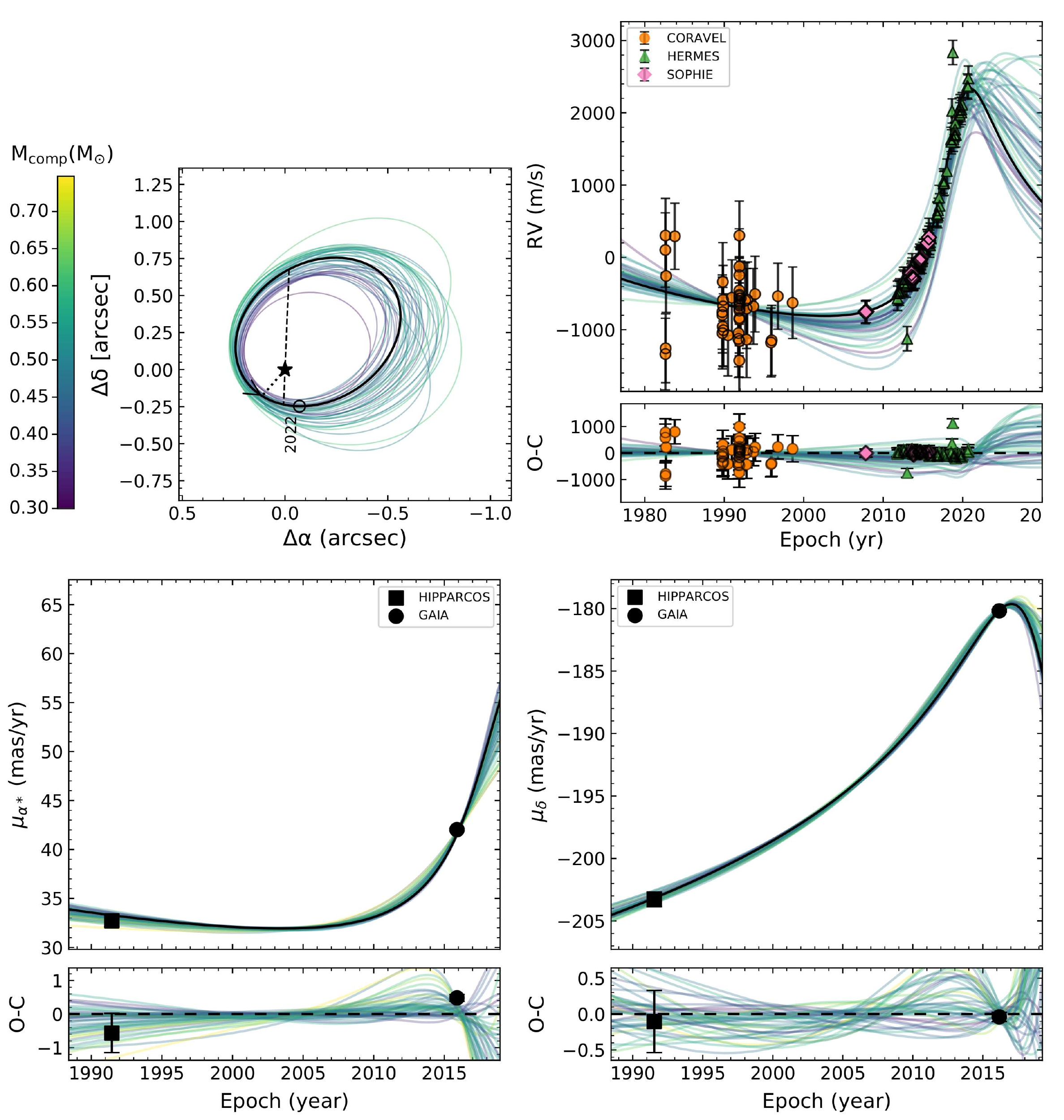}
\caption{\label{Fig:HD2454all} \textsc{Orvara} results for the main-sequence Ba star HD\,2454. \textbf{Top:} astrometric and spectroscopic orbits. The RV plot includes radial-velocity measurements from CORAVEL (orange circles), SOPHIE (pink diamonds), and HERMES (green triangles). \textbf{Bottom:} Hipparcos and Gaia proper motions. In all plots, the best-fitting orbit is plotted as a black thick line, while 40 other well-fitting orbits are included and colour-coded as a function of the companion mass.}
\end{figure*}

\begin{figure*}
\centering
\includegraphics[width=0.7\textwidth]{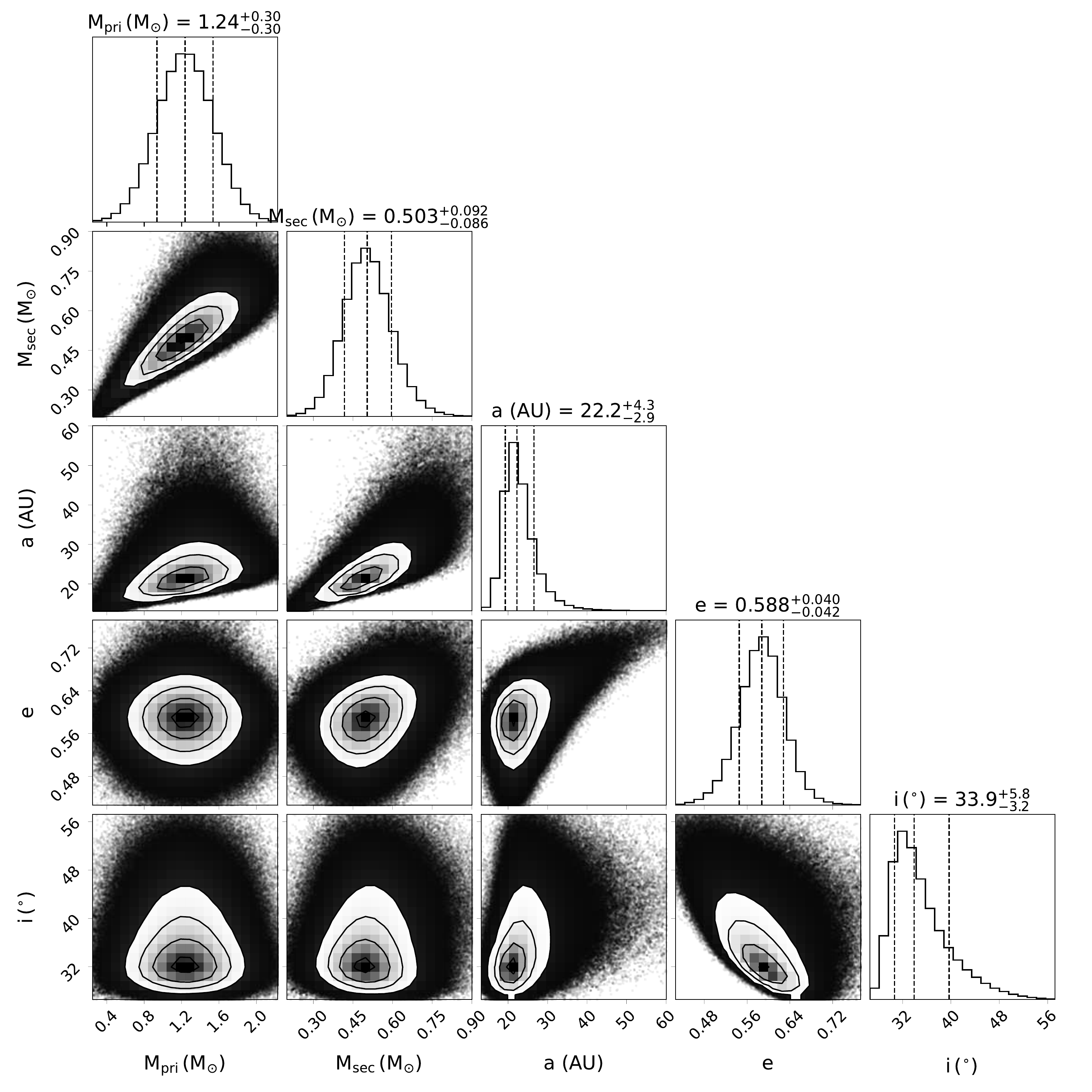}
\caption{\label{Fig:HD2454corner} Corner plot of some derived parameters for HD\,2454 including mass of the two stars, the semimajor axis, the eccentricity, and orbital inclination.}
\end{figure*}


Table \ref{table:results} lists the obtained astrometric-spectroscopic orbital parameters, the best-fitting WD masses, the $\chi^2$ of the best fit, and the HGCA $\chi^2$ values discussed in Sect. \ref{sec:HGCA}. To make the table easier to read, we assume that the error bars we obtained from the MCMC fit are symmetric and listed only the largest value. This means that in some cases, the table lists an overestimated uncertainty in one of the two directions. The $\chi^2$ values are an overall absolute astrometric $\chi^2$, computed adding the $\chi^2$ for the Hipparcos proper motions ($\chi^2_{\rm H}$), the $\chi^2$ for the long-term Hipparcos-Gaia proper motions ($\chi^2_{\rm HG}$), and the $\chi^2$ for the Gaia proper motions ($\chi^2_{\rm G}$). \textsc{orvara} uses RV jitter terms such that the reduced $\chi^2$ of the RV fit is 1, so we did not take it into account to evaluate the goodness of the fit.

The table is ordered based on the orbital period, with the systems with the longest periods first. This way, we can notice that all the systems with periods longer than $\sim3$\,years have significant astrometric accelerations according to their HGCA $\chi^2$ values, while most of the systems below that threshold do not. Finding such a clear threshold in a sample of confirmed binaries is an indication of the type of systems that the HGCA can help identify.

In addition to the table and in order to illustrate and discuss how the results that we get from \textsc{orvara} look like, we include the results for the main-sequence Ba star HD\,2454 in Fig. \ref{Fig:HD2454all}. HD\,2454 was first identified as a Ba dwarf by \cite{Tomkin89}, and \cite{North00} confirmed its binarity even though they did not have enough data to estimate its orbital period. More recently, \cite{Gray11} found direct evidence of the presence of a WD companion in the system thanks to the Galaxy Evolution Explorer (\textit{GALEX}; \citealt{GALEX}) UV observations and, since 2011, HD\,2454 has been part of the long-period binary monitoring program carried out with the HERMES spectrograph (see Sect. \ref{sec:RVdata}). In spite of having almost three decades of RV data between the CORAVEL and HERMES measurements, \cite{Escorza19} were not able to constrain the orbit either. However, combining all these RV data points with the Hipparcos and Gaia information, we can finally estimate the orbital elements of HD\,2454 as well as the mass of its WD companion.

Fig. \ref{Fig:HD2454all} shows, on the top left panel, the astrometric orbit of HD\,2454, including the predicted position of the companion on the scheduled date of Gaia DR3. The best-fitting orbit is plotted as a black thick line, while 40 other well-fitting orbits are colour-coded as a function of the companion mass. On the top right panel, we show the RV curve of HD\,2454. For this target we had CORAVEL (orange circles), SOPHIE (pink diamonds), and HERMES (green triangles) RV data. The plot shows that leaving the RV offsets between instruments completely free produces families of solutions with similar orbits and masses but different RV offsets (displaced vertically in the RV plot). This is especially noticeable in cases like this one, where no data sets covers even half an orbit. We want to note that even though we left the RV offsets free in most cases, we always made sure that the best-fitting solution required reasonable values and, especially in the CORAVEL-HERMES case, that these values were close to the values obtained by \cite{Jorissen19} and \cite{Escorza19}.

The two bottom panels of Fig. \ref{Fig:HD2454all} show the fit to the proper motions in the right ascension (left) and declination (right) directions, as measured by Hipparcos (squared data point) and Gaia (circular data point). All the data sets included in the figures were fitted at the same time, and the plotted models are the same in all plots. Finally, Fig. \ref{Fig:HD2454corner} shows the one and two-dimensional projections of the posterior probability distributions of the masses of the two components in the system and a few orbital parameters (semi-major axis, eccentricity, and inclination) from the joint RV and astrometric MCMC computations. This corner plot shows that the two masses are correlated, and that the semimajor axis is also correlated with the total mass of the system. These correlations are even stronger for other targets.

We have included in Appendix \ref{App} figures similar to Figs. \ref{Fig:HD2454all} and \ref{Fig:HD2454corner} for all the targets in our sample. Additionally, an individual case of study of a Ba dwarf using the same method was presented in \cite{Escorza22IAU}.

\subsection{Spectroscopic orbital parameters}\label{sec:orbparams}

Even though the main goal of this work was deriving the masses of the WD companions of all these Ba stars, an important additional result of this new method are the new orbits of HD\,2454 and BD-11$^{\rm o}$3853, which could not be constrained before, as well as the improved orbits of a few other long-period systems. When comparing the orbital periods obtained using \textsc{orvara} to those presented in \cite{Jorissen19}, \cite{Escorza19} and \cite{North20}, which were obtained by fitting only the RV data, we get a very tight relation. The purely spectroscopic parameters and the new parameters are consistent with each other within error bars in almost all cases, and we discuss the exceptions below.

\subsubsection{HD\,218356}
\begin{figure}
\centering
\includegraphics[width=0.49\textwidth]{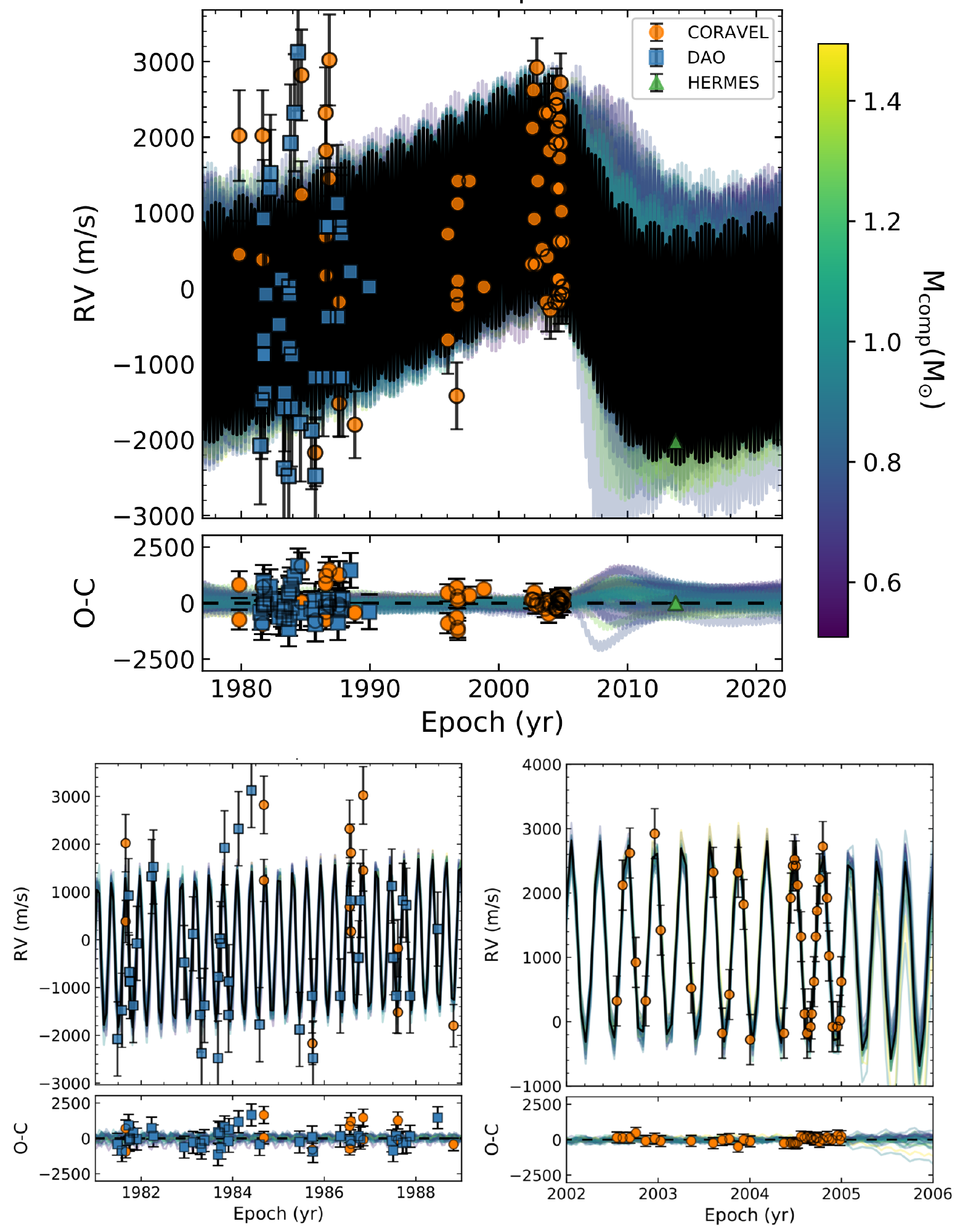}
\caption{\label{Fig:HD218356-RV} Best fitting models to the RV data of HD\,218356}
\end{figure}

Our first orbital fit for this system converged to a period of more than 40 years, while the period published by \cite{Griffin06} and \cite{Jorissen19} for HD\,218356 was 111 days. No third object has been detected in this system in the past, but the mild s-process enhancement in the visible star has been flagged as surprising given the close orbit. We performed a three-body fit, setting strong constraints on the inner orbit using the published spectroscopic parameters, and we succeeded to recover the orbital parameters of two companions, confirming that HD\,218356 is actually a triple system with a third companion in a much longer orbit than the published period. The orbital parameters of the system are included in Table \ref{paramsHD218356} and the combined RV fit can be seen in Fig. \ref{Fig:HD218356-RV}. In order to test the significance of this detection, we compared the Akaike Information Criterion (AIC) of the two- and three-component models using the {\texttt radvel} package \citep{Fulton2018}. We found a $\Delta$AIC of 439 favouring the three-component model. Given the masses of the two companions, we expect the WD that polluted the Ba star to be in the outer orbit. This would also explain the mild s-process enhancement reported for HD\,218356. We included the corner plots with the parameters of both orbits in Appendix \ref{AppHD218356}. Only the outer orbit information is listed together with the other WD orbits in Table \ref{table:results}.

\subsubsection{HD\,201657}
Our orbit fit for HD\,201657 converged to twice the published orbital period and to a much more eccentric orbit. The astrometric data favours the longer orbit, and the RV data is not very constraining since we have only 15 CORAVEL points and one HERMES point. However, given the eccentricity-period diagram of Ba stars, the orbit published by \cite{Jorissen19}, the least eccentric of the two, is the most likely. We attempted to recover this orbit in order to check the quality of such a fit and calculate the WD companion mass by including an orbital eccentricity prior of $0.15 \pm 0.15$. We recovered \cite{Jorissen19}'s orbital solution, although with a slightly higher $\chi^2$ for the astrometric data. Since we considered this solution more likely for a Ba star, we listed this orbit in Table \ref{table:results}, but we show both fits and corner plots in Appendix \ref{AppHD201657}. More HERMES data would be essential to solve this case.

\subsection{Astrometric orbital parameters}

\begin{figure}
\centering
\includegraphics[width=0.49\textwidth]{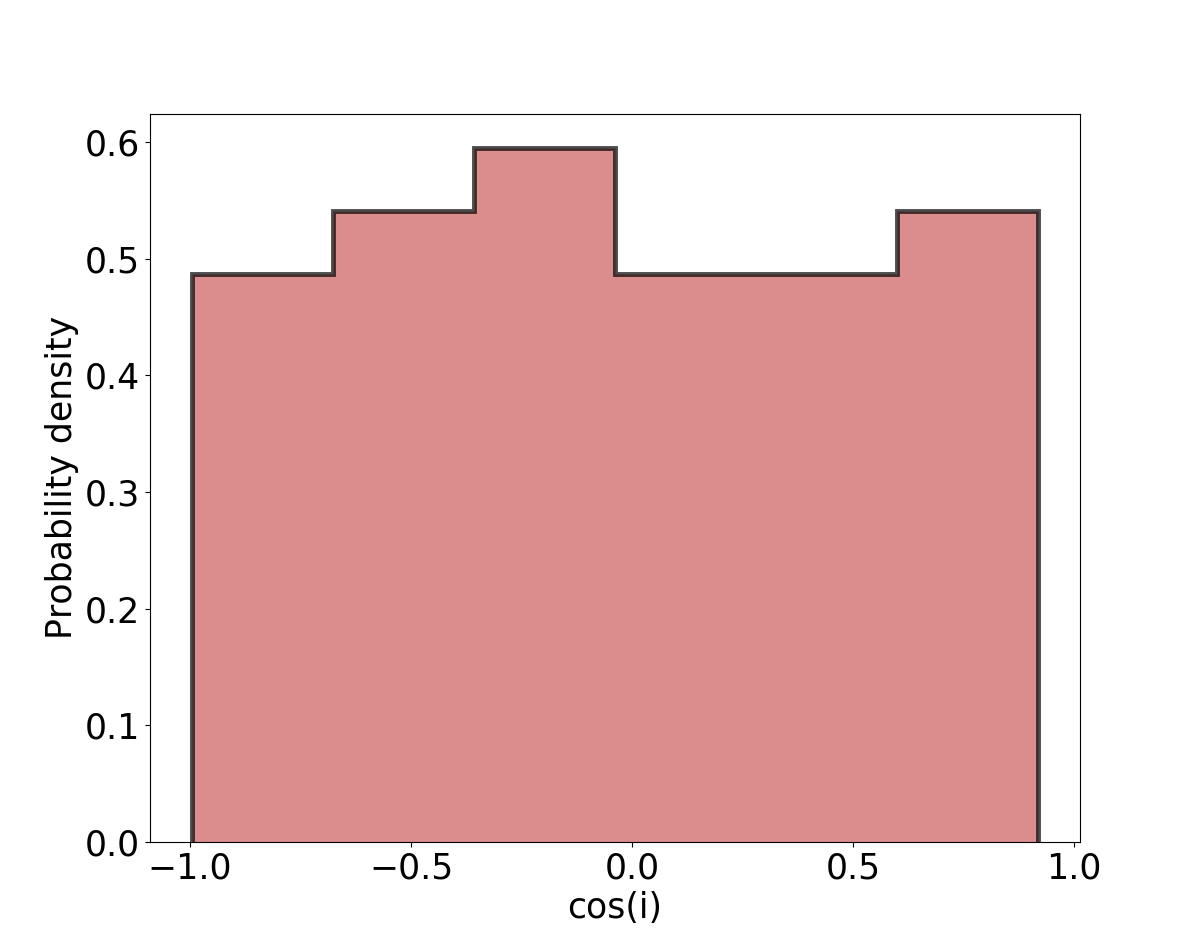}
\caption{\label{Fig:inclinations} Distribution of $\cos i$ where $i$ are the orbital inclinations of the Ba star systems. Bin-width chosen to roughly follow the Freedman–Diaconis rule \citep{Freedman81}.}
\end{figure}

Finally, in addition to the new and improved orbital parameters, this method provided us with orbital inclinations for all these Ba star systems. Fig. \ref{Fig:inclinations} shows the distribution of the obtained $\cos({i})$ values. This distribution should be flat if we could assume our sample of binaries is randomly distributed on the sky, and even though we only have 60 systems, the distribution is compatible with a uniform one. We performed a Kolmogorov-Smirnov (KS) test \citep[e.g.][]{Press86}, and we obtained p-values higher than 0.8 when comparing our $\cos({i})$ distribution with uniform distributions of the same sample size.

The new orbital parameters are also compatible with the astrometric parameters published in Gaia DR3 for the ten targets available in their catalogue. Concerning the periods, all Gaia DR3 values are consistent with our values within 2$\sigma$. The largest difference is found for HD\,221531, for which Gaia DR3 published a period of 1668~$\pm$~135~days, about 260~days longer than our period. The Gaia DR3 time span is about 1000 days \citep{GaiaDR3binaries}, while our data covers a few decades in most cases. Hence, we think that our method is more reliable to obtain the periods of long-period binaries.
The eccentricities are compatible as well, without significant exceptions, and finally, we used the Thiele-Innes elements published in the Gaia DR3 catalogue and followed \cite{Halbwachs22} to compute the orbital inclinations of these systems from the Gaia DR3 data. The Gaia DR3 inclinations are also compatible with the inclinations we obtained with our full RV+astrometric model within 1.2 times our $\sigma$. As discussed above, the HGCA is not very constraining for systems with periods below about 3 years, so while we think our method is better to determine the orbital periods of Ba stars, the Gaia DR3 inclinations are probably of better quality than ours for the shorter-period systems. When the epoch astrometry of the Gaia mission is published, we will be able to combine these data with all our other data sets and improve our results for the shortest period systems.

\renewcommand{\arraystretch}{1.4}
\begin{table}[t]
\begin{center}
\caption{Orbital parameter of the triple system HD\,218356}\label{paramsHD218356}
\vspace{1mm}
\begin{tabular}{l c c}
\hline
{\bf Parameter} & {\bf Inner orbit} & {\bf Outer  orbit}\\
\hline
\textbf{Period, $P$ [days]} & 111.15$^{+0.03}_{-0.03}$ & 15194$^{+2600}_{-1600}$\\
\textbf{Eccentricity, $e$} & 0.072$^{+0.048}_{-0.045}$ & 0.39$^{+0.13}_{-0.12}$\\
\textbf{T. of periastron, $T_{0}$ [HJD]} & 2455289$^{+15}_{-85}$ & 2469014$^{+2800}_{-2800}$\\
\textbf{Semimajor axis, $a$ [AU]} & 0.79$^{+0.10}_{-0.08}$ & 22.1$^{+3.6}_{-2.8}$\\
\textbf{Arg. of periastron, $\omega$ [$^{\circ}$]} & 55$^{+270}_{-37}$ & 73$^{+21}_{-24}$\\
\textbf{Ascending node, $\Omega$ [$^{\circ}$]} & 90$^{+60}_{-62}$ & 153$^{+14}_{-17}$\\
\textbf{Inclination [$^{\circ}$]} & 90$^{+42}_{-41}$ & 157$^{+4}_{-5}$\\
\textbf{Companion mass [$M_{\odot}$]} & 0.13$^{+0.06}_{-0.03}$ & 0.85$^{+0.25}_{-0.18}$\\
\hline
\end{tabular}
\end{center}
\end{table}

\subsection{White Dwarf masses}\label{sec:WDmasses}

\begin{figure}
\centering
\includegraphics[width=0.49\textwidth]{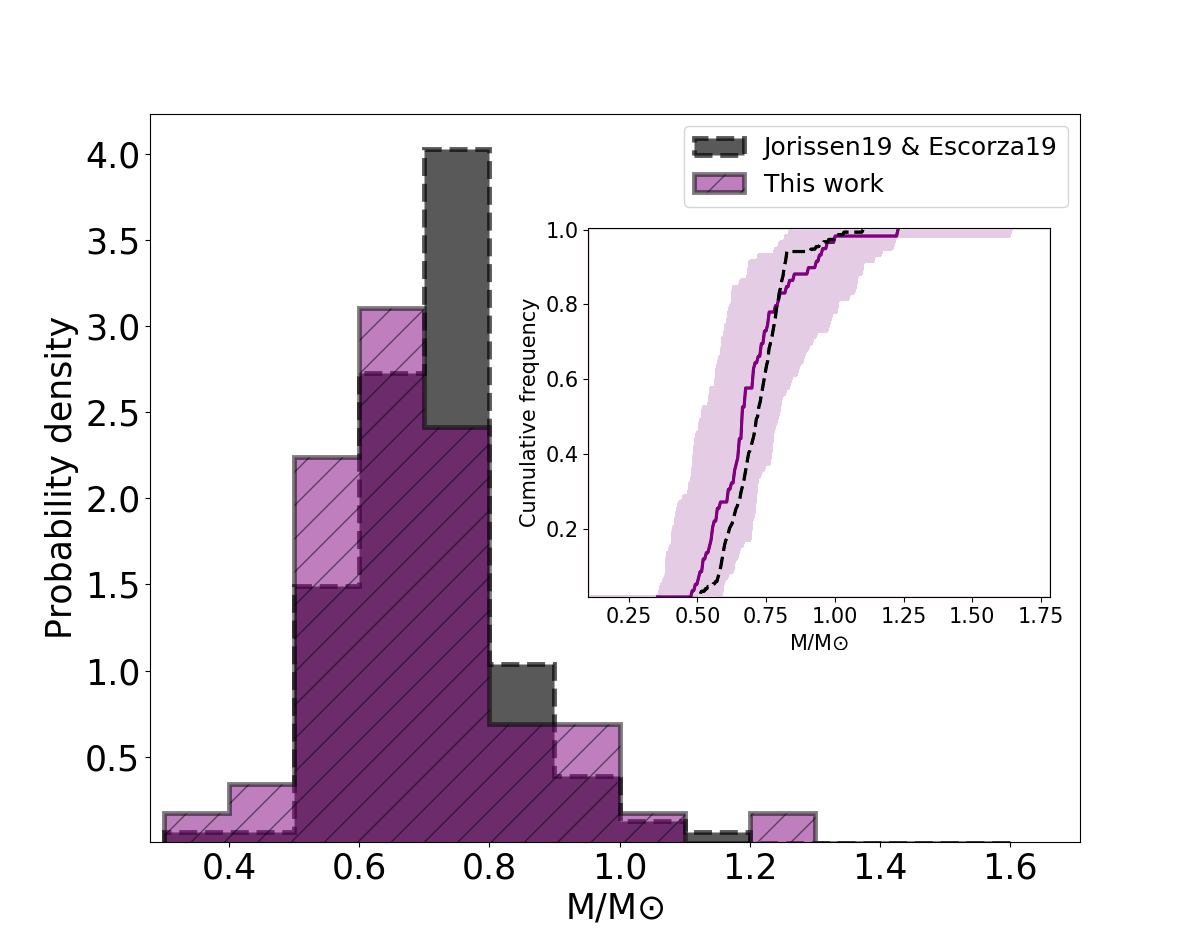}
\caption{\label{Fig:WD} Mass distribution of WD companions of Ba stars. The purple histogram corresponds to the WD masses obtained for this publication, while the black histogram includes the results published in \cite{Jorissen19} and \cite{Escorza19EWASS} assuming a narrow distribution of $Q$ and $M_{\rm WD}$. The bin-width was chosen to roughly follow the Freedman–Diaconis rule \citep{Freedman81}. The insert in the figure shows the cumulative frequency of the same two samples, including a 1-$\sigma$ envelope for our results.}
\end{figure}

\begin{figure}
\centering
\includegraphics[width=0.49\textwidth]{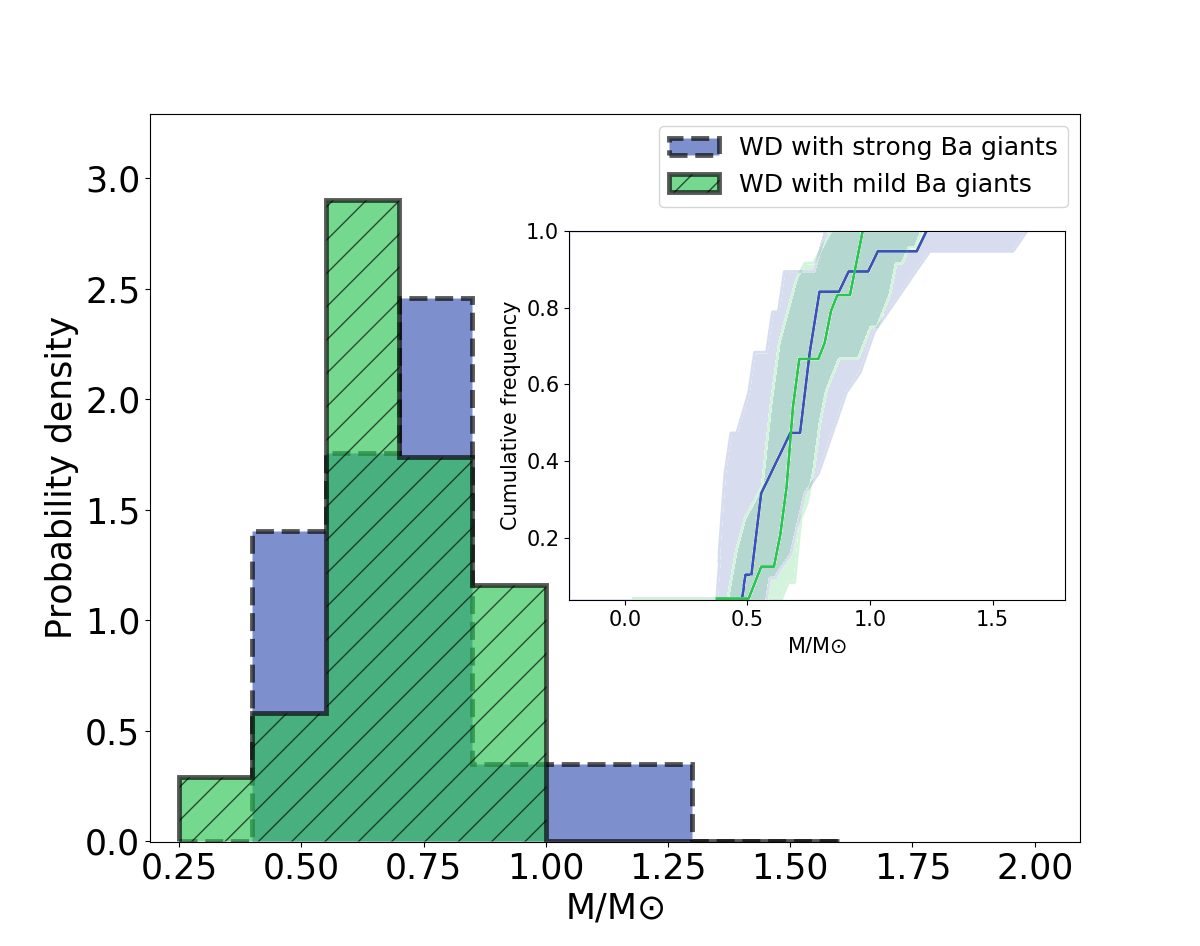}
\caption{\label{Fig:WDgiants} Mass distribution of WD companions of strong (blue) and mild (dashed green) Ba giants. The insert in the figure shows the cumulative frequency of the same two samples, including a 1-$\sigma$ envelope for our results.}
\end{figure}

Table \ref{table:results} lists the masses we obtained for the companions to all the Ba stars in our sample, and Fig. \ref{Fig:WD} shows the distribution of these masses as a purple dashed histogram. Also in Fig. \ref{Fig:WD} we compare this new distribution to the distribution obtained by \cite{Jorissen19} and \cite{Escorza19EWASS} for the same stars, which is drawn in black. The insert in the figure shows the cumulative frequency of the two distributions, including an envelope with the $1-\sigma$ uncertainty for our distribution, which also envelopes the old distribution. We obtained a p-value of 0.010 on a KS test, which is not low enough to reject the null hypothesis. The two distributions are not statistically different.


In Fig. \ref{Fig:WDgiants}, we plot the mass distributions of the companions to strong and mild Ba giants, separately. As mentioned in the introduction, this distinction is made based on the abundance ratios [La/Fe] and [Ce/Fe] and on the Ba index introduced by \cite{Warner1965}. We do not include the pre-RGB stars in this comparison, because the distinction between strong and mild enhancement has not been as clearly established as it has for the giants. We note that the WDs occupying the high-mass tail belong to systems with strong Ba giants. However, we performed a KS test, and we obtained a p-value of 0.45, meaning that we cannot reject that the two samples are drawn from the same distribution. The cumulative distributions plotted in the insert also show that taking the $1-\sigma$ uncertainty into account, the distributions are not very different.

There are a few individual systems that appeared as clear outliers or that even have WDs with unphysical masses. These are briefly discussed below. 

\subsubsection{The least massive WDs: HD\,18182 and HD\,95241}

There are two systems for which our simulations converged to very low WD masses. These are HD\,18182 and HD\,95241. The fit we achieved for the former is less than ideal (see Figs. \ref{Fig:HD18182} and \ref{Fig:HD18182corner}), and even though the mass is small, taking the error bars into account, the value is compatible with an average WD in our sample. The CORAVEL RV data is not very constraining and the HERMES points, being of much higher quality, still fall on the same range of orbital phases, covering in total less than half of the orbit. Additionally, the Hipparcos and Gaia proper motions in the right ascension direction are very similar, not adding strong constraints to the fit either. This WD mass should be taken with caution.

The fit for HD\,95241, on the other hand, is significantly better. We used 97 RV points that cover very well the whole orbit (see Fig. \ref{Fig:HD95241}) and obtained clean and symmetric posterior distributions (see Fig. \ref{Fig:HD95241corner}). Of course, $M_{\rm Ba}$ and $M_{\rm WD}$ are very strongly correlated, so if the $M_{\rm Ba}$ prior was incorrect, too small in this case, it would directly affect $M_{\rm WD}$. The mass of HD\,95241 was determined by \cite{Escorza19} by comparing the location of the star on the HR diagram with STAREVOL \citep{Siess00,SiessArnould08} evolutionary tracks. The stellar parameters were determined from HERMES high-resolution high-signal-to-noise spectra and are in agreement with other studies \citep[e.g.][]{Takeda07,PASTEL16}. However, HD\,95241 was flagged as a mild Ba dwarf by \cite{Edvardsson93} having only a marginal overabundance of s-process elements with respect to iron. Other Ba dwarf candidates of their sample have been proven to be wrongly flagged. Most of them are likely single stars \citep{Escorza19}. It is possible that HD\,95241 has a low-mass companion that is not a WD, and if it is a WD, its AGB progenitor was not massive enough to reach the thermally pulsing AGB phase and produce s-process elements. HD\,95241 is likely not a Ba star and will be removed from further analysis.

\subsubsection{The most massive WDs: HD\,49641 and HD\,31487}

\begin{figure}
\centering
\includegraphics[width=0.49\textwidth]{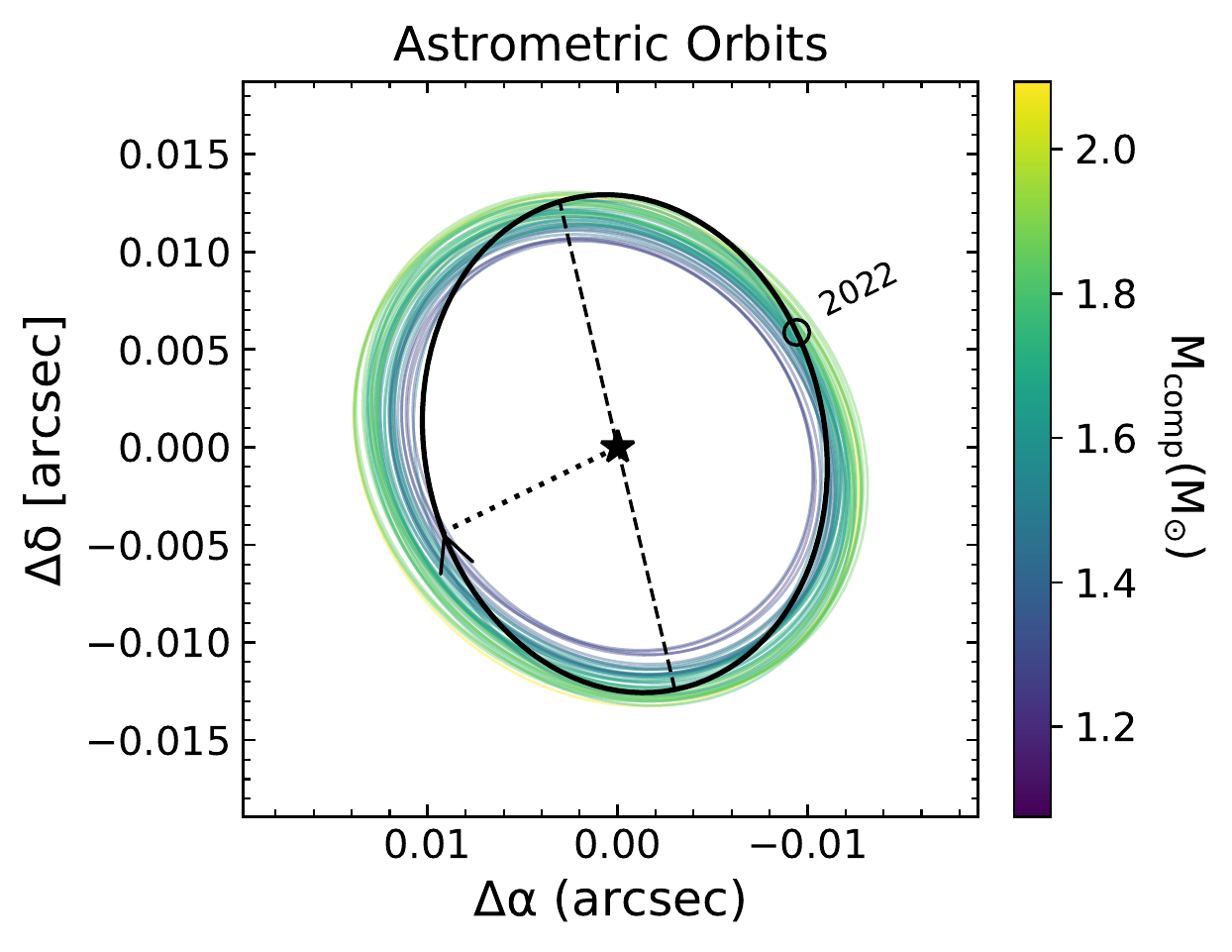}
\caption{\label{Fig:astrom-HD31487} Projection on the sky of HD\,31487.}
\end{figure}

On the high-mass end of the distributions, there are two systems with WD masses clearly outlying from the initial mass distribution ($M_{\rm WD} \geq 1.2$~\Msun). These are HD\,49641, with $M_{\rm WD} = 1.2 \pm 0.4$~\Msun, and HD\,31487, with $M_{\rm WD} = 1.59 \pm 0.22$~\Msun. The fit for HD\,49641 is not very good, because the available RV data was scarce and old, so one should take this WD mass with caution, but the fits for HD\,31487 seems reliable, including a clean result for the orbital projection on the sky (see Fig. \ref{Fig:astrom-HD31487}). In order to try to explain this last mass, one could again try to invoke a wrong $M_{\rm Ba}$ prior. We used the primary mass determined by \cite{Karinkuzhi18}. The primary mass listed by \cite{Jorissen19} is not in agreement with \cite{Karinkuzhi18}'s within error bars, but we decided to use the latter after studying their HR diagram (their Fig. 16). In any case, \cite{Jorissen19}'s mass is higher, and would result in a higher companion mass. \cite{Karinkuzhi18}'s value seems reasonable given the location of the star on the HR diagram, and it is a very average value for giant Ba stars. Additionally, there is no big discrepancy between the parallaxes published in the different Gaia Data Releases. While a wrong parallax could have led to a wrong luminosity, hence mass, determination, we have no good reason to doubt this mass. From the posterior distributions and 1D projections shown in Fig. \ref{Fig:HD31487corner}, one can see that a significantly lower $M_{\rm Ba}$ could lower $M_{\rm WD}$ within the Chandrasekhar limit \citep[about 1.4~\Msun;][]{Chandrasekhar1939}, but that the dynamics of this system do not favour a secondary mass below $\sim$1.2~\Msun.

Only with the dynamical information that we currently have, it is difficult to confirm that this 'massive companion' is a single object, and not a close pair formed, for example, by a faint main-sequence star and a WD (see \citealt{vandenheuvel2020} for an example of such a situation). The strong s-process enhancement strongly suggests that there is a WD in the system, but since we cannot be certain of its mass, HD\,31487 will be removed from further discussion.

\section{Discussion}\label{sec:discussion}
\subsection{Mass distributions}

The mass distribution that we obtained for the WD companions of Ba stars is compatible with current estimates for field WD masses. The average mass of DA WDs (WDs with only Balmer lines in their spectra) is about 0.60~\Msun, while that of DB WDs (WDs with no H or metals lines in their spectra, only helium lines) is 0.68~\Msun \citep{Kleinman13}. The weighted average of our mass distribution is 0.65~\Msun, after removing the two targets mentioned in Sect. \ref{sec:WDmasses}. There is a high-mass tail present in the mass distribution of WDs orbiting Ba giant that \cite{Jorissen19} and \cite{Escorza19EWASS} already discussed (see also Fig. \ref{Fig:WD}).

In order to evaluate if $Q = M^{3}_{\rm WD}/(M_{\rm Ba} + M_{\rm WD})^{2}$ is constant, we computed this value for all our targets and present the average and the standard deviation for each one of the three subsamples separately in Table \ref{table:Qvalues}. The new distributions are marginally different to literature $Q$ distributions \citep[see Table 1 in][]{Escorza19EWASS}. We obtained p = 0.048 for the strong Ba giants, p = 0.035 for the mild Ba giants and p = 0.012 for the Ba dwarfs, when we performed KS tests. The main difference is that the new distributions are not as narrow as obtained in the past when modelling $f(m) = Q \sin^3i$, with $f(m)$ being the spectroscopic mass function. In order to check if this is caused by the fact that the individual inclination uncertainties play a role now, while an inclination distribution was assumed in the past, we calculated new $Q$ distributions removing the 10 and 20\% systems with larger uncertainties. All the observed distributions are broader than the literature ones, but not significantly different.

In Table \ref{table:Qvalues}, we have also included the average and standard deviations of the current mass ratios of the three Ba star subsamples. The two subsamples of giants show closer values, with the average mass ratio of strong Ba giants being slightly higher than that of mild Ba giants, in agreement with what \cite{Jorissen19} reported. We perform a KS test and obtained a p-value of 0.015, which is not low enough to statistically confirm that this difference. The average mass ratio of Ba dwarfs is much higher, but the currently known Ba dwarfs are significantly less massive than the giants \citep{Escorza19}, then accounting for this result.

\begin{table}
\caption{Average and standard deviations of the $Q$-values, with $Q = M^{3}_{\rm WD}/(M_{\rm Ba} + M_{\rm WD})^{2}$, and the mass ratios of strong Ba giants, mild Ba giants and pre-RGB Ba stars.}\label{table:Qvalues}
\centering
\begin{tabular}{|l|c|c|}
\hline       
\textbf{Ba type} & \textbf{$Q$-value} & \textbf{Mass ratio ($q$)}\\
\hline 
Strong Ba giant & 0.054 $\pm$ 0.022 & 0.37 $\pm$ 0.09\\
Mild Ba giant & 0.036 $\pm$ 0.019 & 0.28 $\pm$ 0.08\\
Ba dwarf & 0.091 $\pm$ 0.035 & 0.60 $\pm$ 0.14\\
\hline
\end{tabular}
\end{table}

\subsection{A comment on nucleosynthesis predictions}

It is difficult to make a direct correlation between the WD companion mass and the s-process enhancement of the Ba star because many parameters, apart from the WD progenitor mass, strongly affect the final Ba star abundances and the unknowns are still stronger than the observational constraints \citep[see][for a study of abundances in individual Ba giant systems]{Cseh21}. For example, the efficiency of the mass transfer and the dilution factor, the ratio between the accreted mass and the mass in the Ba star envelope over which this is mixed in, are major uncertainties in our understanding of the formation of Ba stars and will directly affect the final s-process enhancement \cite[e.g.][]{Stancliffe21}. Of course, the efficiency of the s-process of nucleosynthesis in the interiors of AGB stars, which strongly depends on the mass and the metallicity of the star itself, is also a key parameter in order to explain a possible correlation between WD mass and Ba enhancement\cite[e.g.][]{Busso01, Karakas2016,VanderSwaelmen17}. Additionally, even the number of thermal pulses and third dredge-ups experienced by the AGB star before the mass transfer episode took place will have an effect on the final s-process abundance pattern \citep[e.g.][]{Shetye18}, as well as mixing and diffusion below the AGB star's convective envelope will \citep[e.g.][]{GorielySiess18}.

Standard stellar-evolution models do not predict solar-metallicity low-mass AGB stars undergo third dredge-ups \citep[e.g.][]{Cristallo15, Karakas2016}. This limit can go down to 1 \Msun at lower metallicities \citep[e.g.][]{Stancliffe05, Lugaro12, Fishlock14}. However, including different additional effects in the models can help, for example, \cite{Weiss09} showed that including some overshooting below the convective pulse, their models could make a 1 \Msun AGB star undergo third dredge-ups. Additionally, \cite{Shetye19, Shetye21} found several low-mass AGB stars currently undergoing third dredge-ups and their models succeeded to reproduce the s-process overabundance including diffusive mixing at the bottom of the stellar envelope. Additionally, according to several studies, the AGB stars that polluted Ba stars need to have masses below 3 \Msun to be able to reproduce their abundance ratios with models \citep{Lugaro03, Lugaro12, Lugaro16, Cseh18, Karinkuzhi18}.

\begin{figure}
\centering
\includegraphics[width=0.49\textwidth]{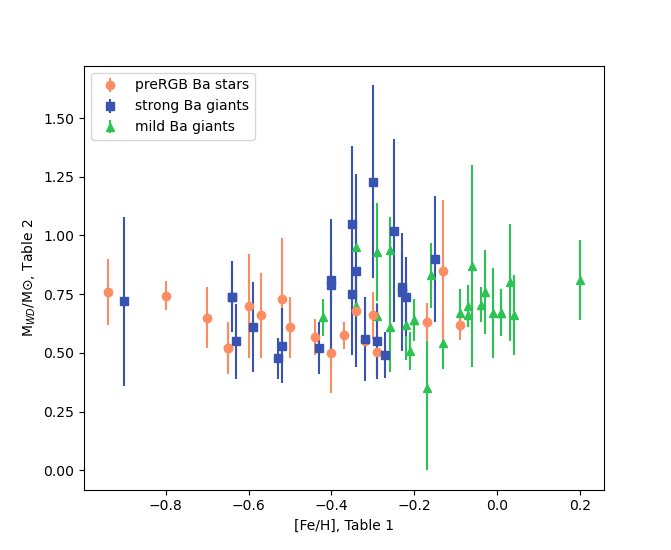}
\caption{\label{fig:FeH-WD} Relation between the metallicity (Table \ref{table:targets}) and the obtained WD masses (Table \ref{table:results}) for the preRGB Ba stars (orange circles), the strong Ba giants (blue squares) and the mild Ba giants (green triangles) in our sample.}
\end{figure}

Figure \ref{fig:FeH-WD} shows the relation between the metallicity (listed in Table \ref{table:targets}) and the obtained WD masses (Table \ref{table:results}) for the preRGB Ba stars (orange circles), the strong Ba giants (blue squares) and the mild Ba giants (green triangles) in our sample. The figure shows an expected correlation between the Ba-type and the metallicity, caused by the fact that the efficiency of the s-process in AGB stars decreases as the metallicity increases \citep[e.g.][]{Cseh18, Jorissen19}. However, there is no obvious correlation between the WD mass and the metallicity, even though the AGB mass directly affects the s-process efficiency as well. The least massive WDs are in systems with [Fe/H]~$< -0.1$, in agreement with the models, and the most massive WDs accompany Ba giants of [Fe/H] between $-$0.4 and $-$0.2, with the three most massive WDs being in a strong Ba star systems.

Among our sample of 58 systems (after having removed (HD\,95241 and HD\,31487 from the WD sample), we do not find Ba stars with unexpectedly low mass companions. As discussed in Sect. \ref{sec:WDmasses}, the companion mass for HD\,18182 should be taken with caution, but all other Ba star systems have WDs of around or more massive than 0.5~\Msun, meaning that their progenitors were AGB stars of around or more massive than 1~\Msun.  Note that to make such a statement, one needs to rely on initial-final mass relationships (IFMR). We used as a reference the relation published by \cite{Marigo20, Marigo22}. Using the same relation, we can claim that a fraction of the AGB stars that polluted our sample of Ba stars were more massive than the expected 3~\Msun limit, since we found that several WDs have masses around or higher than 0.8~\Msun. This is the case even taking into account the kink that \cite{Marigo20, Marigo22} find for WDs of about 0.70 -- 0.75~\Msun with carbon AGB progenitors. Most IFMRs \citep[e.g.][]{Weidemann00, Kalirai08, Williams09, Andrews15, Cummings16, El-Badr18} flatten at around $M_{\rm WD} \sim 0.8$~\Msun, making stars with a wide range of initial masses accumulate at that WD mass. However, their progenitors are expected to have initial masses in the range between 3.5 and 5.5~\Msun, hence more massive than what the Ba stars abundance ratios seem to indicate. 

The presence of these massive WDs orbiting around both strongly and mildly polluted Ba stars presents important constraints, as well as an interesting challenge, for evolutionary and nucleosynthesis models. Future studies of these systems following the line presented by \cite{Stancliffe21} or \cite{Cseh21}, but using these new WD masses, might be able to tell us new things about AGB stars. We note that our error bars are significant and that these statements blur if we consider two or three sigma uncertainties. This will improve when we have NSS parallaxes to obtain more accurate Ba star masses and Gaia astrometric epochs to improve the RV+astrometry fit. Direct imaging observations could also help constrain the longest-period systems better (see Sect. \ref{sec:sphereplans}).

\section{Future observational prospects: direct imaging of these white dwarfs with SPHERE}\label{sec:sphereplans}
\begin{figure*}
\centering
\includegraphics[width=0.91\textwidth]{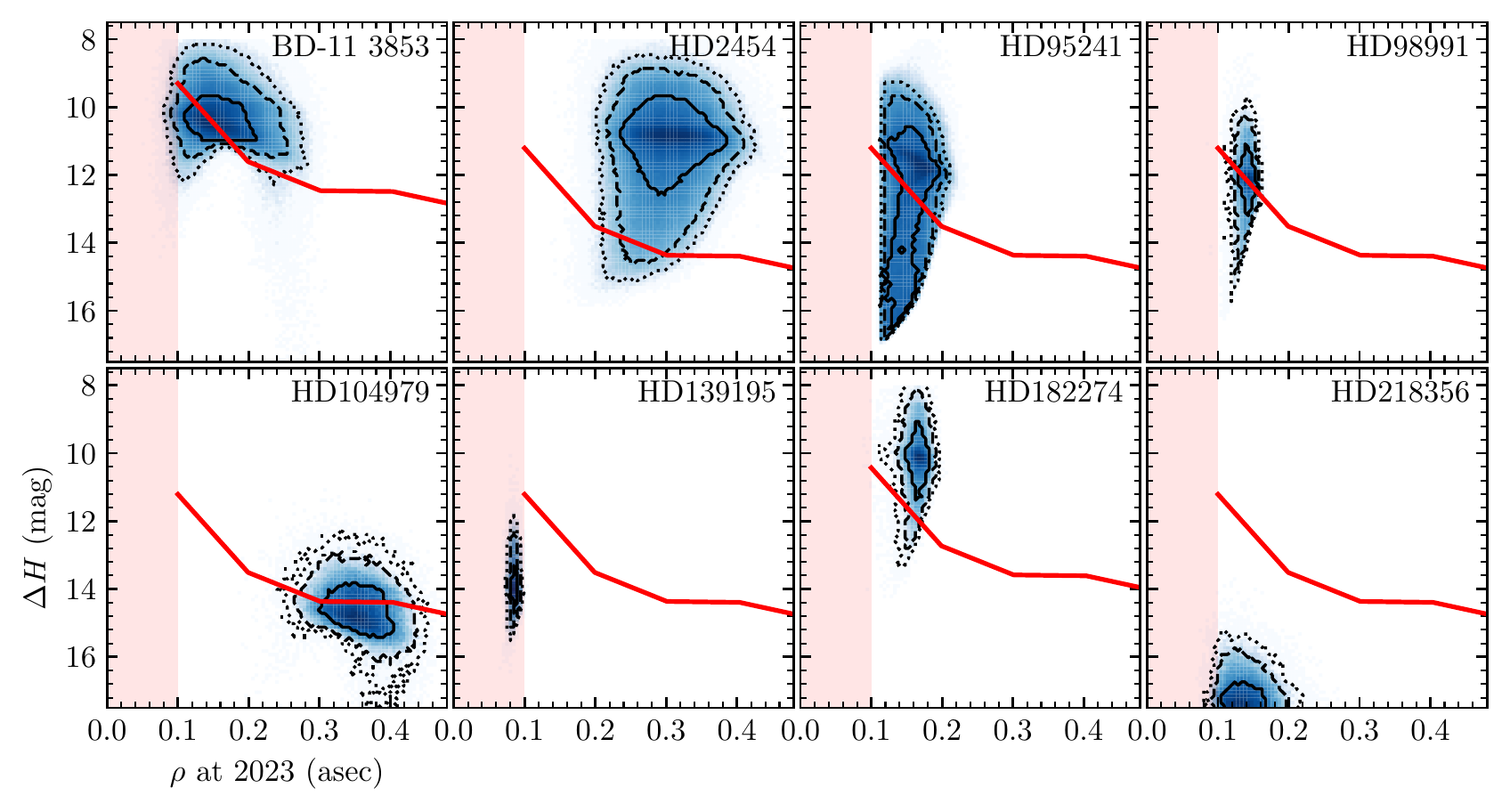}
\caption{\label{fig:sphere}Predicted angular separation in 2023 and contrast of the eight systems with median apoastron distances of $>100$\,mas. Contours indicate 1, 2, and 3$\sigma$ credible regions. The predicted SPHERE contrast is given by the solid red line, and the red shaded region corresponds to the inner working angle of the instrument. Six of the systems are amenable to direct detection in the near future.}
\end{figure*}

The nearby ($d\lesssim 100$\,pc) Ba stars that are host to long-period ($P\gtrsim 10$\,yr) companions are suitable candidates for high-contrast imaging observations to spatially resolve the companion. These observations would provide relative astrometric and photometric measurements between the WD and the Ba star host. A single measurement of the instantaneous angular separation between the components would constrain both the total semi-major axis, and thus the total system mass, and the inclination (unless the observation occurred when the companion was crossing the line of nodes). Photometric measurements of the companion could be used to estimate the bolometric luminosity of the companion which, in conjunction with the mass, can be compared to WD cooling models to estimate the age of the companion (see e.g. \citealt{Bonavita20} for the discovery and analysis of a WD companion around a K-type star with SPHERE and e.g. \citealt{Gratton21} for the study of a sample of Sirius-like systems, long-period main-sequence + white dwarf binaries).

We assessed the feasibility of spatially resolving the companion by comparing the predicted angular separation and flux ratio between the WD companion and the Ba host star to the performance of the VLT/SPHERE instrument \citep{SPHERE}. We filtered the sample to exclude systems with a median apoastron distance within 100\,mas, calculated from the MCMC samples described in Section~\ref{sec:methods}. For these systems, the companion would always be within the inner working angle of the instrument, and impossible to resolve with SPHERE. This filter resulted in a sub-sample of eight systems for which the companion will be at a projected separation of $\rho>100$\,mas at some point in its orbit.

The feasibility of direct detection also depends on the flux ratio between the WD companion and Ba star host. We estimate the {\it H}-band flux ratio for each MCMC sample using pure-hydrogen (DA) atmosphere mass-luminosity relations from \cite{HolbergBergeron06}. We assigned an age to each MCMC sample at random from a uniform distribution between $10^6$ and $10^{10}$\,yr to account for the unknown age of the WD. The model grid was linearly interpolated in $(\log t, M)$ to extract an absolute {\it H}-band magnitude. This was converted into a flux ratio using the parallax from the MCMC sample and the apparent {\it H}-band magnitude of the Ba star. We assumed the companion has negligible flux in the {\it H}-band relative to the Ba star, such that the catalogue {\it H}-band magnitude of the system can be entirely ascribed to the Ba star.

We converted the orbital elements to the angular separation between the WD and the Ba star host at the epochs 2023, 2024, and 2025 for each MCMC sample. The predicted angular separation and flux ratio for each sample was then compared to the SPHERE contrast curve given in \cite{Wahhaj21}. We accounted for the degradation in contrast performance for fainter stars \citep[e.g.][]{Jones22} by scaling the contrast curve by the square root of the {\it H}-band flux ratio between the Ba star host and HR 8799, the star observed by \cite{Wahhaj21} to measure the contrast curve. 

The predicted separations in 2023 and $H$-band contrast for each of the eight systems are shown in Figure~\ref{fig:sphere}. There are six systems with a non-negligible probability of detection at this epoch; the others are too faint to be detected given the expected contrast curve. The majority of these systems exhibit a strong correlation between the separation at the 2023 epoch and the mass of the WD companion. This can partly be explained by the constraint provided by a direct measurement of the semi-major axis of the system, leading to a much more precise measurement of the total mass of the system.

\section{Summary and conclusions}\label{sec:conclusions}
The WD companions of Ba stars contain important information about the formation of these chemically peculiar stars, about the binary interaction processes that these systems underwent in the past, and about the nucleosynthesis processes that took place inside their AGB progenitors. However, they are cool, dim, and generally not detected by direct methods, so they have not been studied in detail in the past. A few absolute masses had been determined before this work by combining the spectroscopic orbital parameters of these systems with Hipparcos astrometric data. However, most published masses for WD companions of Ba stars were computed by making assumptions on the relation between the masses of the two stellar components in these systems or on their orbital inclinations.

In this work, we used the software package \textsc{orvara} to combine radial-velocity data, Hipparcos and Gaia positions and proper motions through the Hipparcos-Gaia Catalogue of Accelerations, and astrometric epoch measurements from the Hipparcos mission, and determine the astrometric orbital parameters of 60 stars flagged as Ba dwarfs or giants. Using this method, we could constrain the orbits of two long-period systems that could not have been constrained before with RV data only, and we improved the orbital solution of a few other systems. Orbital inclinations were also determined for the first time for many of these systems, and finally, including a prior on the Ba star masses, we also derived the mass of the secondary stars in these systems. Finally, we discovered that HD\,218356, one of the shortest period Ba star systems known, is actually a triple system. We determined the parameters of both the inner and outer orbits and the masses of the two components, and it is very likely that the WD companion that polluted HD\,218356 is in the outer orbit, explaining the mild s-process enhancement of the Ba giant. 

The WD mass distribution presented in this work includes all systems published by \cite{Jorissen19}, \cite{Escorza19} and \cite{North20} that had a single-star Hipparcos solution and that were not confirmed triples. This mass distribution is compatible with field WD mass distributions and with those published before for Ba stars. The distribution extends to high WD masses, higher than expected by theoretical models of the s-process of nucleosynthesis that have focused on reproducing the abundance ratios measured on Ba star atmospheres. This work brings new observational constraints for these models and an interesting challenge to our understanding of the formation of Ba stars.

In order to look at Ba stars with new eyes, we plan future direct imaging observations of six of the longest-period systems with SPHERE. On the one hand, this data will provide us with a measurement of the instantaneous angular separation between the components of the system, partially breaking the total mass - semimajor axis correlation and helping us get more accurate masses. On the other hand, we will be able to estimate the bolometric luminosity of the WD, which combined with its mass, can be compared to WD cooling models to estimate the age of these systems. \\

\begin{acknowledgements}
The authors thank Prof. Dr. Alain Jorissen and Dr. Henri Boffin for the enriching discussions and Prof. Dr. Hans Van Winckel for providing us with a few new, unpublished HERMES RV points for our targets. We also want to thank the referee, Dr. Carine Babusiaux, for helping us to improve this manuscript.
We are grateful to all observers of the HERMES consortium for their time dedicated to the Mercator-HERMES long-term binary-monitoring program. The HERMES spectrograph is supported by the Fund for Scientific Research of Flanders (FWO), Belgium, the Research Council of KU Leuven, Belgium, the Fonds National de la Recherche Scientifique (F.R.S.-FNRS), Belgium, the Royal Observatory of Belgium, the Observatoire de Genève, Switzerland and the Thüringer Landessternwarte Tautenburg, Germany.
This publication includes data retrieved from the SOPHIE and the ELODIE archives at Observatoire de Haute-Provence (OHP), available at \url{http://atlas.obs-hp.fr/sophie} and \url{http://atlas.obs-hp.fr/elodie}, respectively.
This work makes use of the "Synthetic Colors and Evolutionary Sequences of Hydrogen- and Helium-Atmosphere White Dwarfs" hosted at \url{http://www.astro.umontreal.ca/~bergeron/CoolingModels} and of the SIMBAD database, operated at CDS, Strasbourg, France.
Last but not least, this work has made use of data from the European Space Agency (ESA) mission Gaia (\url{https://www.cosmos.esa.int/gaia}), processed by the Gaia Data Processing and Analysis Consortium (DPAC, \url{https://www.cosmos.esa.int/web/gaia/dpac/consortium}). Funding for the DPAC has been provided by national institutions, in particular the institutions participating in the Gaia Multilateral Agreement.
\end{acknowledgements}

\bibliographystyle{aa}
\bibliography{references}

\clearpage

\onecolumn 
\begin{appendix}
\section{RV curves, proper motions and corner plots}\label{App}

\begin{figure*}[h]
\centering
\includegraphics[width=0.85\textwidth]{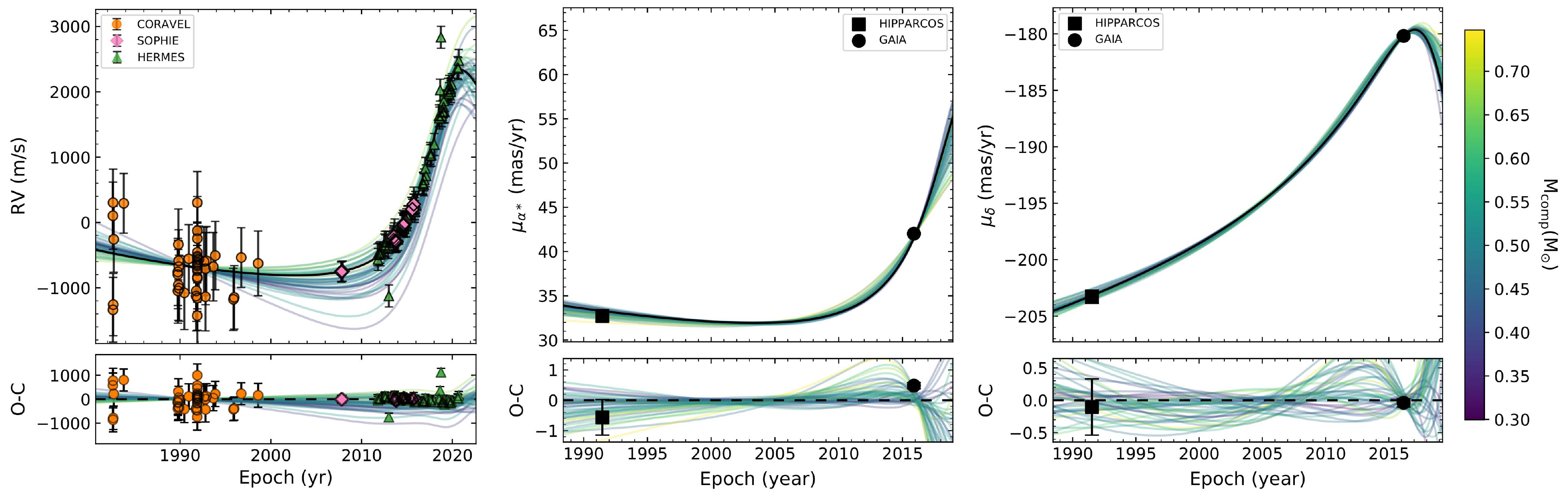}
\caption{\label{Fig:HD2454} RV curve and proper motions of HD\,2454}
\end{figure*}
\begin{figure*}[h]
\centering
\includegraphics[width=0.85\textwidth]{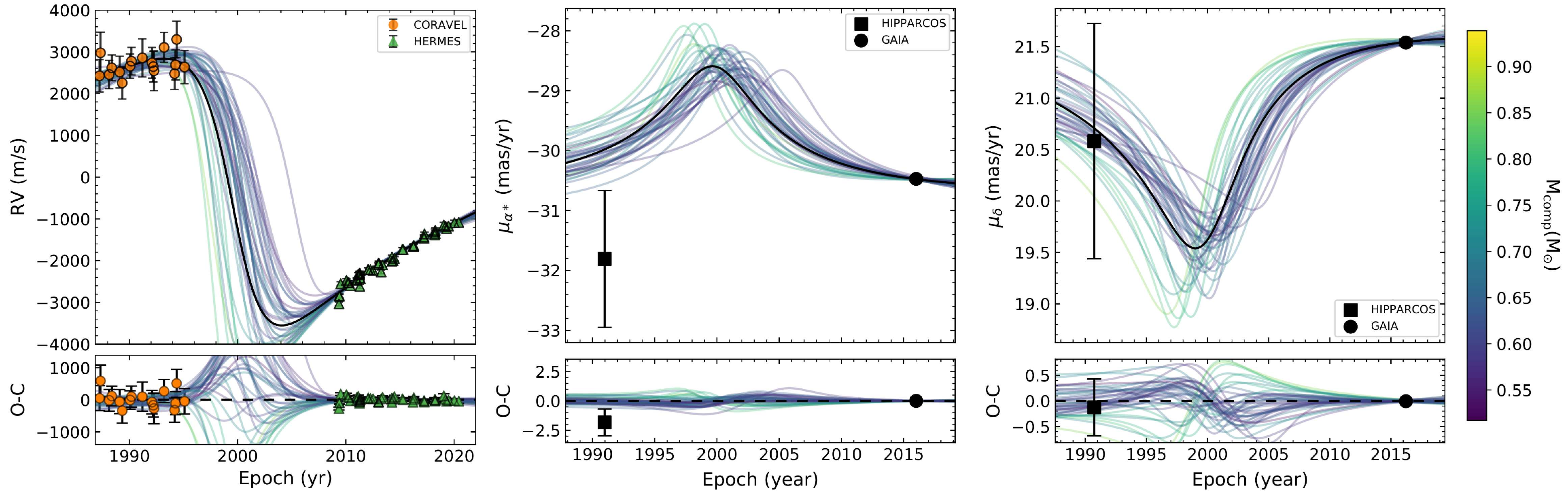}
\caption{\label{Fig:HD119185} RV curve and proper motions of HD\,119185. We used a fixed RV offset of 500 m/s \citep{Jorissen19}.}
\end{figure*}

\begin{figure}
\begin{minipage}{6.5cm} 
\includegraphics[scale=0.28]{Corner_HD2454.pdf}
\caption{\label{Fig:BD-HD2454corner2} Corner plot of HD\,2454}
\end{minipage}
\hspace{3cm} 
\begin{minipage}{6.5cm} 
\includegraphics[scale=0.28]{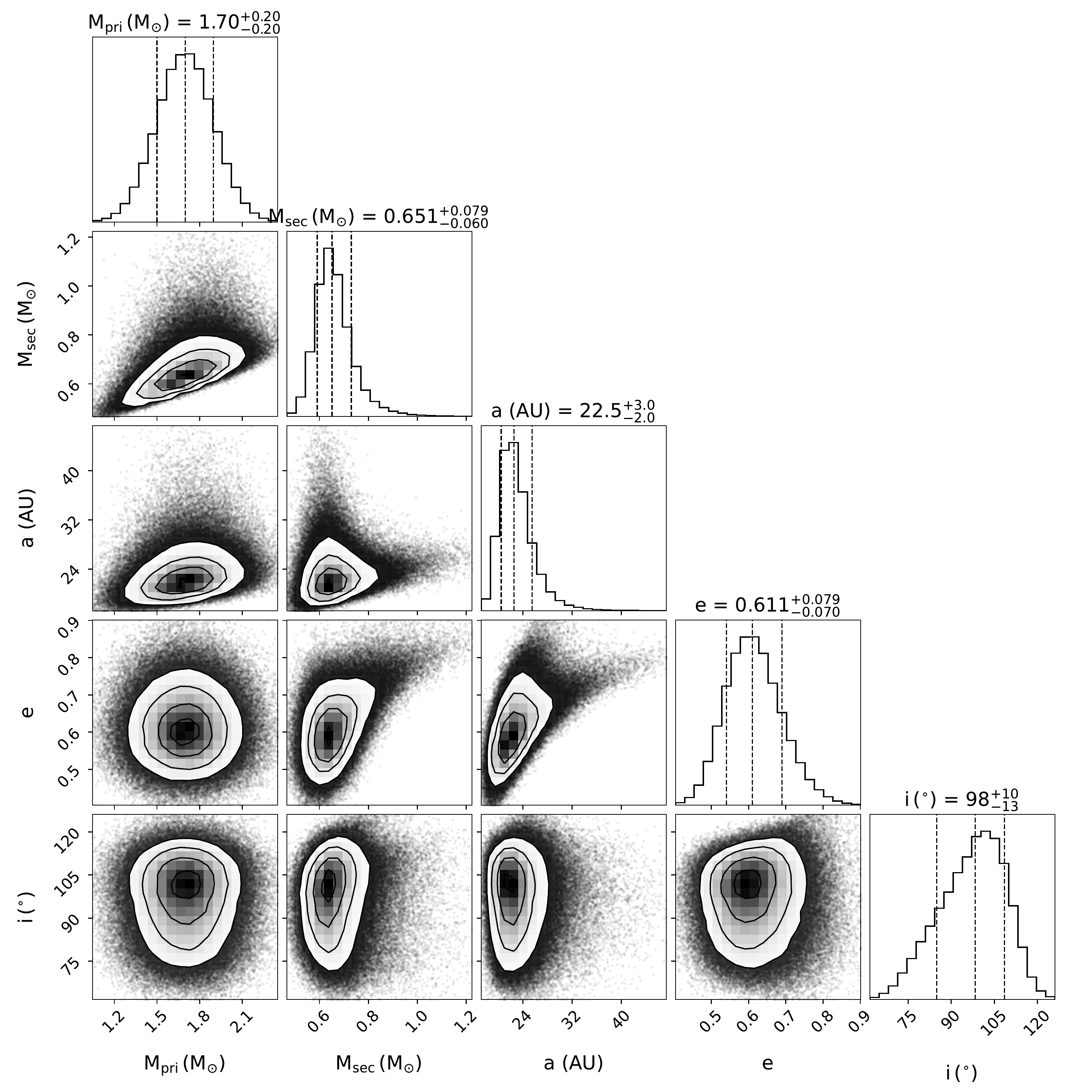}
\caption{\label{Fig:HD119185corner} Corner plot of HD\,119185}
\end{minipage} 
\end{figure} 

\begin{figure*}[t]
\centering
\includegraphics[width=\textwidth]{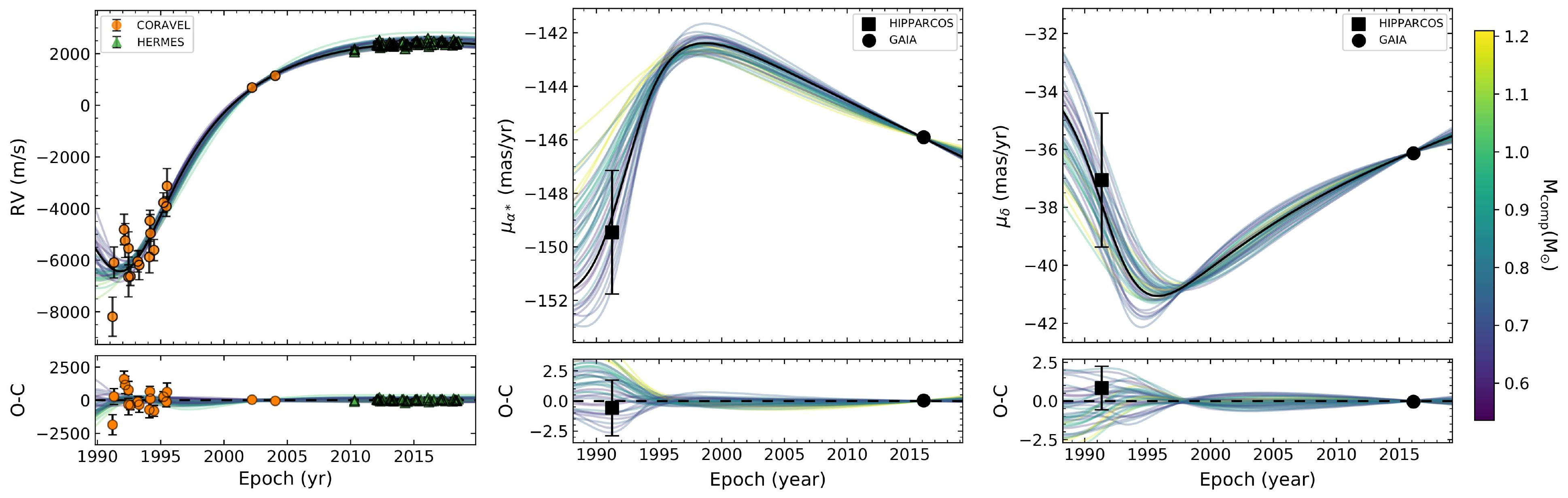}
\caption{\label{Fig:BD-11_3853} RV curve and proper motions of BD-11$^{o}$3853}
\end{figure*}
\begin{figure*}
\centering
\includegraphics[width=\textwidth]{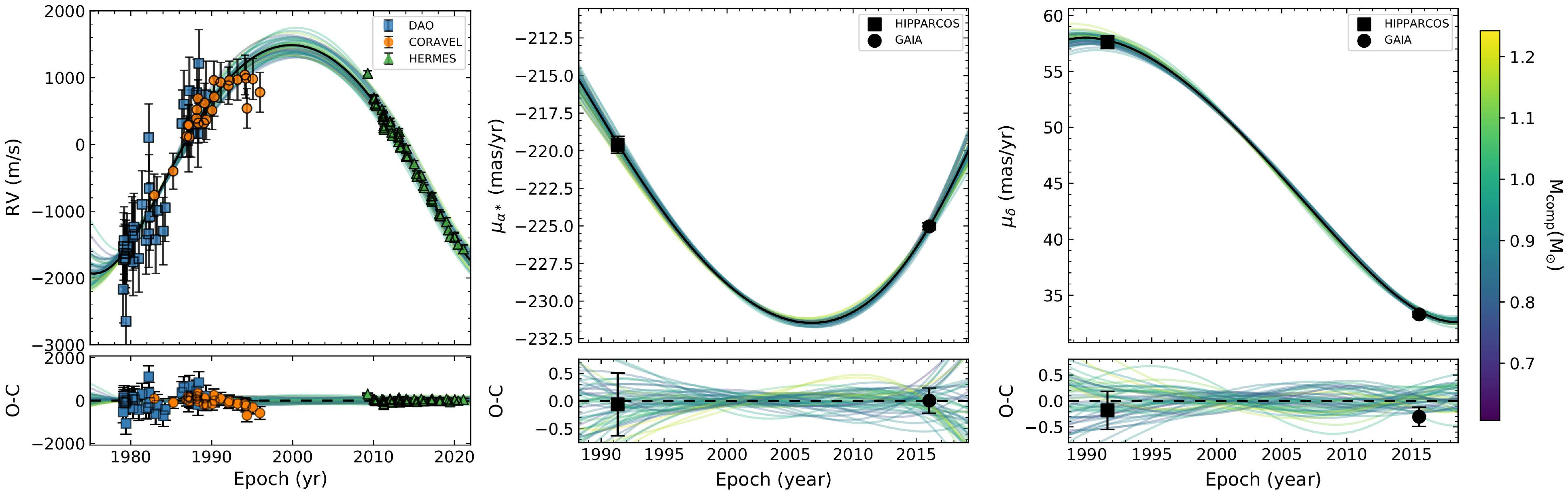}
\caption{\label{Fig:HD104979} RV curve and proper motions of HD\,104979}
\end{figure*}

\begin{figure}
\begin{minipage}[l]{6cm} 
\includegraphics[scale=0.3]{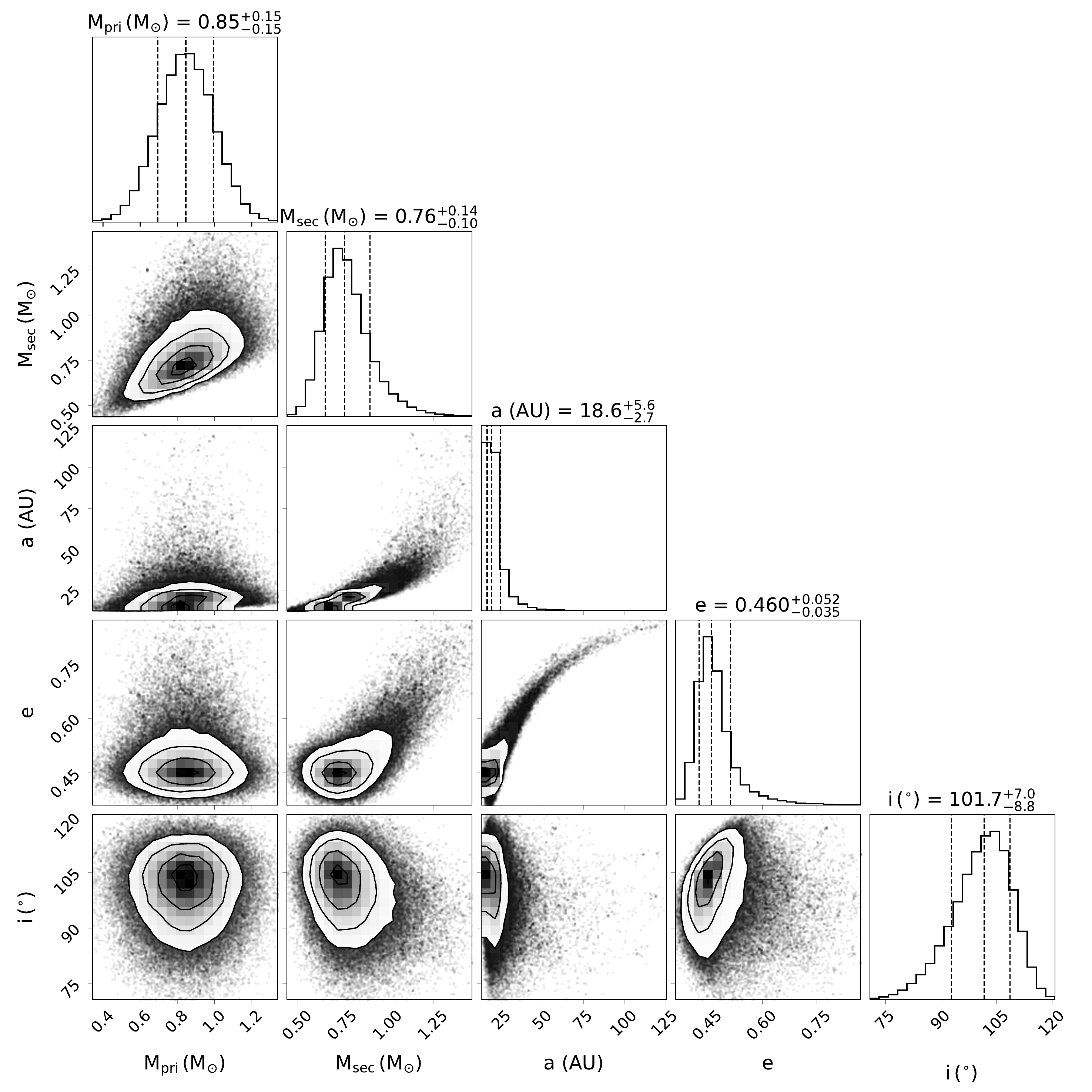}
\caption{\label{Fig:BD-11_3853corner} Corner plot of BD-11$^{o}$3853}
\end{minipage} 
\hspace{3cm} 
\begin{minipage}[r]{6cm} 
\includegraphics[scale=0.3]{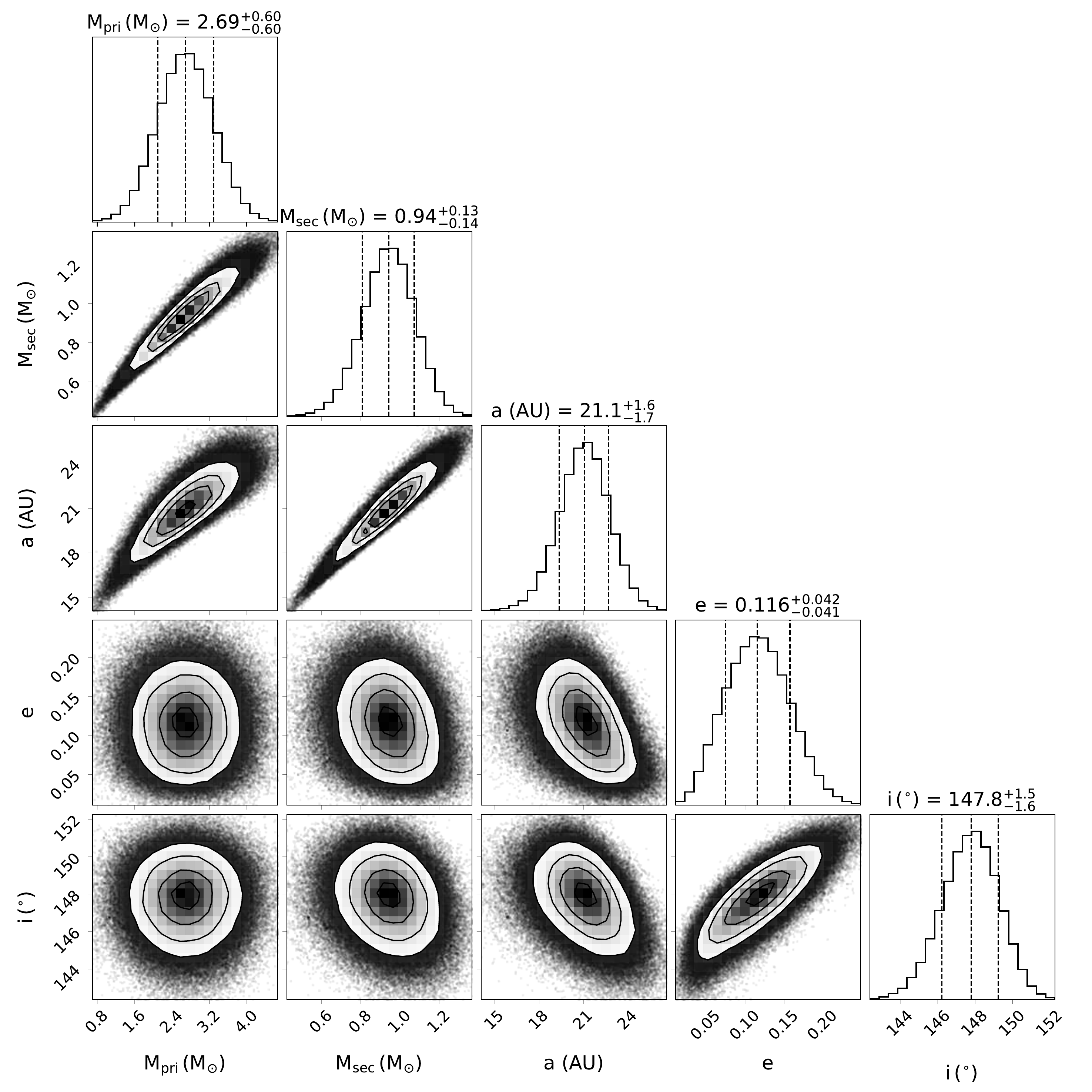}
\caption{\label{Fig:HD104979corner} Corner plot of HD\,104979}
\end{minipage} 
\end{figure}

\begin{figure*}[t]
\centering
\includegraphics[width=\textwidth]{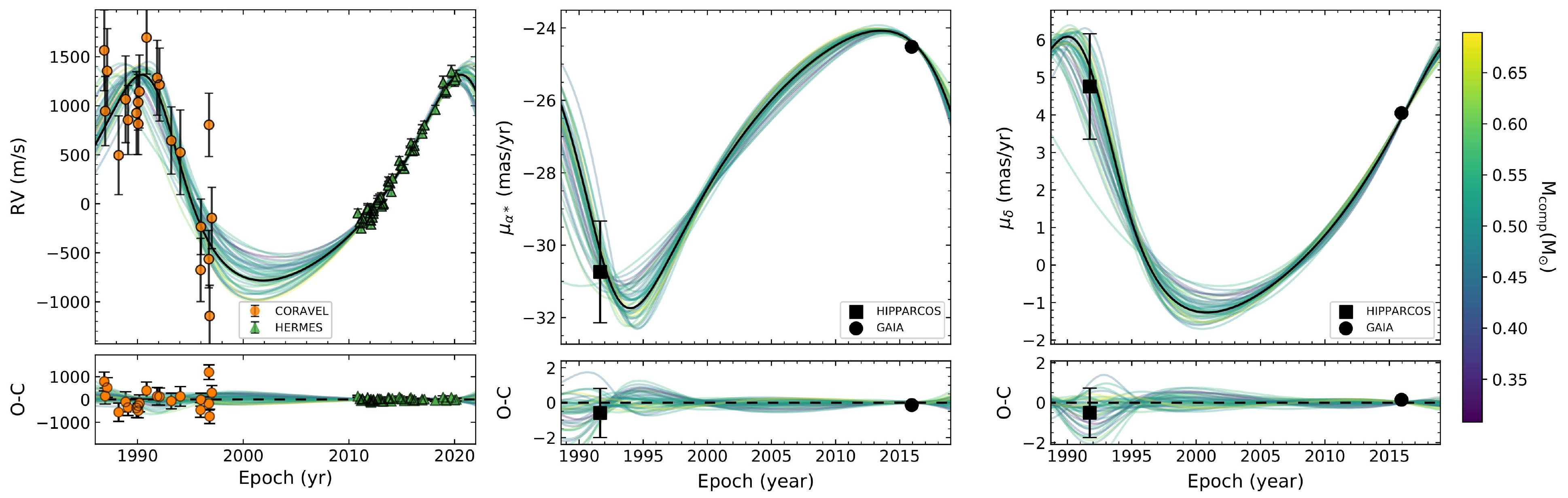}
\caption{\label{Fig:HD51959} RV curve and proper motions of HD\,51959. We used a fixed RV offset of 500 m/s \citep{Jorissen19}.}
\end{figure*}
\begin{figure*}
\centering
\includegraphics[width=\textwidth]{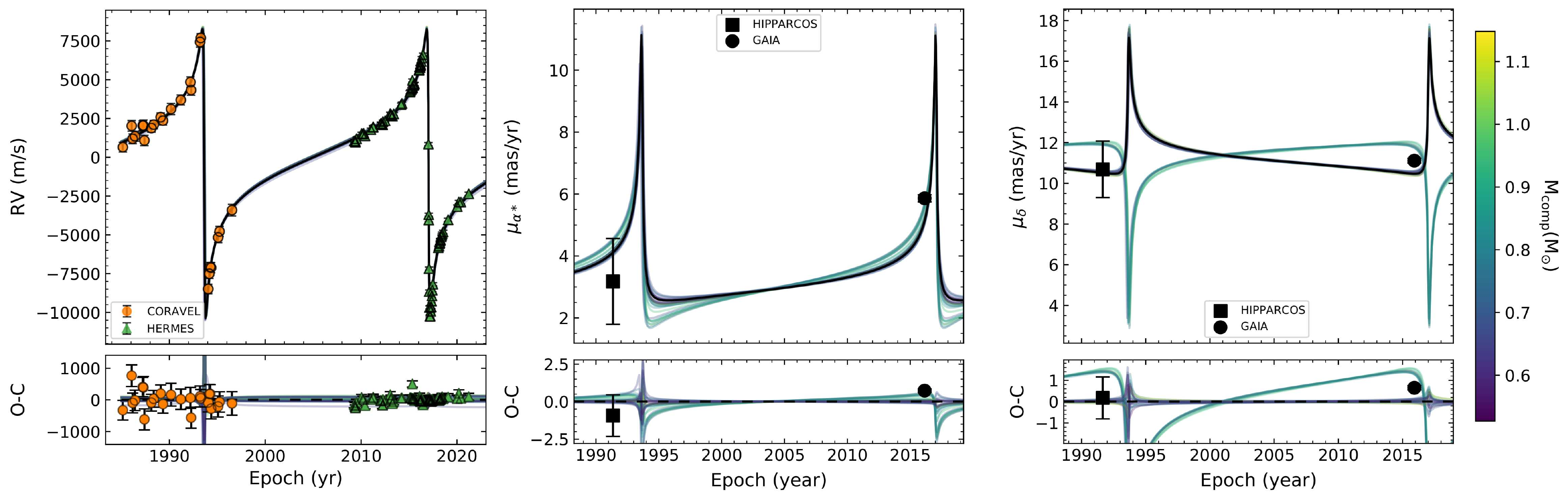}
\caption{\label{Fig:HD123949} RV curve and proper motions of HD\,123949}
\end{figure*}
\vspace{1mm}

\begin{figure}
\begin{minipage}[l]{6cm} 
\includegraphics[scale=0.3]{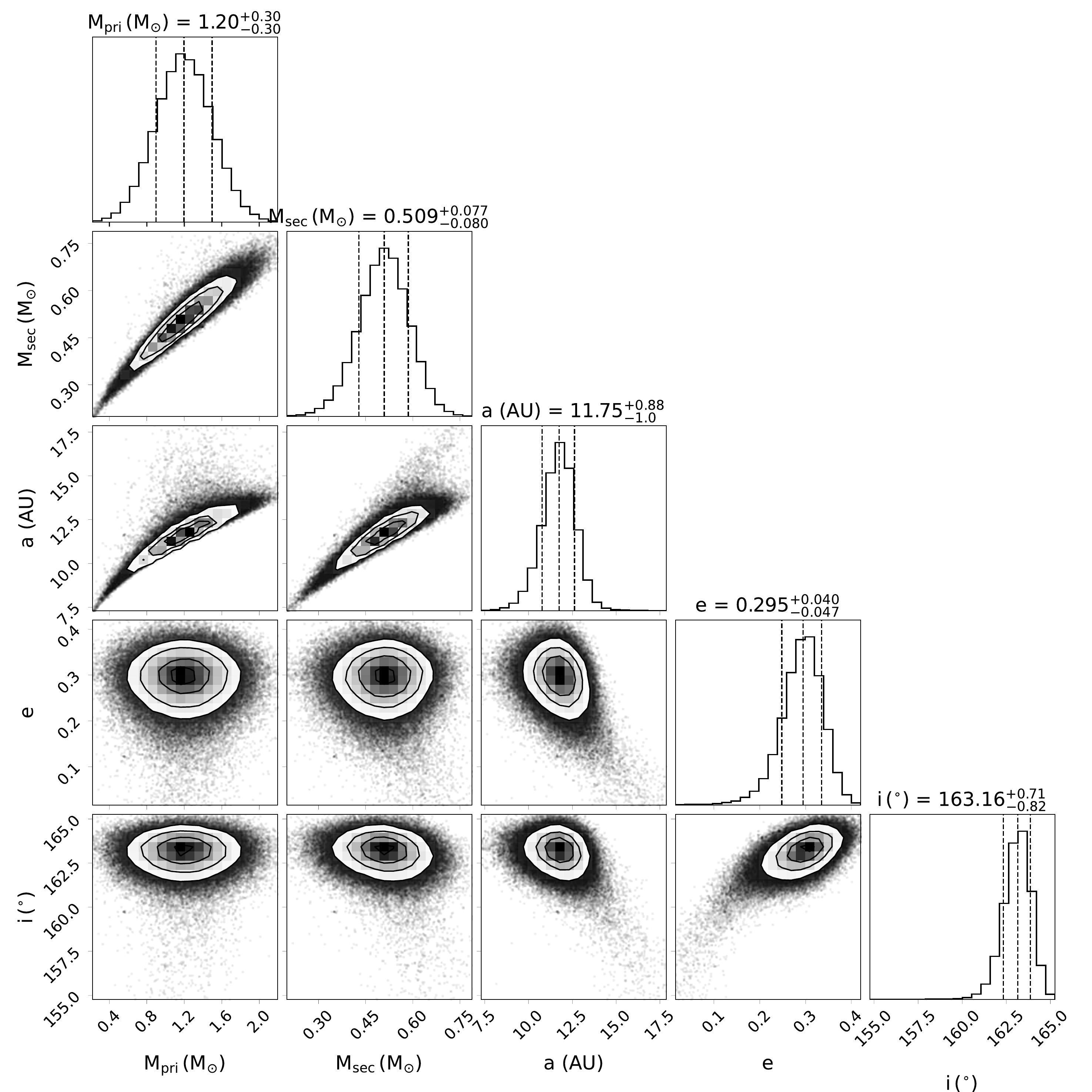}
\caption{\label{Fig:HD51959corner} Corner plot of HD\,51959}
\end{minipage} 
\hspace{3cm} 
\begin{minipage}[r]{6cm} 
\includegraphics[scale=0.3]{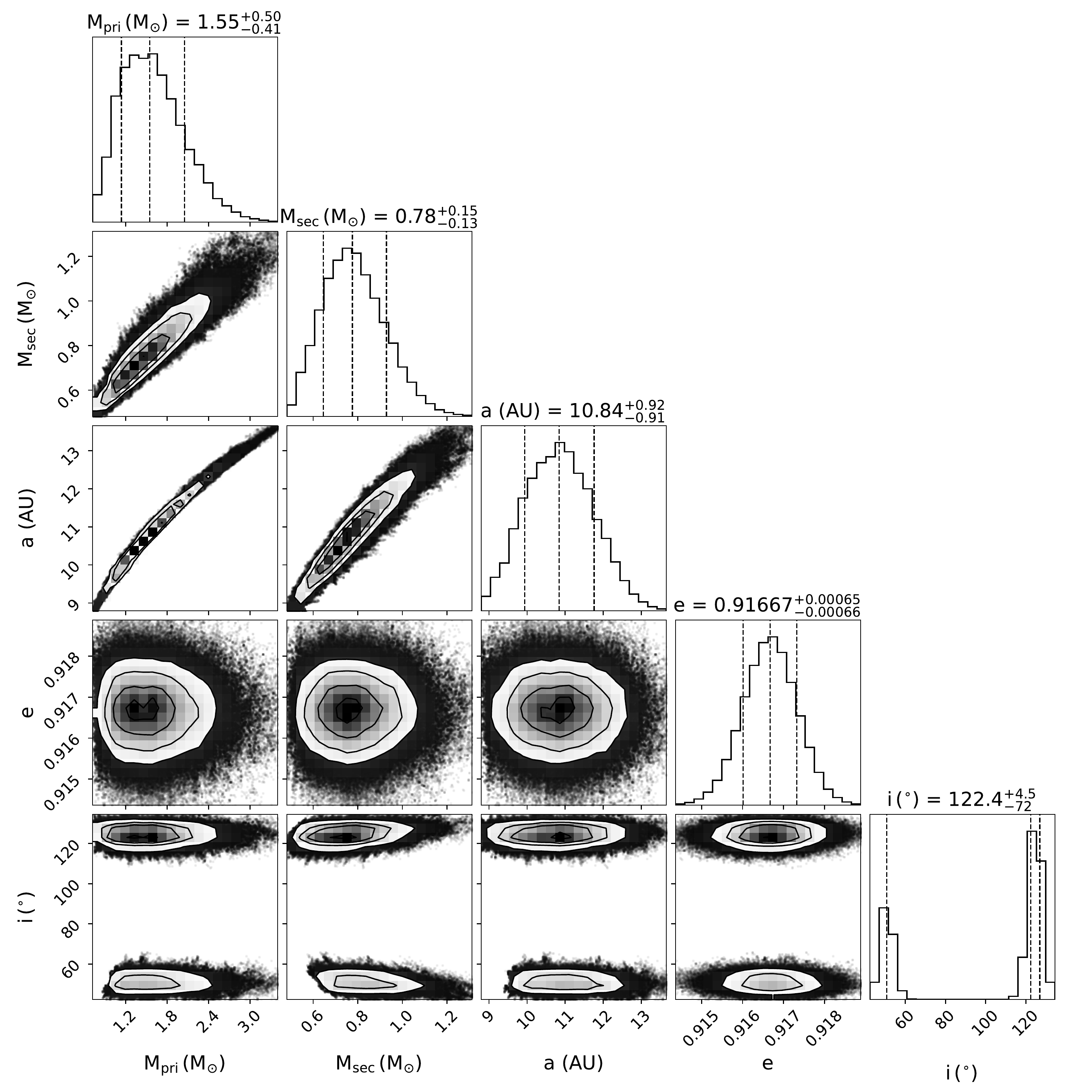}
\caption{\label{Fig:HD123949corner} Corner plot of HD\,123949}
\end{minipage} 
\end{figure}

\begin{figure*}[t]
\centering
\includegraphics[width=\textwidth]{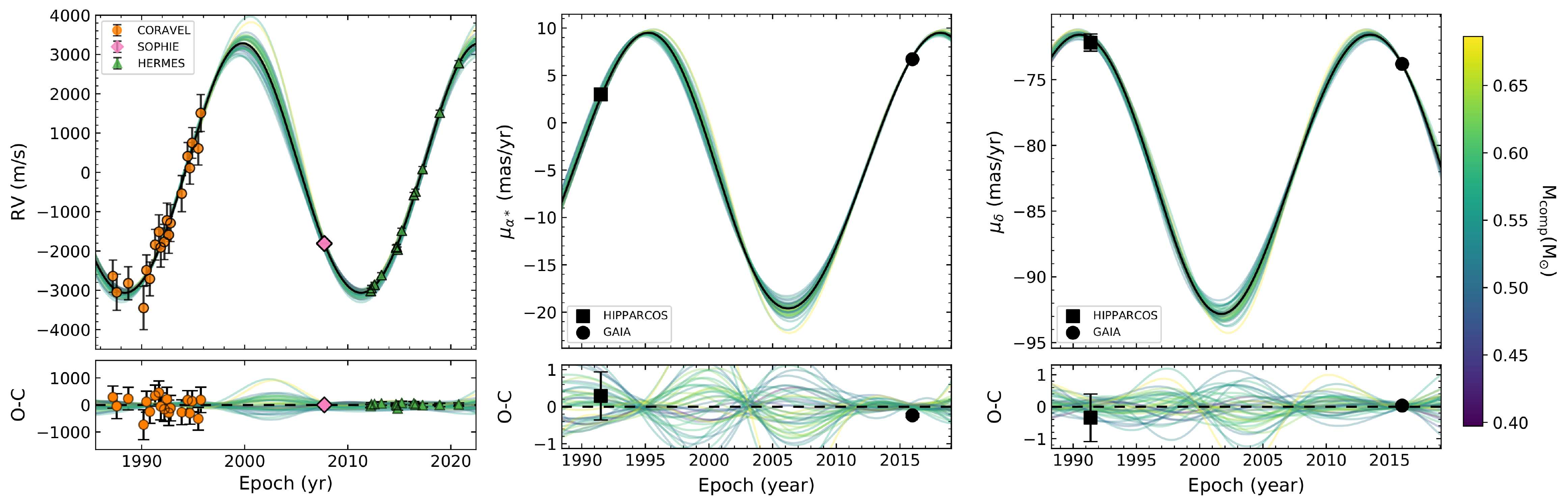}
\caption{\label{Fig:HD182274} RV curve and proper motions of HD\,182274}
\end{figure*}
\begin{figure*}
\centering
\includegraphics[width=\textwidth]{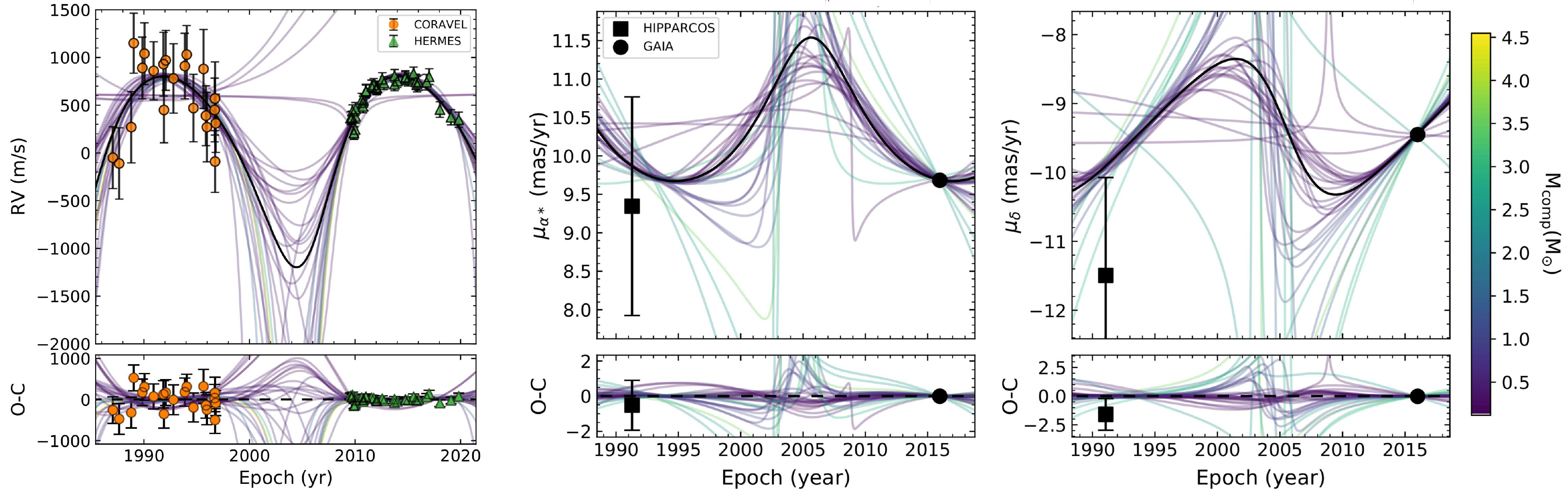}
\caption{\label{Fig:HD18182} RV curve and proper motions of HD\,18182. We used a fixed RV offset of 500 m/s \citep{Jorissen19}.}
\end{figure*}
\vspace{1mm}
\begin{figure}
\begin{minipage}[l]{6cm} 
\includegraphics[scale=0.3]{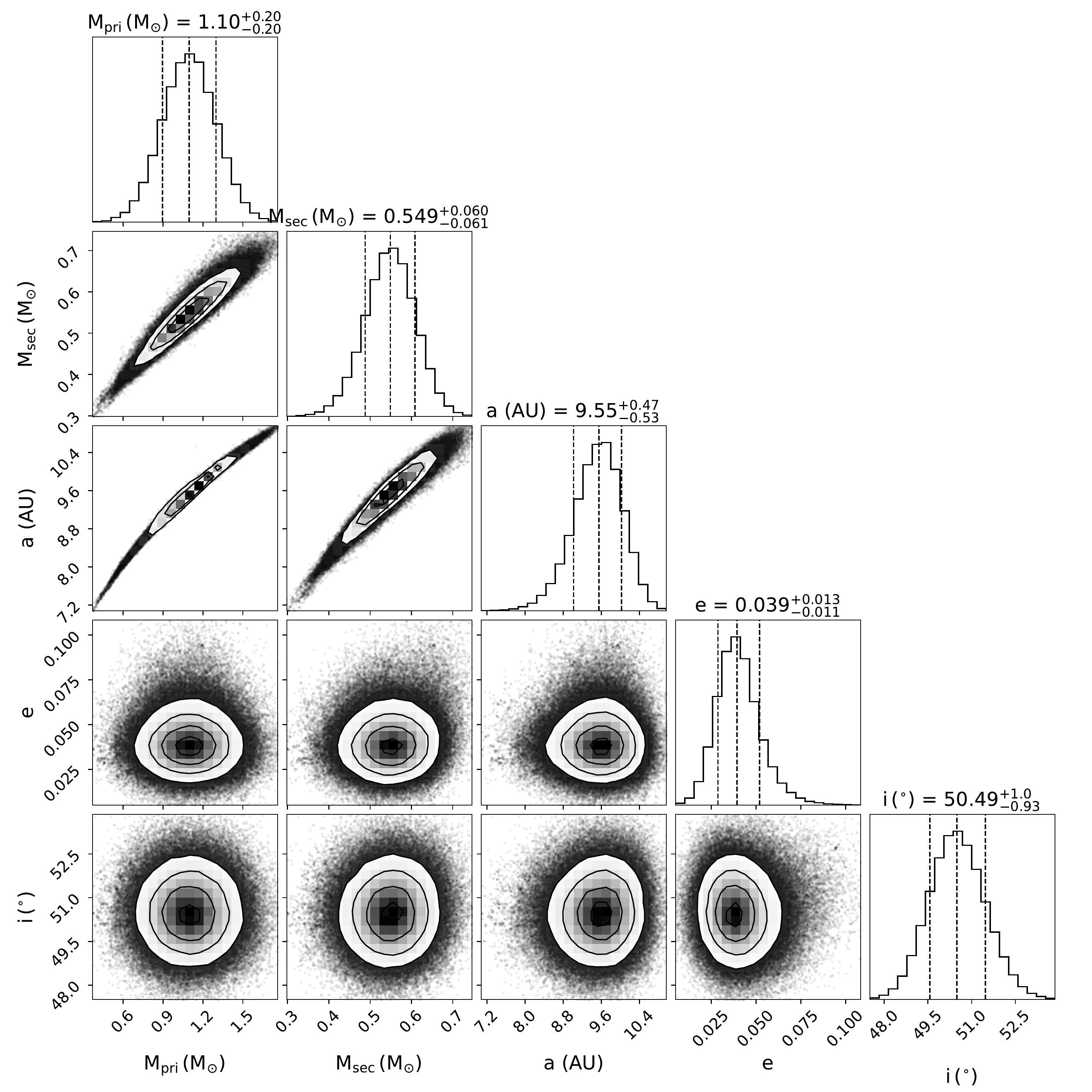}
\caption{\label{Fig:HD182274corner} Corner plot of HD\,182274}
\end{minipage} 
\hspace{3cm} 
\begin{minipage}[r]{6cm} 
\includegraphics[scale=0.3]{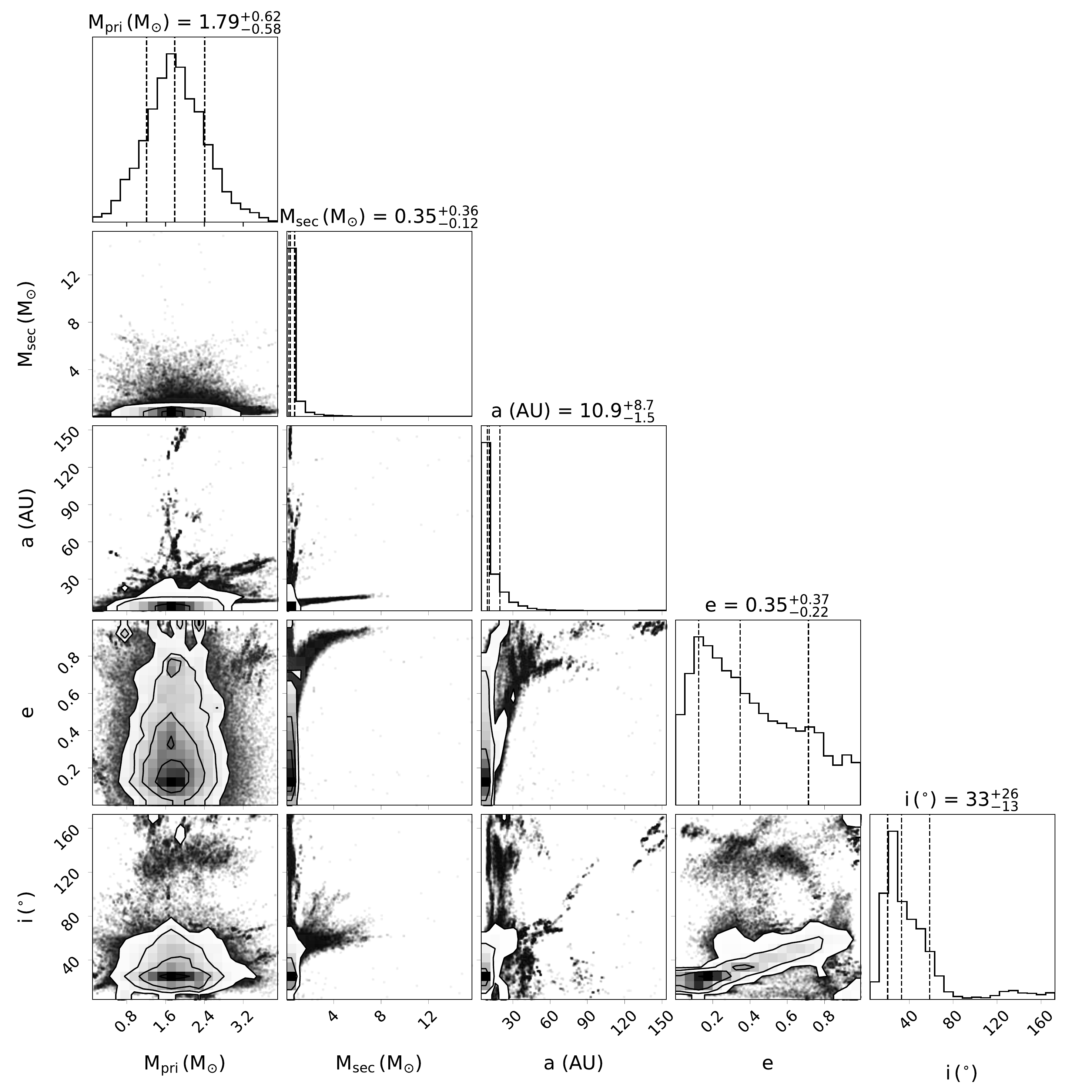}
\caption{\label{Fig:HD18182corner} Corner plot of HD\,18182}
\end{minipage} 
\end{figure}

\begin{figure*}[t]
\centering
\includegraphics[width=\textwidth]{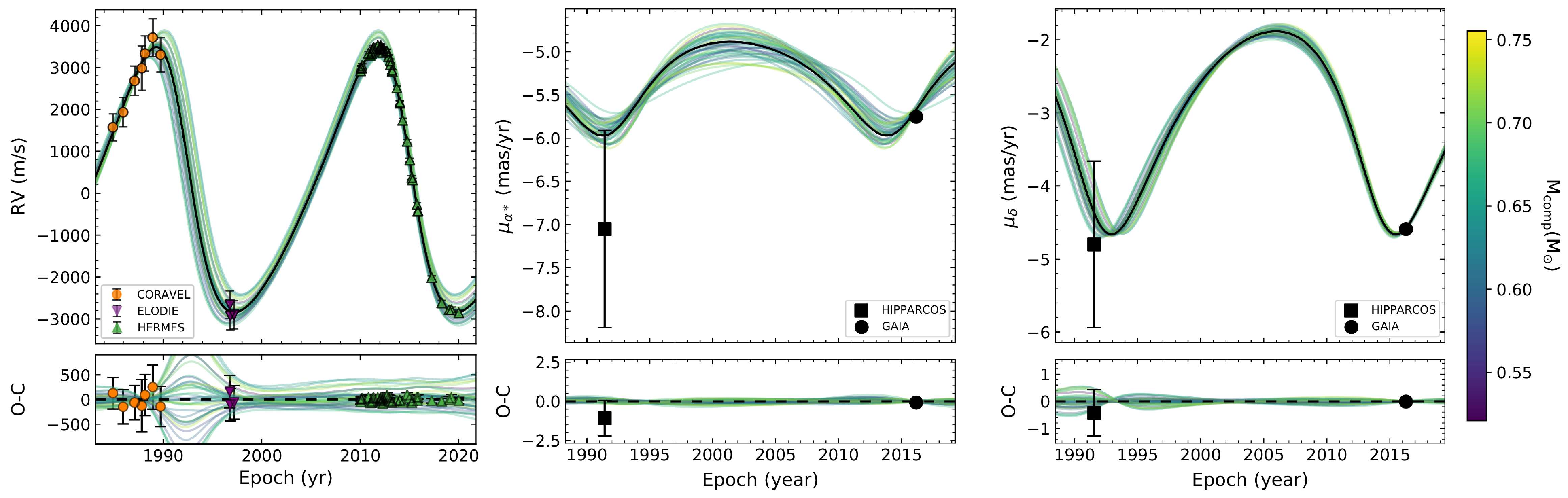}
\caption{\label{Fig:HD53199} RV curve and proper motions of HD\,53199.}
\end{figure*}
\begin{figure*}
\centering
\includegraphics[width=\textwidth]{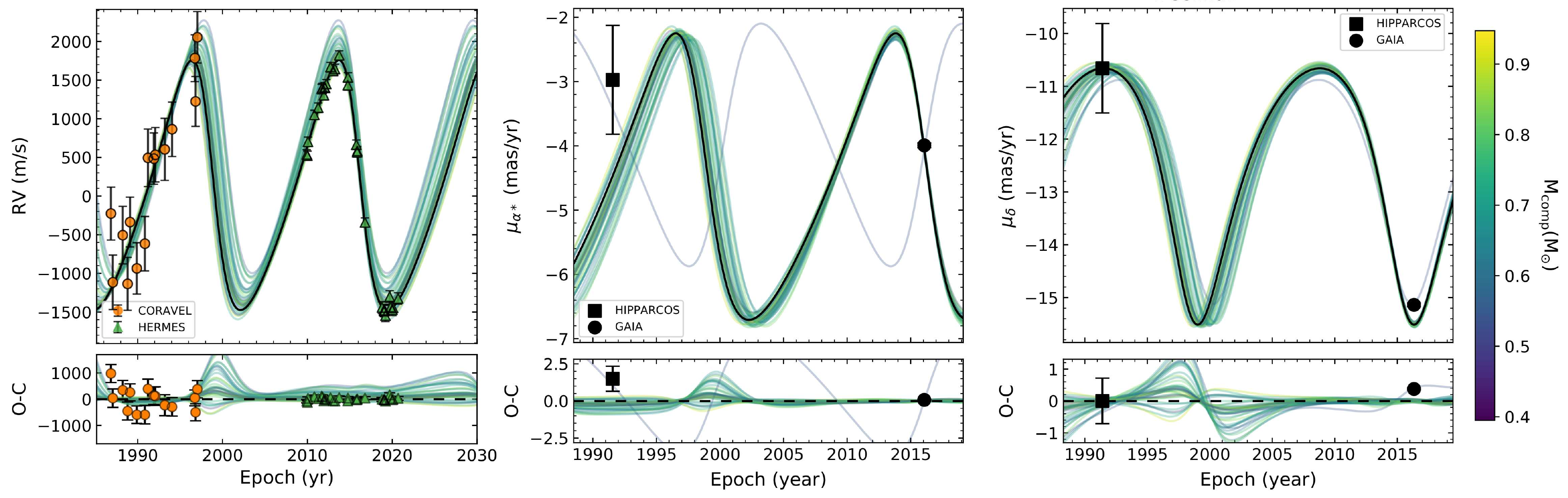}
\caption{\label{Fig:HD40430} RV curve and proper motions of HD\,40430}
\end{figure*}
\vspace{1mm}
\begin{figure}
\begin{minipage}[l]{6cm} 
\includegraphics[scale=0.3]{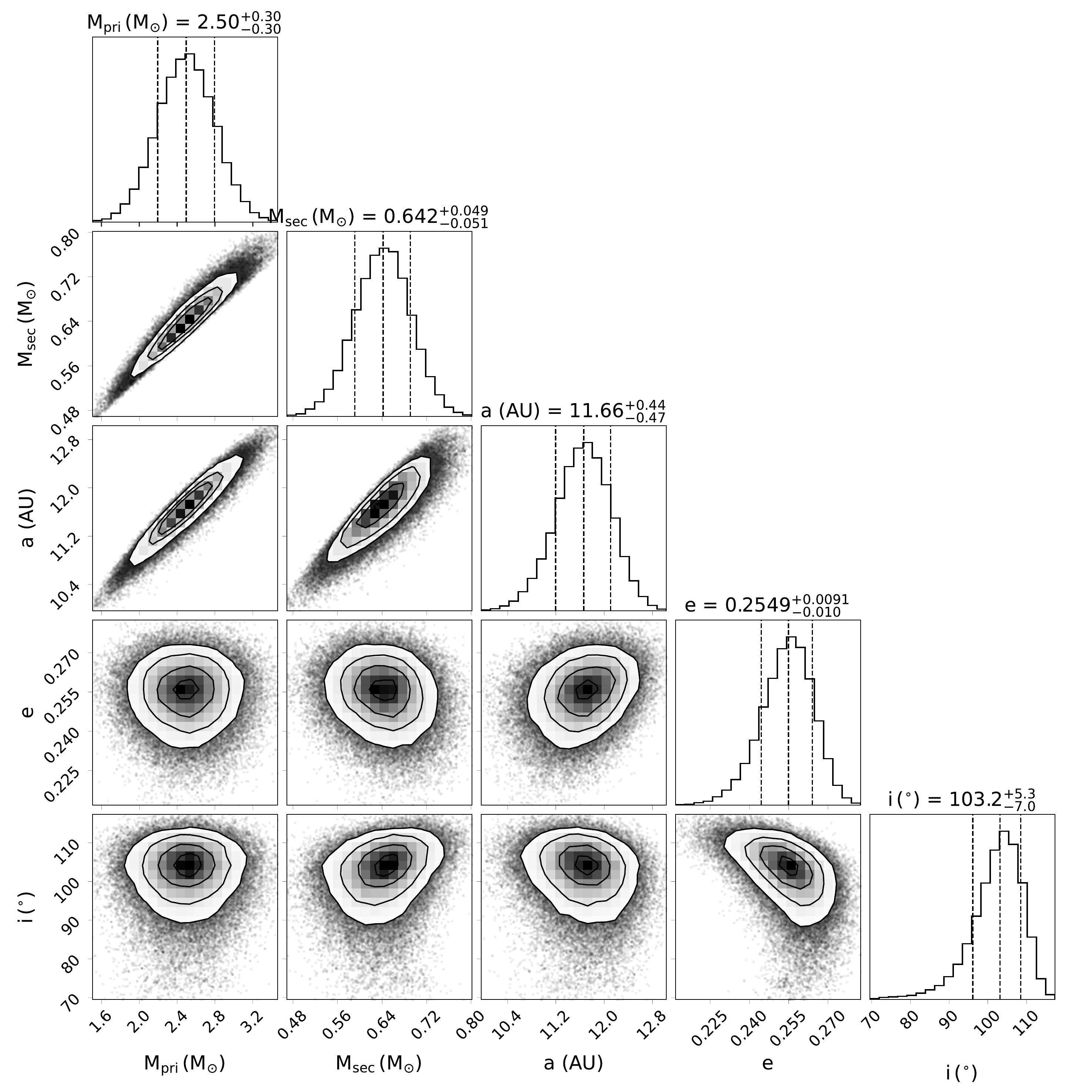}
\caption{\label{Fig:HD53199corner} Corner plot of HD\,53199}
\end{minipage} 
\hspace{3cm} 
\begin{minipage}[r]{6cm} 
\includegraphics[scale=0.3]{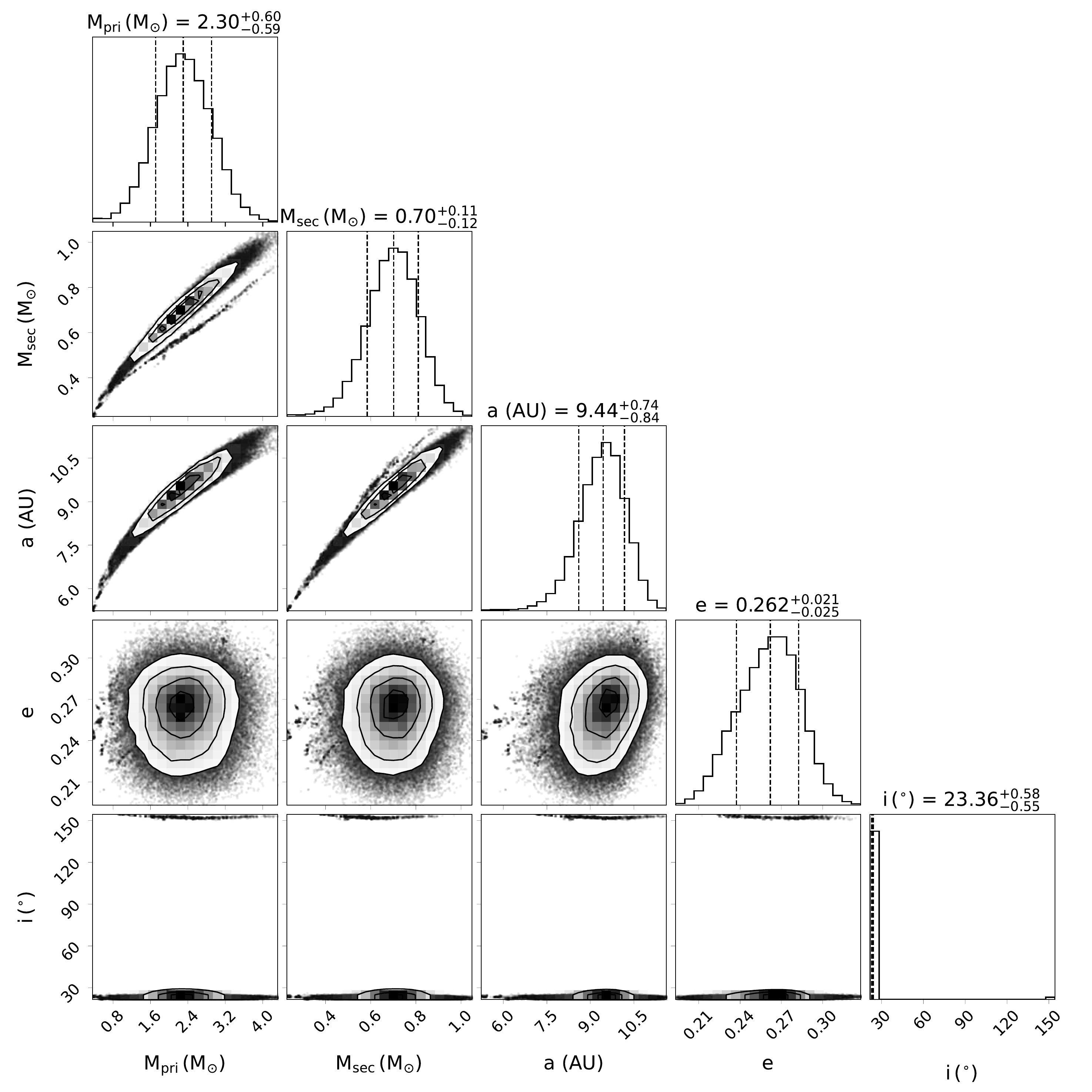}
\caption{\label{Fig:HD40430corner} Corner plot of HD\,40430}
\end{minipage} 
\end{figure} 

\begin{figure*}[t]
\centering
\includegraphics[width=\textwidth]{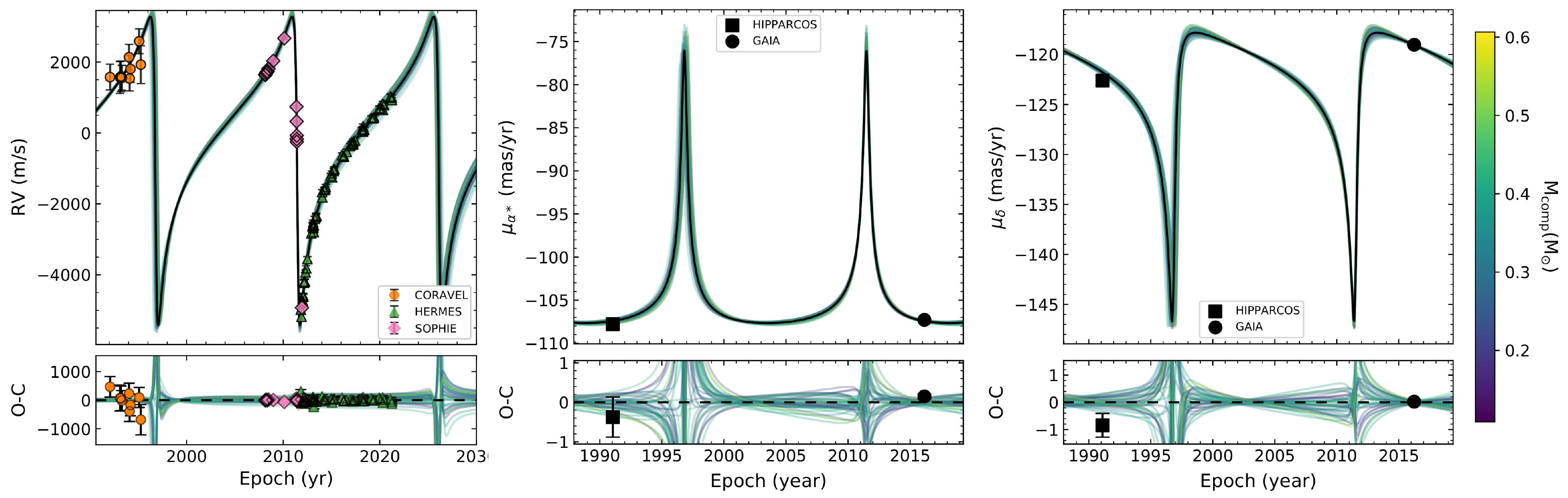}
\caption{\label{Fig:HD95241} RV curve and proper motions of HD\,95241}
\end{figure*}
\begin{figure*}
\centering
\includegraphics[width=\textwidth]{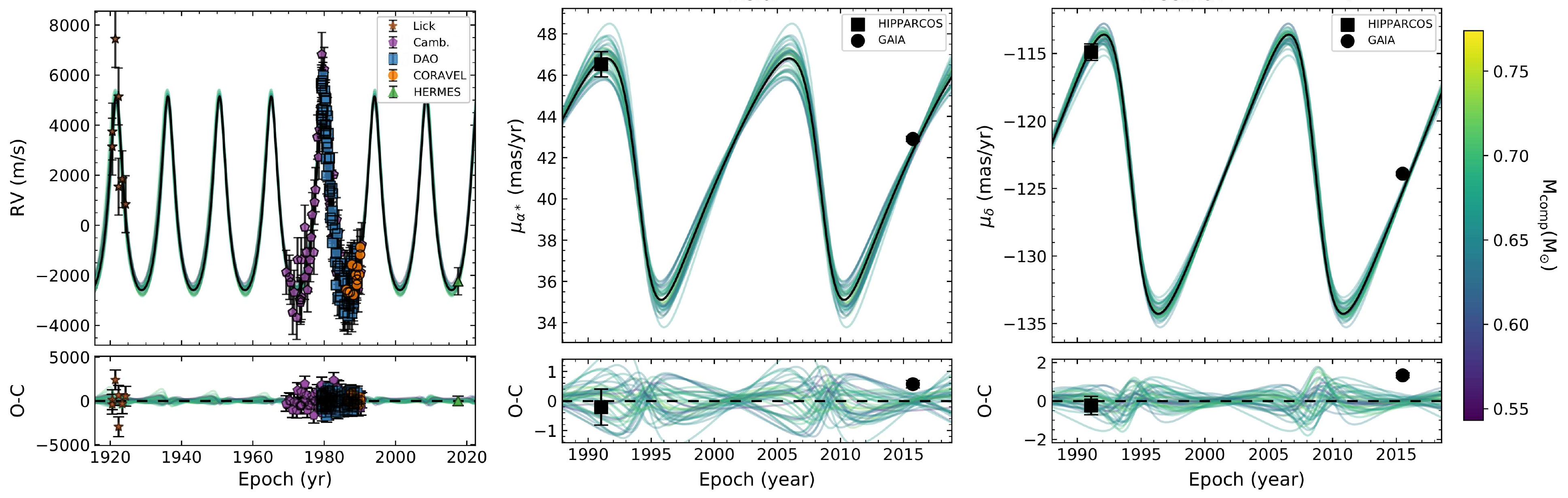}
\caption{\label{Fig:HD139195} RV curve and proper motions of HD\,139195}
\end{figure*}
\vspace{1mm}
\begin{figure}
\begin{minipage}[l]{6cm} 
\includegraphics[scale=0.3]{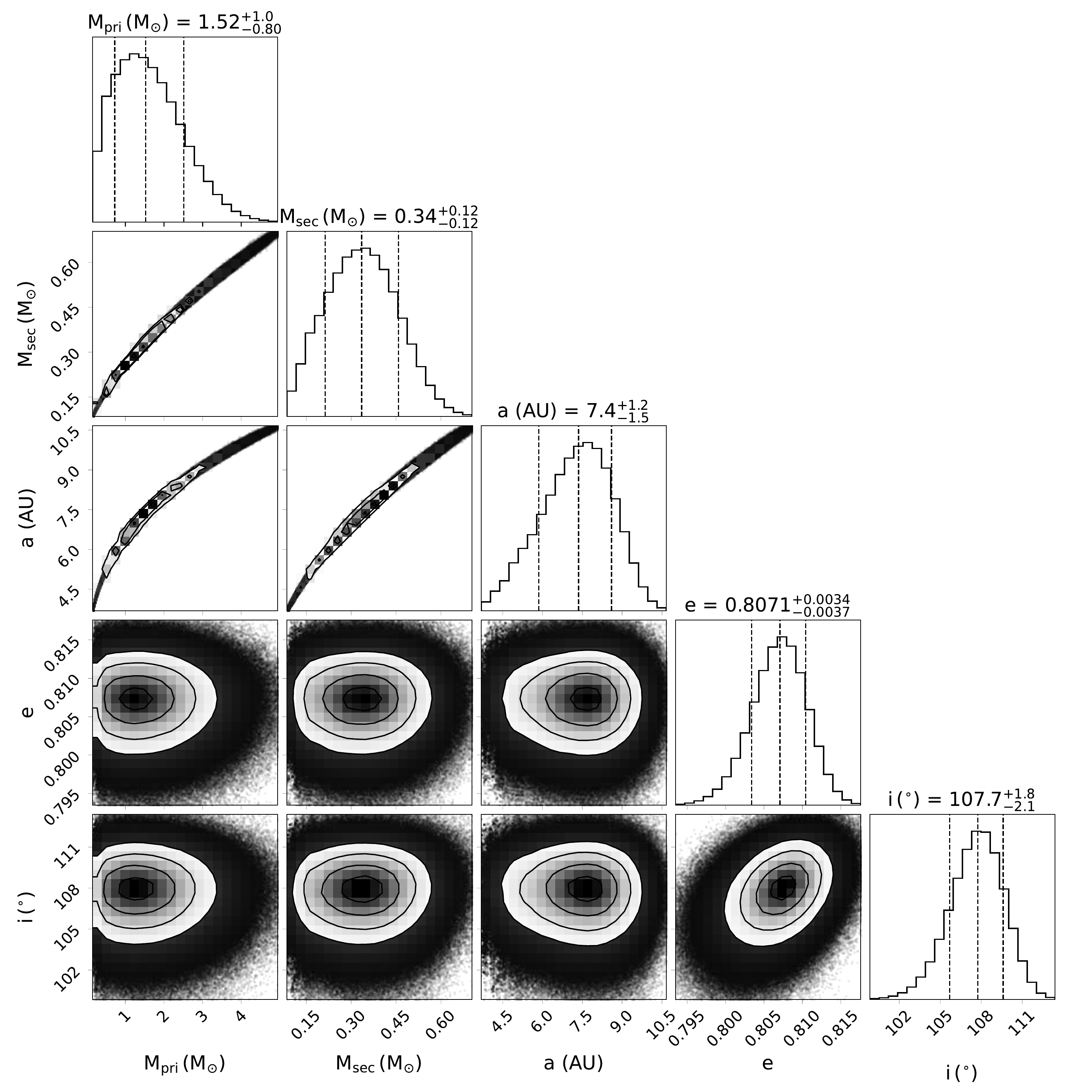}
\caption{\label{Fig:HD95241corner} Corner plot of HD\,95241}
\end{minipage} 
\hspace{3cm} 
\begin{minipage}[r]{6cm} 
\includegraphics[scale=0.3]{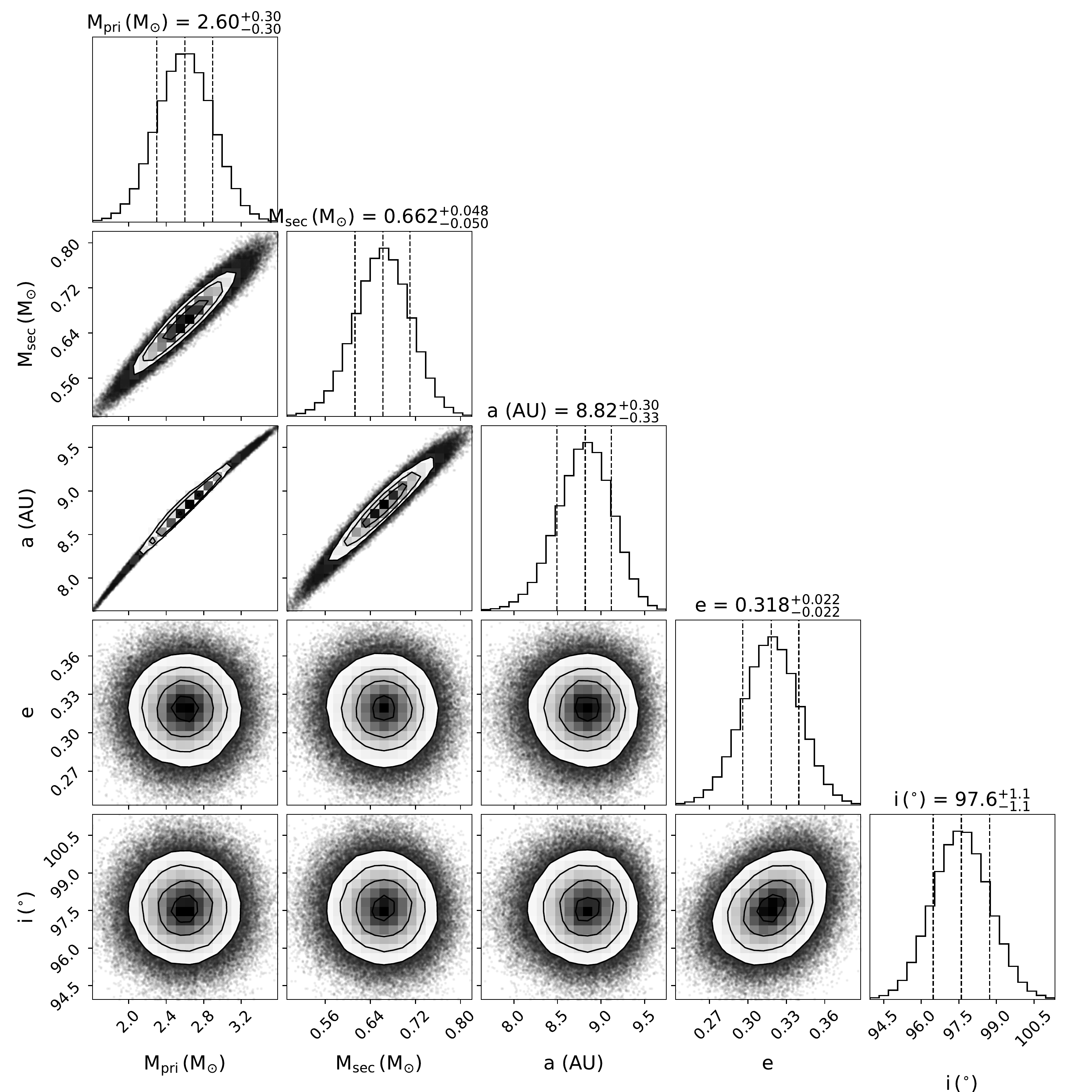}
\caption{\label{Fig:HD139195corner} Corner plot of HD\,139195}
\end{minipage} 
\end{figure}

\begin{figure*}[t]
\centering
\includegraphics[width=\textwidth]{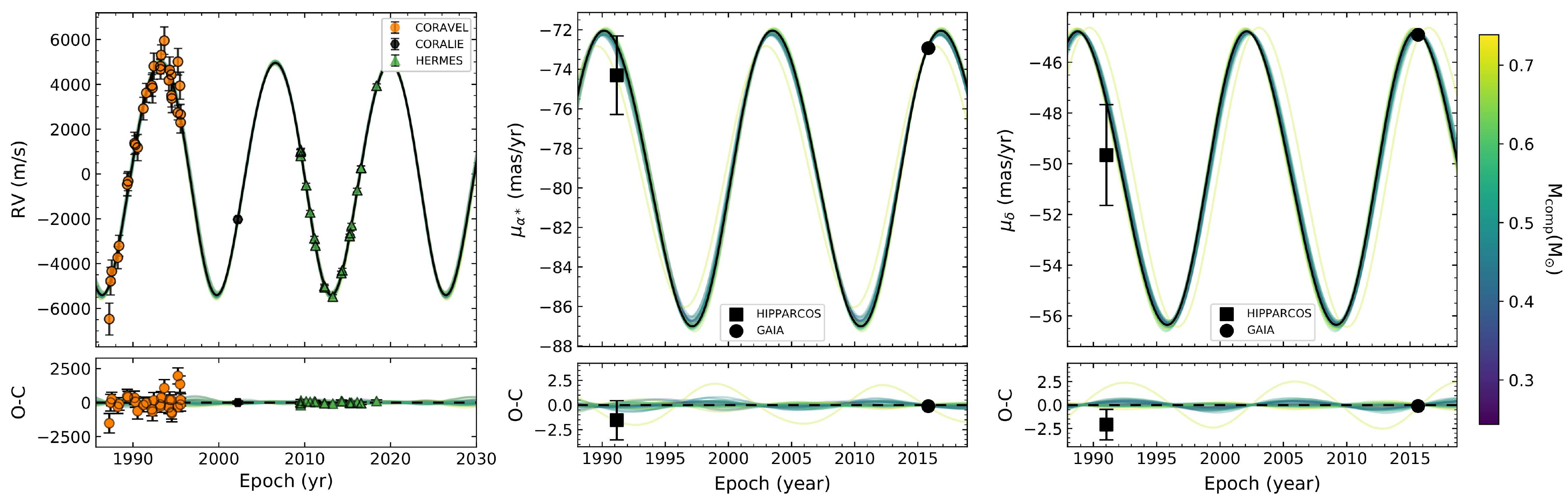}
\caption{\label{Fig:BD-10_4311} RV curve and proper motions of BD-10$^{\rm o}$4311}
\end{figure*}
\begin{figure*}
\centering
\includegraphics[width=\textwidth]{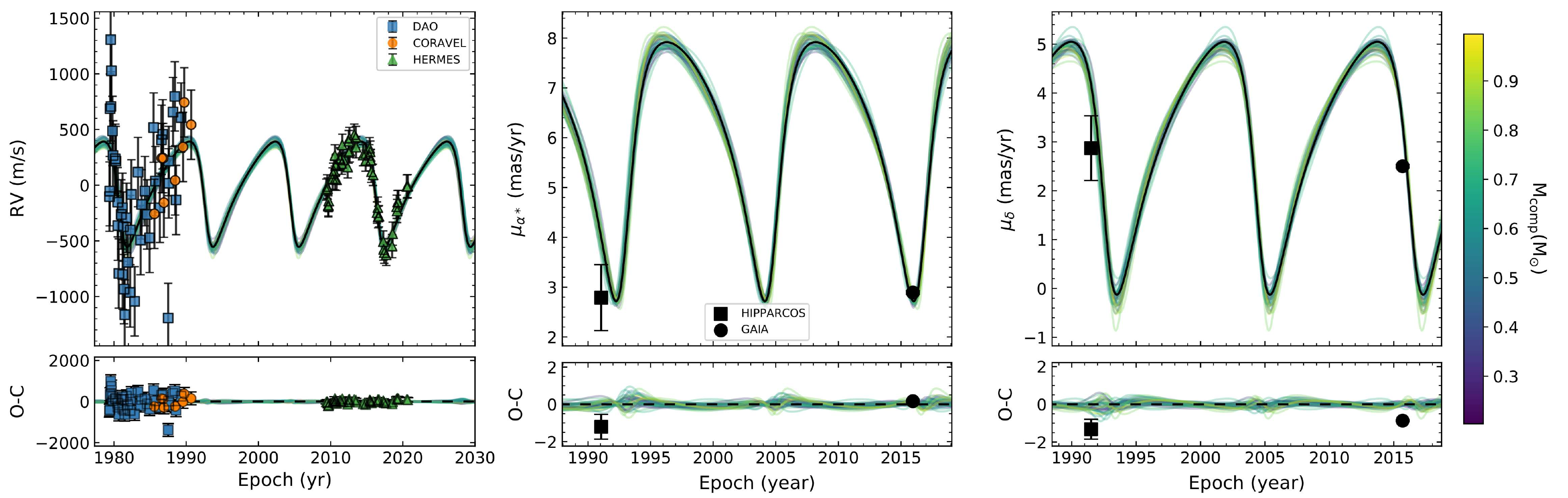}
\caption{\label{Fig:HD183915} RV curve and proper motions of HD\,183915}
\end{figure*}
\vspace{1mm}
\begin{figure}
\begin{minipage}[l]{6cm} 
\includegraphics[scale=0.3]{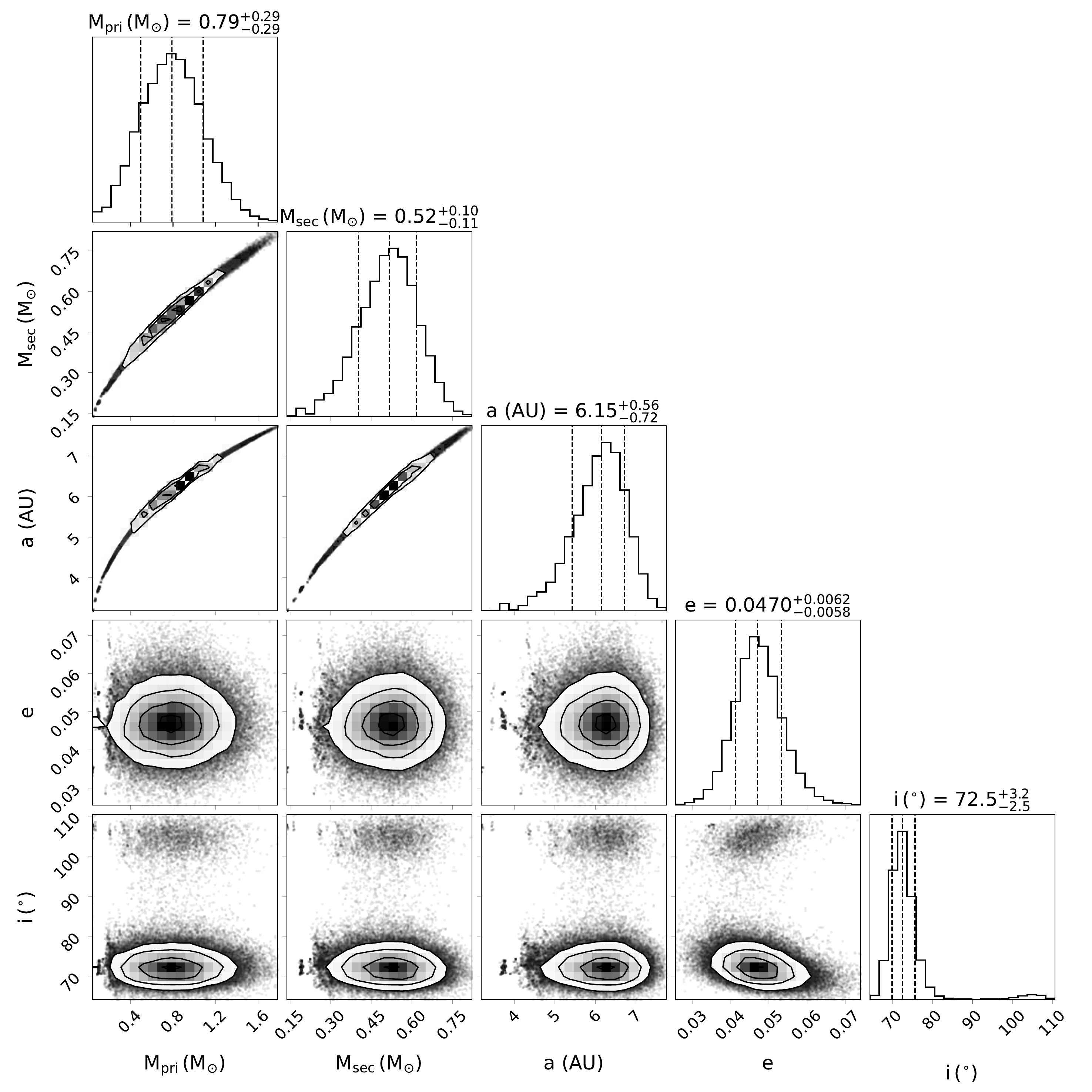}
\caption{\label{Fig:BD-10_4311corner} Corner plot of BD-10$^{\rm o}$4311}
\end{minipage} 
\hspace{3cm} 
\begin{minipage}[r]{6cm} 
\includegraphics[scale=0.3]{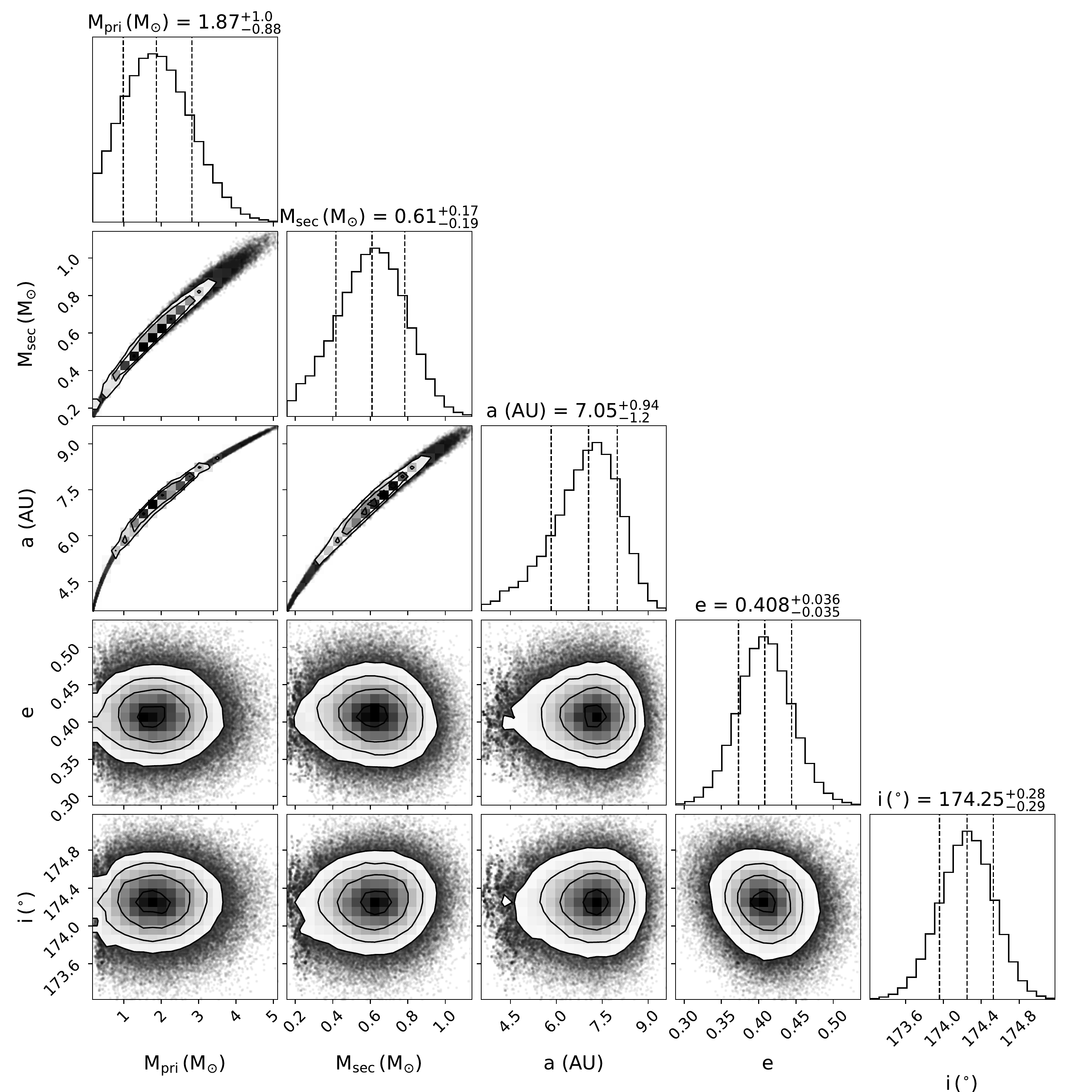}
\caption{\label{Fig:HD183915corner} Corner plot of HD\,183915}
\end{minipage} 
\end{figure}

\begin{figure*}[t]
\centering
\includegraphics[width=\textwidth]{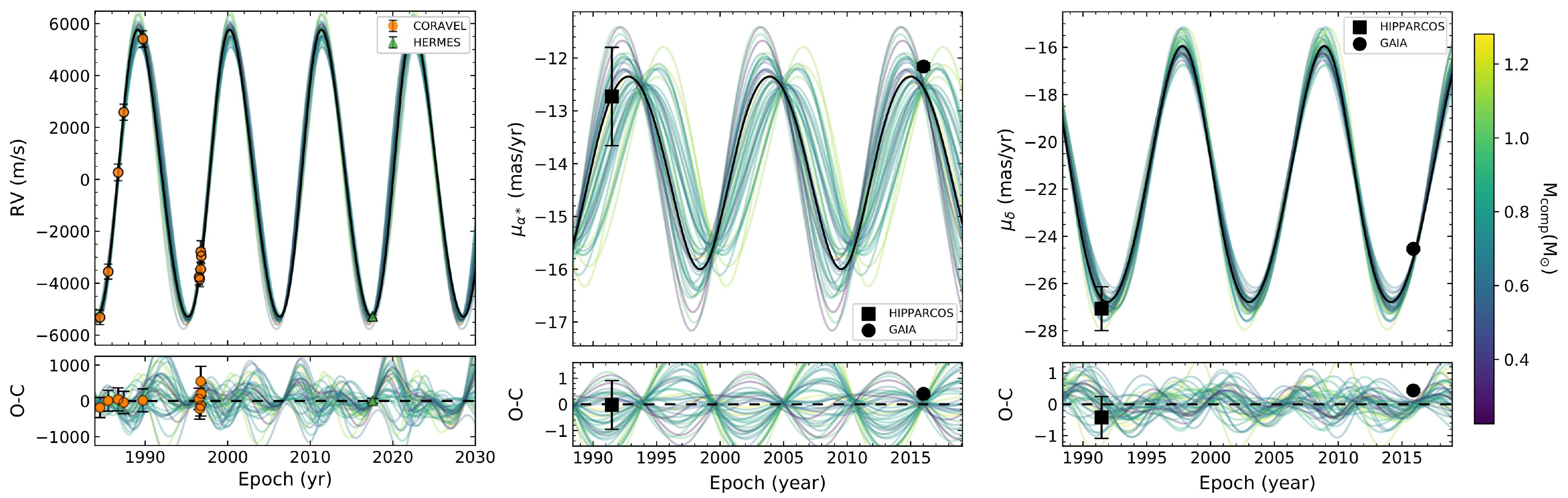}
\caption{\label{Fig:HD180622} RV curve and proper motions of HD\,180622. We used a fixed RV offset of 500 m/s \citep{Jorissen19}.}
\end{figure*}
\begin{figure*}
\centering
\includegraphics[width=\textwidth]{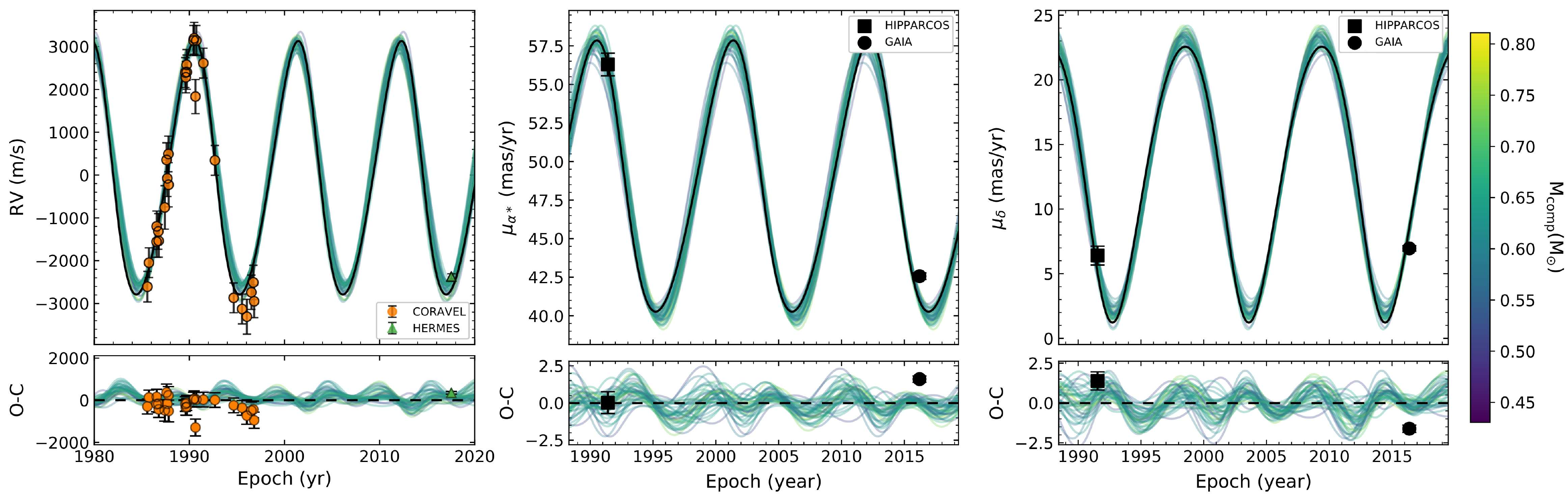}
\caption{\label{Fig:HD216219}RV curve and proper motions of HD\,216219. We used a fixed HERMES-CORAVEL RV offset of 186 m/s \cite{Escorza19}}
\end{figure*}
\vspace{1mm}
\begin{figure}
\begin{minipage}[l]{6cm} 
\includegraphics[scale=0.3]{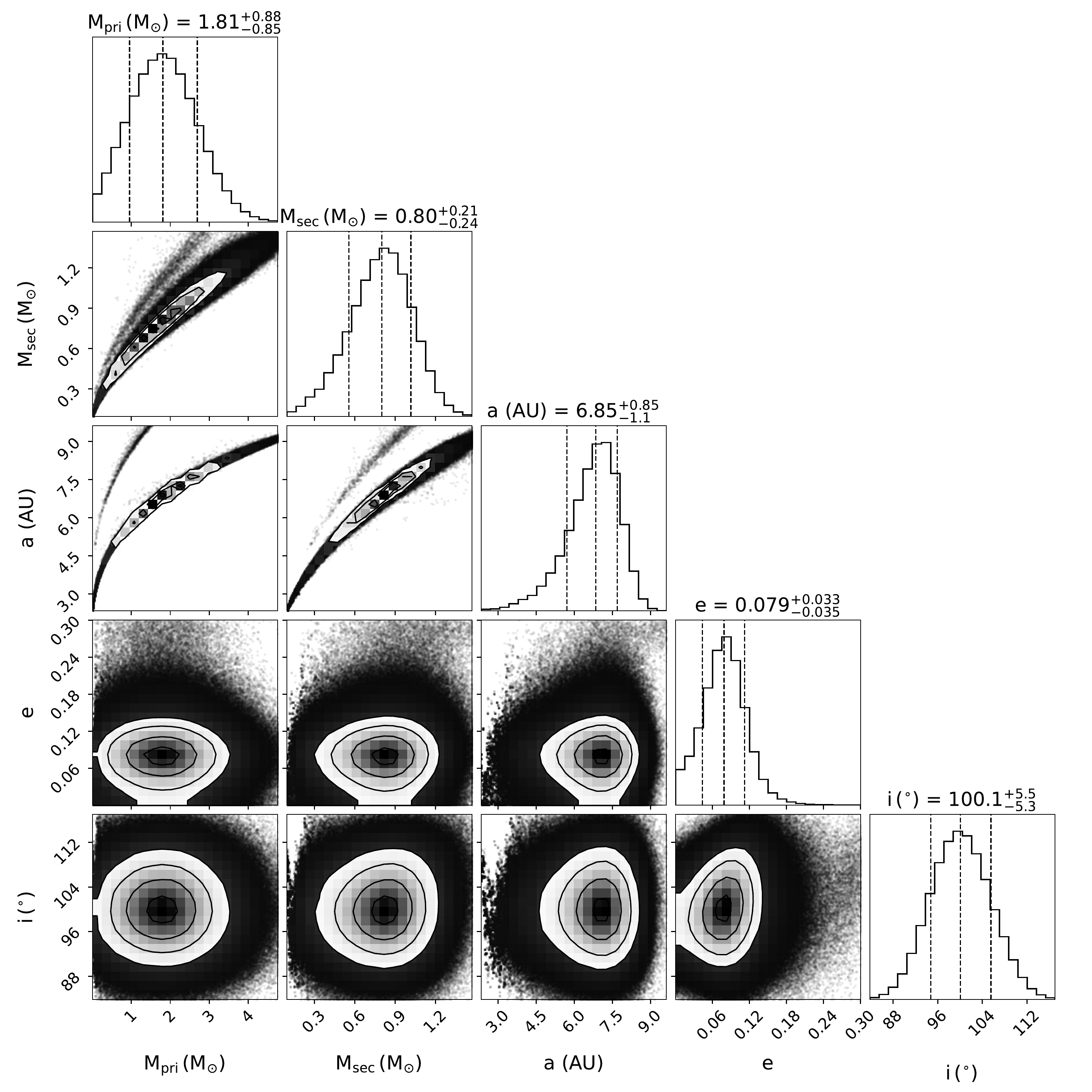}
\caption{\label{Fig:HD180622corner} Corner plot of HD\,180622}
\end{minipage} 
\hspace{3cm} 
\begin{minipage}[r]{6cm} 
\includegraphics[scale=0.3]{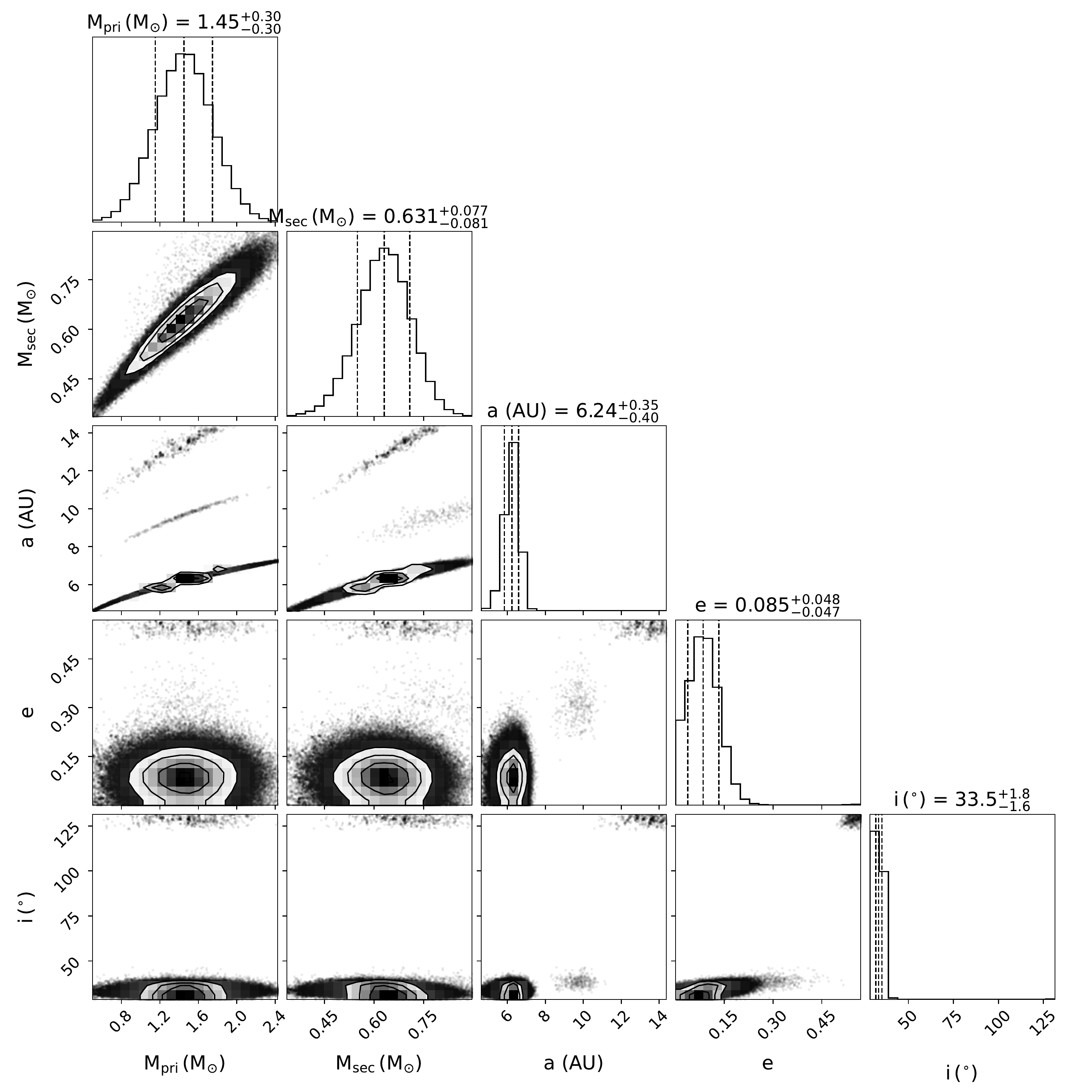}
\caption{\label{Fig:HD216219corner} Corner plot of HD\,216219}
\end{minipage} 
\end{figure}

\begin{figure*}[t]
\centering
\includegraphics[width=\textwidth]{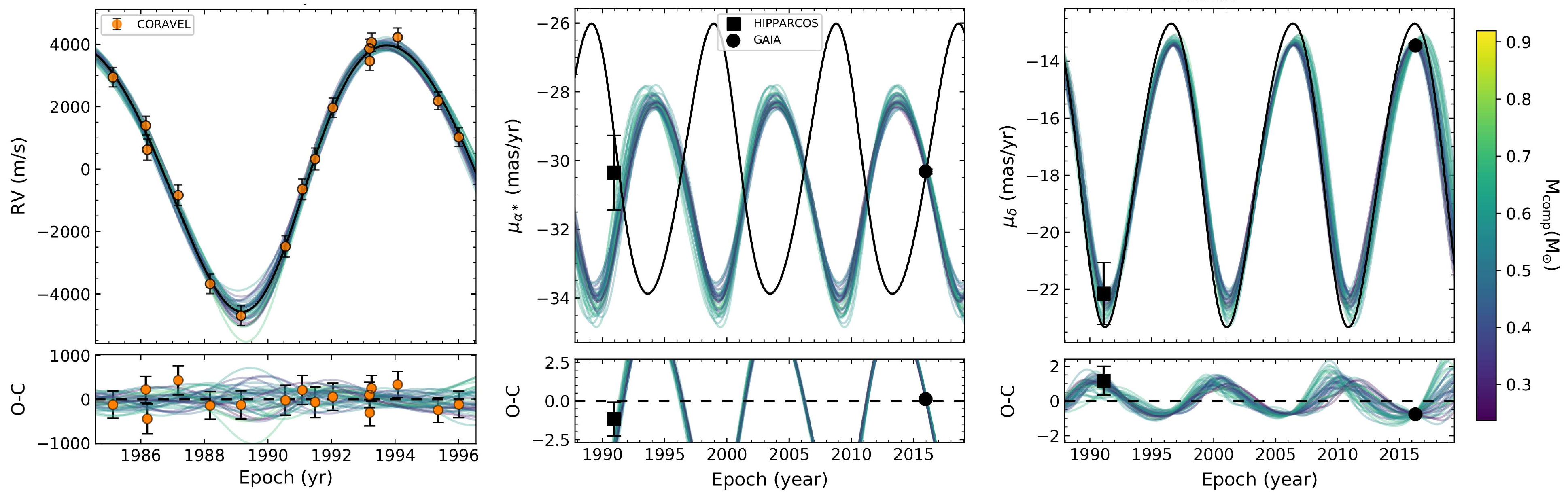}
\caption{\label{Fig:HD107541} RV curve and proper motions of HD\,107541}
\end{figure*}
\begin{figure*}
\centering
\includegraphics[width=\textwidth]{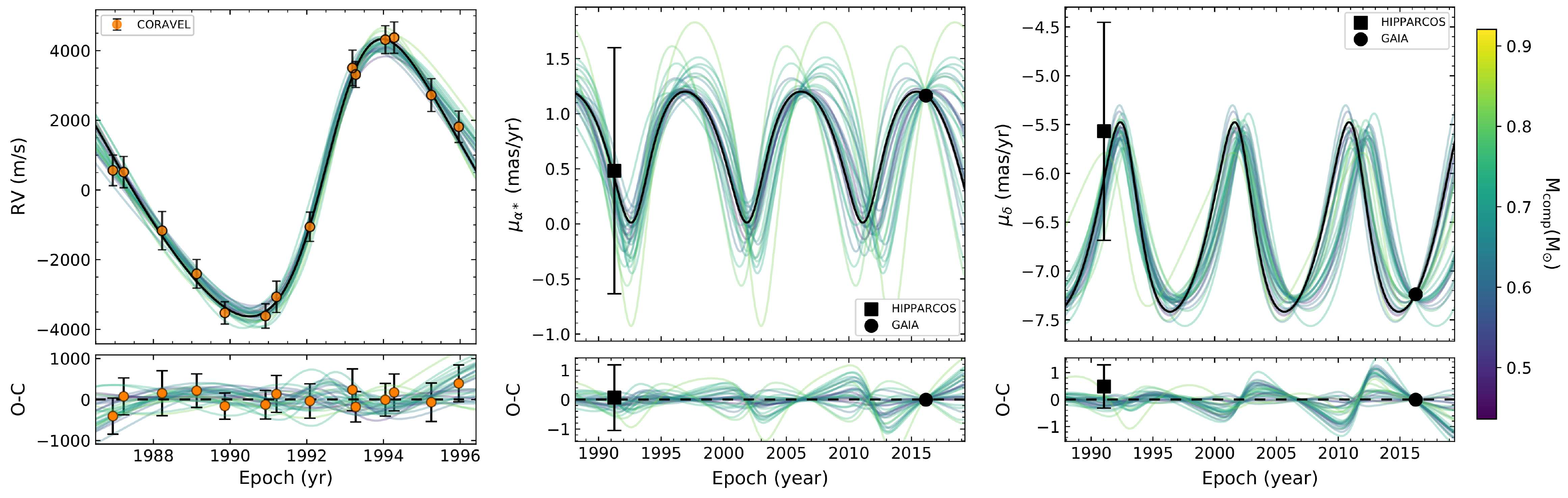}
\caption{\label{Fig:BD-14_2678} RV curve and proper motions of BD-14$^{\rm o}$2678}
\end{figure*}
\vspace{1mm}
\begin{figure}
\begin{minipage}[l]{6cm} 
\includegraphics[scale=0.3]{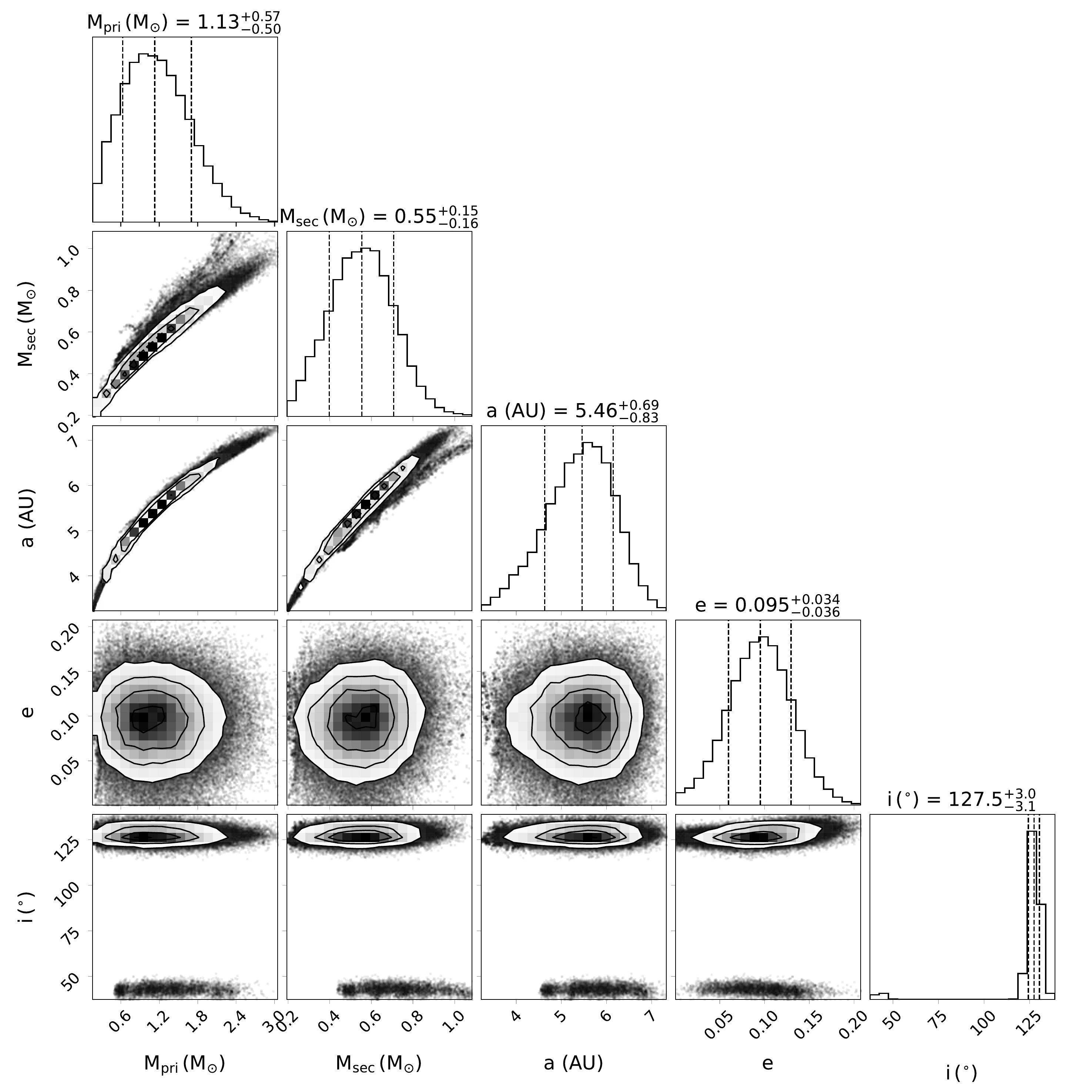}
\caption{\label{Fig:HD107541corner} Corner plot of HD\,107541}
\end{minipage} 
\hspace{3cm} 
\begin{minipage}[r]{6cm} 
\includegraphics[scale=0.3]{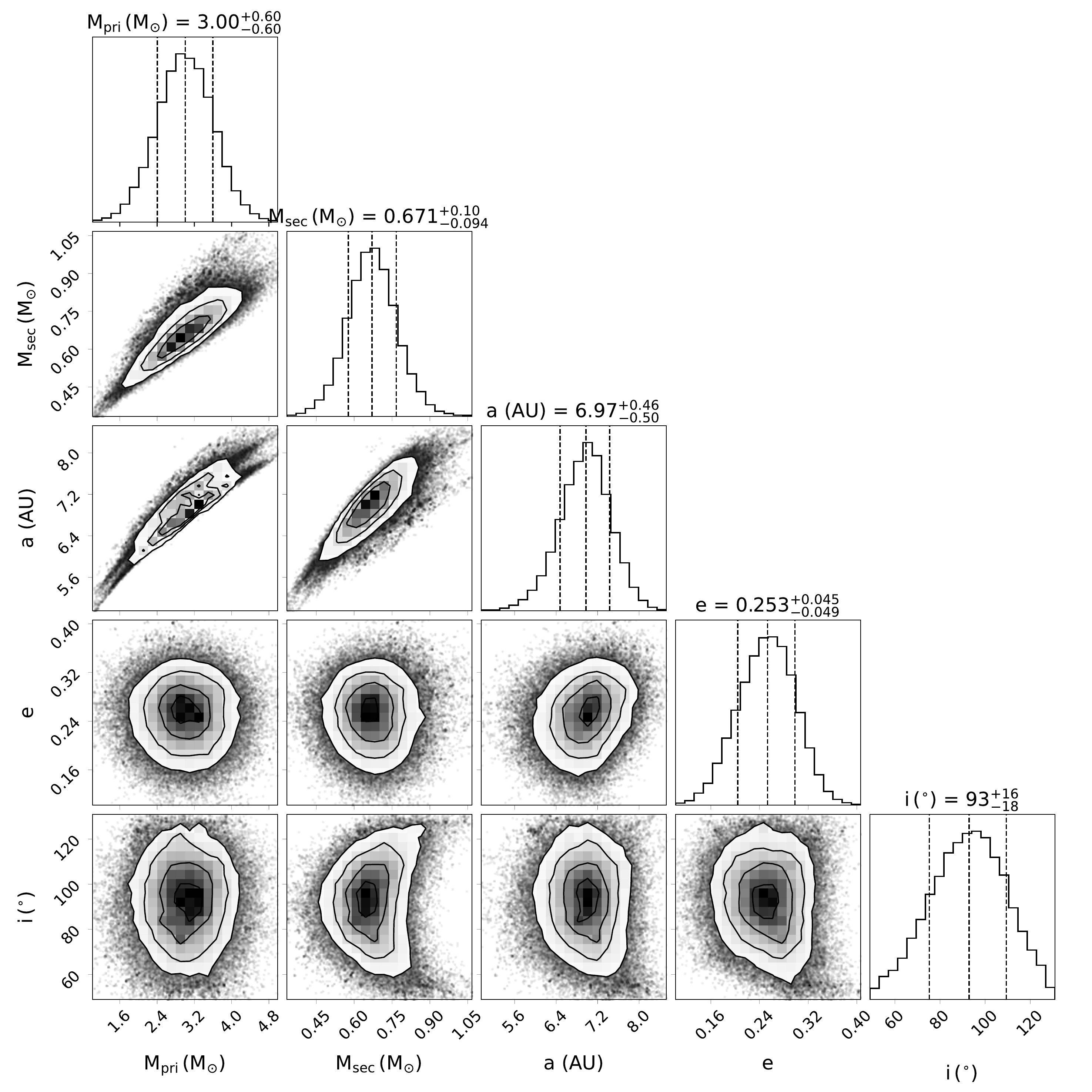}
\caption{\label{Fig:BD-14_2678corner} Corner plot of BD-14$^{\rm o}$2678}
\end{minipage} 
\end{figure}

\begin{figure*}[t]
\centering
\includegraphics[width=\textwidth]{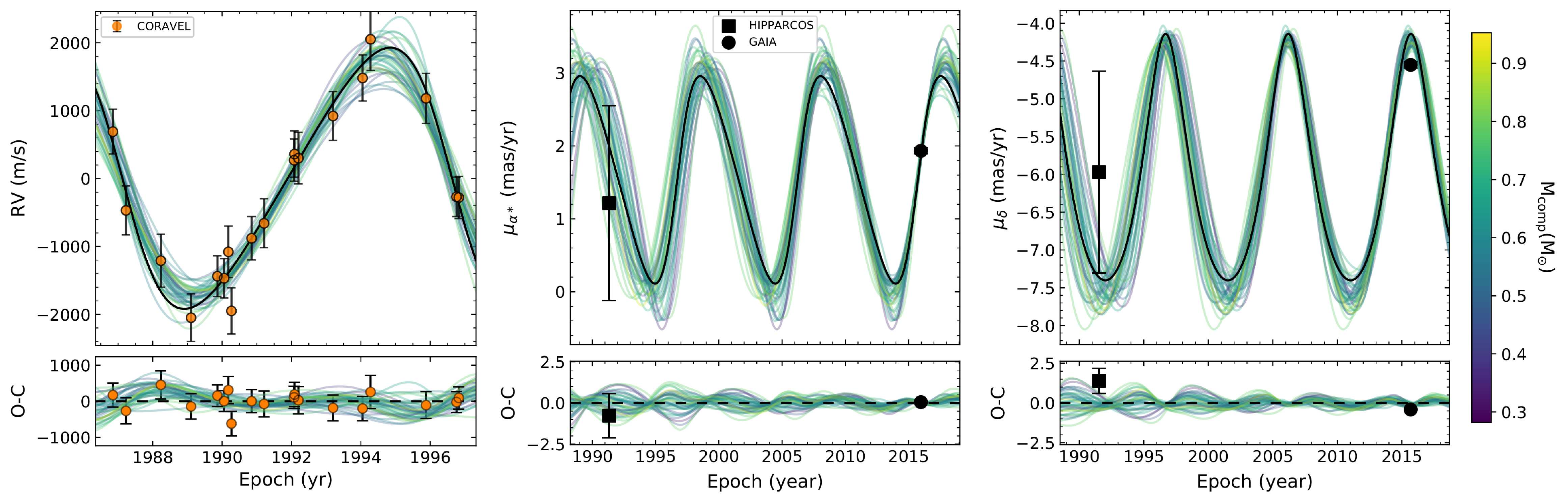}
\caption{\label{Fig:HD59852} RV curve and proper motions of HD\,59852}
\end{figure*}
\begin{figure*}
\centering
\includegraphics[width=1.0\textwidth]{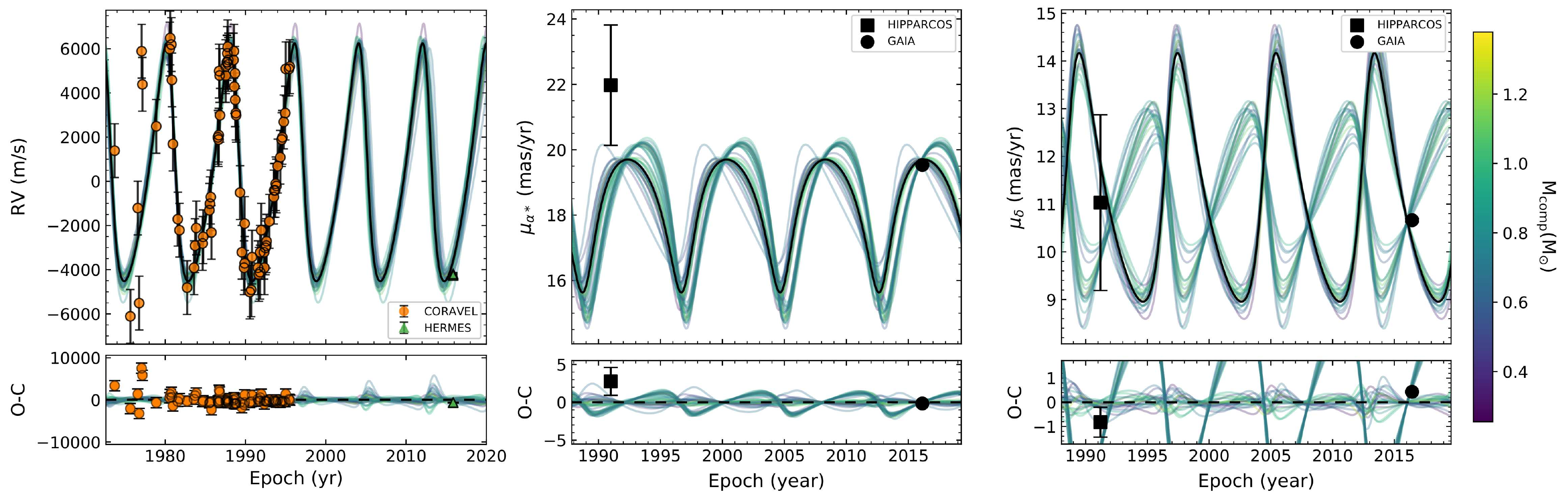}
\caption{\label{Fig:HD201824} RV curve and proper motions of HD\,201824. We used a fixed RV offset of 500 m/s \citep{Jorissen19}.}
\end{figure*}
\vspace{1mm}
\begin{figure}
\begin{minipage}[l]{6cm} 
\includegraphics[scale=0.3]{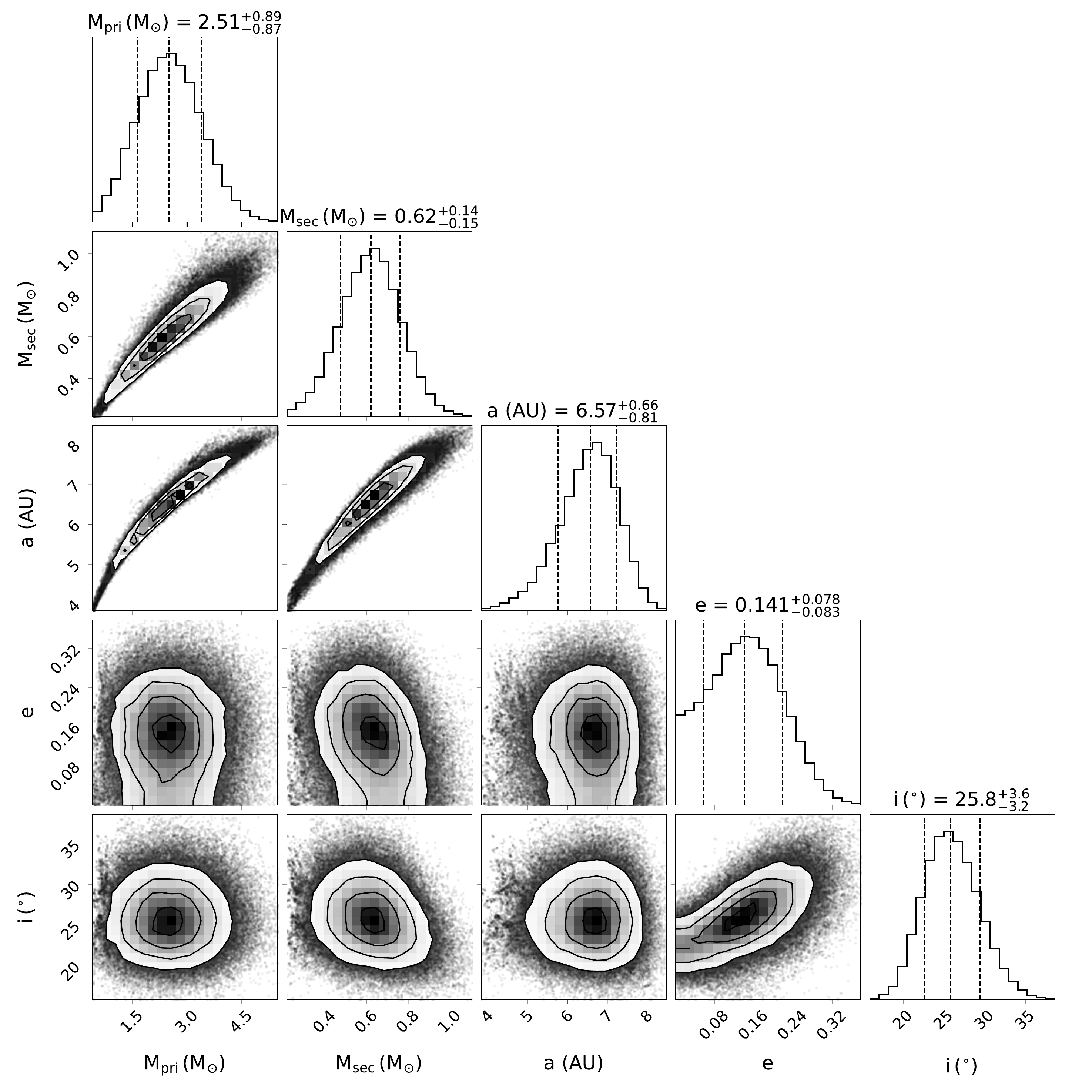}
\caption{\label{Fig:HD59852corner} Corner plot of HD\,59852}
\end{minipage} 
\hspace{3cm} 
\begin{minipage}[r]{6cm} 
\includegraphics[scale=0.3]{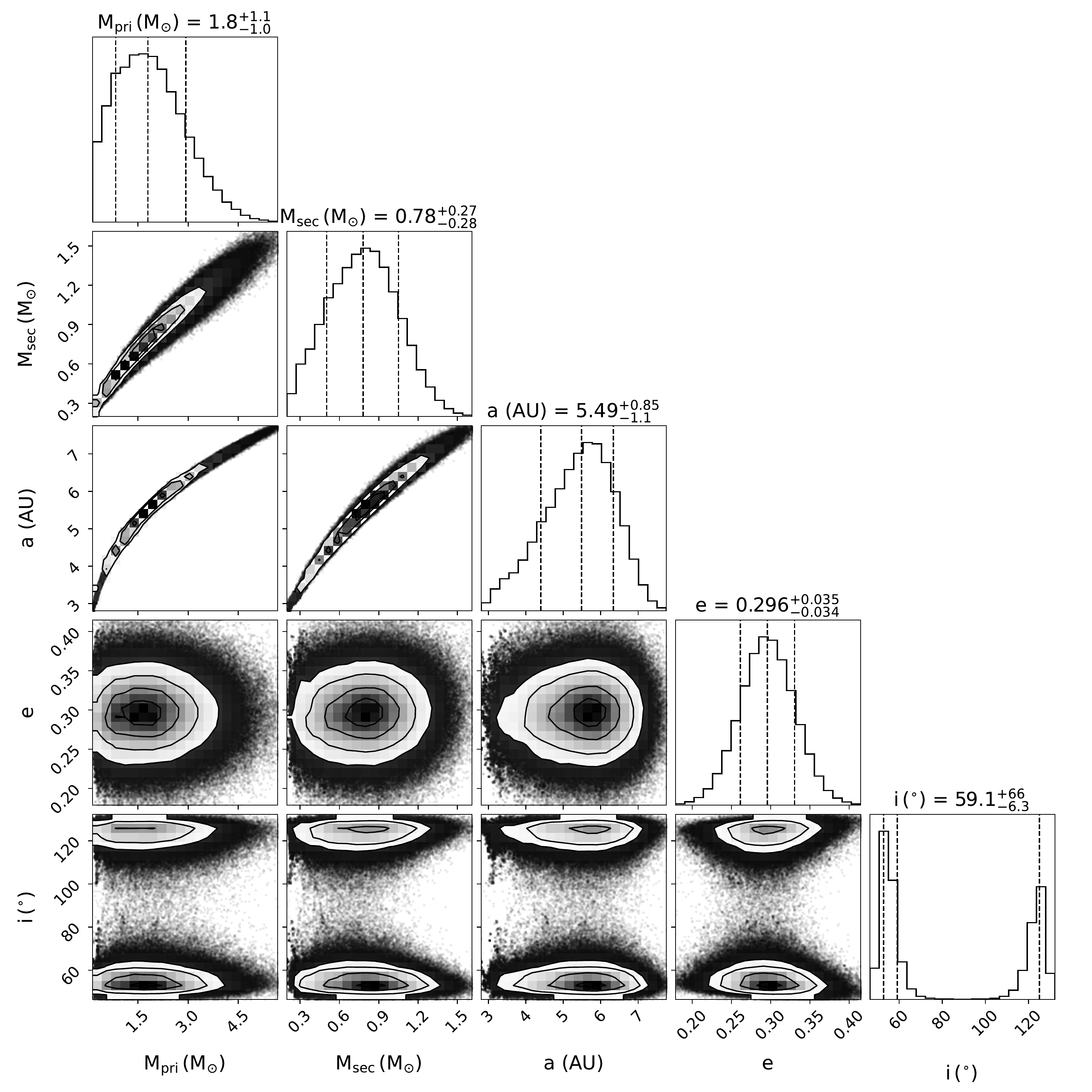}
\caption{\label{Fig:HD201824corner} Corner plot of HD\,201824}
\end{minipage} 
\end{figure}

\begin{figure*}[t]
\centering
\includegraphics[width=1.0\textwidth]{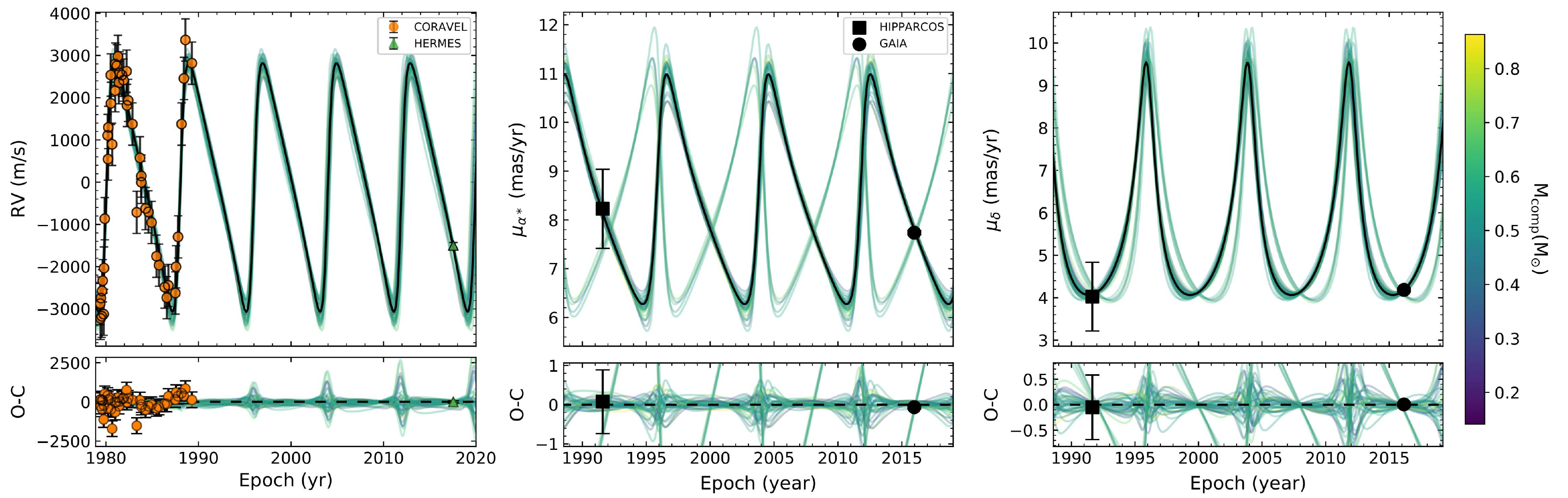}
\caption{\label{Fig:HD178717} RV curve and proper motions of HD\,178717. We used a fixed RV offset of 500 m/s \citep{Jorissen19}.}
\end{figure*}
\begin{figure*}
\centering
\includegraphics[width=1.0\textwidth]{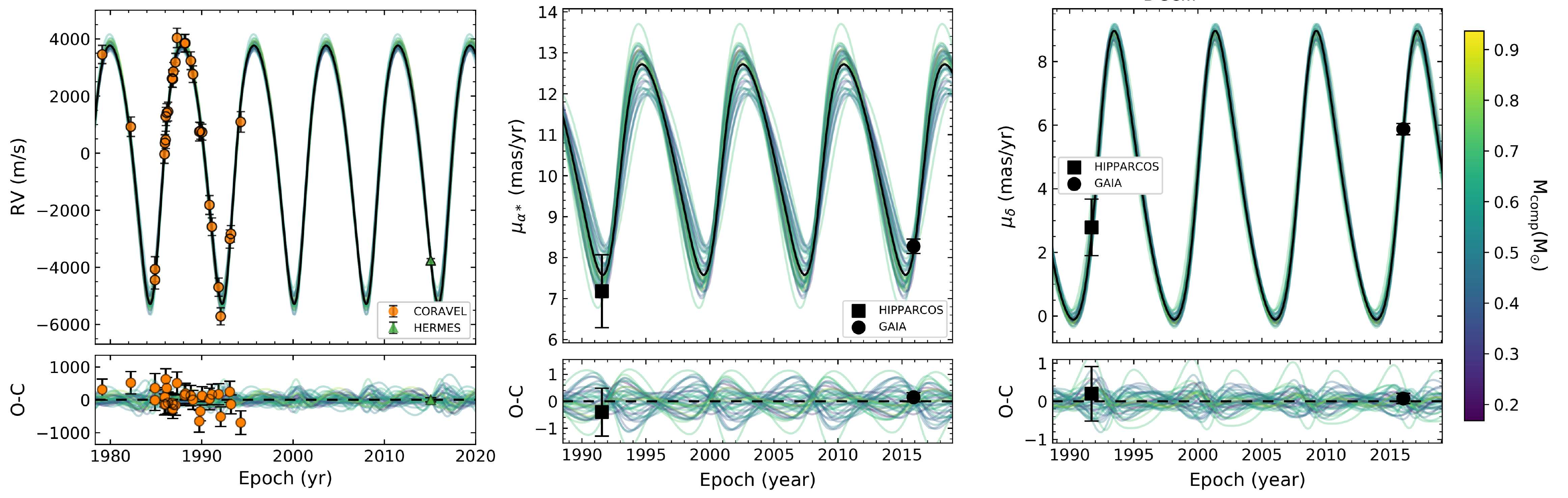}
\caption{\label{Fig:HD50082} RV curve and proper motions of HD\,50082. We used a fixed RV offset of 500 m/s \citep{Jorissen19}.}
\end{figure*}
\vspace{1mm}
\begin{figure}
\begin{minipage}[l]{6cm} 
\includegraphics[scale=0.3]{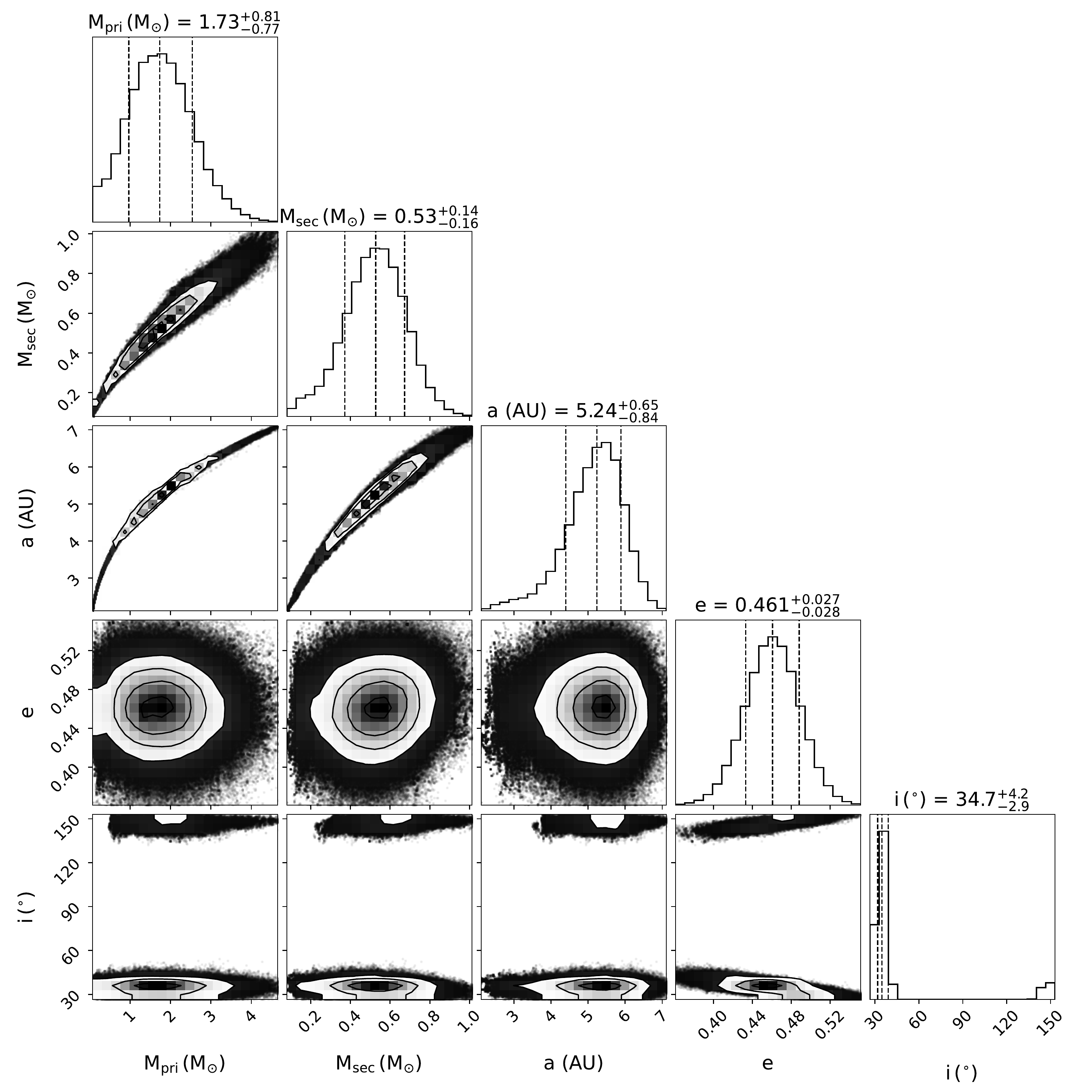}
\caption{\label{Fig:HD178717corner} Corner plot of HD\,178717}
\end{minipage} 
\hspace{3cm} 
\begin{minipage}[r]{6cm} 
\includegraphics[scale=0.3]{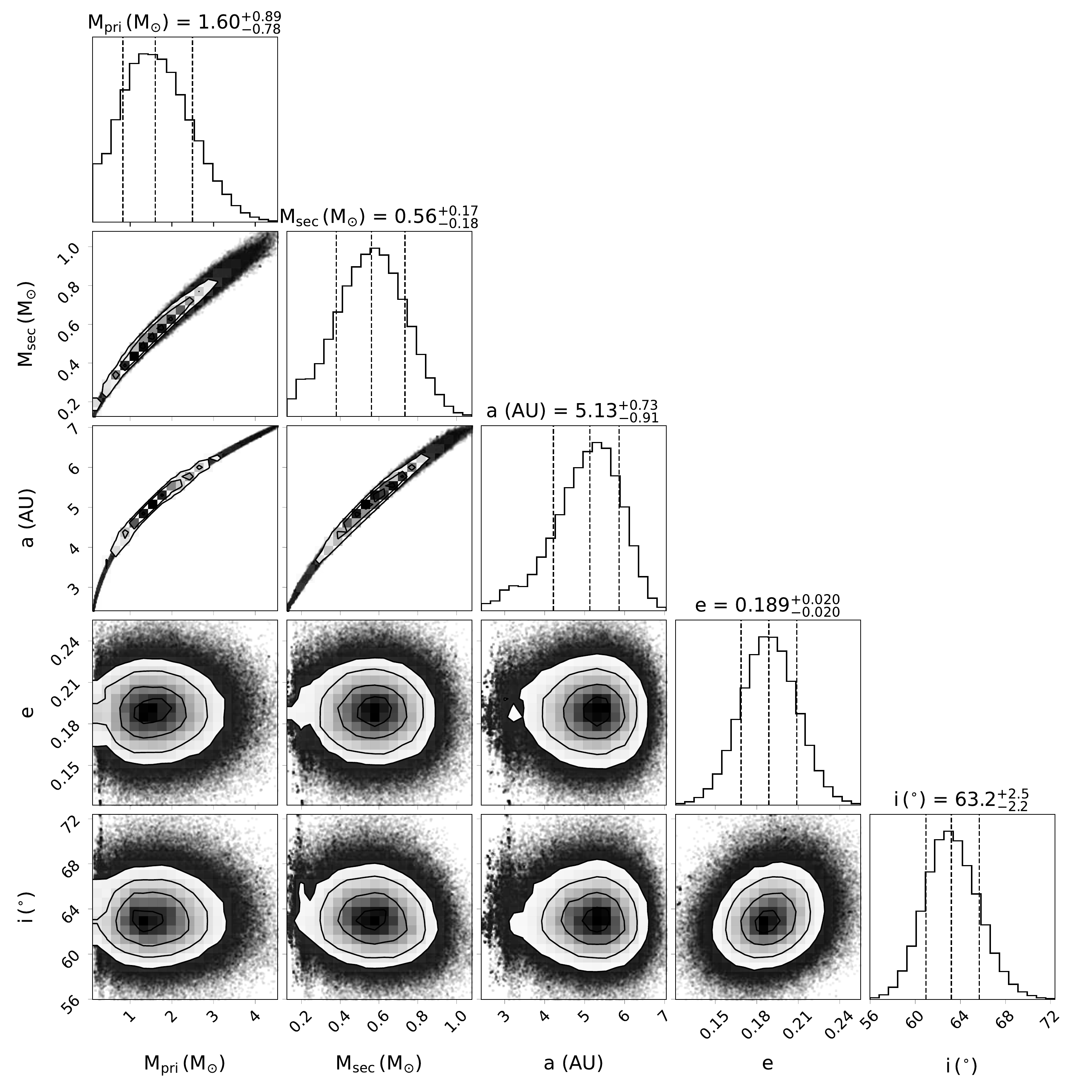}
\caption{\label{Fig:HD50082corner} Corner plot of HD\,50082}
\end{minipage} 
\end{figure}

\begin{figure*}[t]
\includegraphics[width=\textwidth]{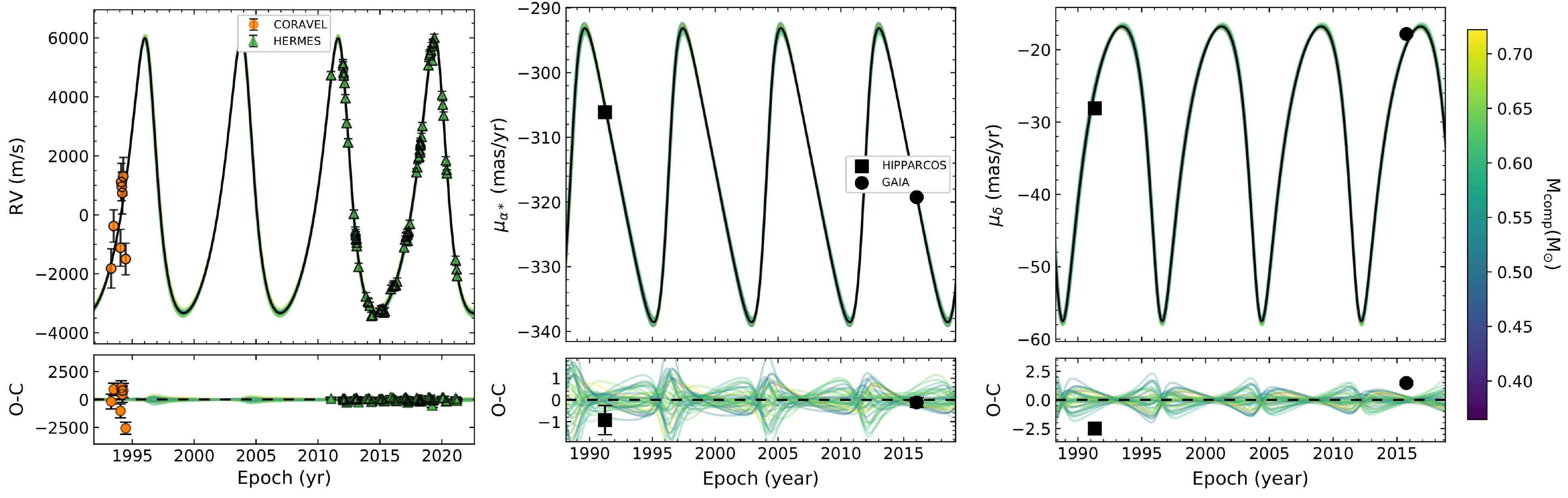}
\caption{\label{Fig:HD98991} RV curve and proper motions of HD\,98991}
\end{figure*}
\begin{figure*}
\includegraphics[width=1.0\textwidth]{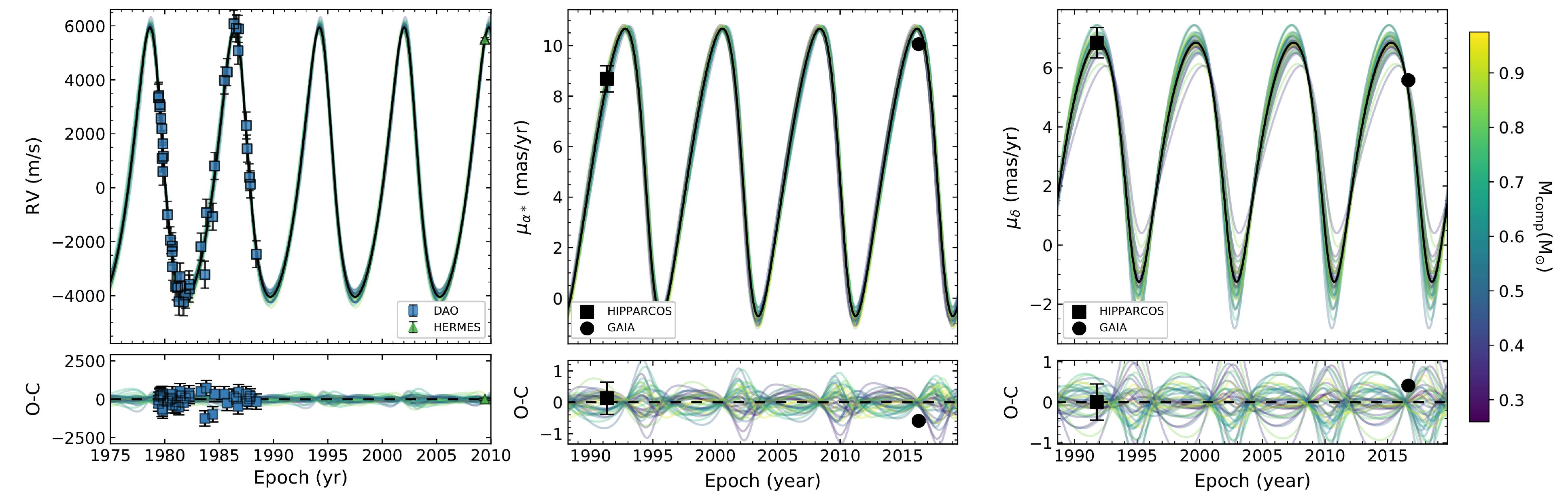}
\caption{\label{Fig:HD205011} RV curve and proper motions of HD\,205011}
\end{figure*}
\begin{figure}
\begin{minipage}[l]{6cm} 
\includegraphics[scale=0.3]{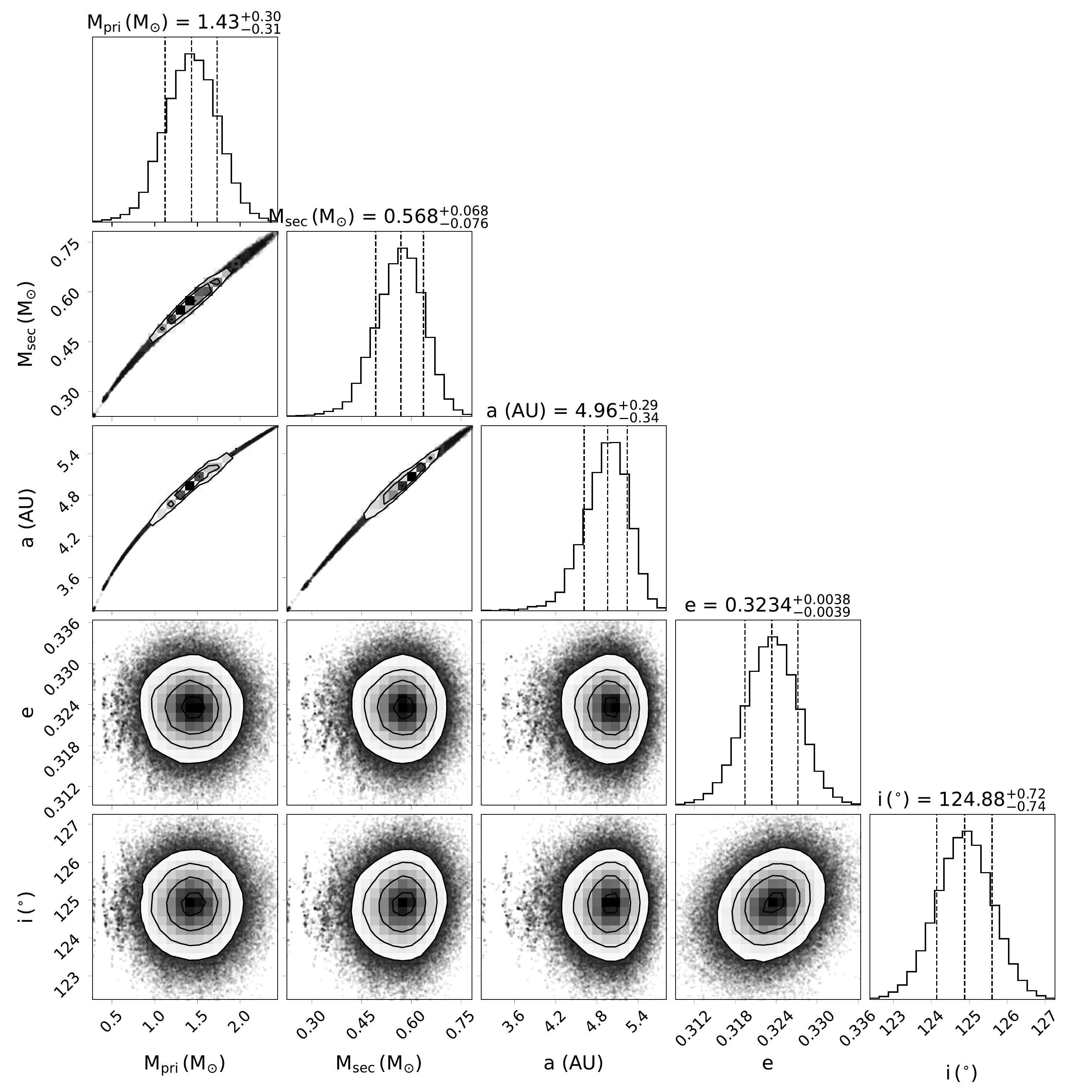}
\caption{\label{Fig:HD98991corner} Corner plot of HD\,98991}
\end{minipage} 
\hspace{3cm} 
\begin{minipage}[r]{6cm} 
\includegraphics[scale=0.3]{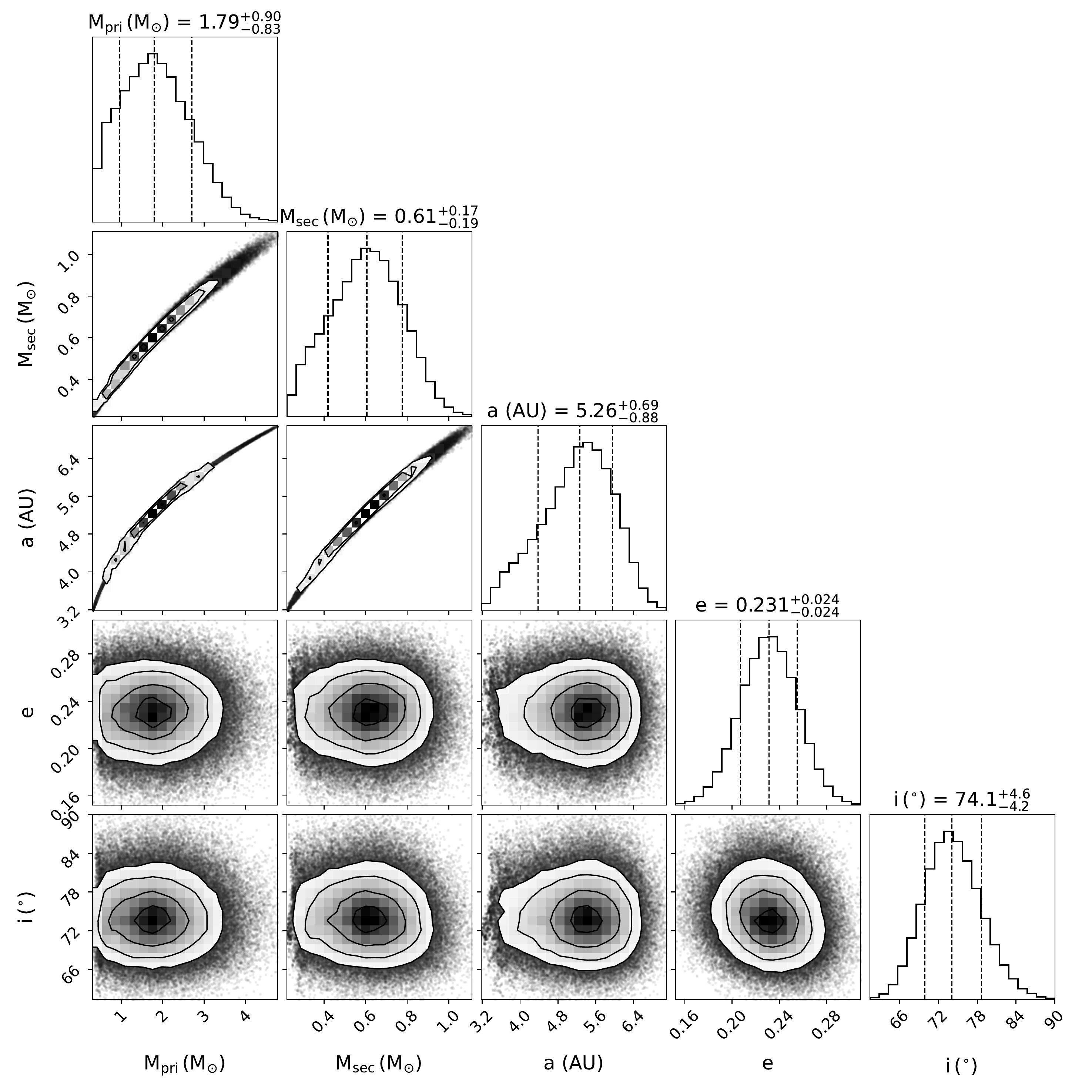}
\caption{\label{Fig:HD205011corner} Corner plot of HD\,205011}
\end{minipage} 
\end{figure}

\begin{figure*}[t]
\includegraphics[width=\textwidth]{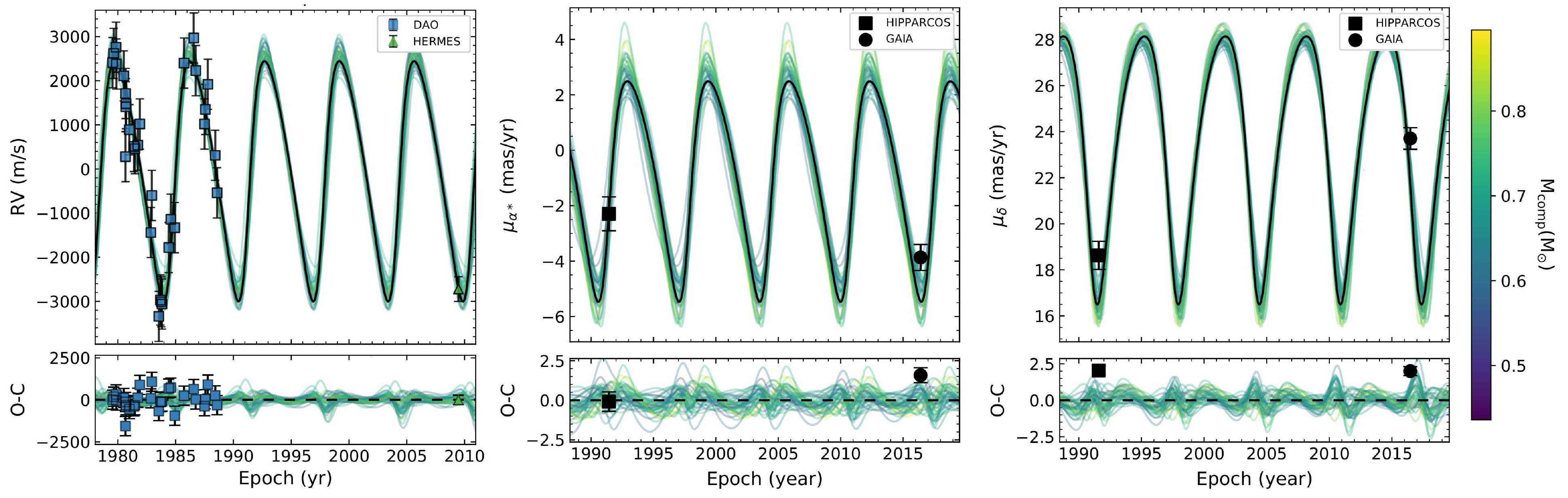}
\caption{\label{Fig:HD204075} RV curve and proper motions of HD\,204075}
\end{figure*}
\begin{figure*}
\includegraphics[width=\textwidth]{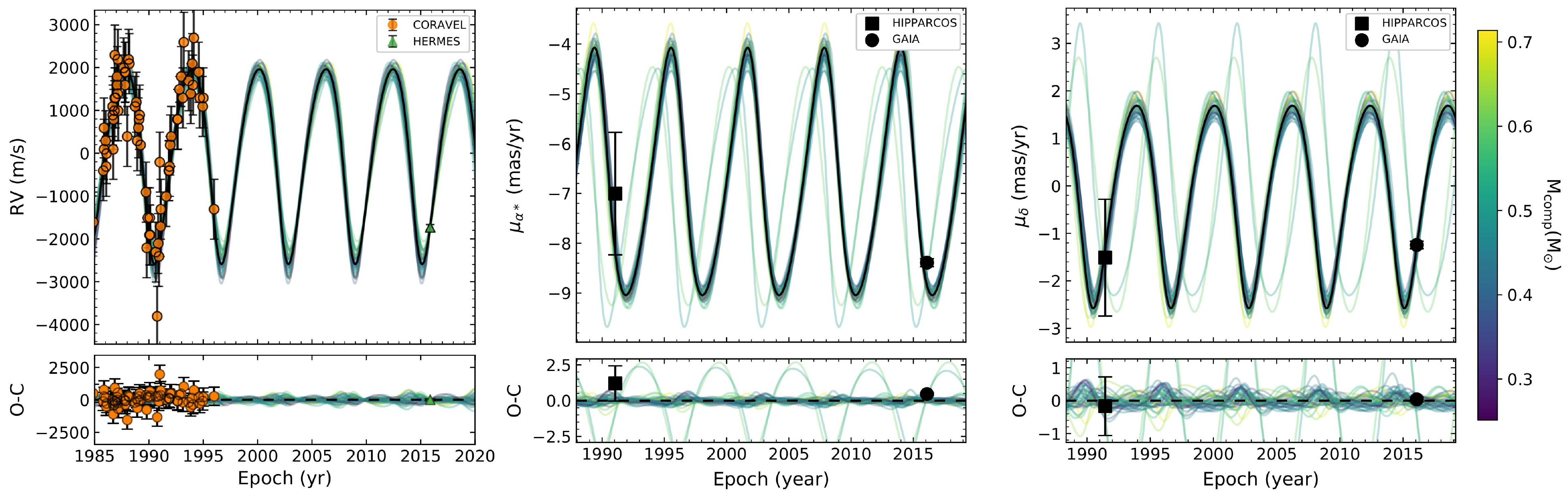}
\caption{\label{Fig:HD20394} RV curve and proper motions of HD\,20394. We used a fixed RV offset of 500 m/s \citep{Jorissen19}.}
\end{figure*}
\begin{figure}
\begin{minipage}[l]{6cm} 
\includegraphics[scale=0.3]{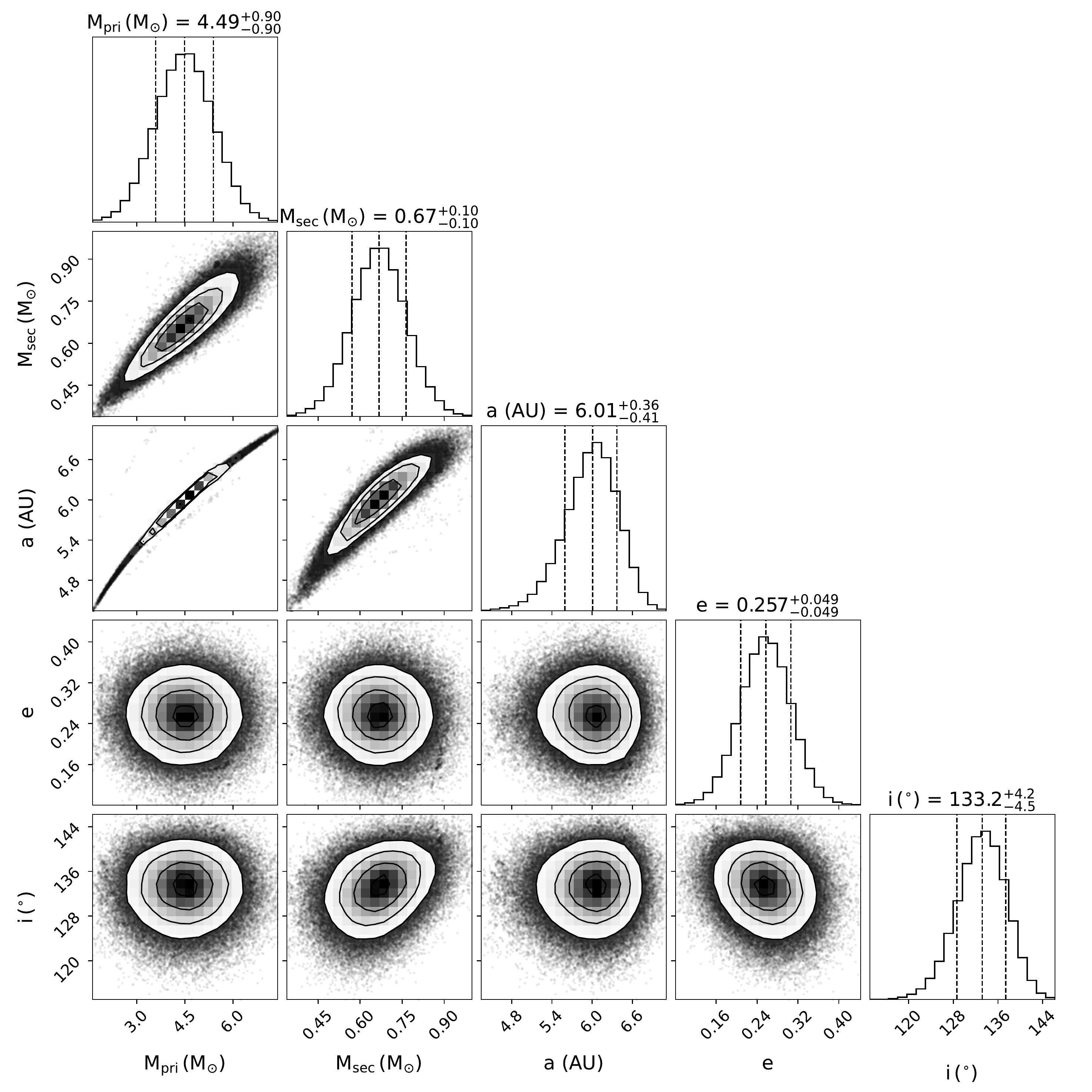}
\caption{\label{Fig:HD204075corner} Corner plot of HD\,204075}
\end{minipage} 
\hspace{3cm} 
\begin{minipage}[r]{6cm} 
\includegraphics[scale=0.3]{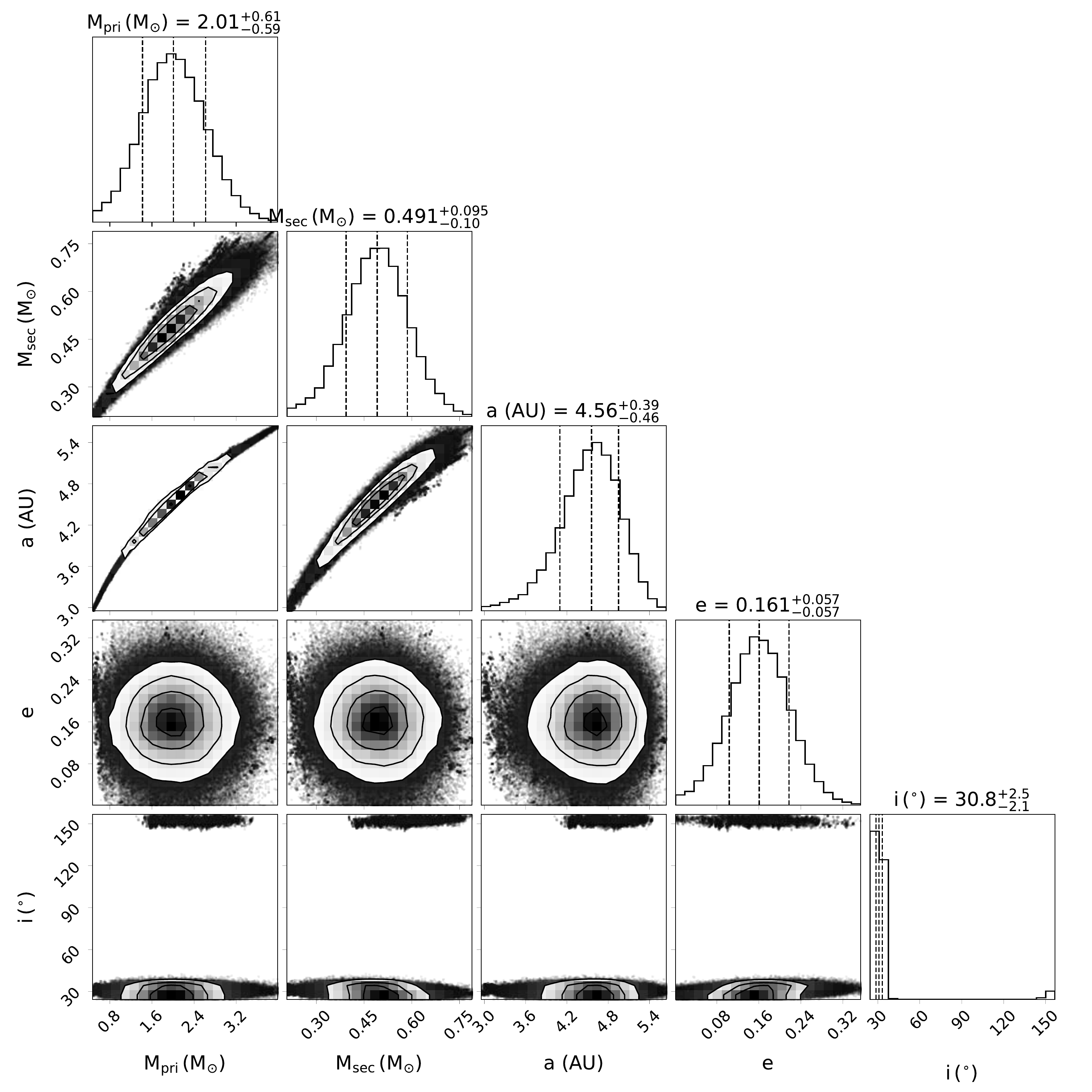}
\caption{\label{Fig:HD20394corner} Corner plot of HD\,20394}
\end{minipage} 
\end{figure}

\begin{figure*}[t]
\centering
\includegraphics[width=\textwidth]{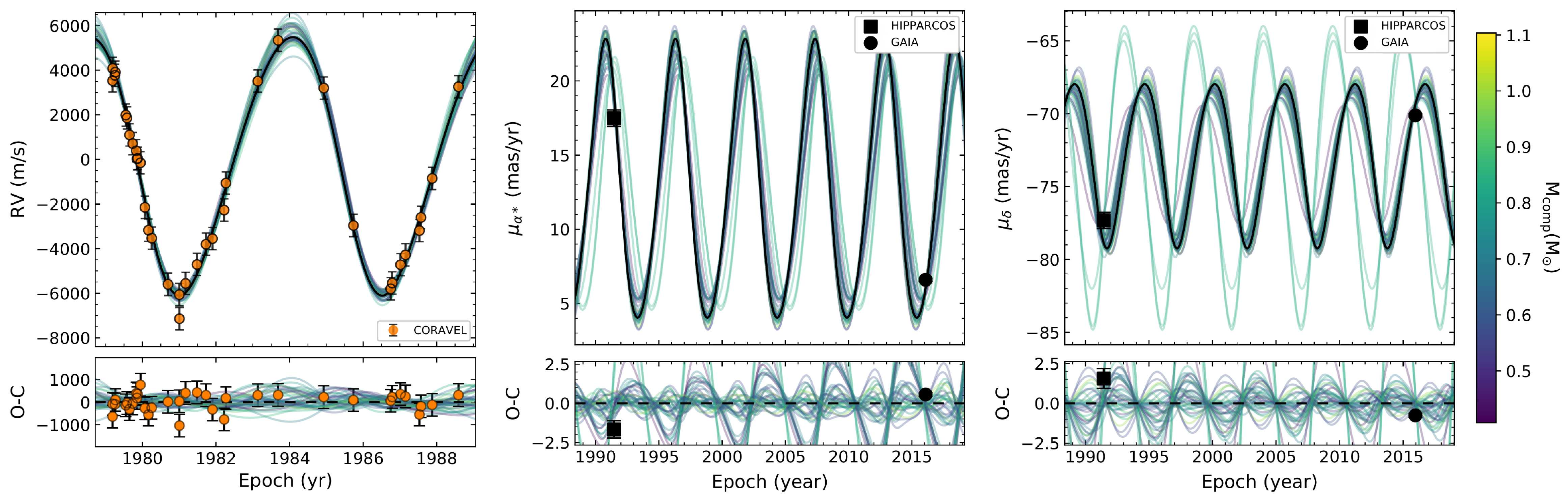}
\caption{\label{Fig:HD16458} RV curve and proper motions of HD\,16458}
\end{figure*}
\begin{figure*}
\centering
\includegraphics[width=\textwidth]{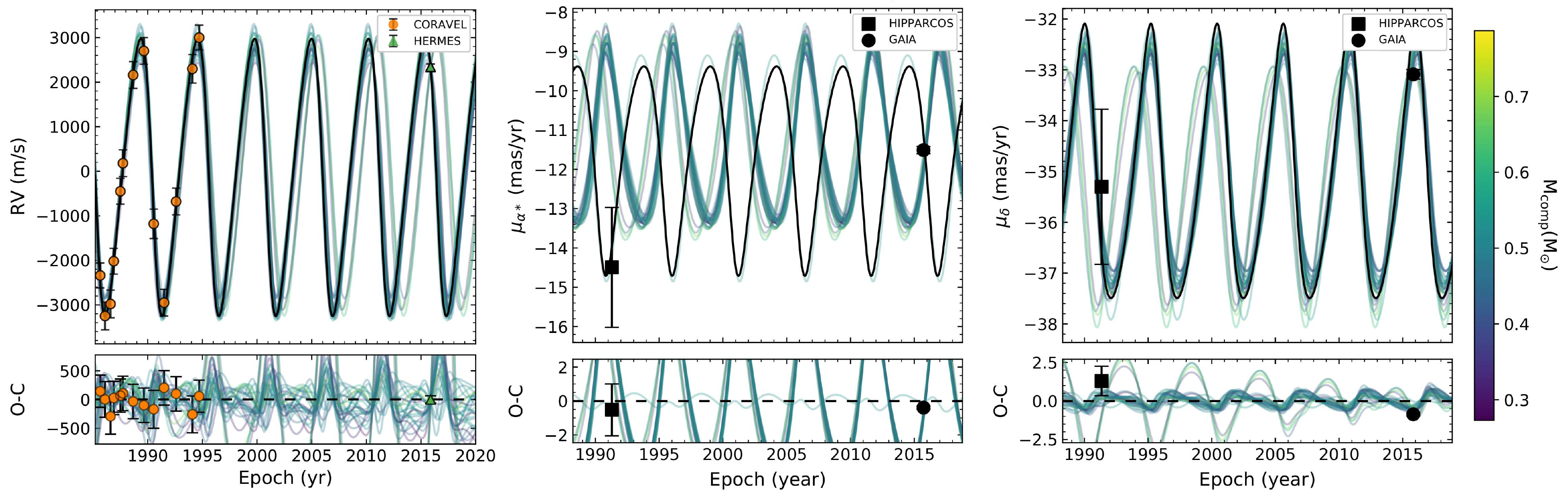}
\caption{\label{Fig:HD5424} RV curve and proper motions of HD\,5424. We used a fixed RV offset of 500 m/s \citep{Jorissen19}.}
\end{figure*}
\begin{figure}
\begin{minipage}[l]{6cm} 
\includegraphics[scale=0.3]{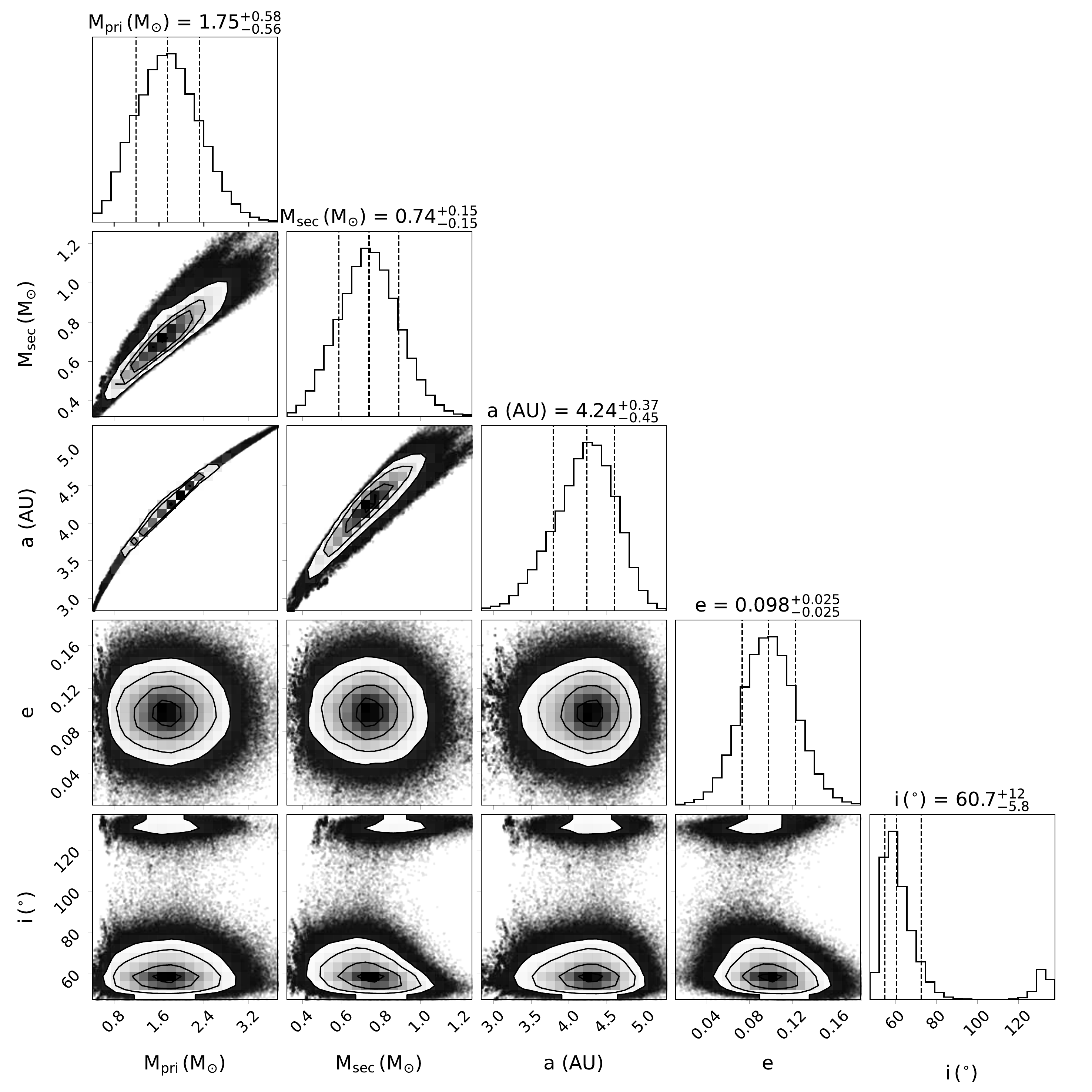}
\caption{\label{Fig:HD16458corner} Corner plot of HD\,16458}
\end{minipage} 
\hspace{3cm} 
\begin{minipage}[r]{6cm} 
\includegraphics[scale=0.3]{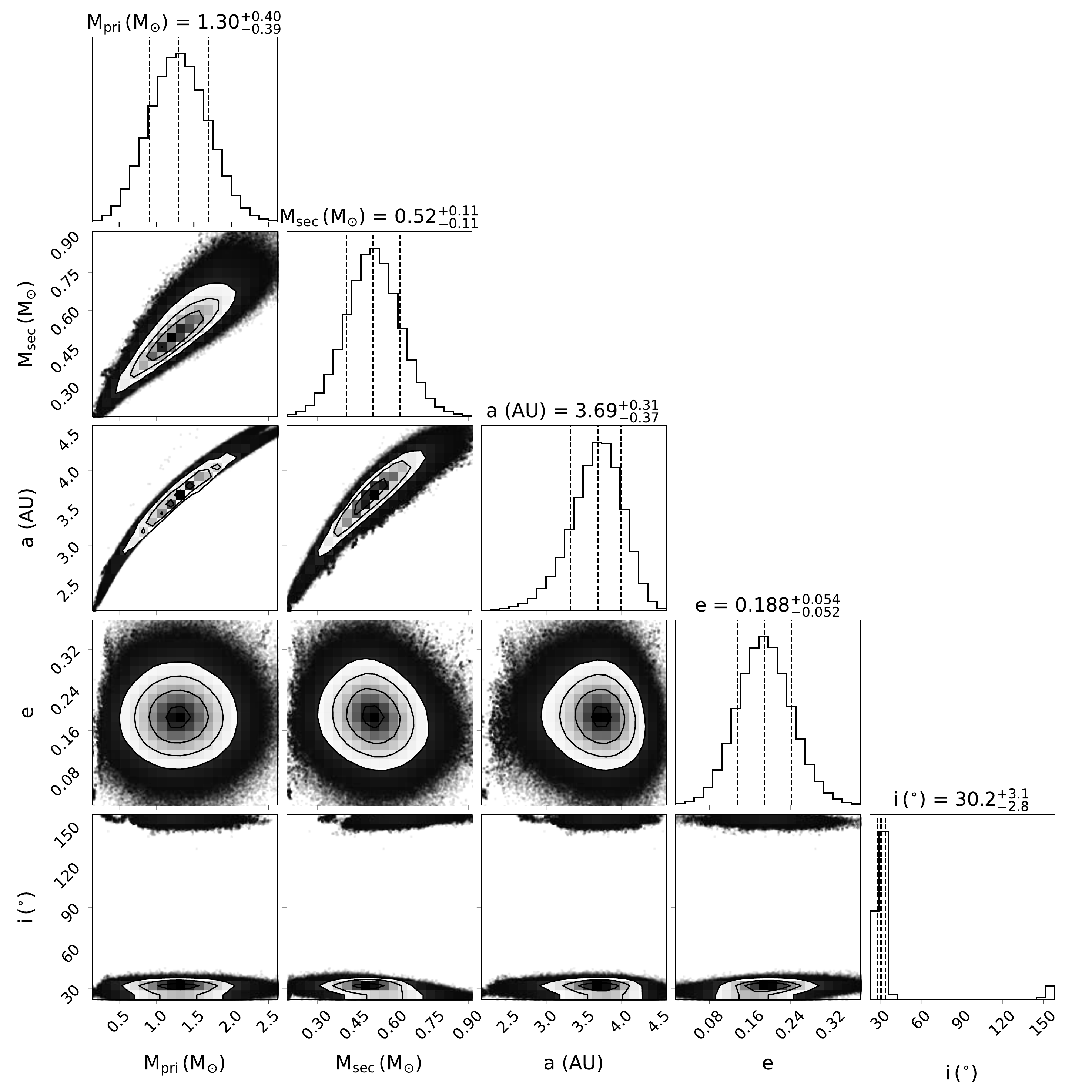}
\caption{\label{Fig:HD5424corner} Corner plot of HD\,5424}
\end{minipage} 
\end{figure}

\begin{figure*}[t]
\centering
\includegraphics[width=\textwidth]{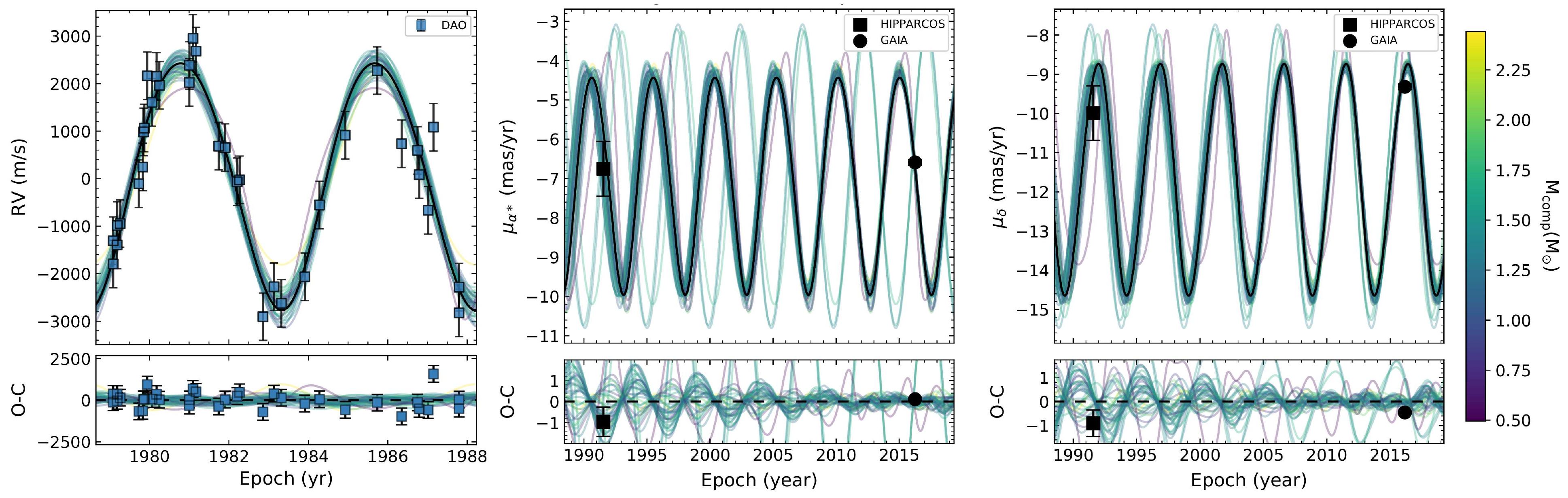}
\caption{\label{Fig:HD49641} RV curve and proper motions of HD\,49641}
\end{figure*}
\begin{figure*}
\centering
\includegraphics[width=\textwidth]{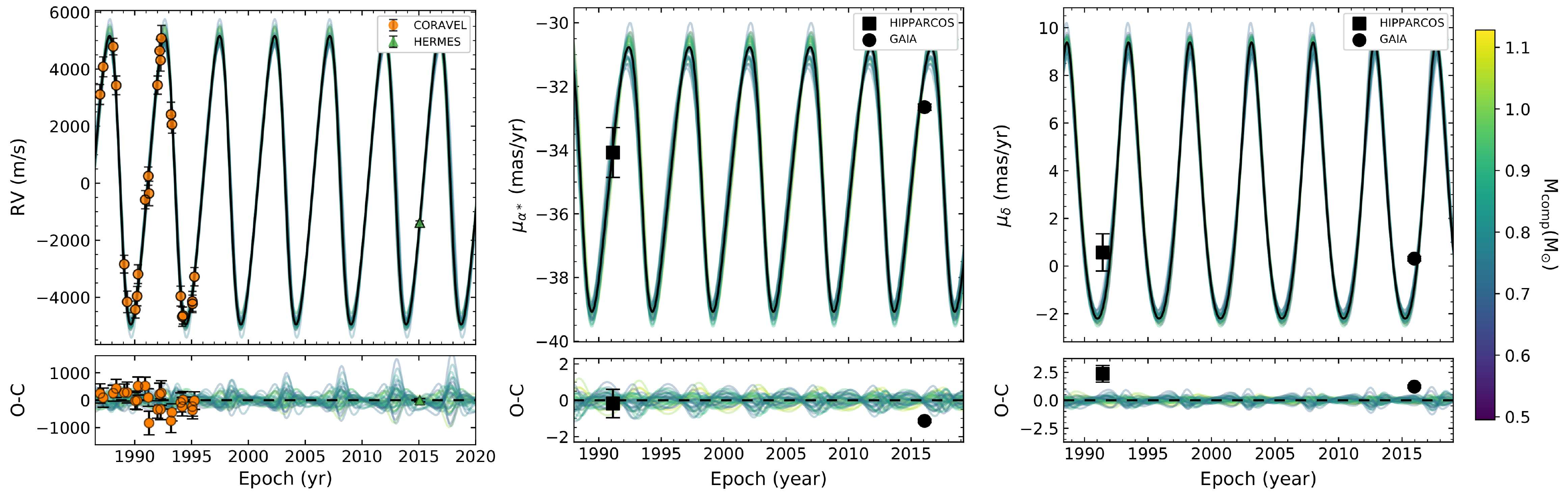}
\caption{\label{Fig:HD91208} RV curve and proper motions of HD\,91208. We used a fixed RV offset of 500 m/s \citep{Jorissen19}.}
\end{figure*}
\begin{figure}
\begin{minipage}[l]{6cm} 
\includegraphics[scale=0.3]{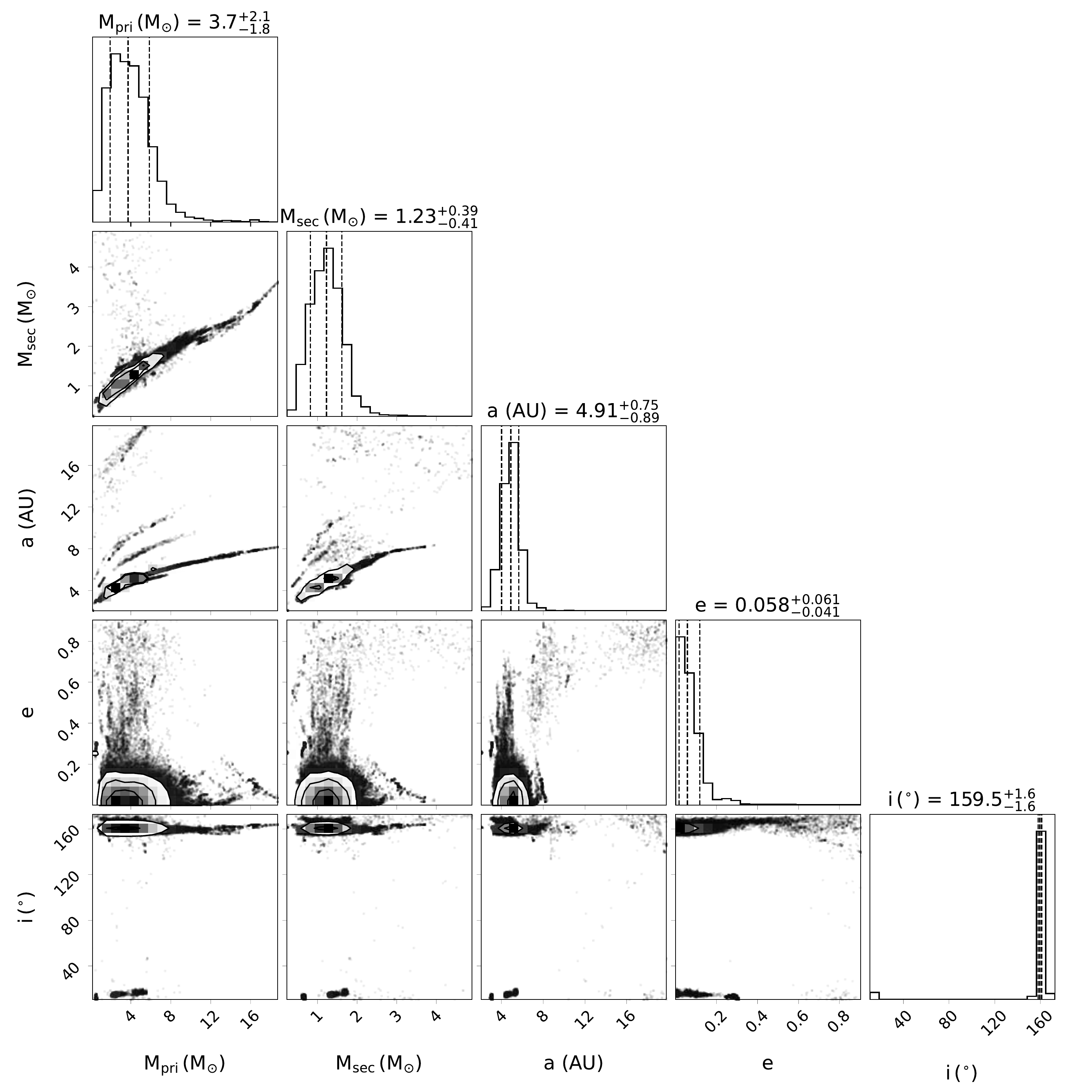}
\caption{\label{Fig:HD49641corner}  Corner plot of HD\,49641}
\end{minipage} 
\hspace{3cm} 
\begin{minipage}[r]{6cm} 
\includegraphics[scale=0.3]{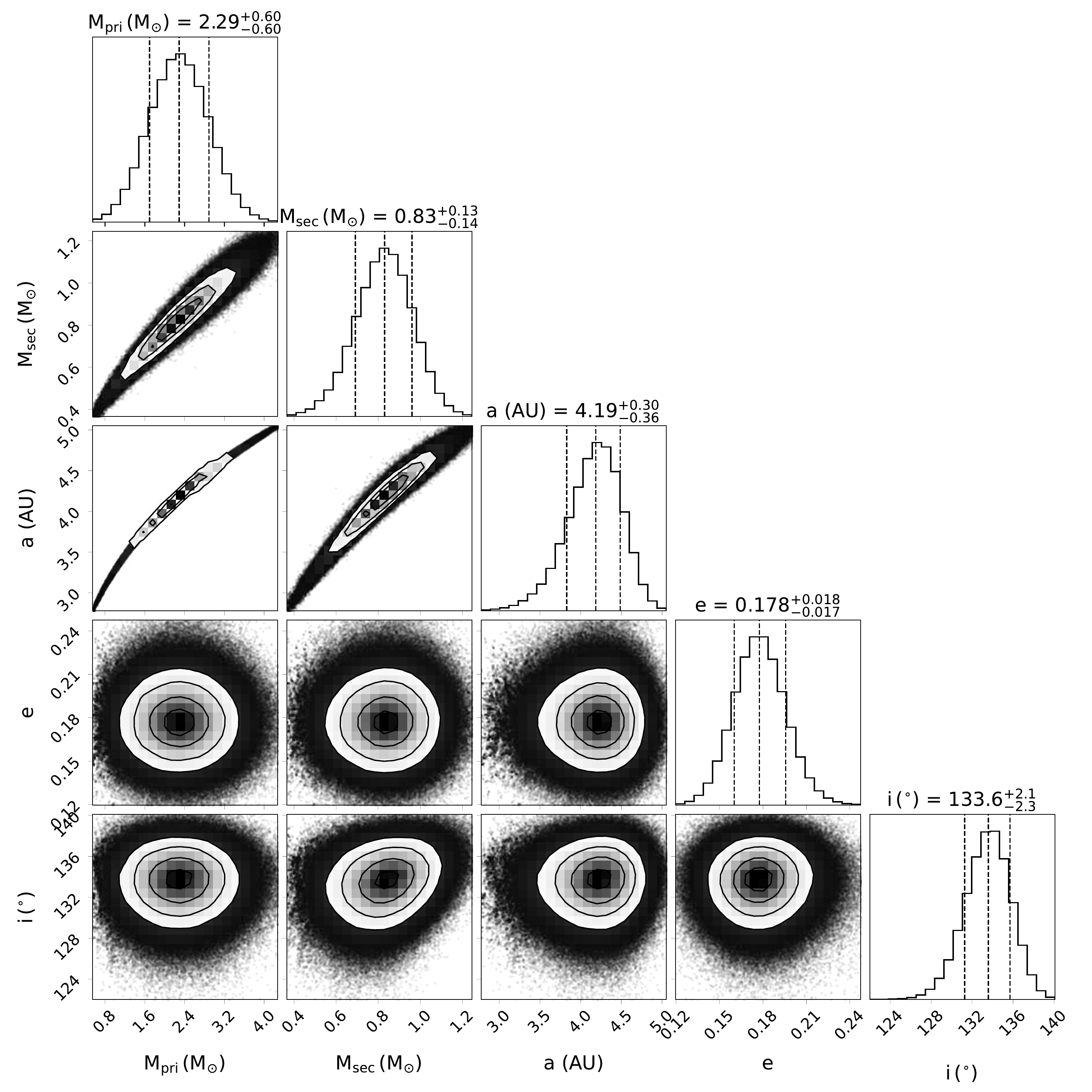}
\caption{\label{Fig:HD91208corner} Corner plot of HD\,91208}
\end{minipage} 
\end{figure}

\begin{figure*}[t]
\centering
\includegraphics[width=\textwidth]{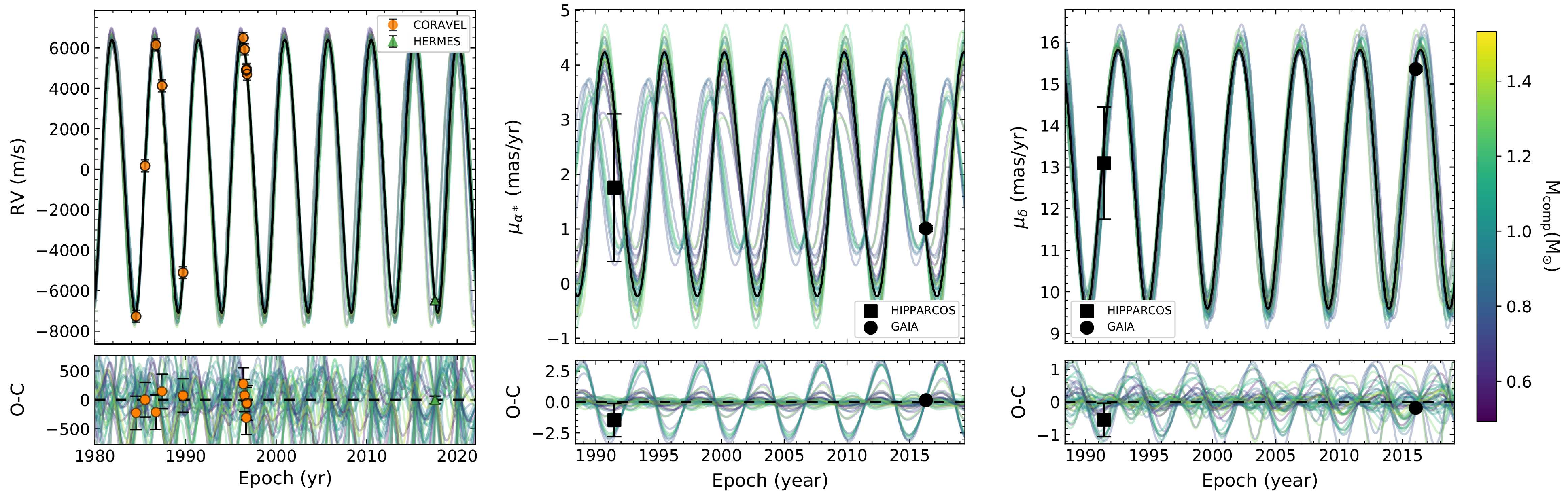}
\caption{\label{Fig:HD200063} RV curve and proper motions of HD\,200063. We used a fixed RV offset of 500 m/s \citep{Jorissen19}.}
\end{figure*}
\begin{figure*}
\centering
\includegraphics[width=\textwidth]{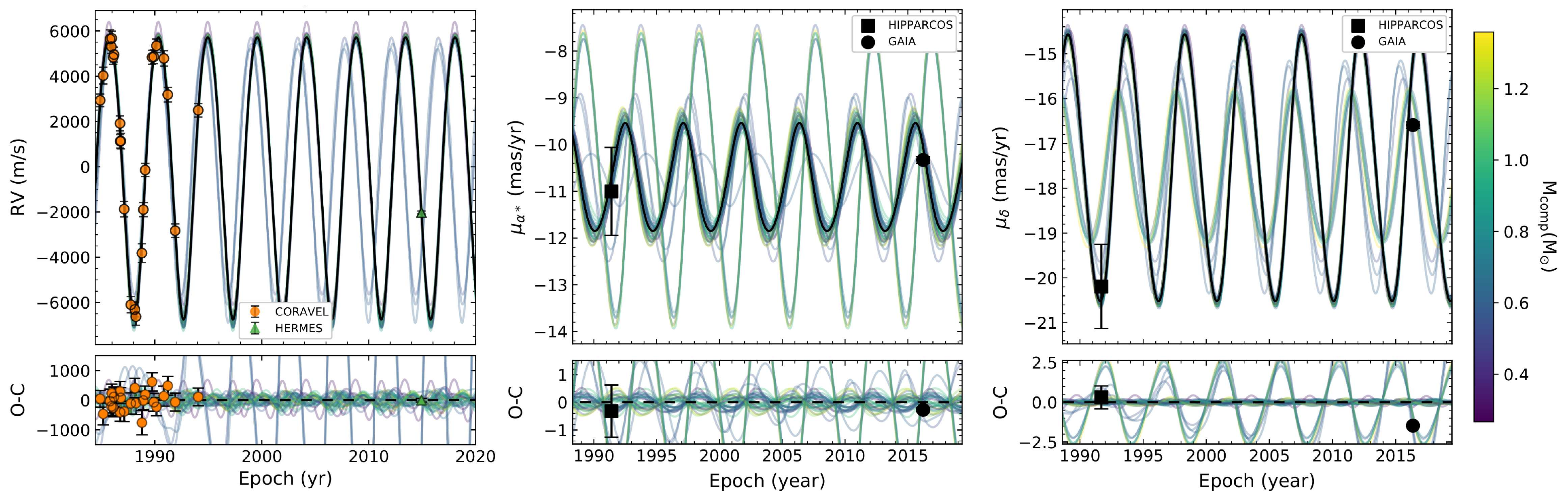}
\caption{\label{Fig:HD43389} RV curve and proper motions of HD\,43389. We used a fixed RV offset of 500 m/s \citep{Jorissen19}.}
\end{figure*}
\begin{figure}
\begin{minipage}[l]{6cm} 
\includegraphics[scale=0.3]{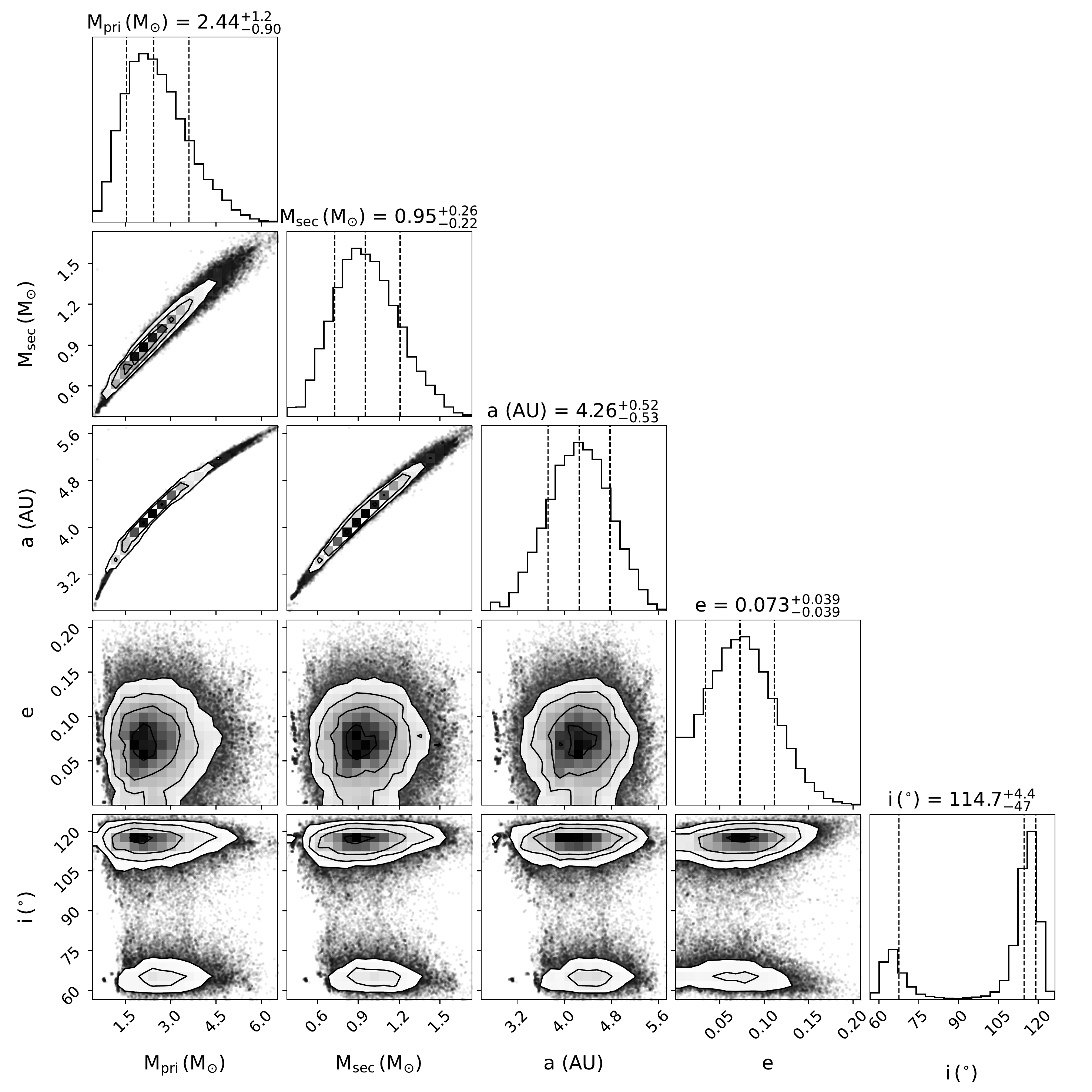}
\caption{\label{Fig:HD200063corner} Corner plot of HD\,200063}
\end{minipage} 
\hspace{3cm} 
\begin{minipage}[r]{6cm} 
\includegraphics[scale=0.3]{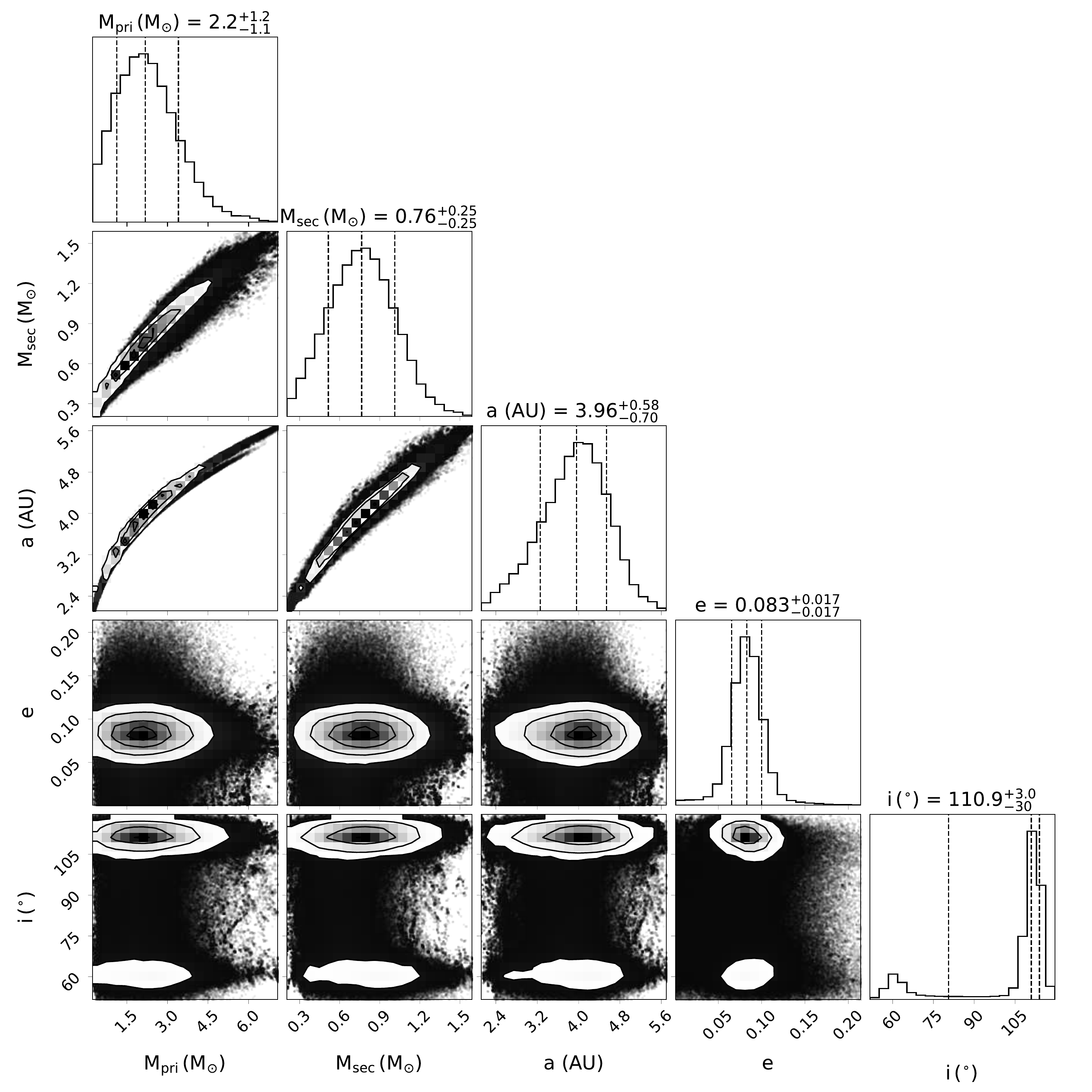}
\caption{\label{Fig:HD43389corner} Corner plot of HD\,43389}
\end{minipage} 
\end{figure}

\begin{figure*}[t]
\centering
\includegraphics[width=\textwidth]{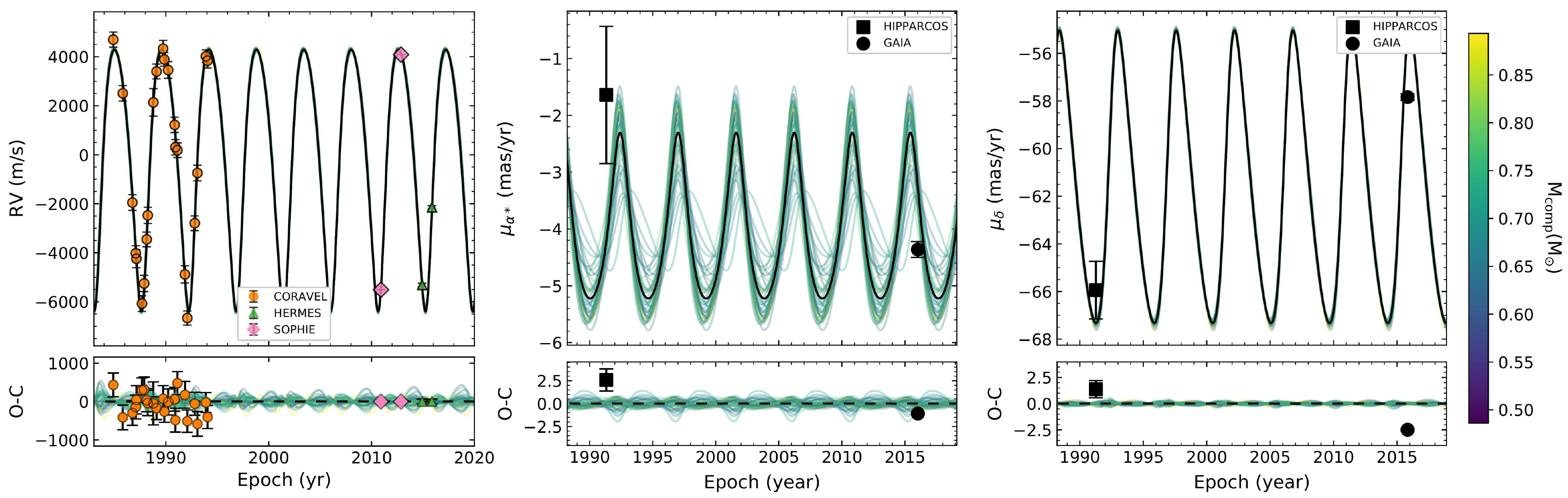}
\caption{\label{Fig:HD27271} RV curve and proper motions of HD\,27271.}
\end{figure*}
\begin{figure*}
\centering
\includegraphics[width=\textwidth]{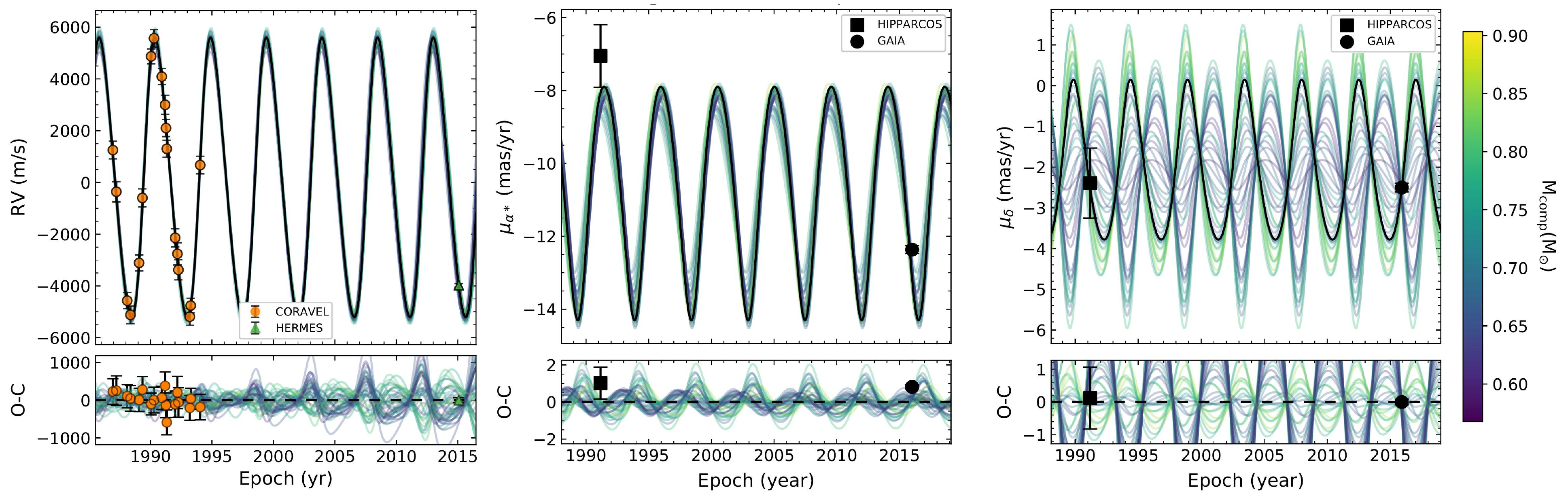}
\caption{\label{Fig:HD95193} RV curve and proper motions of HD\,95193. We used a fixed RV offset of 500 m/s \citep{Jorissen19}.}
\end{figure*}
\begin{figure}
\begin{minipage}[l]{6cm} 
\includegraphics[scale=0.3]{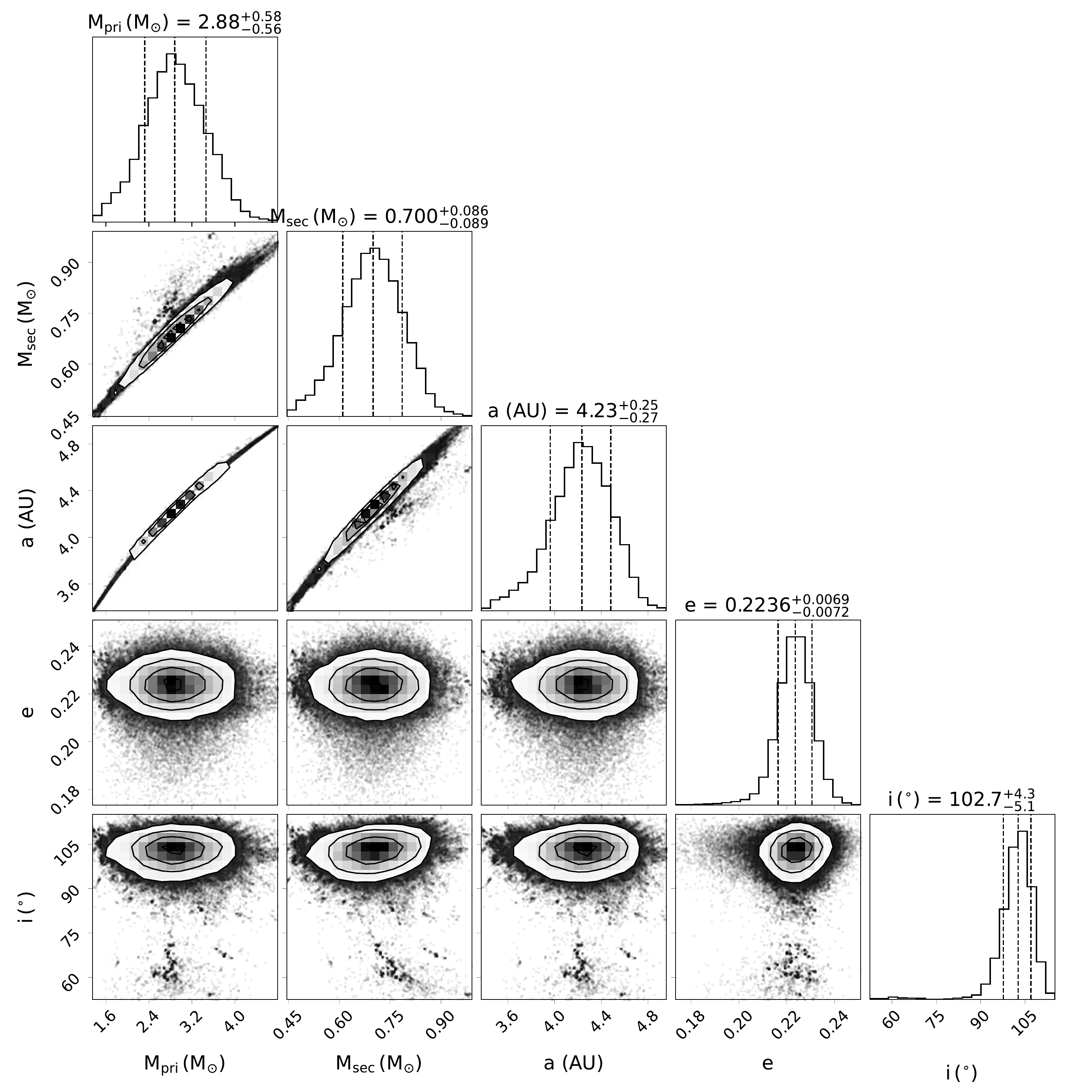}
\caption{\label{Fig:HD27271corner} Corner plot of HD\,27271}
\end{minipage} 
\hspace{3cm} 
\begin{minipage}[r]{6cm} 
\includegraphics[scale=0.3]{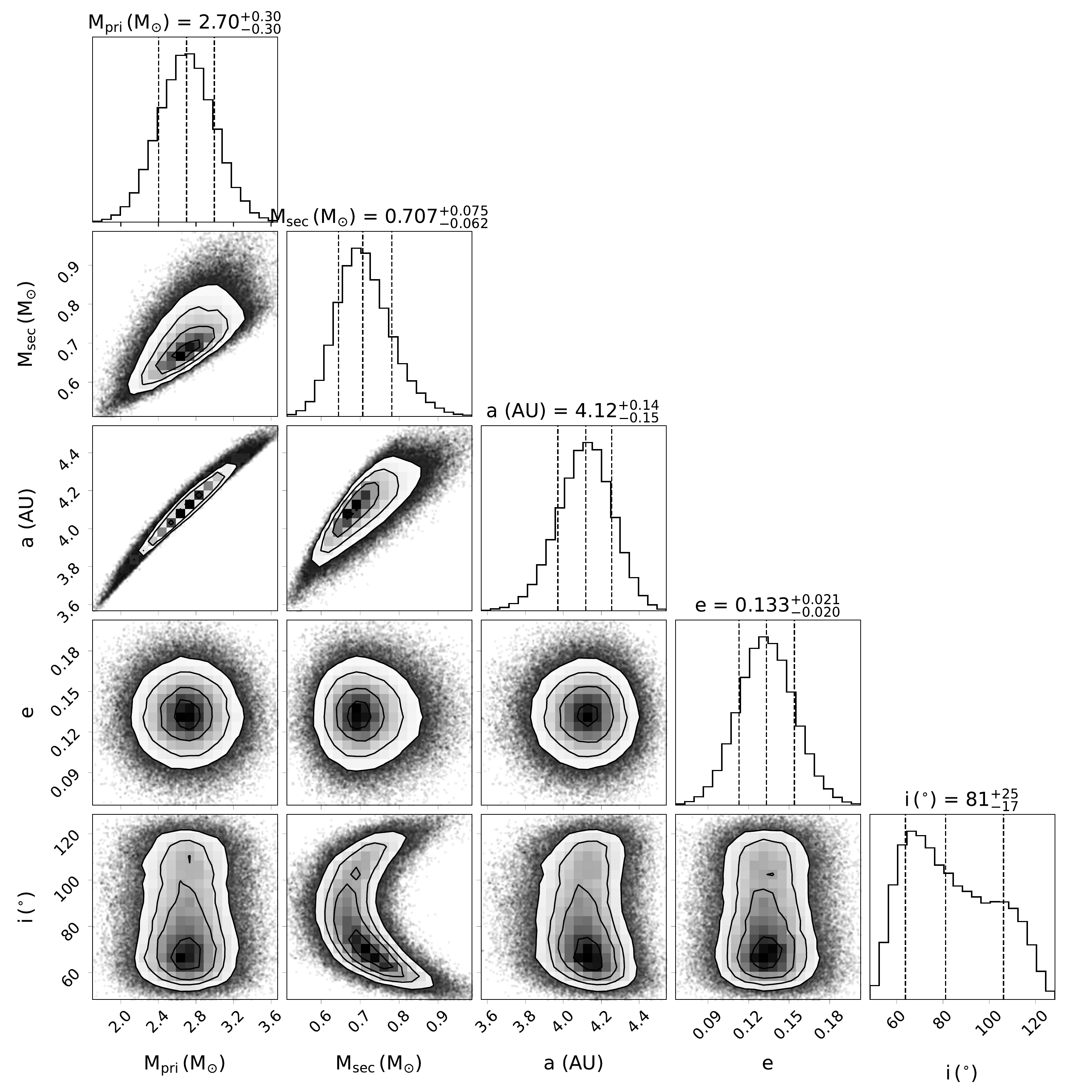}
\caption{\label{Fig:HD95193corner} Corner plot of HD\,95193}
\end{minipage} 
\end{figure}

\begin{figure*}[t]
\centering
\includegraphics[width=\textwidth]{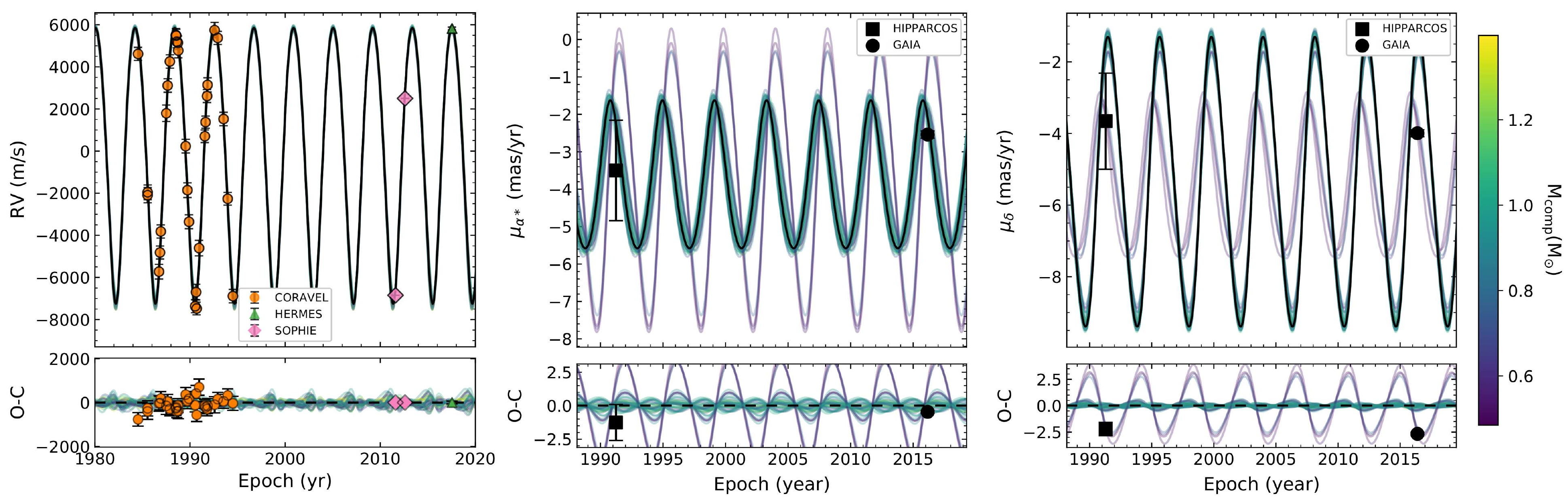}
\caption{\label{Fig:HD210946} RV curve and proper motions of HD\,210946}
\end{figure*}
\begin{figure*}
\centering
\includegraphics[width=\textwidth]{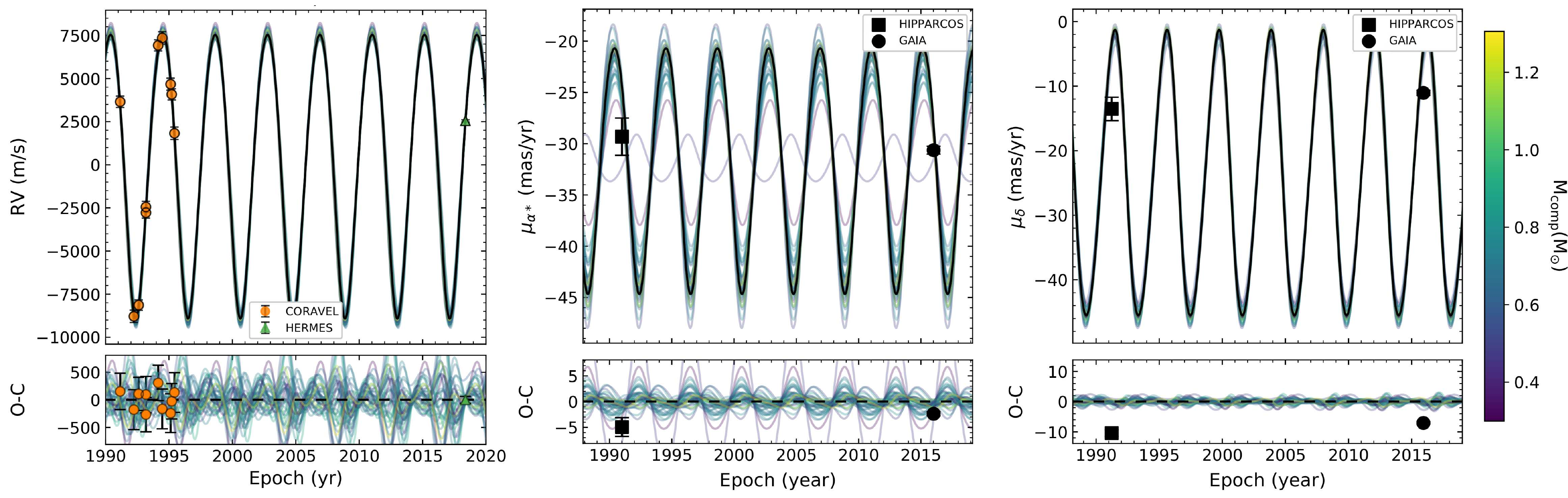}
\caption{\label{Fig:HD127392} RV curve and proper motions of HD\,127392. We used a fixed RV offset of 195 m/s \citep{Escorza19}}
\end{figure*}
\begin{figure}
\begin{minipage}[l]{6cm} 
\includegraphics[scale=0.3]{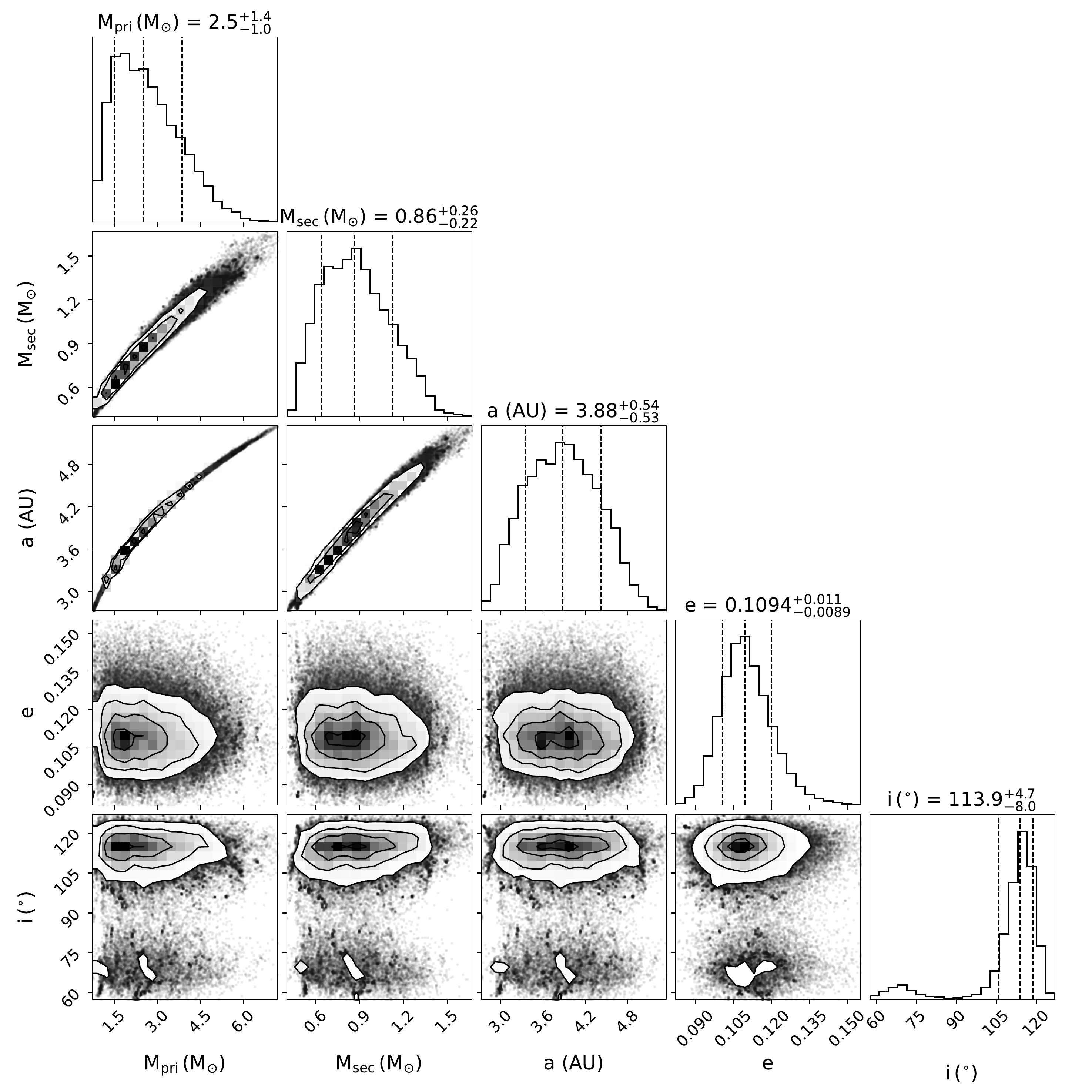}
\caption{\label{Fig:HD210946corner} Corner plot of HD\,210946}
\end{minipage} 
\hspace{3cm} 
\begin{minipage}[r]{6cm} 
\includegraphics[scale=0.3]{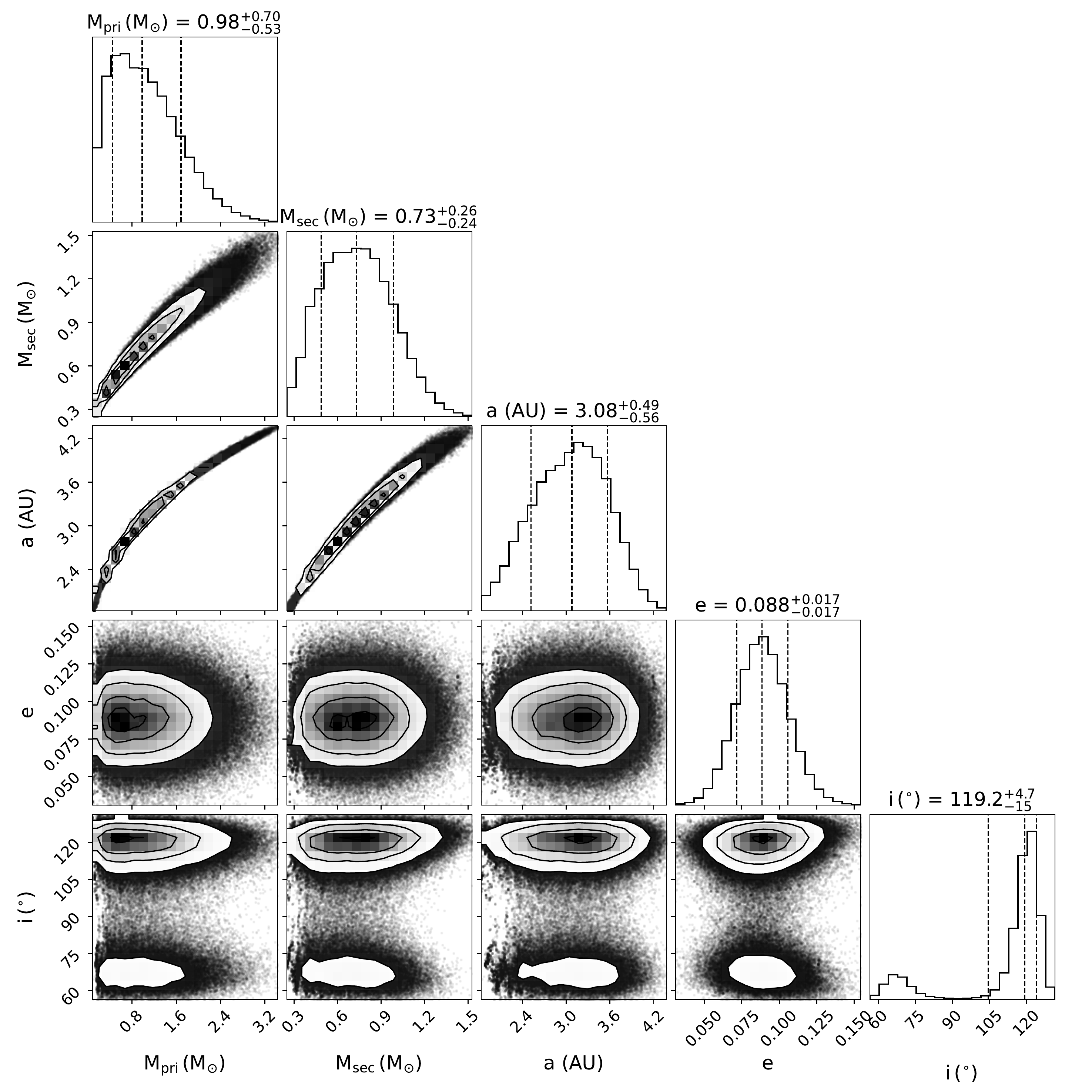}
\caption{\label{Fig:HD127392corner} Corner plot of HD\,127392}
\end{minipage} 
\end{figure}

\begin{figure*}[t]
\centering
\includegraphics[width=\textwidth]{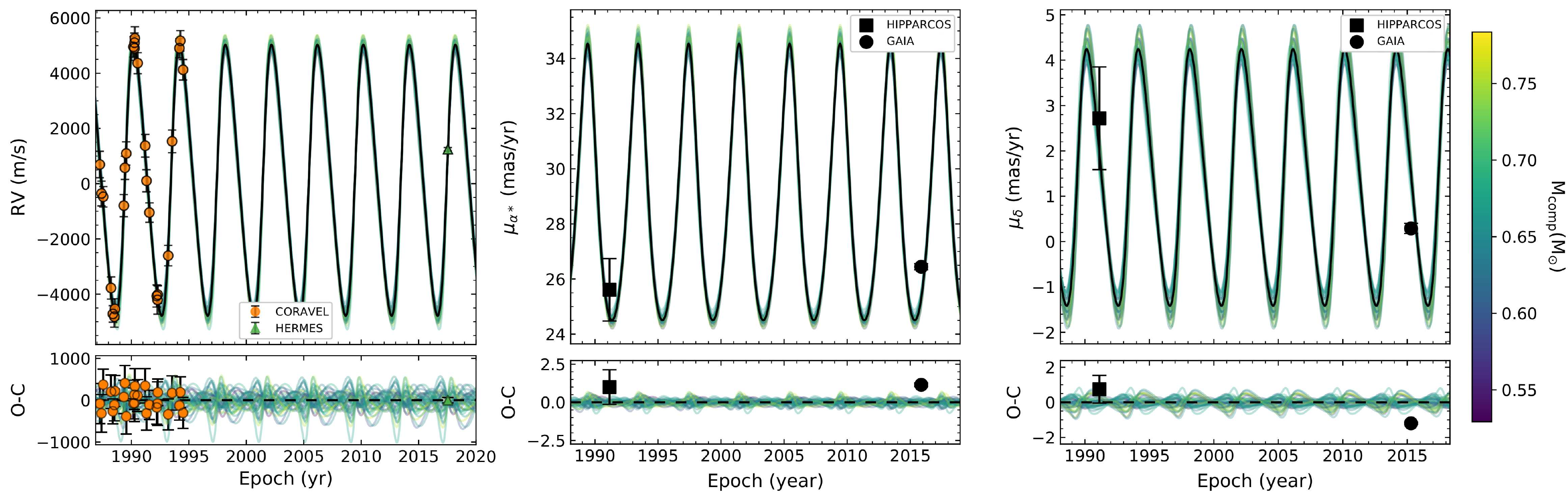}
\caption{\label{Fig:HD143899} RV curve and proper motions of HD\,143899. We used a fixed RV offset of 500 m/s \citep{Jorissen19}.}
\end{figure*}
\begin{figure*}
\centering
\includegraphics[width=\textwidth]{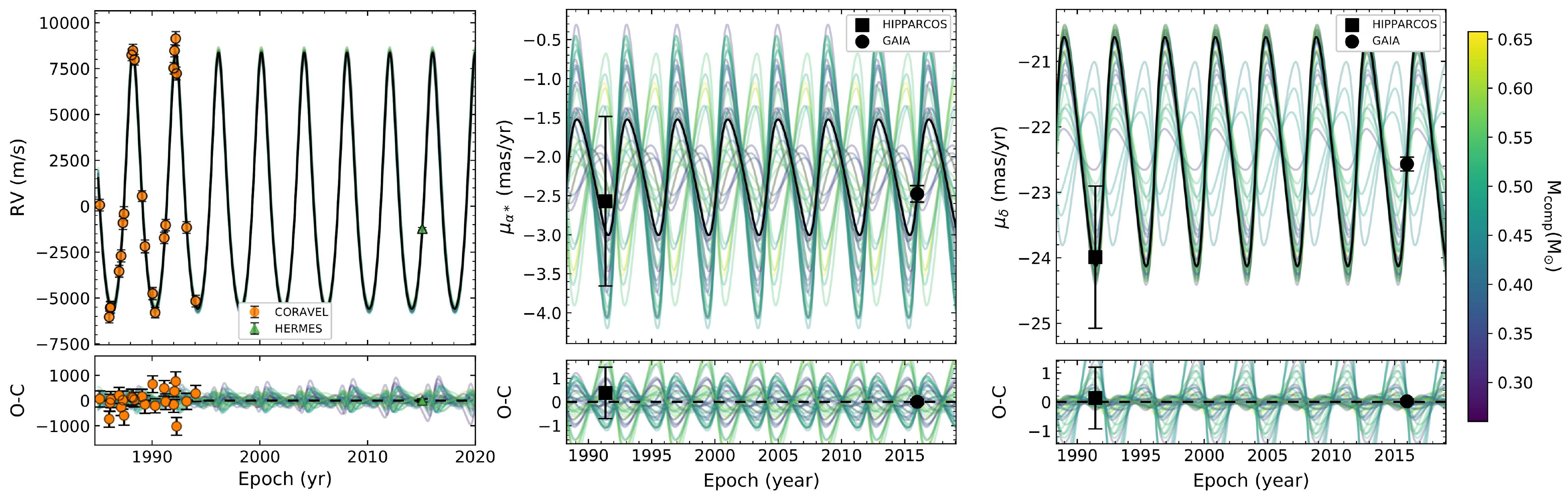}
\caption{\label{Fig:HD88562} RV curve and proper motions of HD\,88562. We used a fixed RV offset of 500 m/s \citep{Jorissen19}.}
\end{figure*}
\begin{figure}
\begin{minipage}[l]{6cm} 
\includegraphics[scale=0.3]{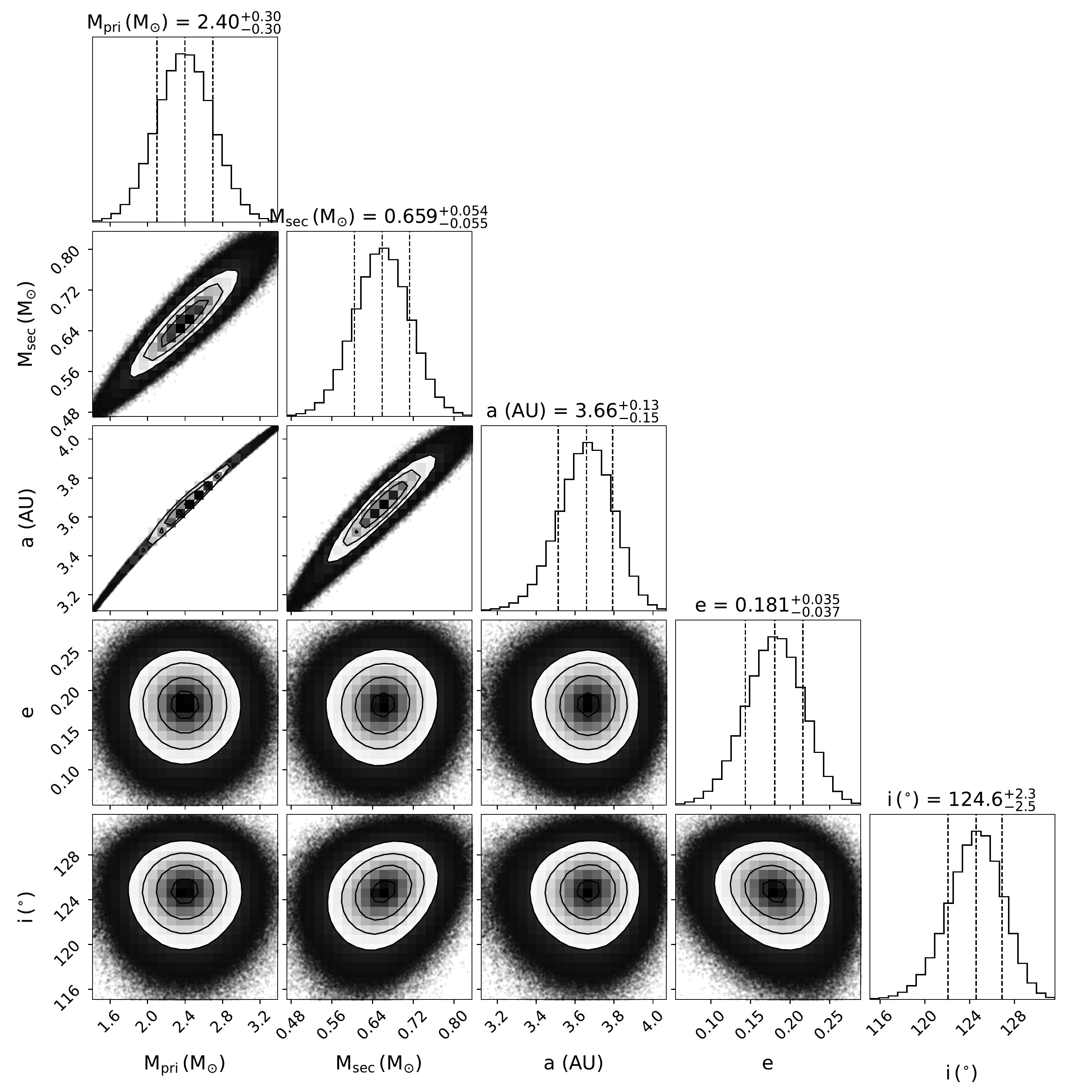}
\caption{\label{Fig:HD143899corner} Corner plot of HD\,143899}
\end{minipage} 
\hspace{3cm} 
\begin{minipage}[r]{6cm} 
\includegraphics[scale=0.3]{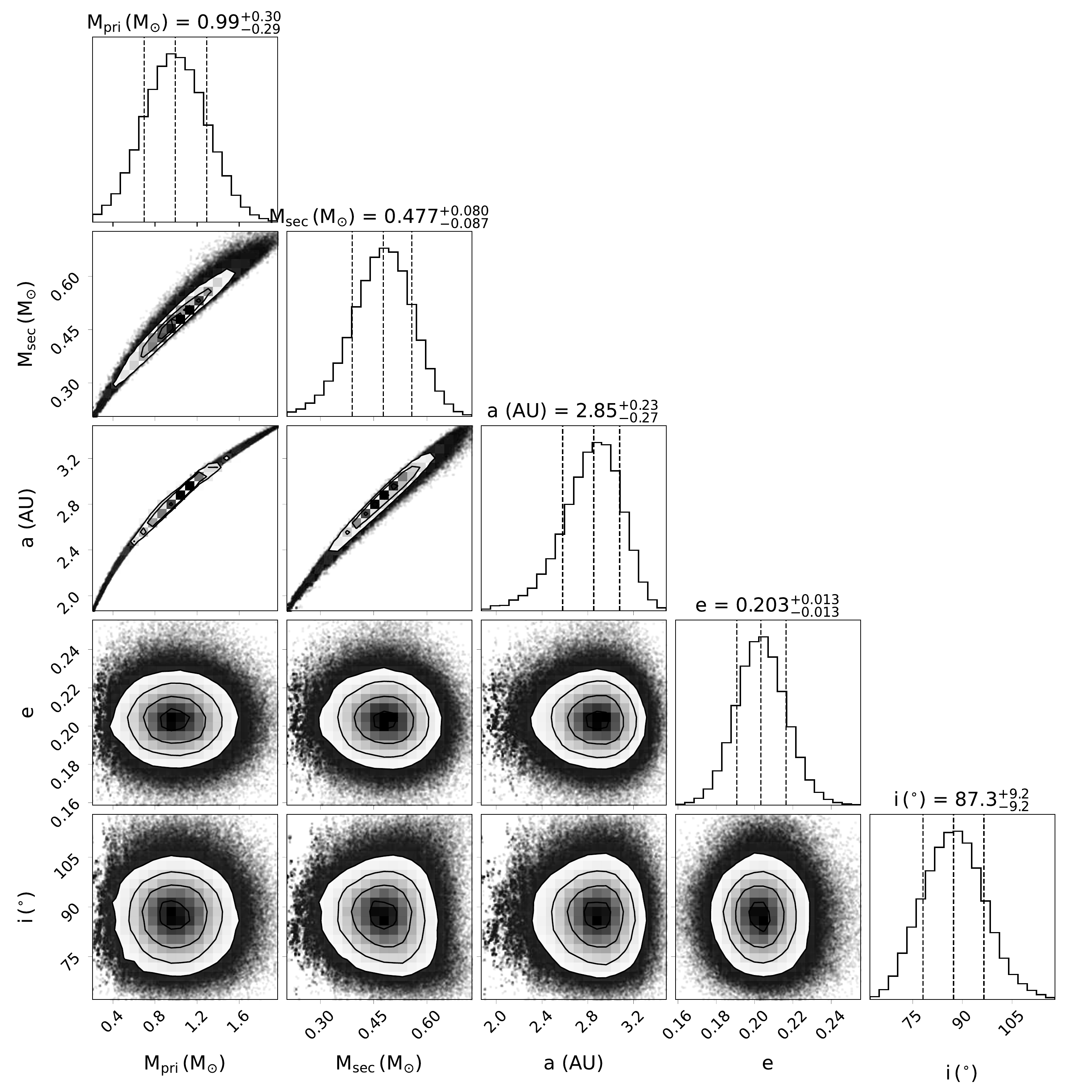}
\caption{\label{Fig:HD88562corner} Corner plot of HD\,88562}
\end{minipage} 
\end{figure}

\begin{figure*}[t]
\centering
\includegraphics[width=\textwidth]{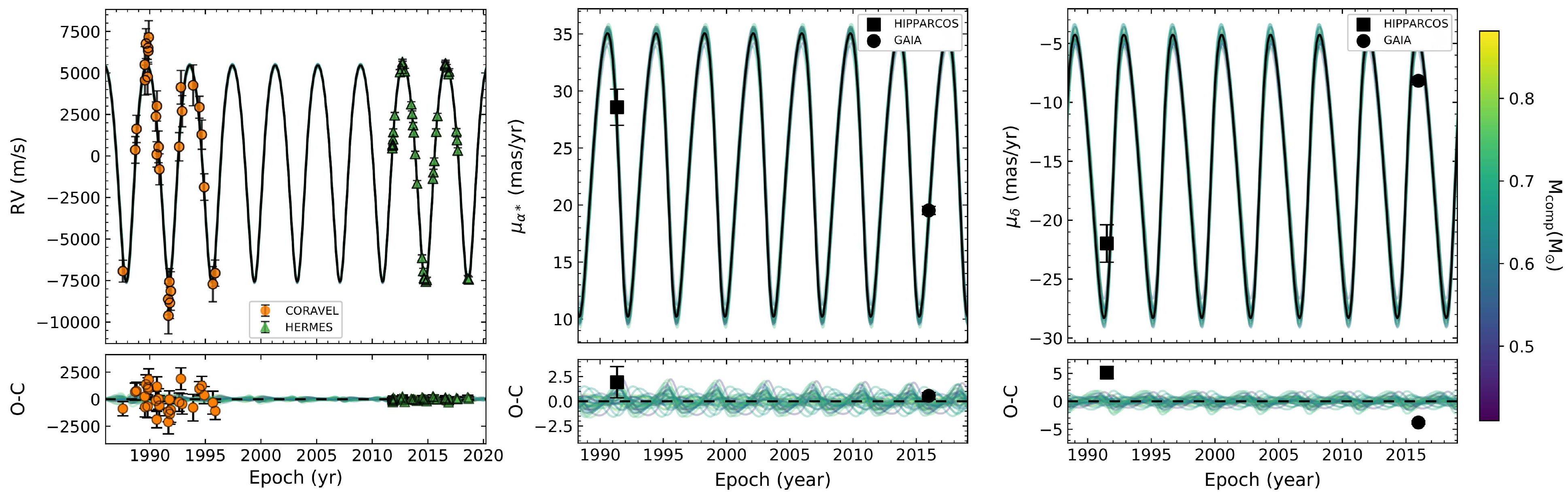}
\caption{\label{Fig:HD221531} RV curve and proper motions of HD\,221531}
\end{figure*}
\begin{figure*}
\centering
\includegraphics[width=\textwidth]{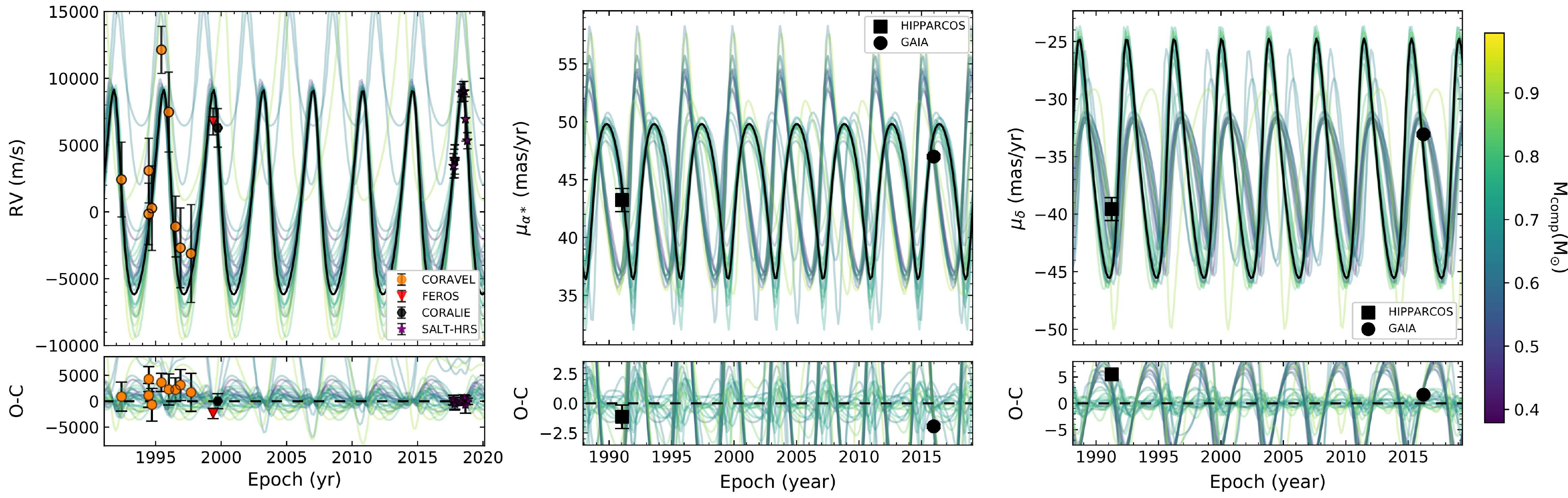}
\caption{\label{Fig:HD202400} RV curve and proper motions of HD\,202400}
\end{figure*}
\begin{figure}
\begin{minipage}[l]{6cm} 
\includegraphics[scale=0.3]{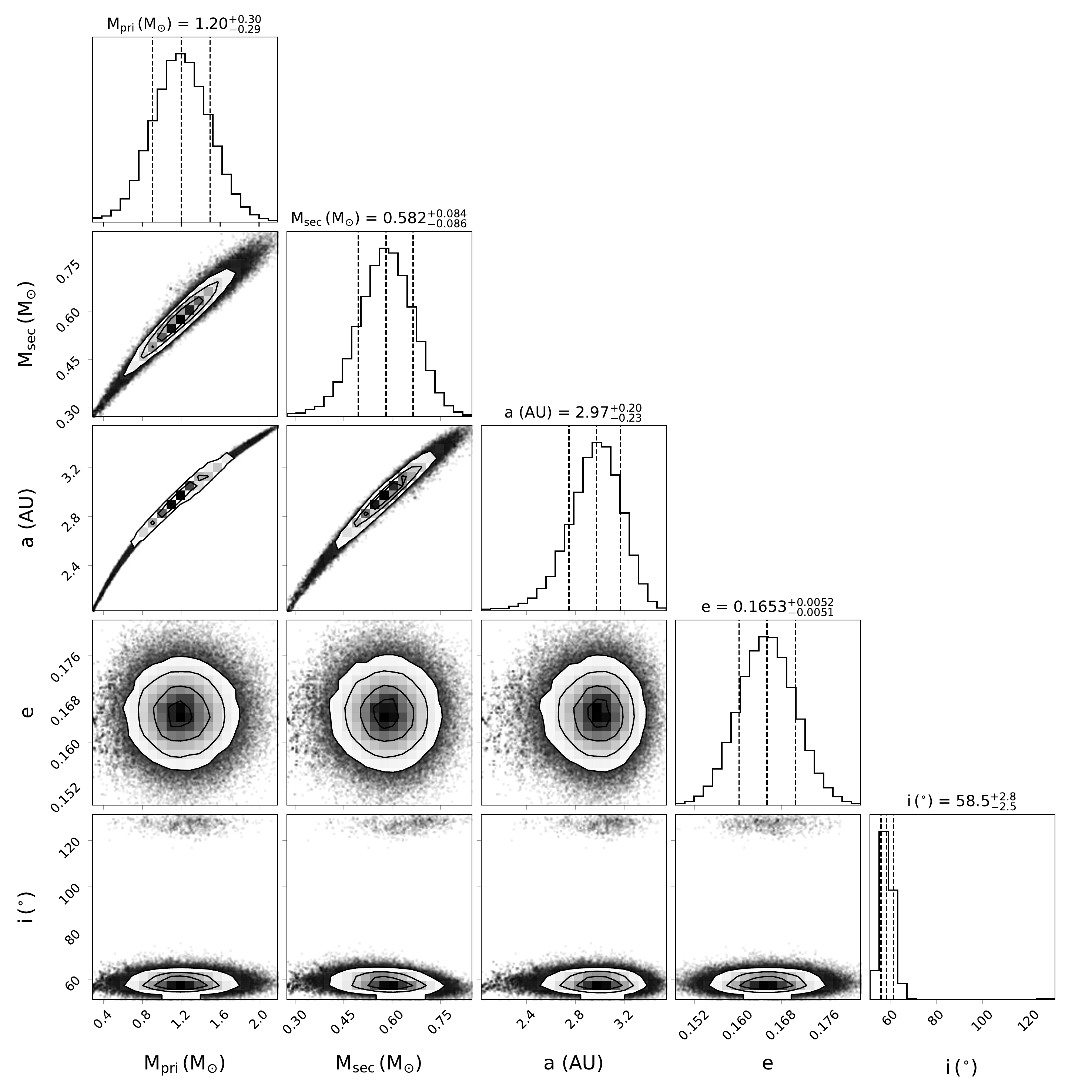}
\caption{\label{Fig:HD221531corner} Corner plot of HD\,221531}
\end{minipage} 
\hspace{3cm} 
\begin{minipage}[r]{6cm} 
\includegraphics[scale=0.3]{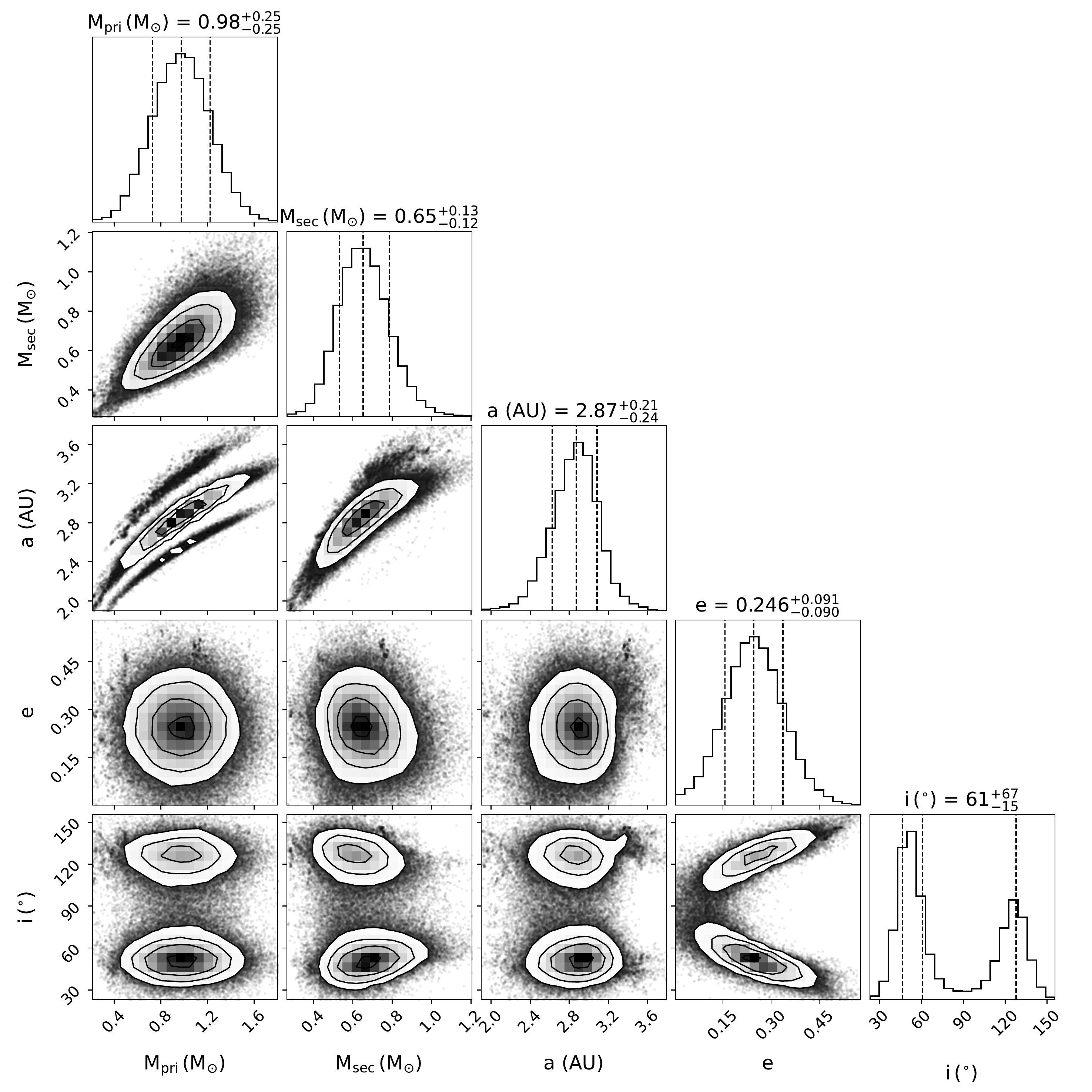}
\caption{\label{Fig:HD202400corner} Corner plot of HD\,202400}
\end{minipage} 
\end{figure}

\begin{figure*}[t]
\centering
\includegraphics[width=\textwidth]{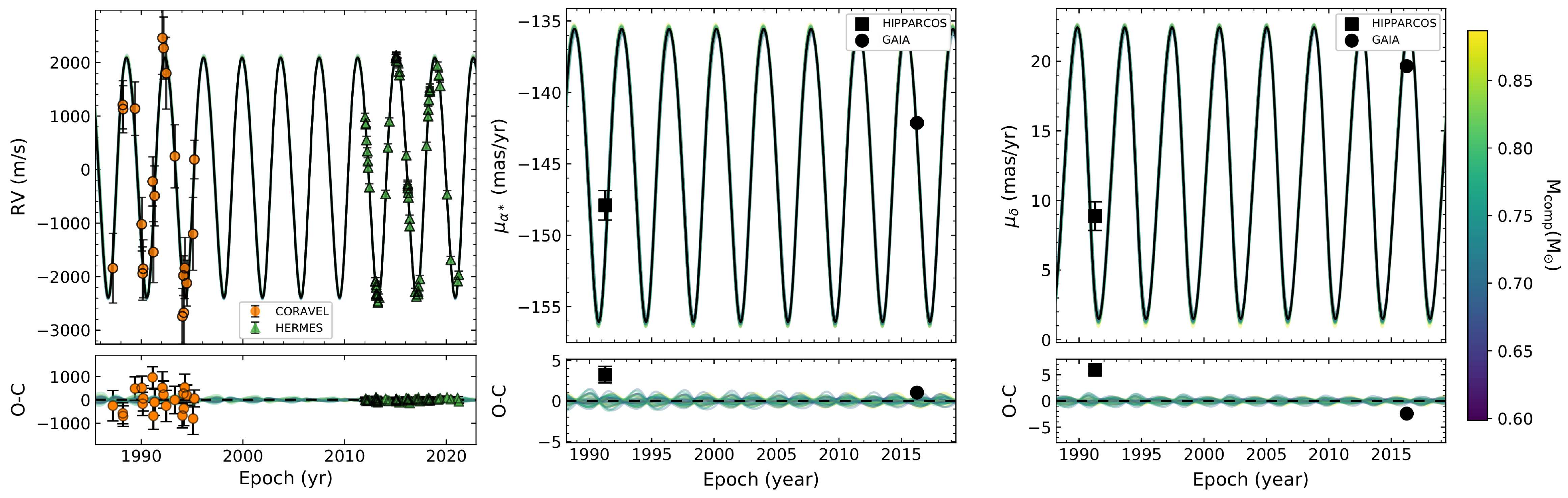}
\caption{\label{Fig:HD107574} RV curve and proper motions of HD\,107574}
\end{figure*}
\begin{figure*}
\centering
\includegraphics[width=\textwidth]{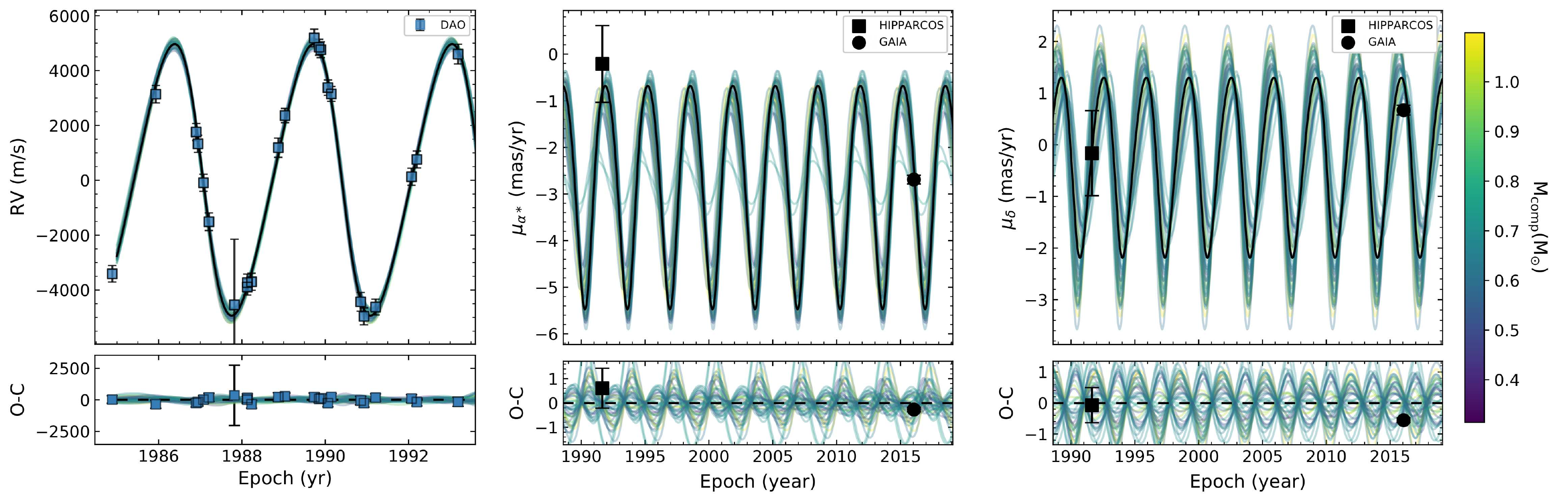}
\caption{\label{Fig:HD58121} RV curve and proper motions of HD\,58121}
\end{figure*}
\begin{figure}
\begin{minipage}[l]{6cm} 
\includegraphics[scale=0.3]{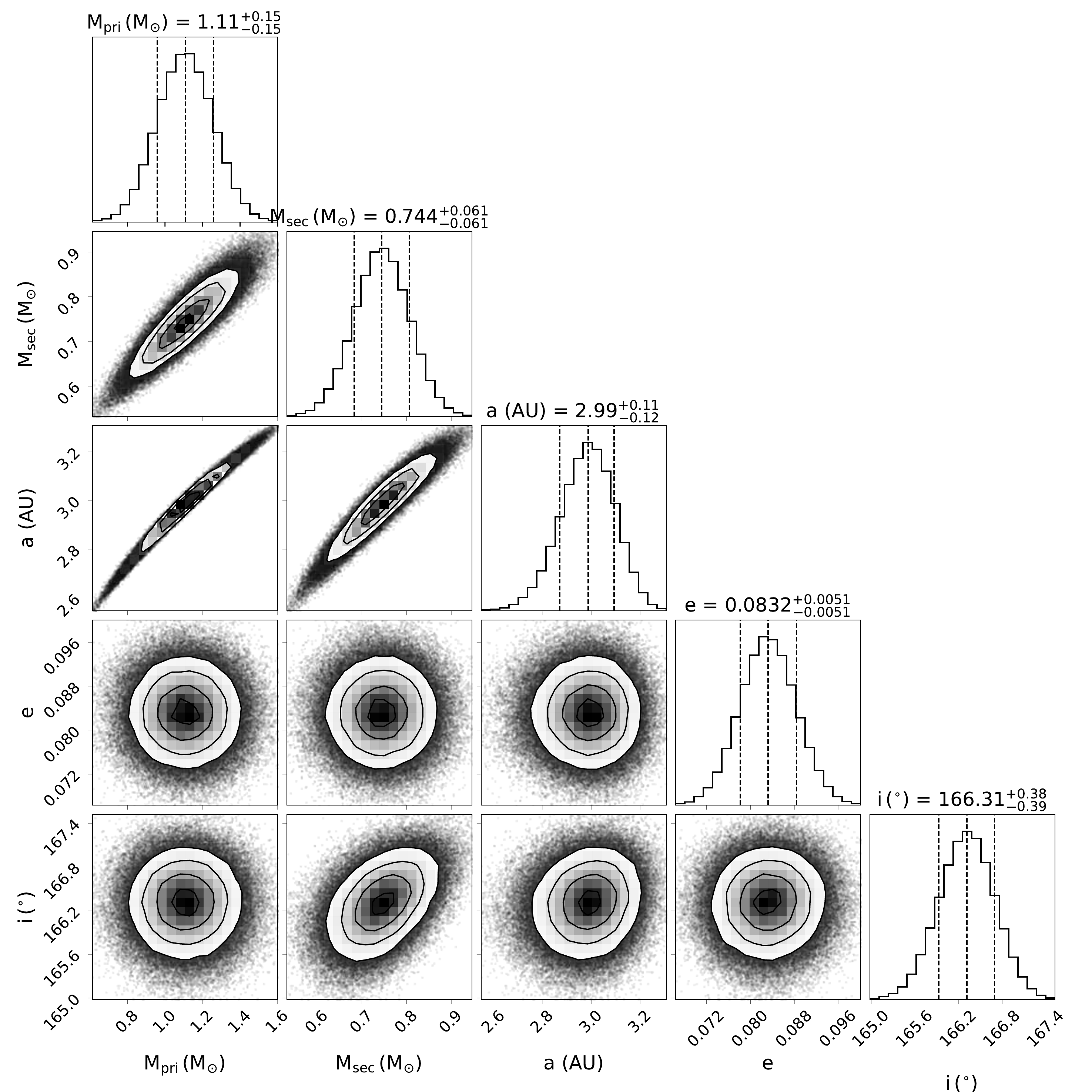}
\caption{\label{Fig:HD107574corner} Corner plot of HD\,107574}
\end{minipage} 
\hspace{3cm} 
\begin{minipage}[r]{6cm} 
\includegraphics[scale=0.3]{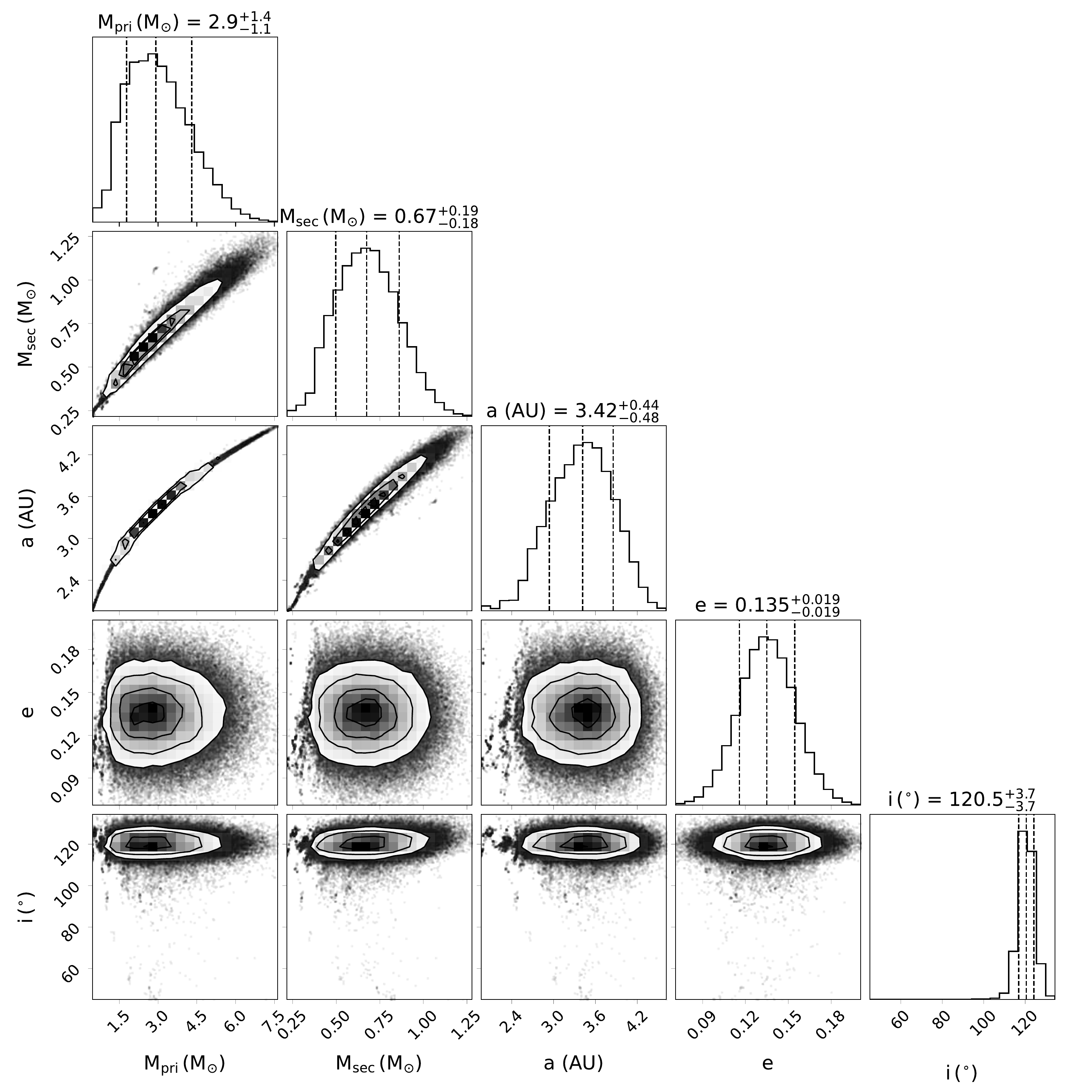}
\caption{\label{Fig:HD58121corner} Corner plot of HD\,58121}
\end{minipage} 
\end{figure}

\begin{figure*}[t]
\centering
\includegraphics[width=\textwidth]{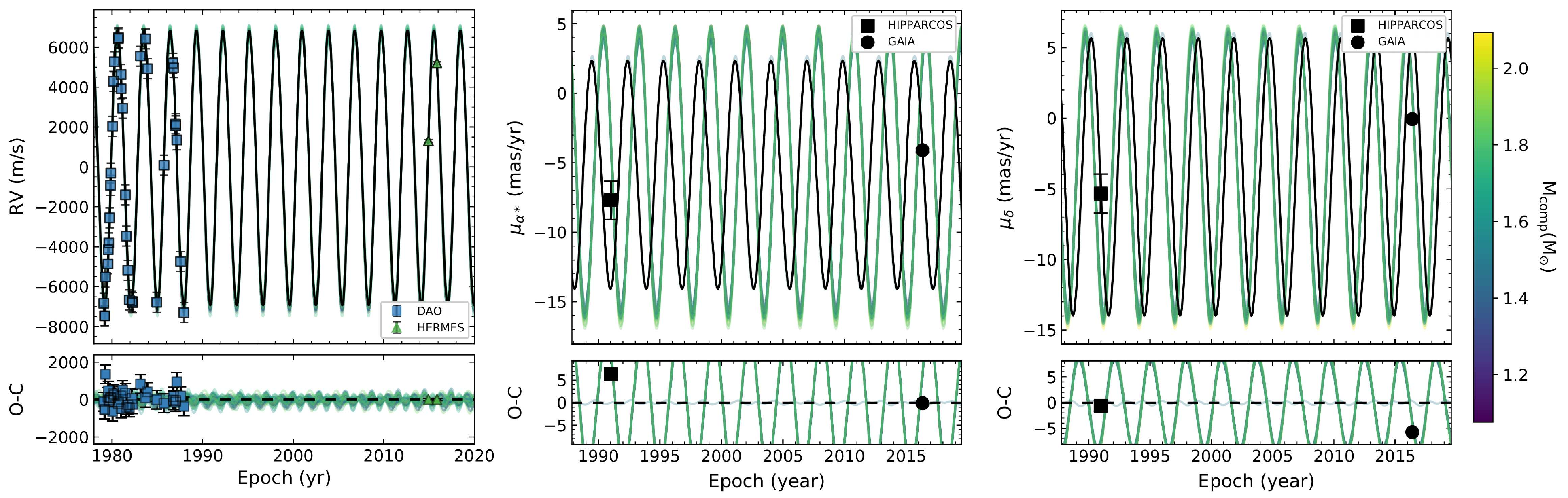}
\caption{\label{Fig:HD31487} RV curve and proper motions of HD\,31487}
\end{figure*}
\begin{figure*}
\centering
\includegraphics[width=\textwidth]{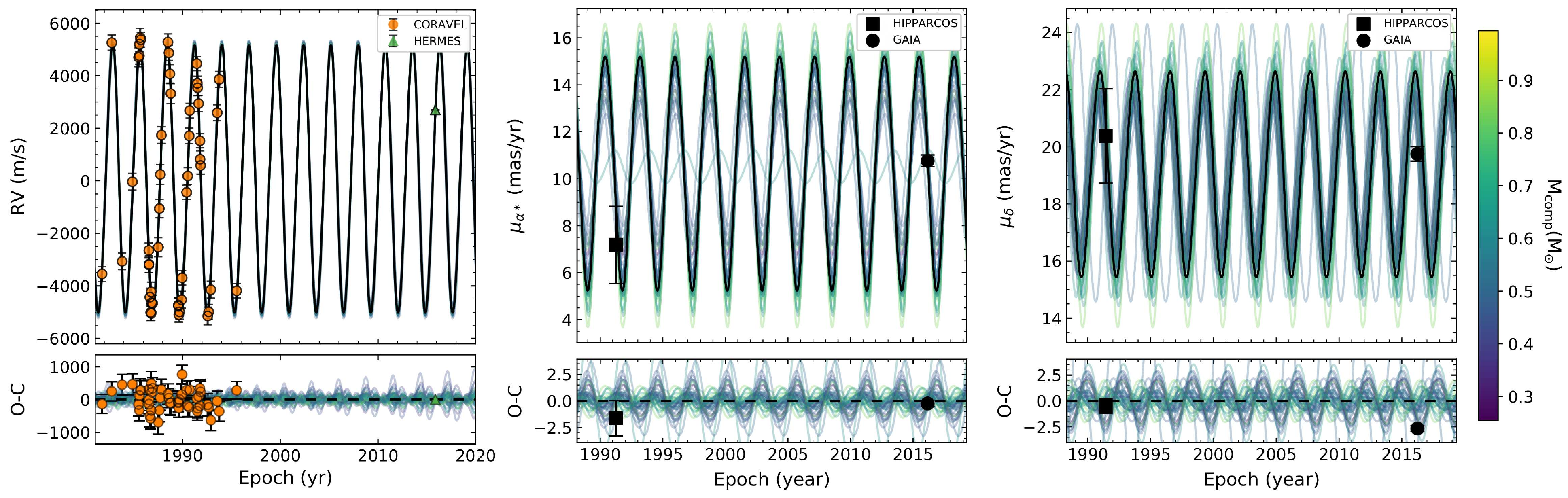}
\caption{\label{Fig:HD211594} RV curve and proper motions of HD\,211594. We used a fixed RV offset of 500 m/s \citep{Jorissen19}.}
\end{figure*}
\begin{figure}
\begin{minipage}[l]{6cm} 
\includegraphics[scale=0.3]{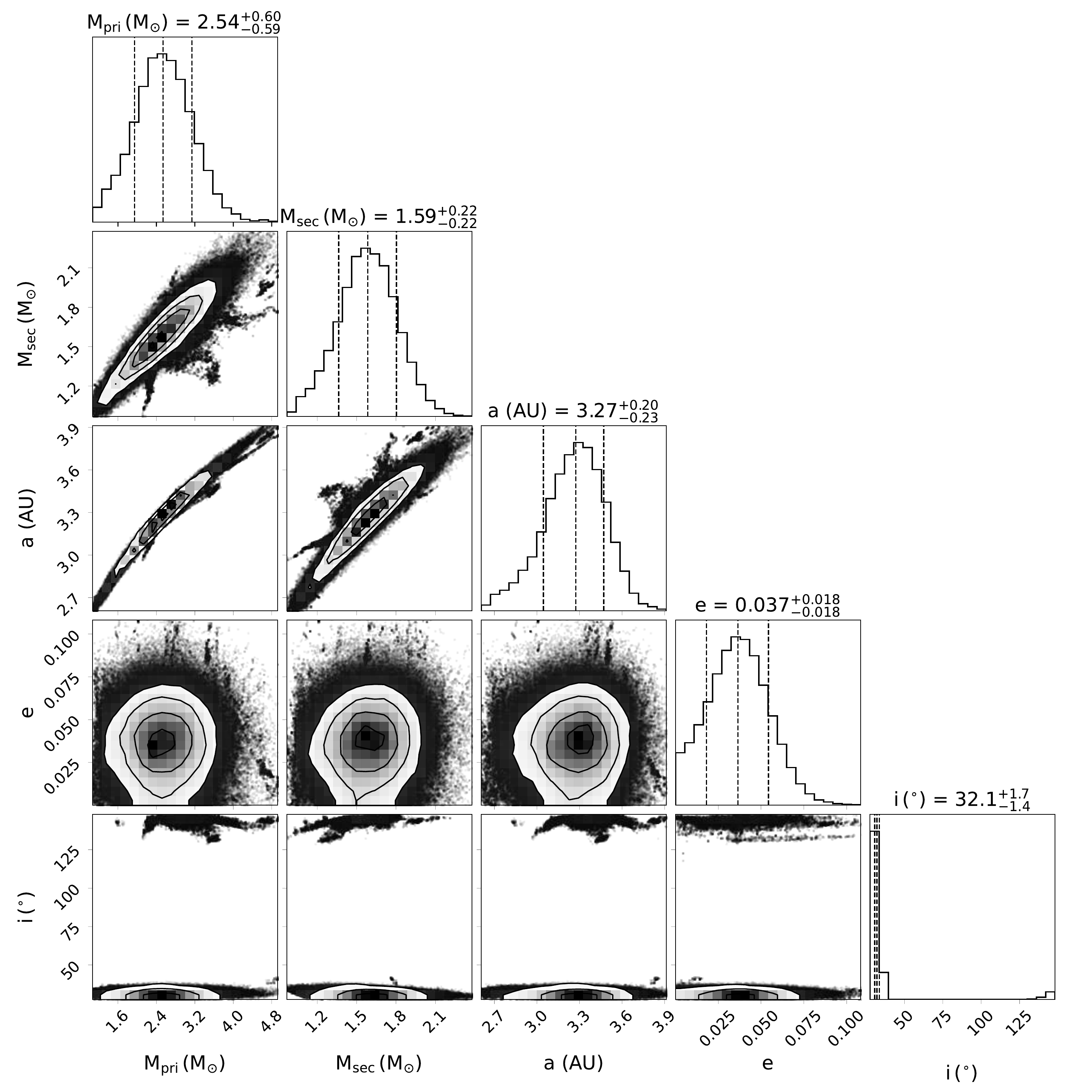}
\caption{\label{Fig:HD31487corner} Corner plot of HD\,31487}
\end{minipage} 
\hspace{3cm} 
\begin{minipage}[r]{6cm} 
\includegraphics[scale=0.3]{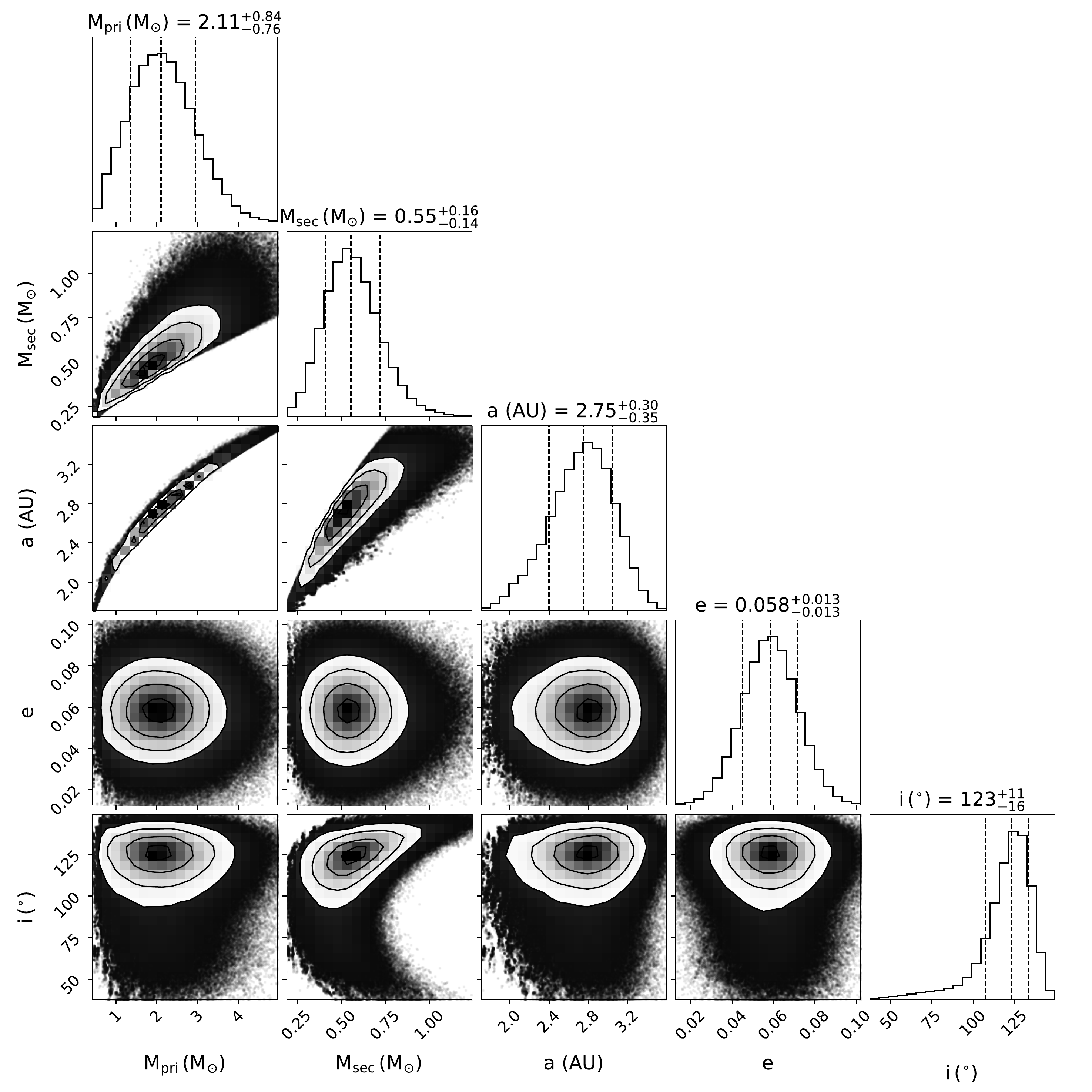}
\caption{\label{Fig:HD211594corner} Corner plot of HD\,211594}
\end{minipage} 
\end{figure}

\begin{figure*}[t]
\centering
\includegraphics[width=\textwidth]{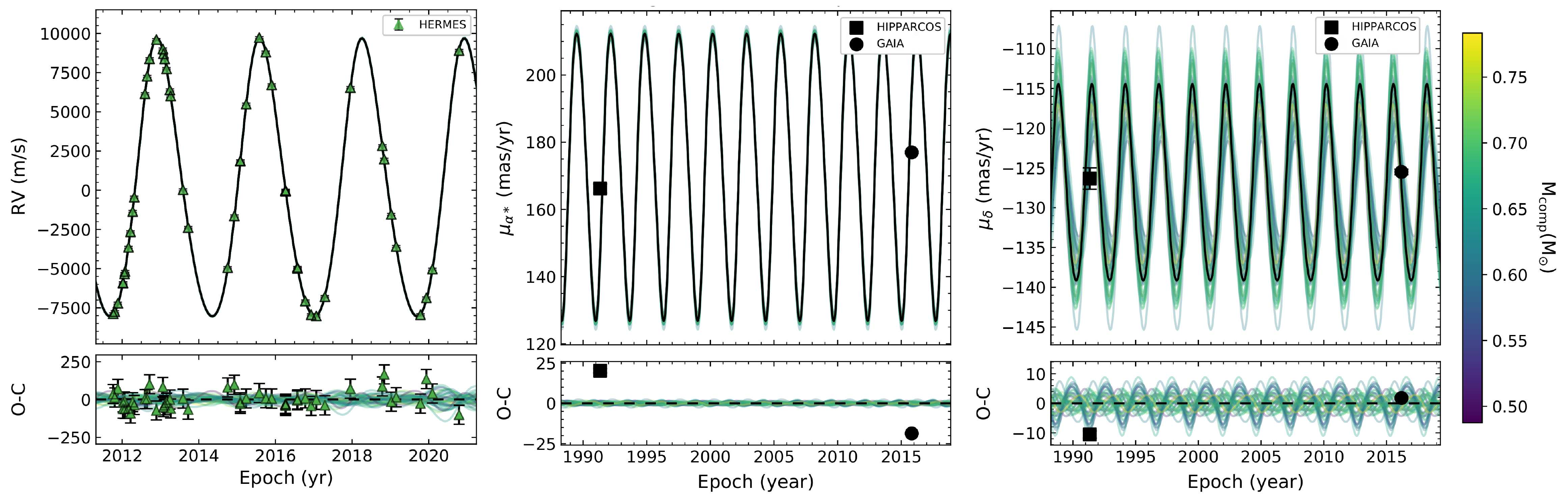}
\caption{\label{Fig:HD34654} RV curve and proper motions of HD\,34654}
\end{figure*}
\begin{figure*}
\centering
\includegraphics[width=\textwidth]{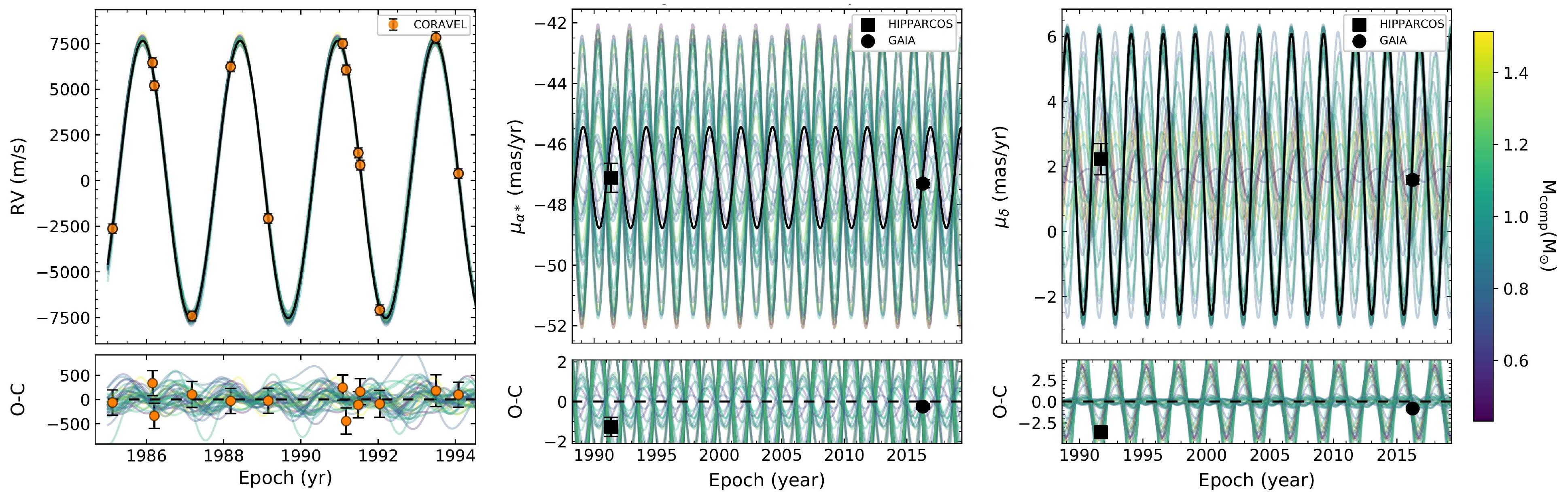}
\caption{\label{Fig:HD92626} RV curve and proper motions of HD\,92626}
\end{figure*}
\begin{figure}
\begin{minipage}[l]{6cm} 
\includegraphics[scale=0.3]{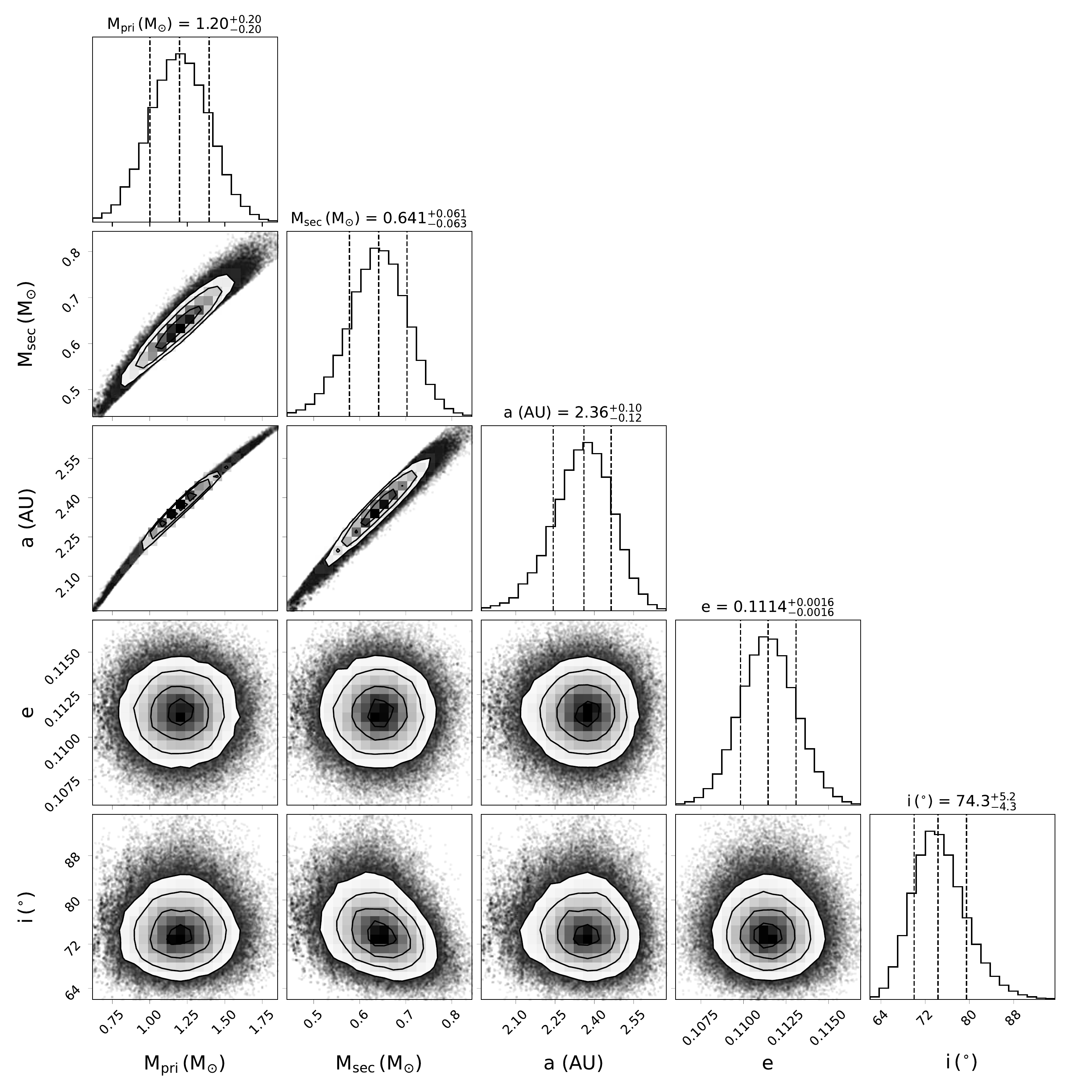}
\caption{\label{Fig:HD34654corner} Corner plot of HD\,34654}
\end{minipage} 
\hspace{3cm} 
\begin{minipage}[r]{6cm} 
\includegraphics[scale=0.3]{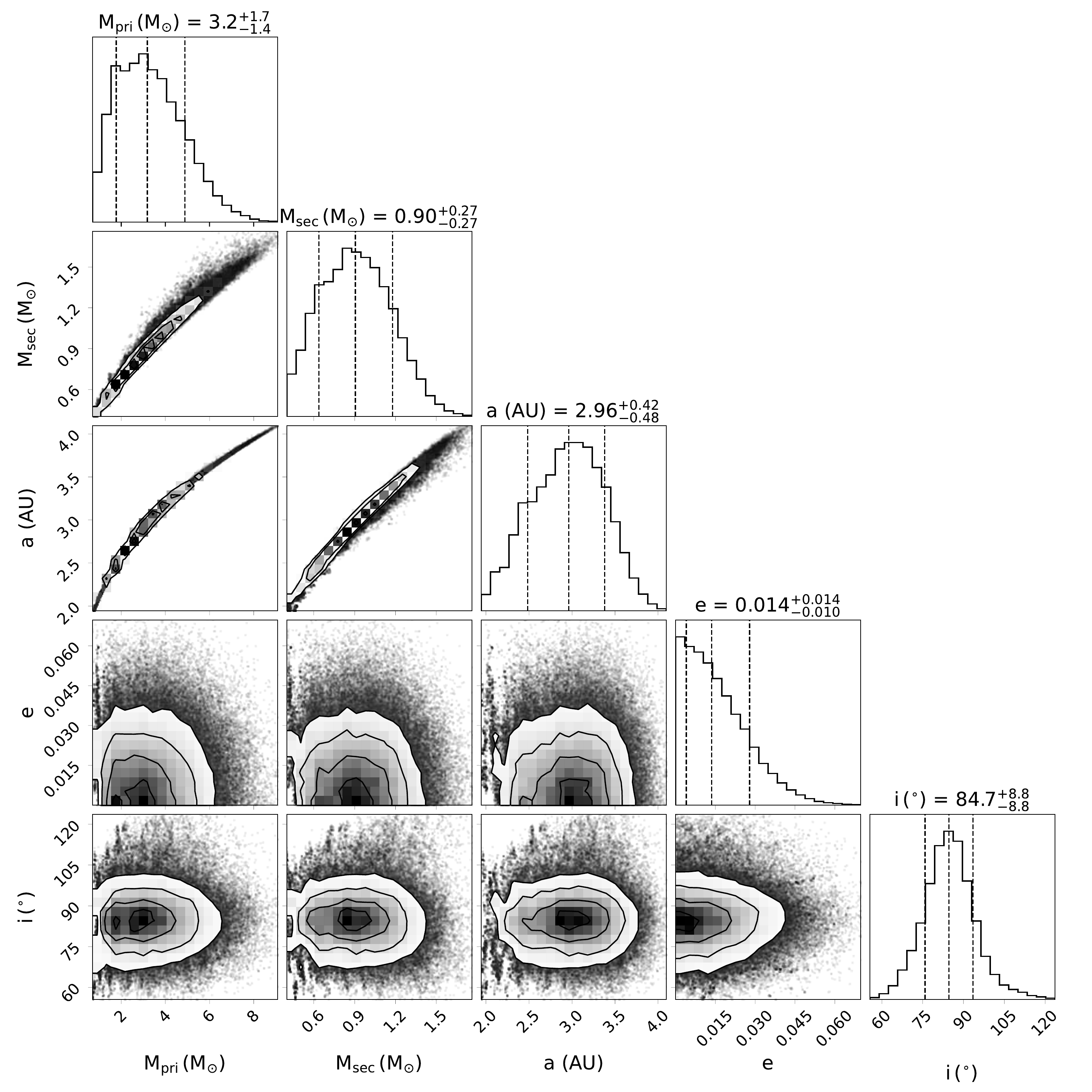}
\caption{\label{Fig:HD92626corner} Corner plot of HD\,92626}
\end{minipage} 
\end{figure}

\begin{figure*}[t]
\centering
\includegraphics[width=\textwidth]{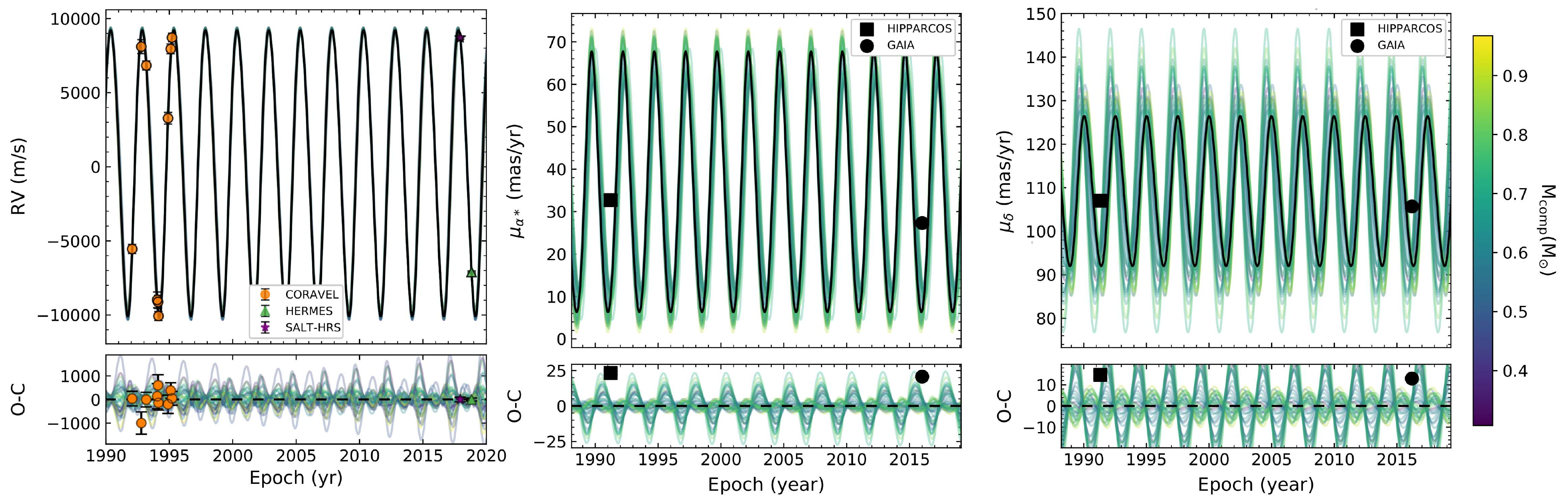}
\caption{\label{Fig:HD50264} RV curve and proper motions of HD\,50264.}
\end{figure*}
\begin{figure*}
\centering
\includegraphics[width=\textwidth]{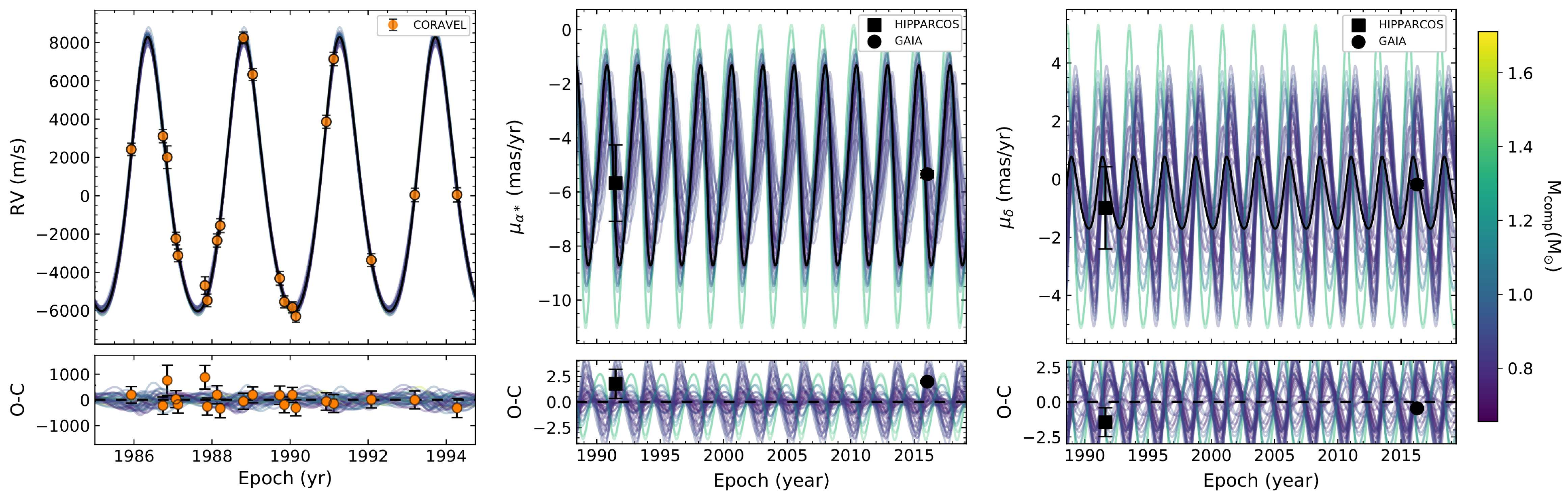}
\caption{\label{Fig:HD49841} RV curve and proper motions of HD\,49841}
\end{figure*}
\begin{figure}
\begin{minipage}[l]{6cm} 
\includegraphics[scale=0.3]{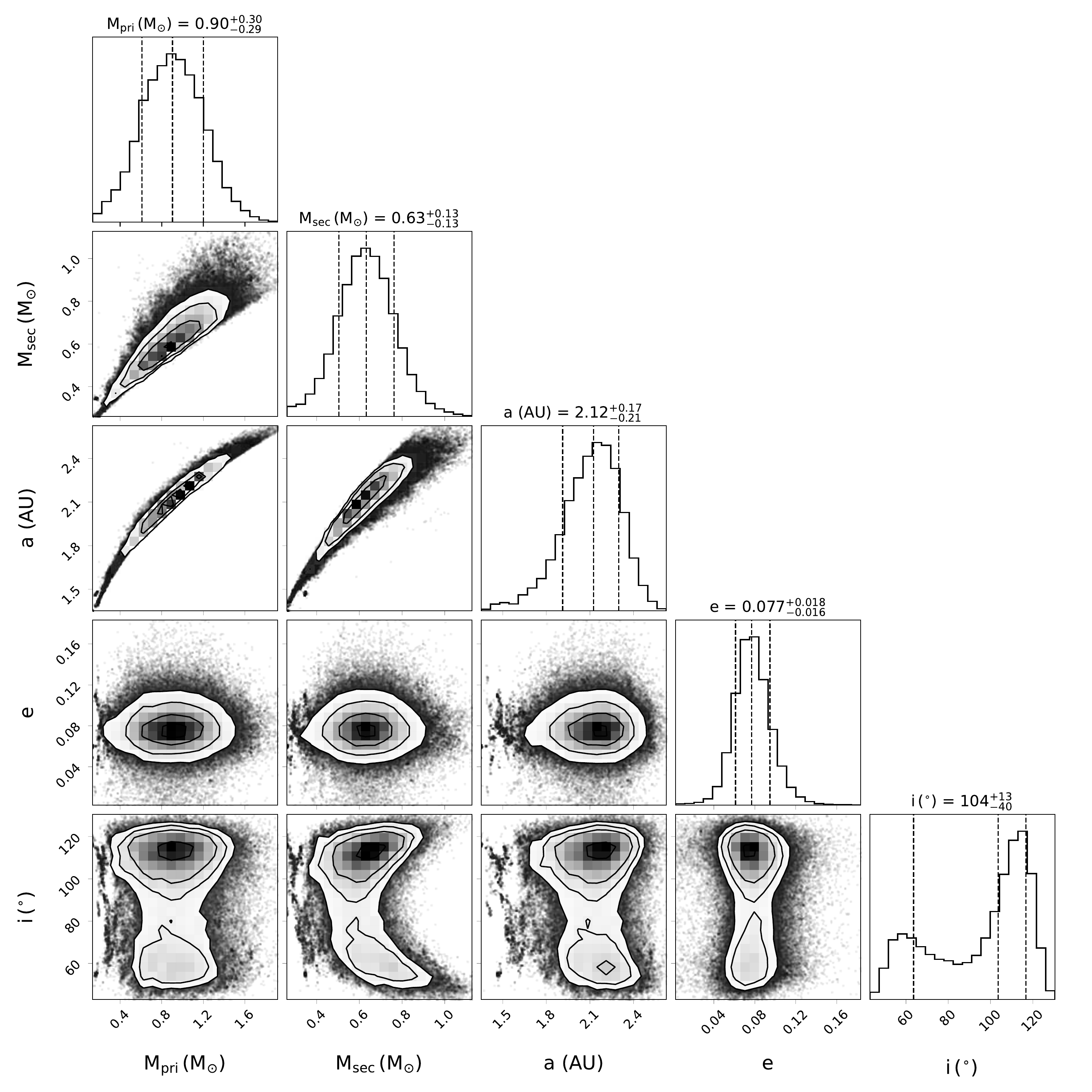}
\caption{\label{Fig:HD50264corner} Corner plot of HD\,50264}
\end{minipage} 
\hspace{3cm} 
\begin{minipage}[r]{6cm} 
\includegraphics[scale=0.3]{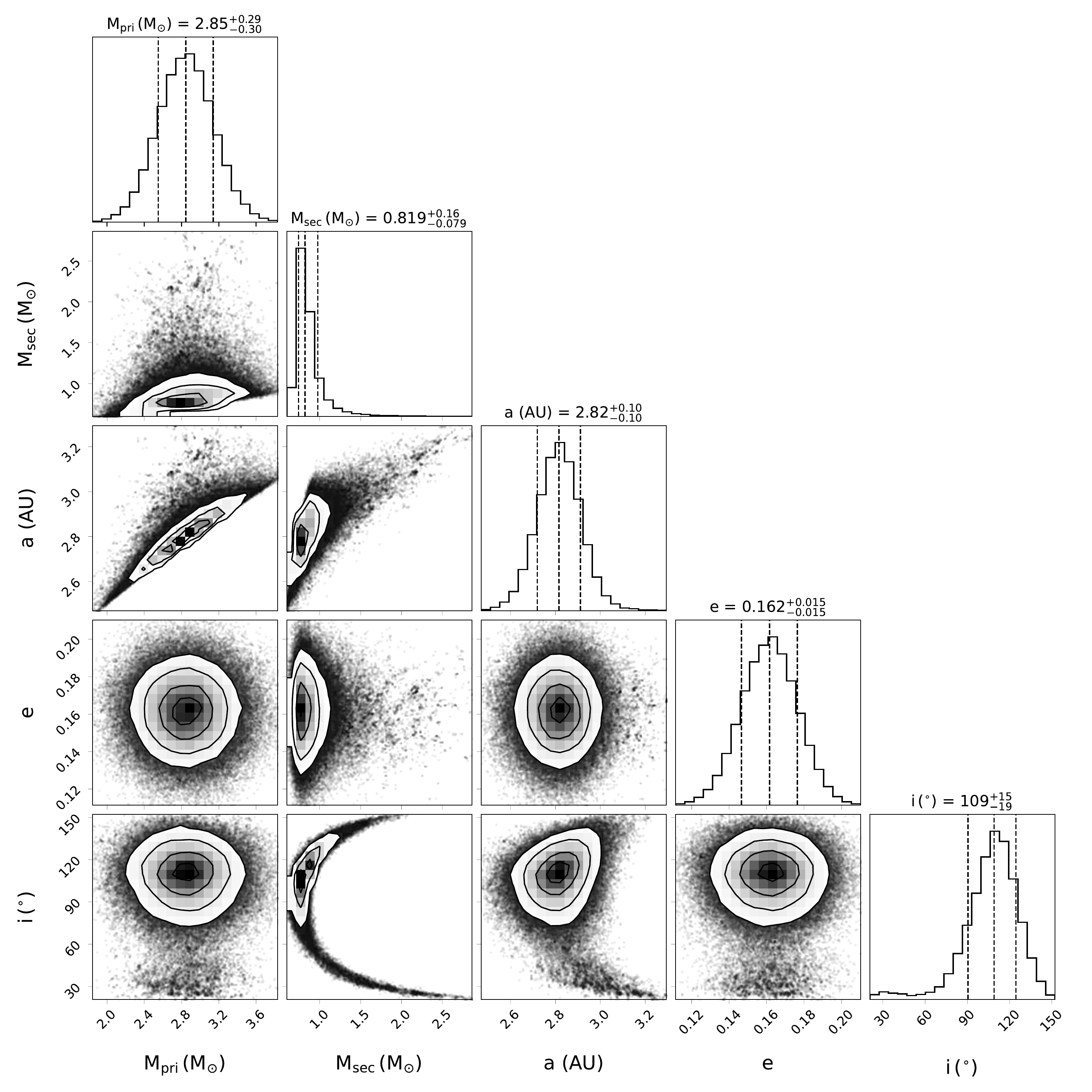}
\caption{\label{Fig:HD49841corner} Corner plot of HD\,49841}
\end{minipage} 
\end{figure}

\begin{figure*}[t]
\centering
\includegraphics[width=\textwidth]{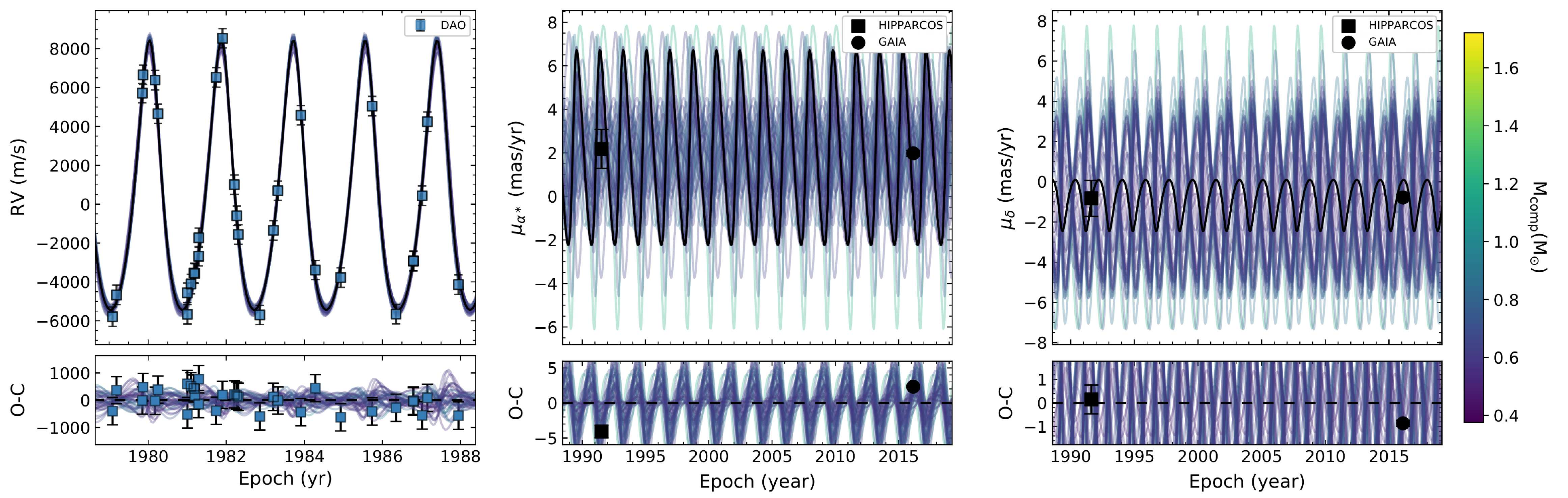}
\caption{\label{Fig:HD58368} RV curve and proper motions of HD\,58368}
\end{figure*}
\begin{figure*}
\centering
\includegraphics[width=\textwidth]{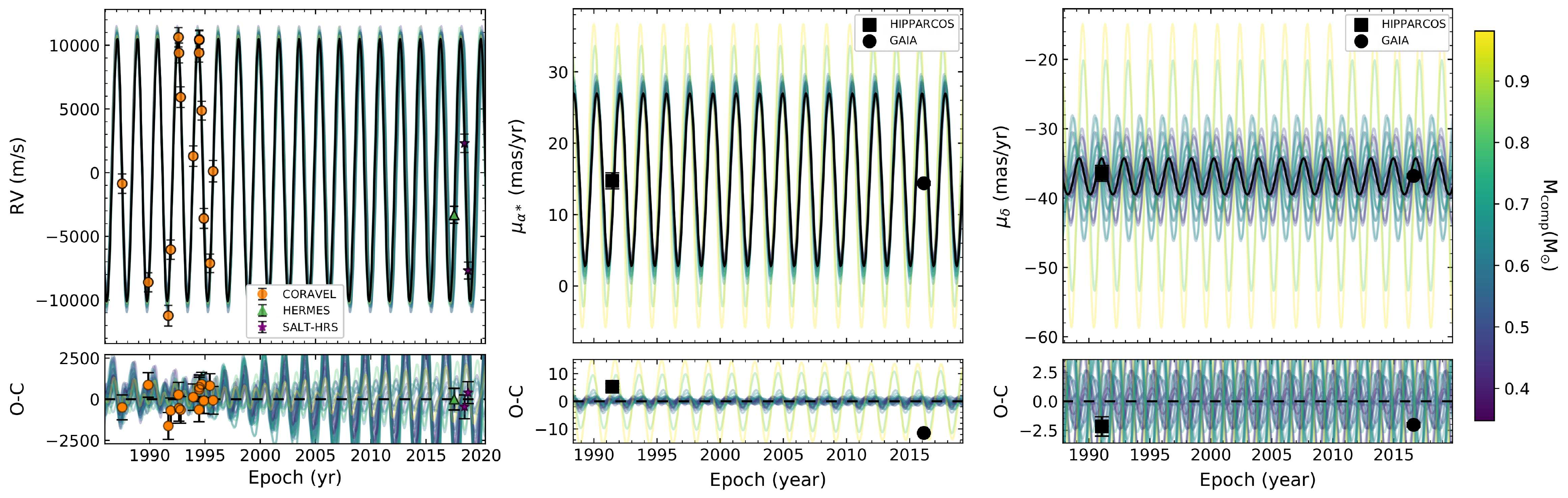}
\caption{\label{Fig:HD207585} RV curve and proper motions of HD\,207585}
\end{figure*}
\begin{figure}
\begin{minipage}[l]{6cm} 
\includegraphics[scale=0.3]{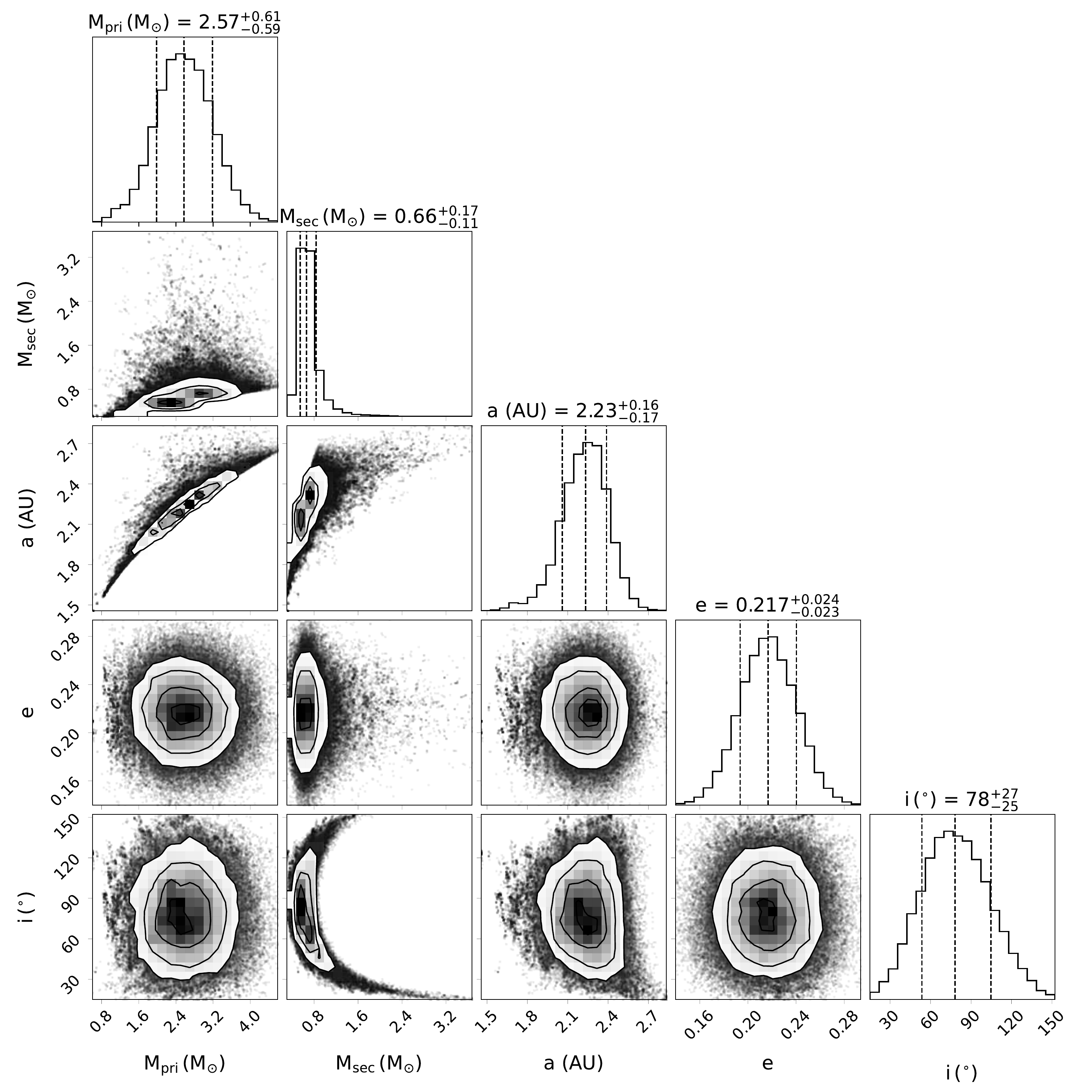}
\caption{\label{Fig:HD58368corner} Corner plot of HD\,58368}
\end{minipage} 
\hspace{3cm} 
\begin{minipage}[r]{6cm} 
\includegraphics[scale=0.3]{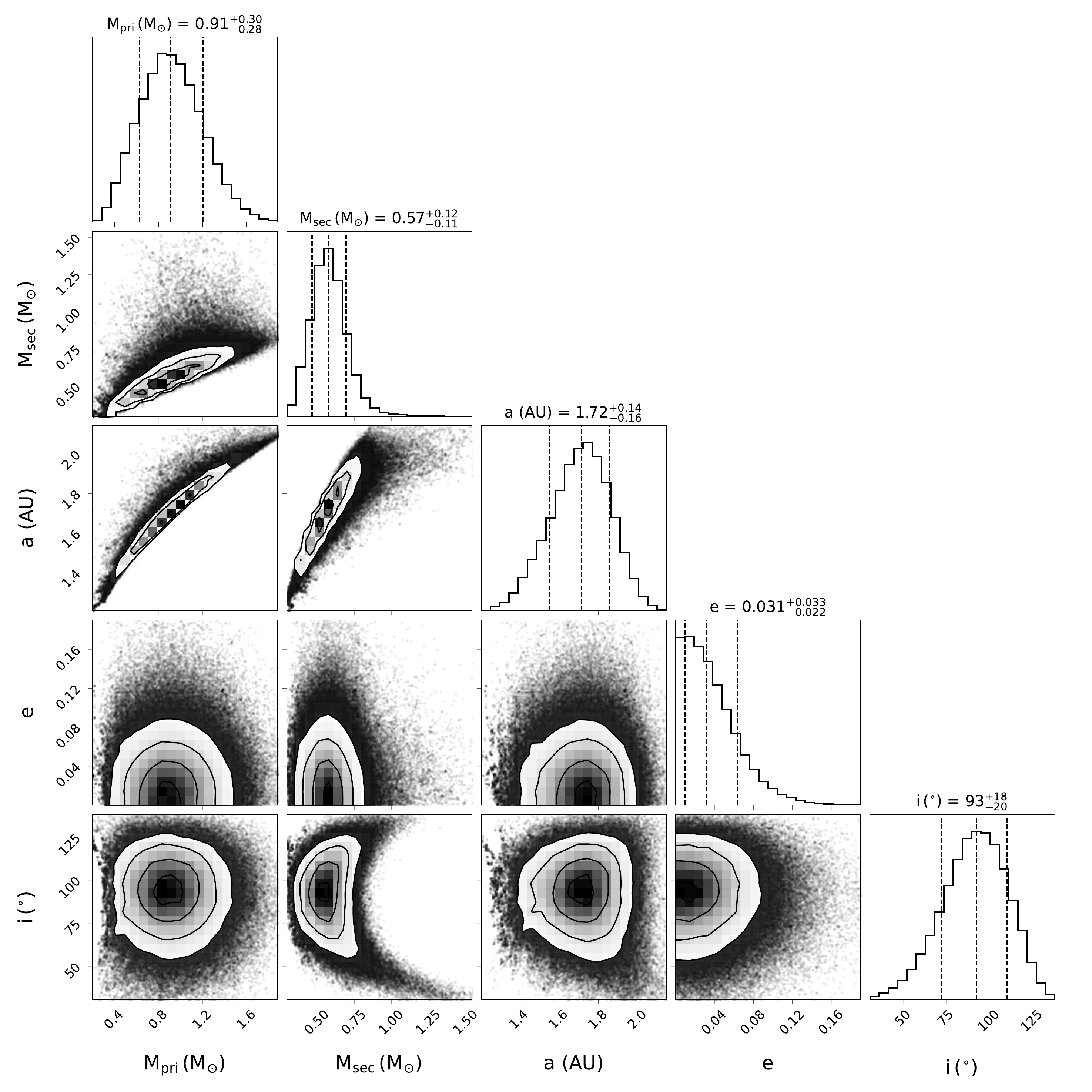}
\caption{\label{Fig:HD207585corner} Corner plot of HD\,207585}
\end{minipage} 
\end{figure}

\begin{figure*}[t]
\centering
\includegraphics[width=\textwidth]{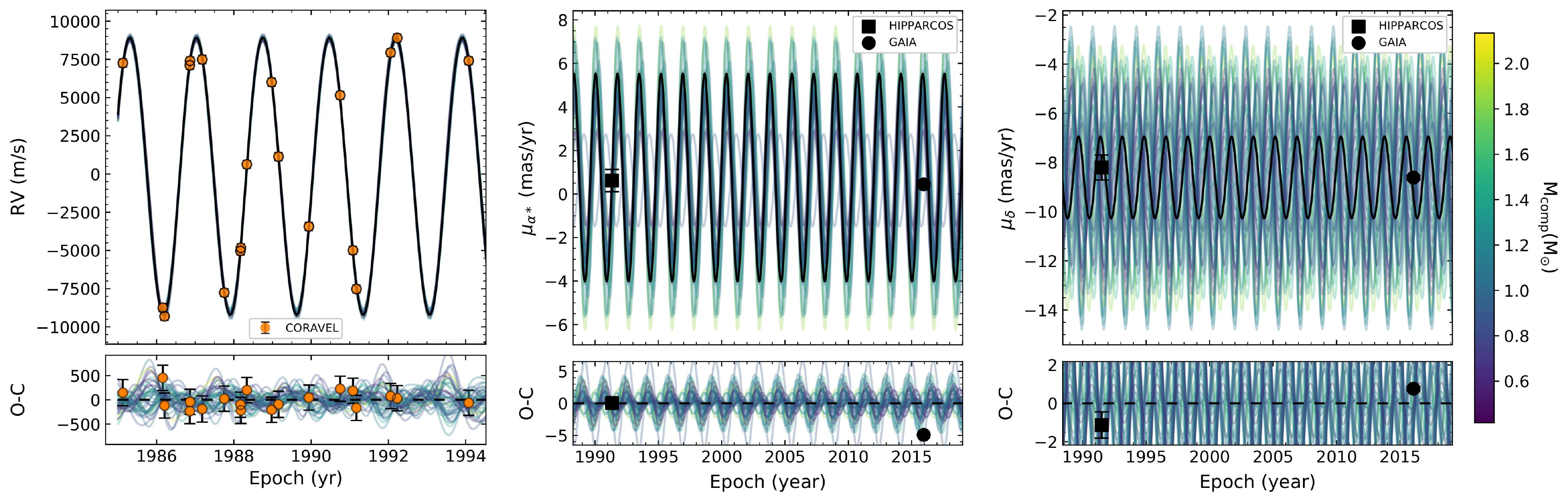}
\caption{\label{Fig:HD44896} RV curve and proper motions of HD\,44896}
\end{figure*}
\begin{figure*}
\centering
\includegraphics[width=\textwidth]{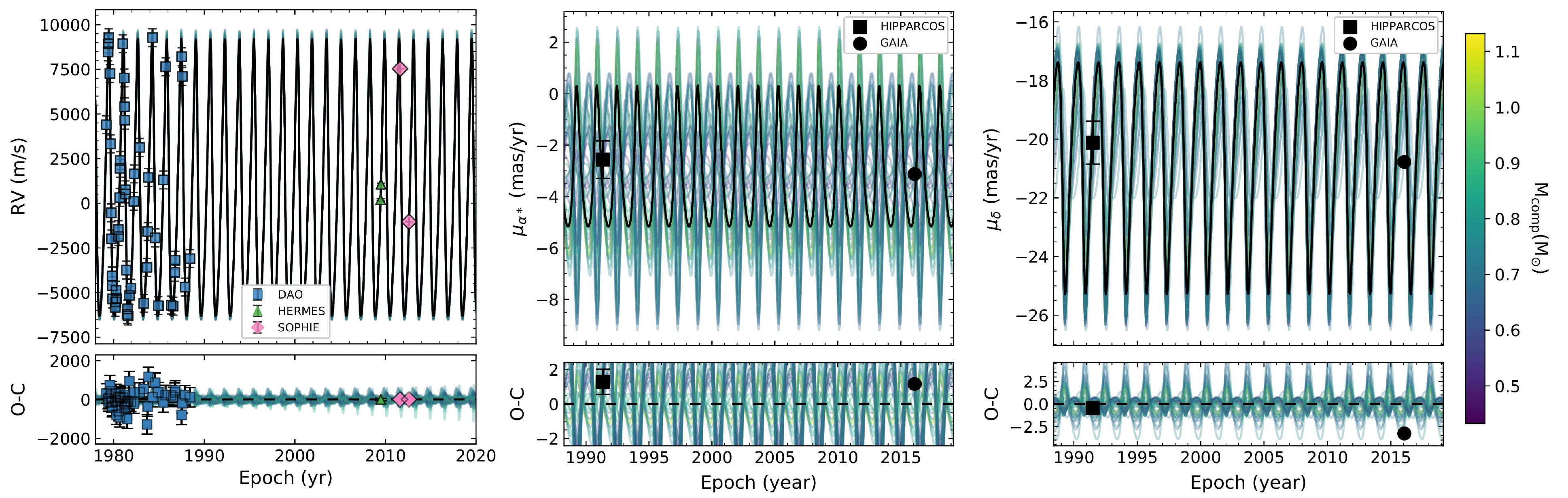}
\caption{\label{Fig:HD199939} RV curve and proper motions of HD\,199939}
\end{figure*}
\begin{figure}
\begin{minipage}[l]{6cm} 
\includegraphics[scale=0.3]{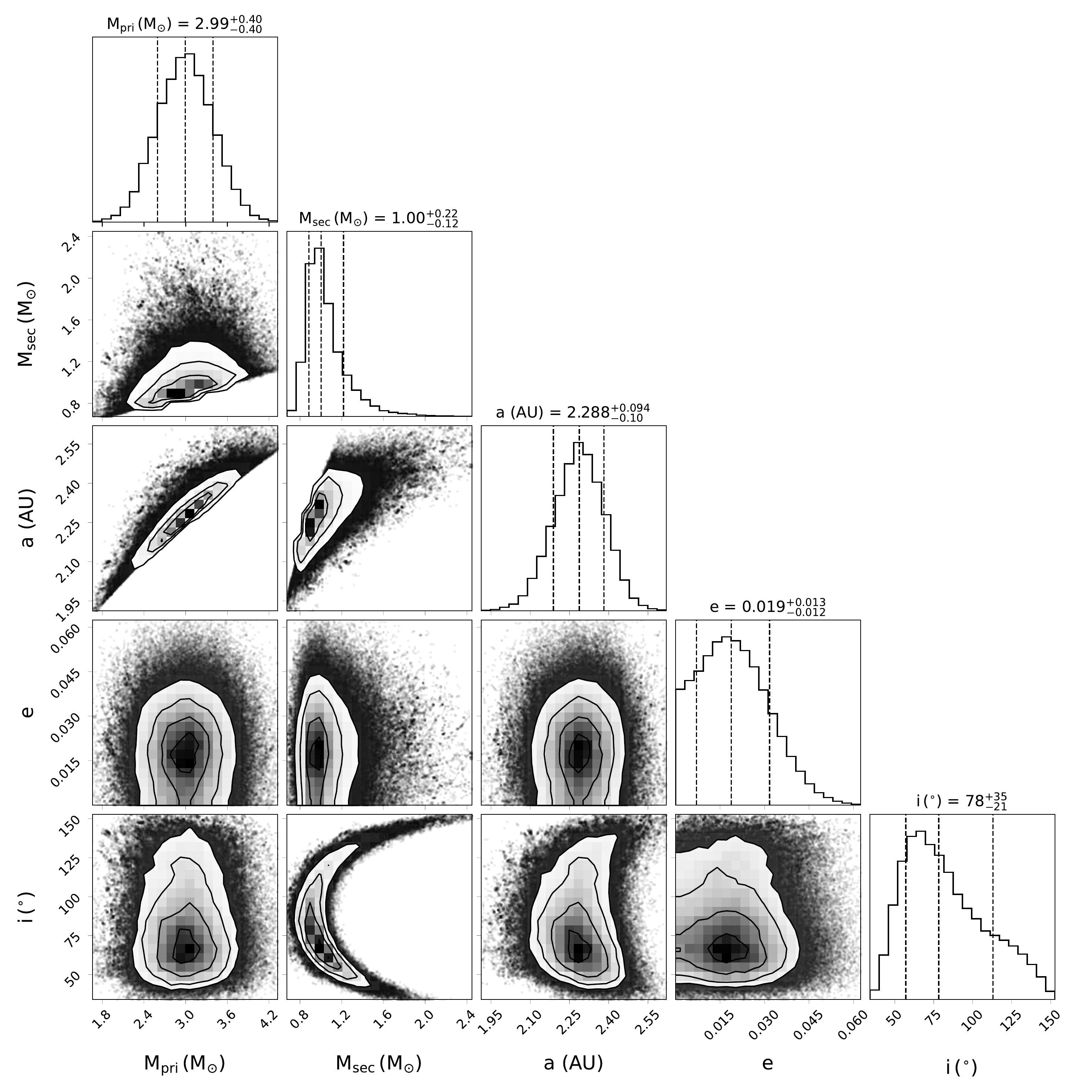}
\caption{\label{Fig:HD44896corner} Corner plot of HD\,44896}
\end{minipage} 
\hspace{3cm} 
\begin{minipage}[r]{6cm} 
\includegraphics[scale=0.3]{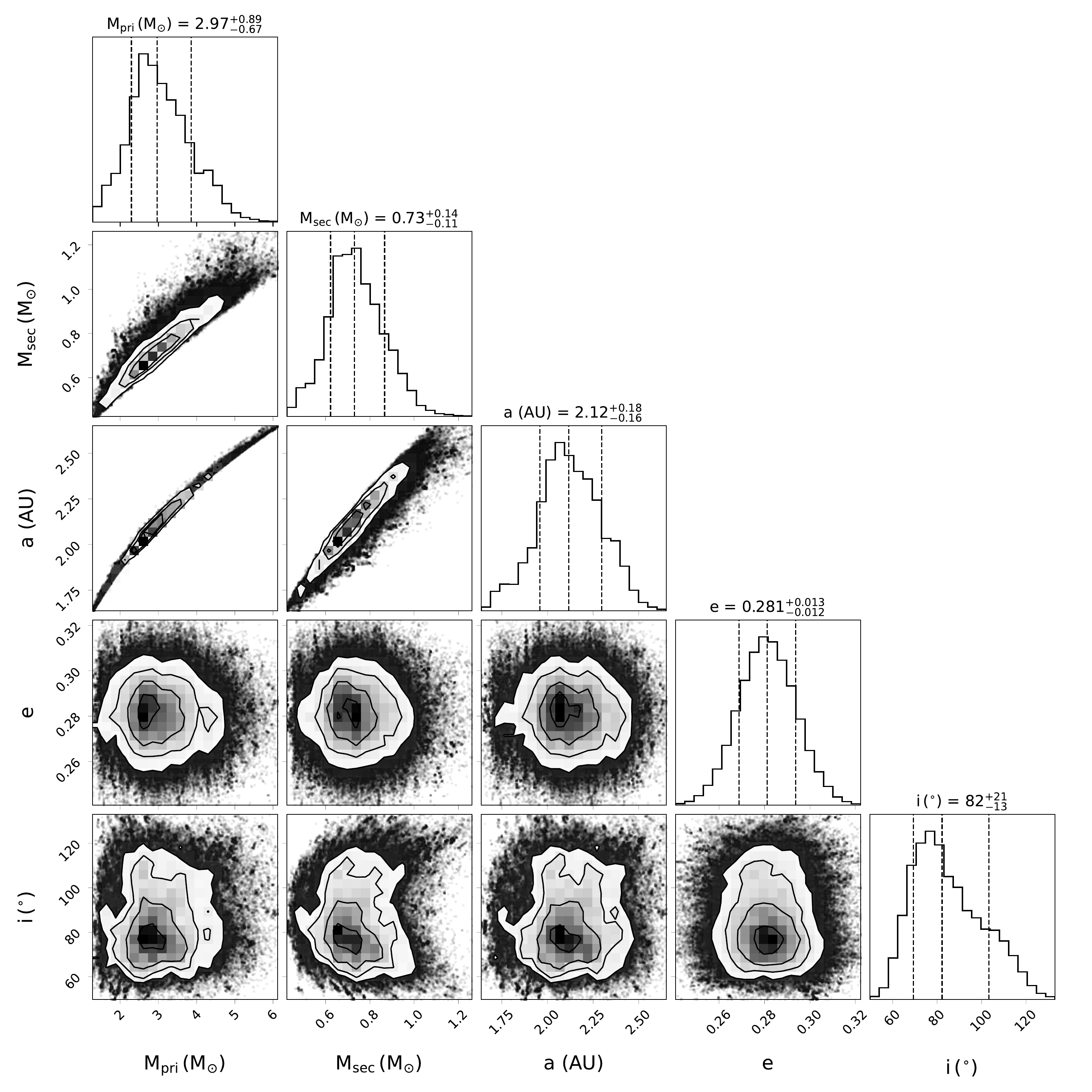}
\caption{\label{Fig:HD199939corner} Corner plot of HD\,199939}
\end{minipage} 
\end{figure}

\begin{figure*}[t]
\centering
\includegraphics[width=\textwidth]{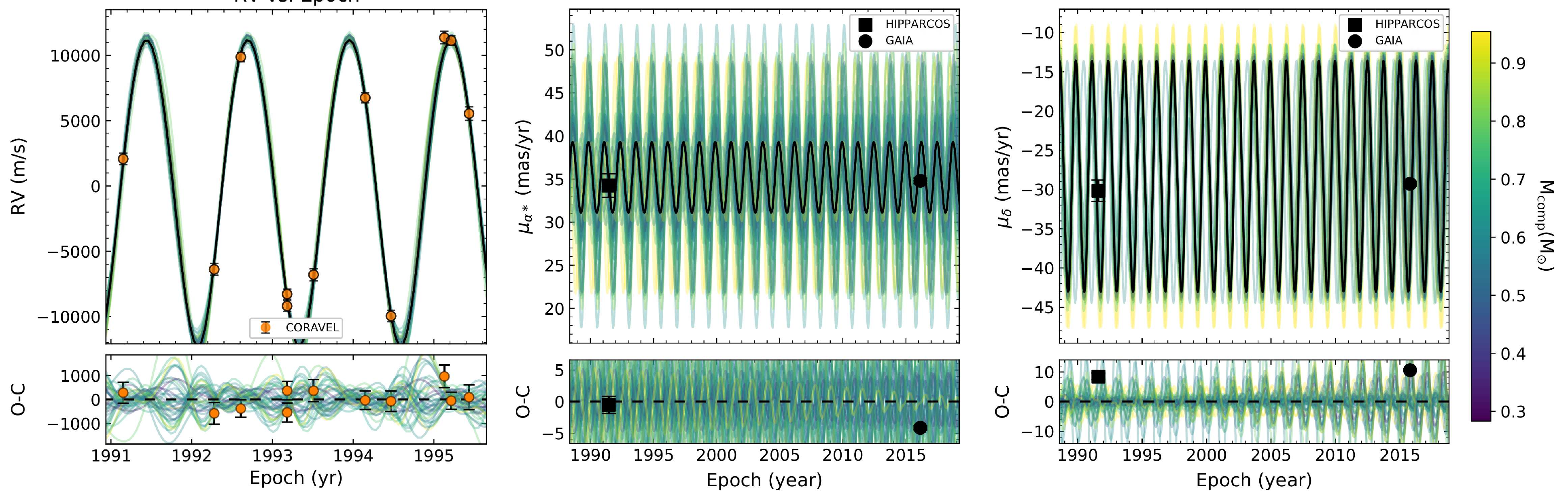}
\caption{\label{Fig:HD123585} RV curve and proper motions of HD\,123585}
\end{figure*}
\begin{figure*}
\centering
\includegraphics[width=\textwidth]{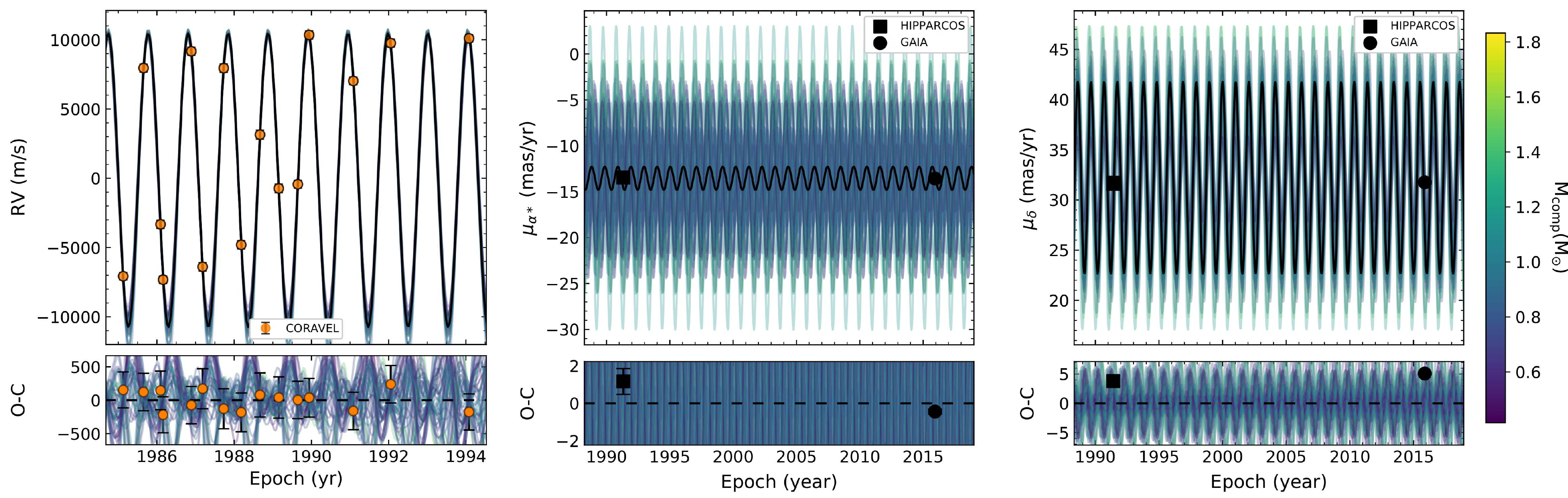}
\caption{\label{Fig:HD24035} RV curve and proper motions of HD\,24035}
\end{figure*}
\begin{figure}
\begin{minipage}[l]{6cm} 
\includegraphics[scale=0.3]{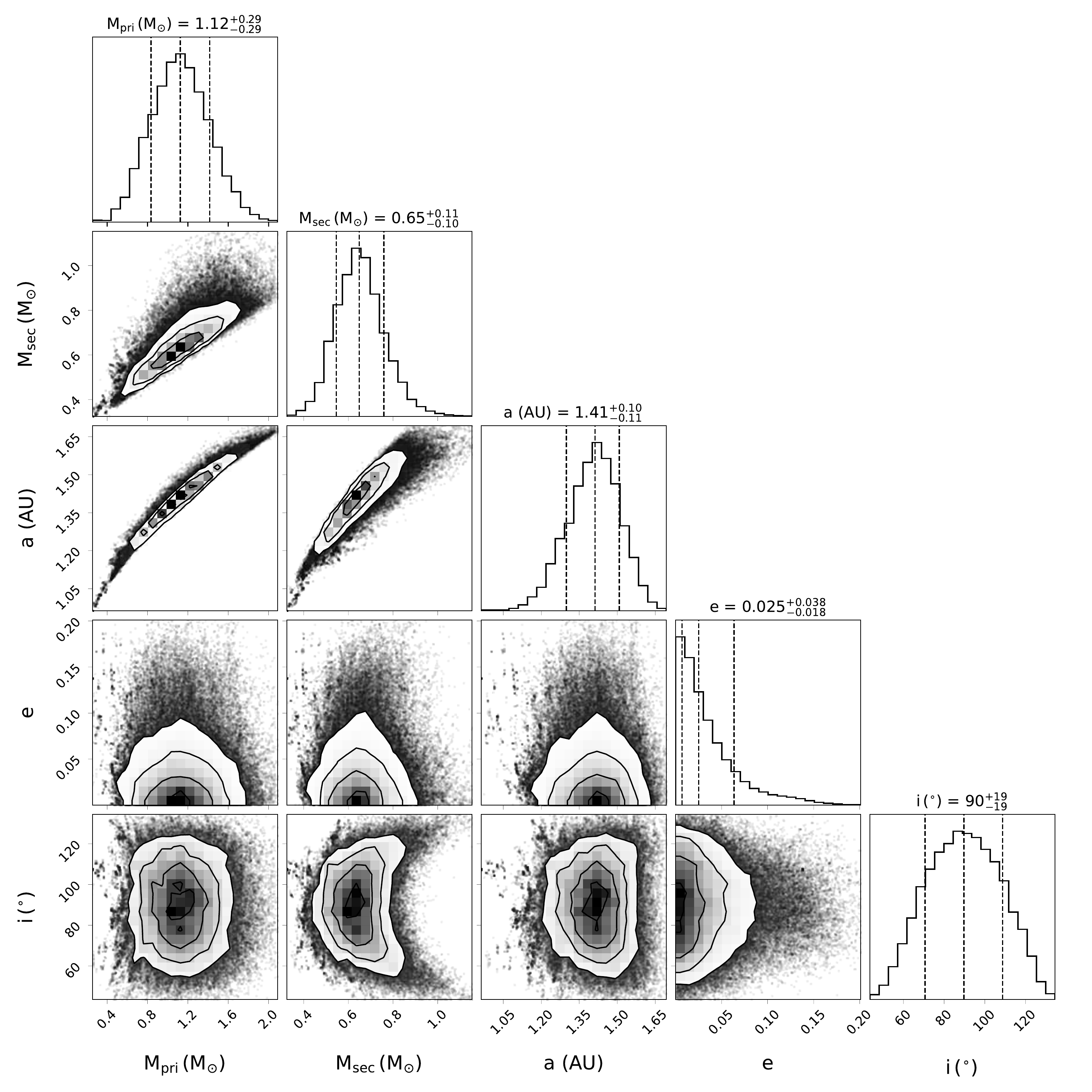}
\caption{\label{Fig:HD123585corner} Corner plot of HD\,123585}
\end{minipage} 
\hspace{3cm} 
\begin{minipage}[r]{6cm} 
\includegraphics[scale=0.3]{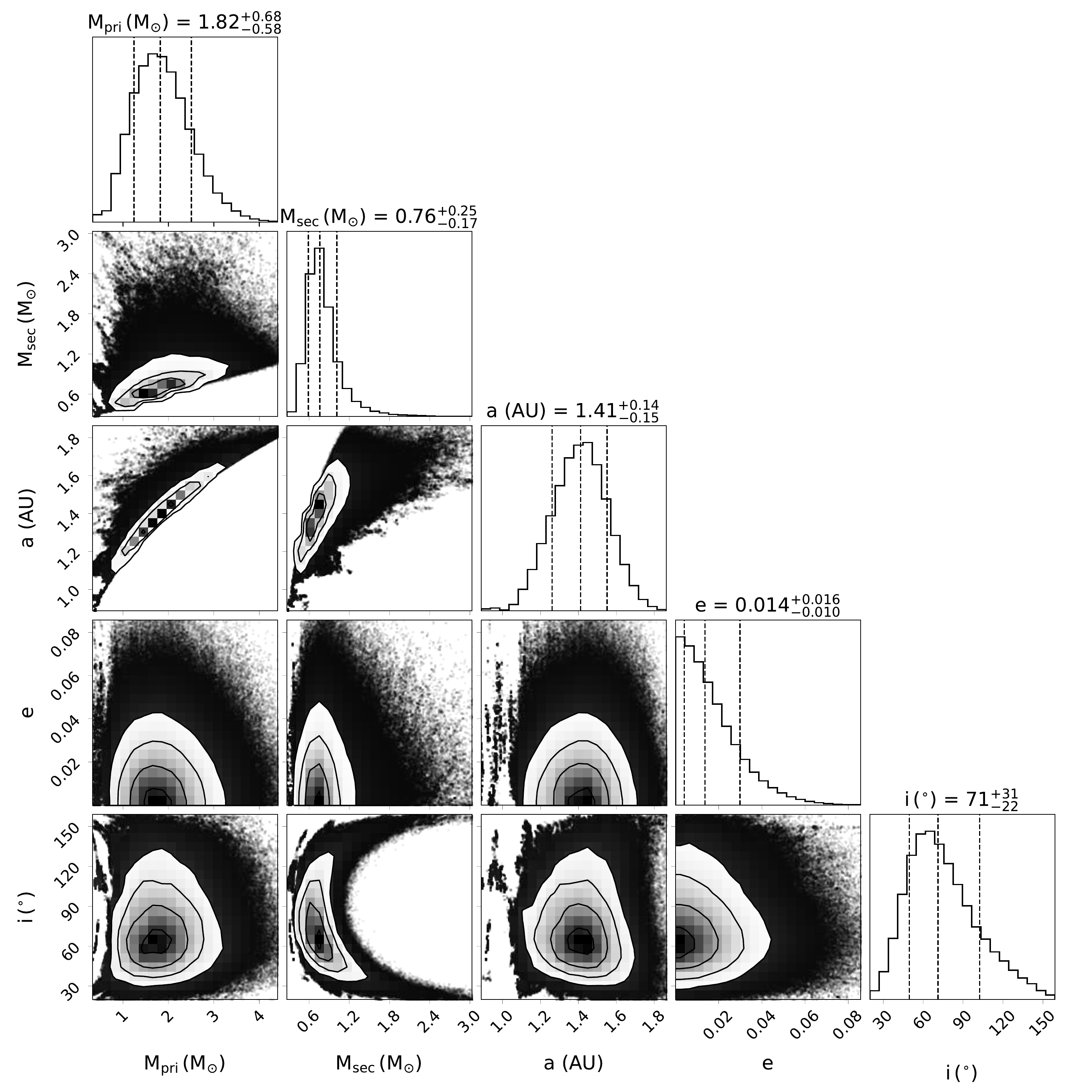}
\caption{\label{Fig:HD24035corner} Corner plot of HD\,24035}
\end{minipage} 
\end{figure}

\begin{figure*}[t]
\centering
\includegraphics[width=\textwidth]{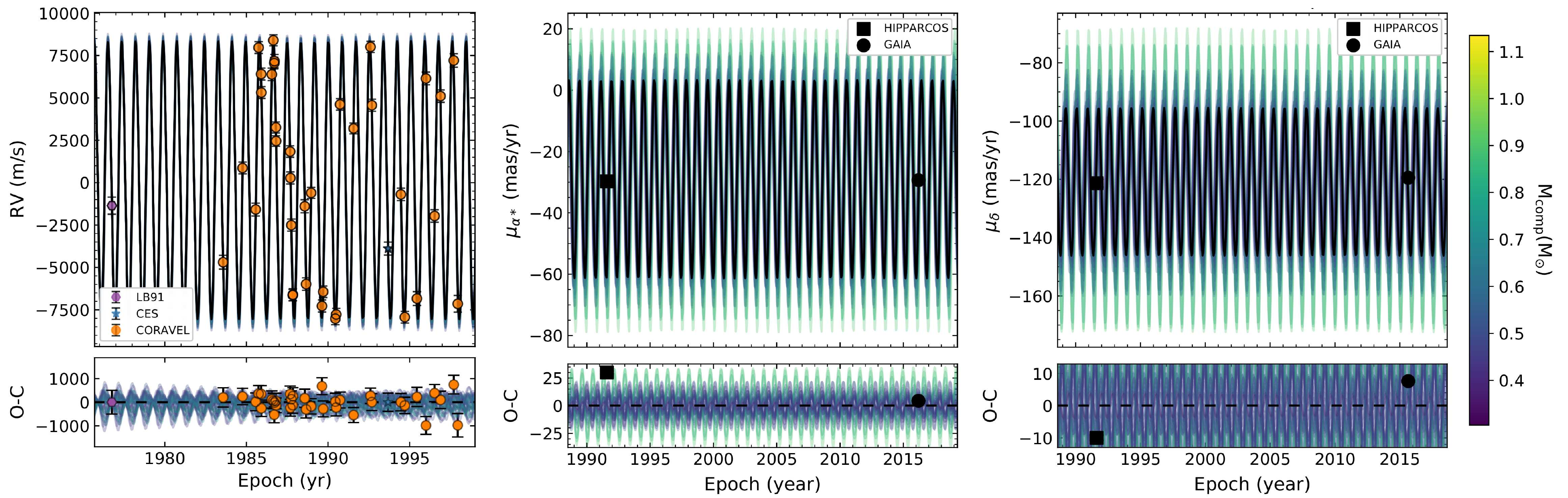}
\caption{\label{Fig:HD224621} RV curve and proper motions of HD\,224621}
\end{figure*}
\begin{figure*}
\centering
\includegraphics[width=\textwidth]{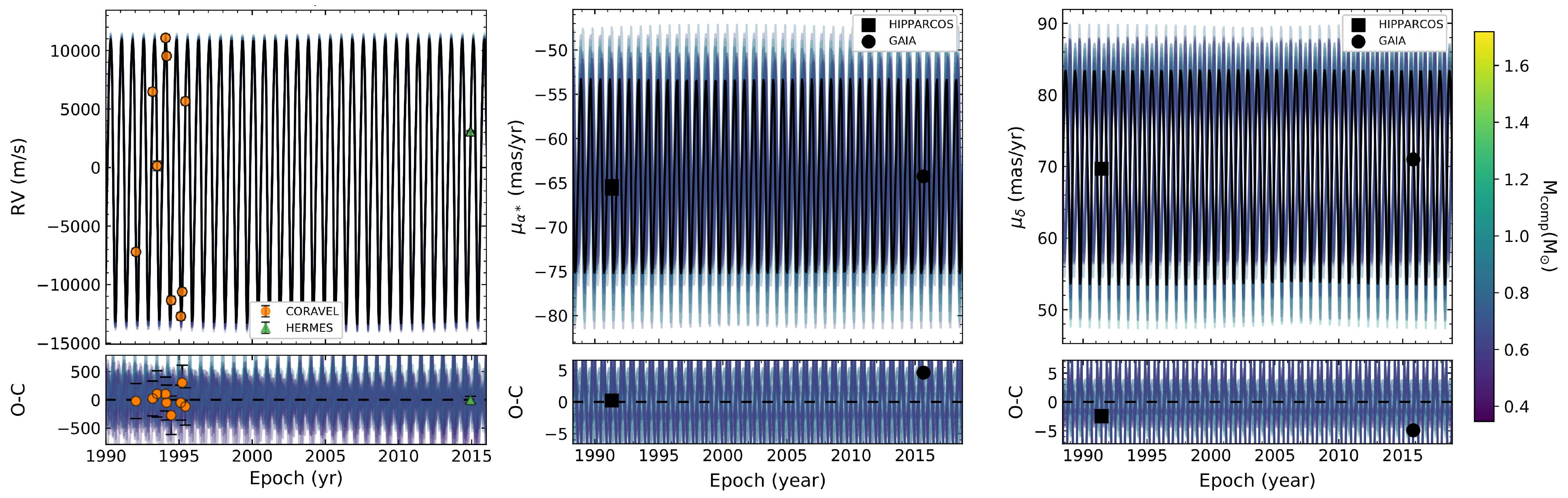}
\caption{\label{Fig:HD87080} RV curve and proper motions of HD\,87080. We used a fixed RV offset of 248 m/s \citep{Escorza19}.}
\end{figure*}
\begin{figure}
\begin{minipage}[l]{6cm} 
\includegraphics[scale=0.3]{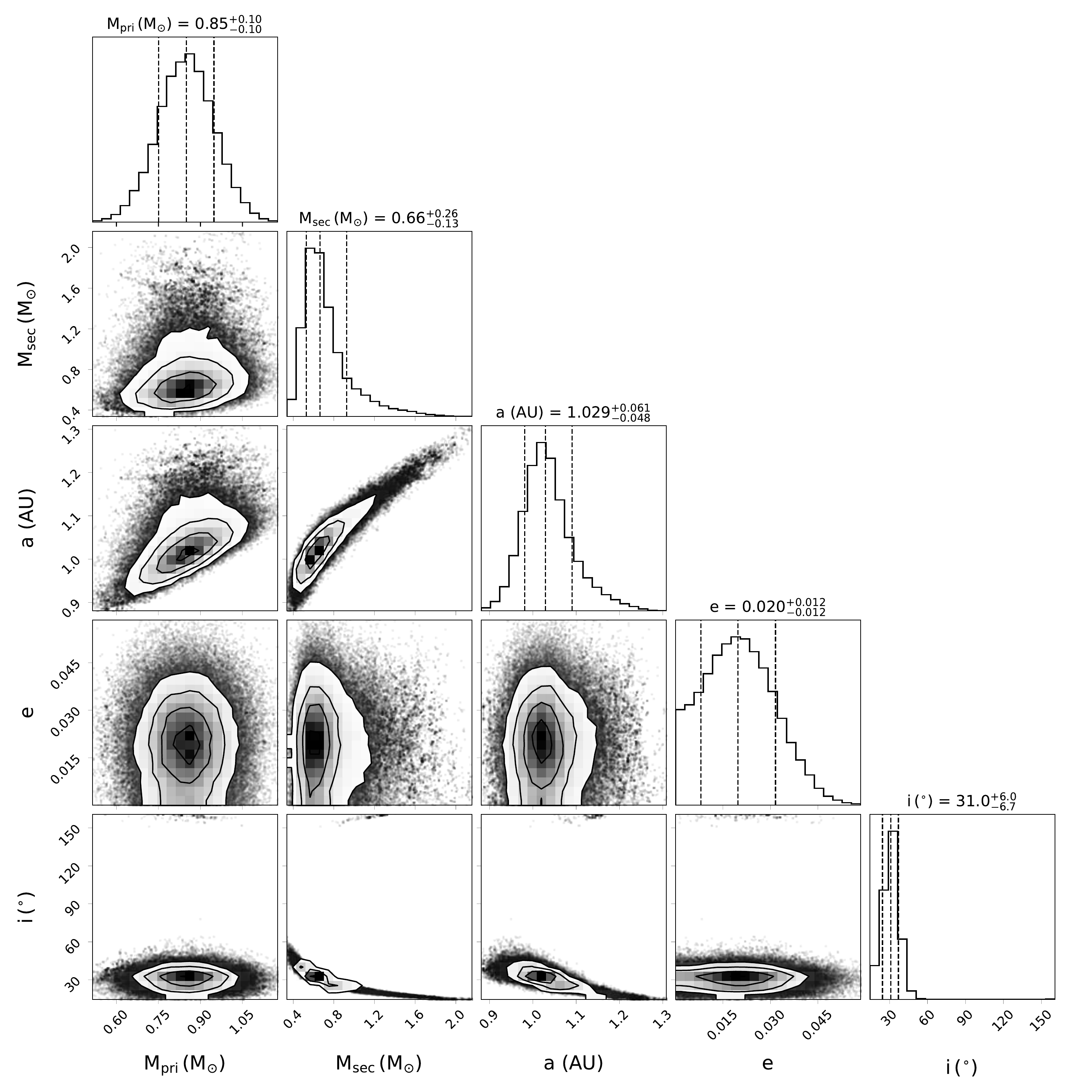}
\caption{\label{Fig:HD224621corner} Corner plot of HD\,224621}
\end{minipage} 
\hspace{3cm} 
\begin{minipage}[r]{6cm} 
\includegraphics[scale=0.3]{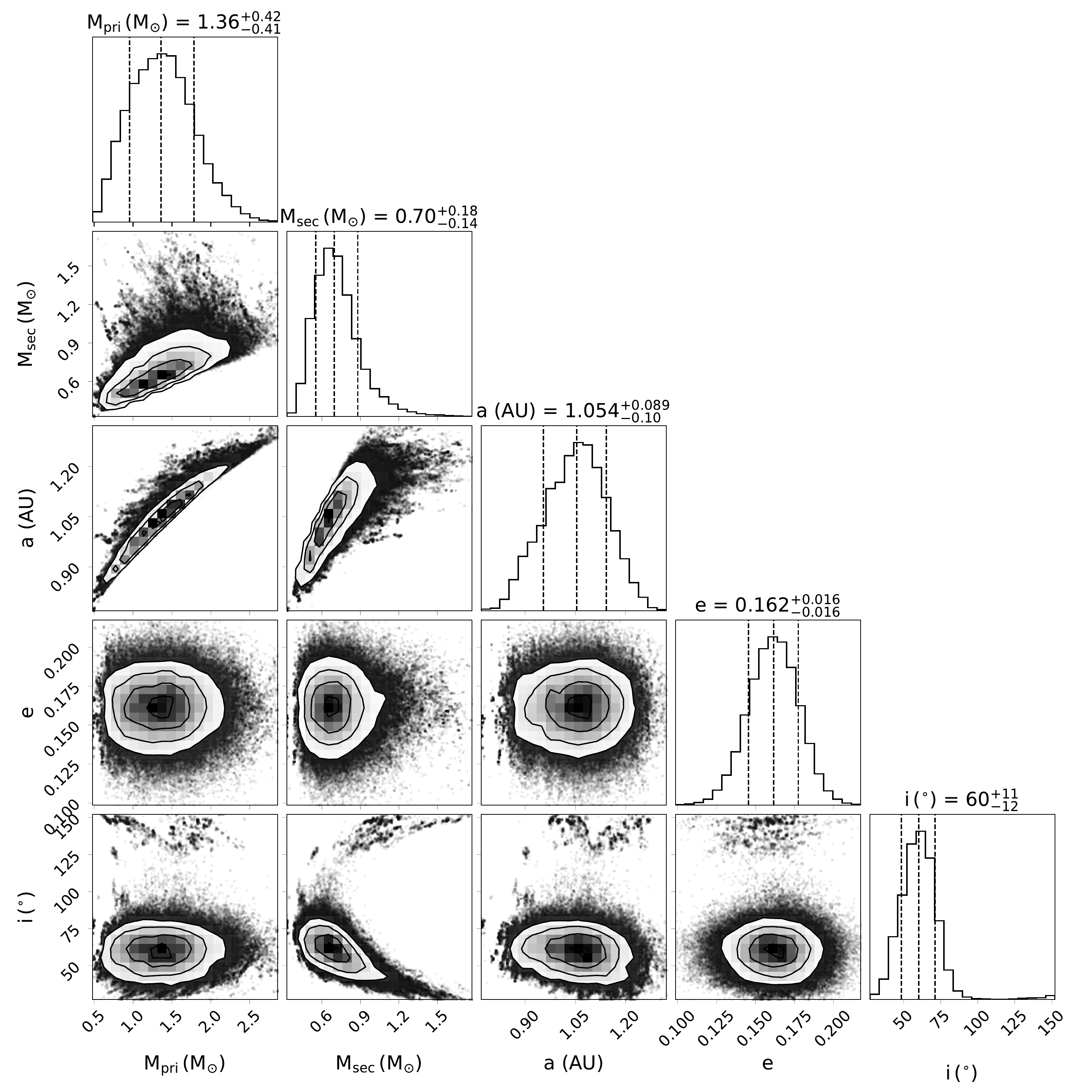}
\caption{\label{Fig:HD87080corner} Corner plot of HD\,87080}
\end{minipage} 
\end{figure}

\begin{figure*}[t]
\centering
\includegraphics[width=\textwidth]{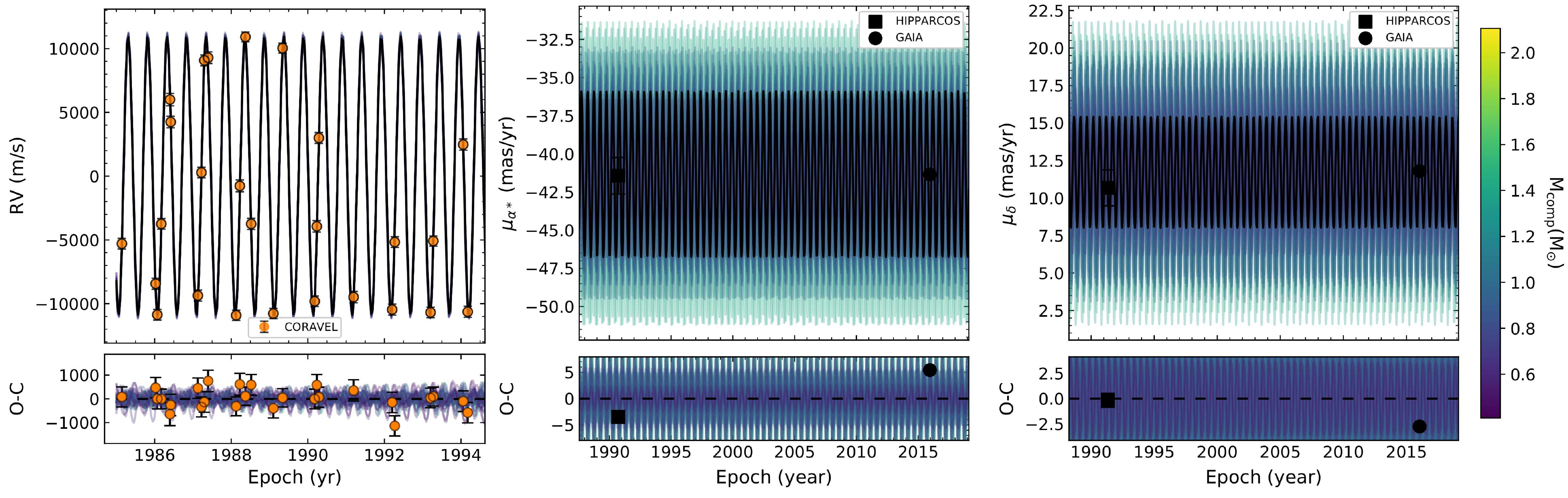}
\caption{\label{Fig:HD121447} RV curve and proper motions of HD\,121447}
\end{figure*}
\begin{figure*}
\centering
\includegraphics[width=\textwidth]{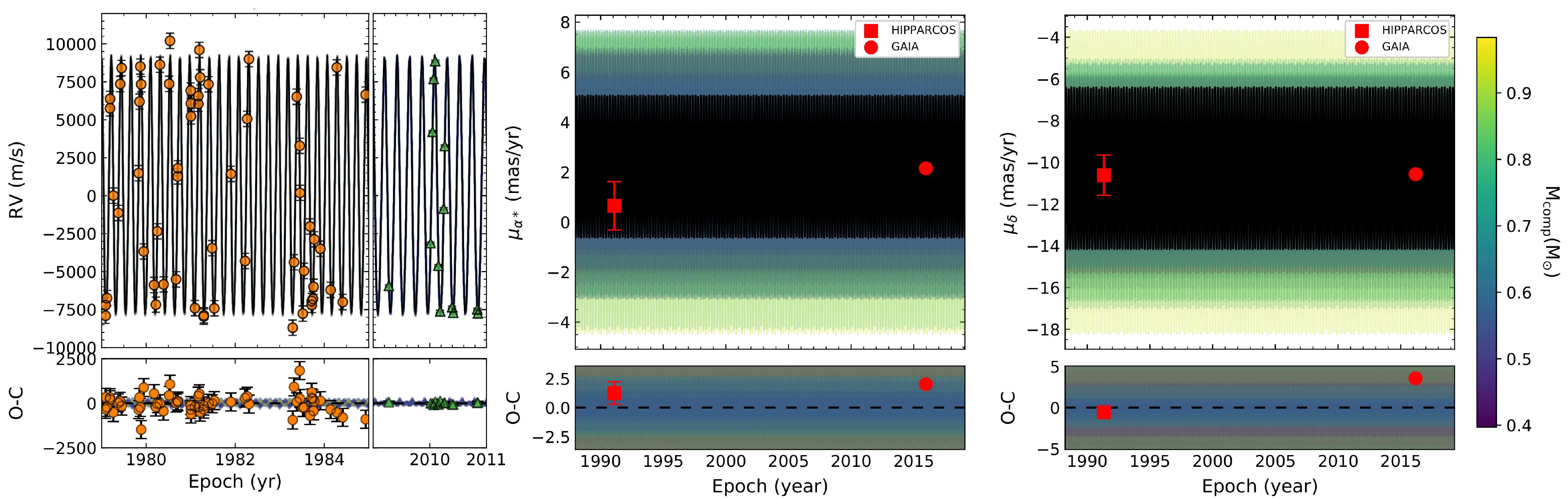}
\caption{\label{Fig:HD77247} RV curve and proper motions of HD\,77247}
\end{figure*}
\begin{figure}
\begin{minipage}[l]{6cm} 
\includegraphics[scale=0.3]{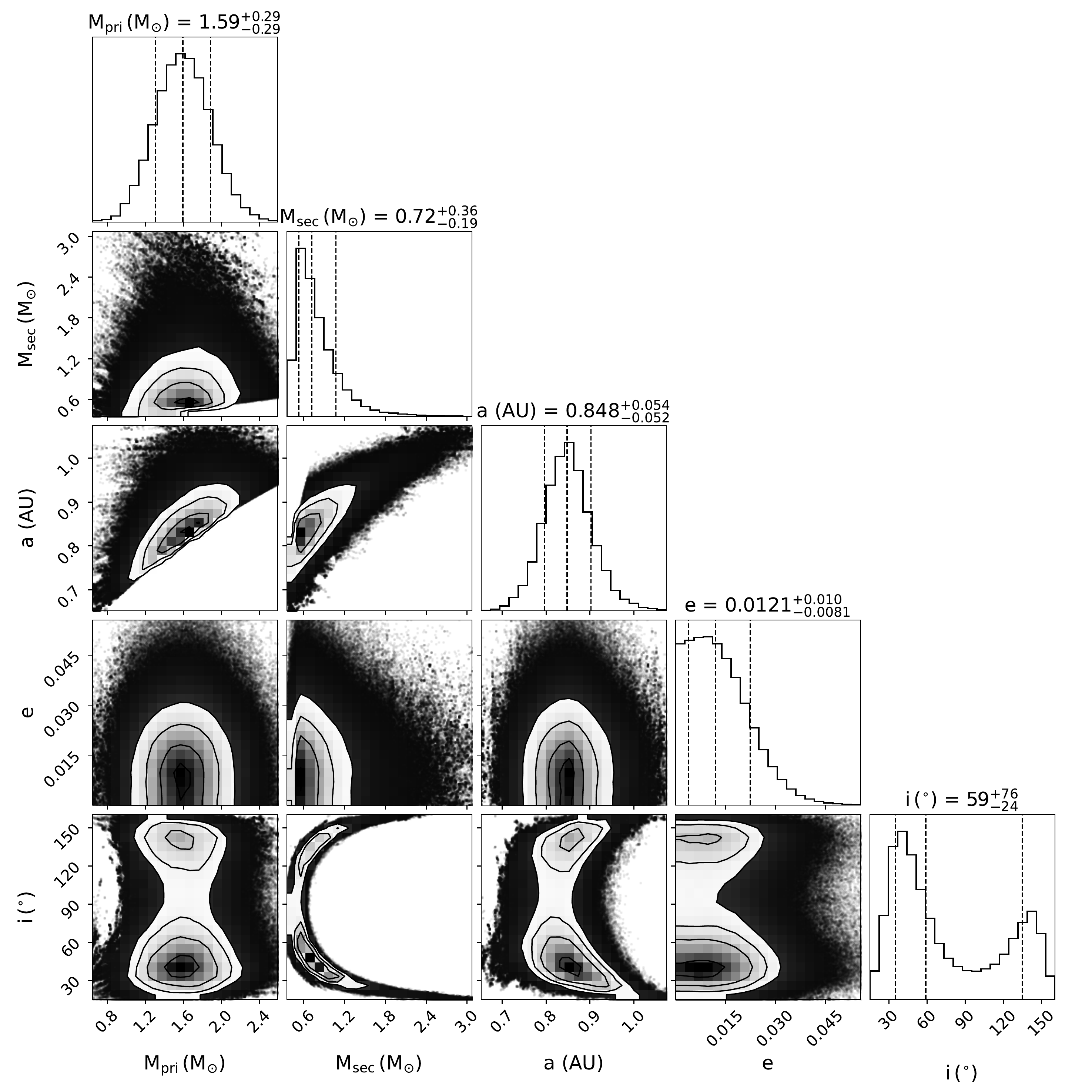}
\caption{\label{Fig:HD121447corner} Corner plot of HD\,121447}
\end{minipage} 
\hspace{3cm} 
\begin{minipage}[r]{6cm} 
\includegraphics[scale=0.3]{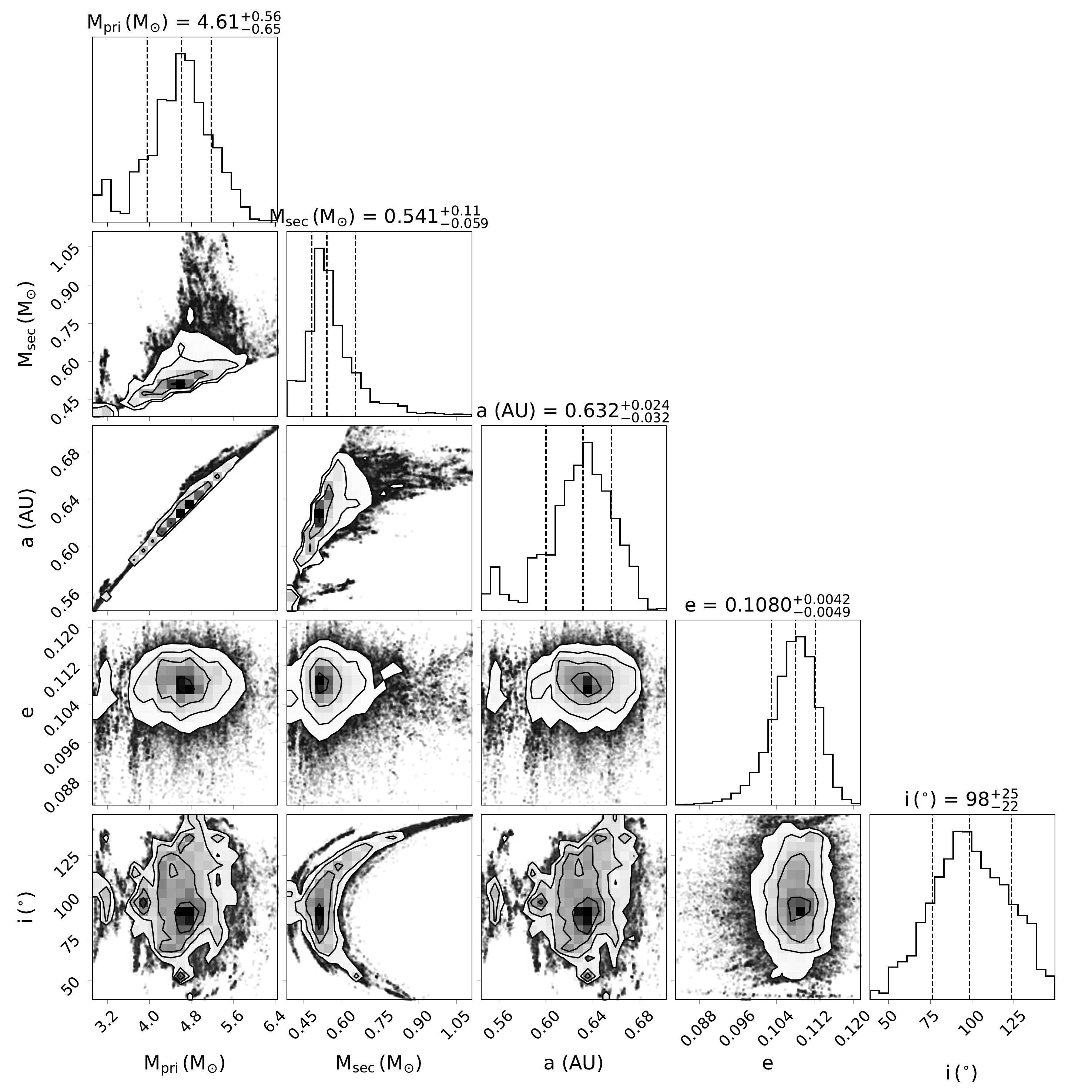}
\caption{\label{Fig:HD77247corner} Corner plot of HD\,77247}
\end{minipage} 
\end{figure}

\FloatBarrier
\section{Corner plots of HD\,218356}\label{AppHD218356}

\begin{figure*}[h!]
\centering
\includegraphics[width=0.49\textwidth]{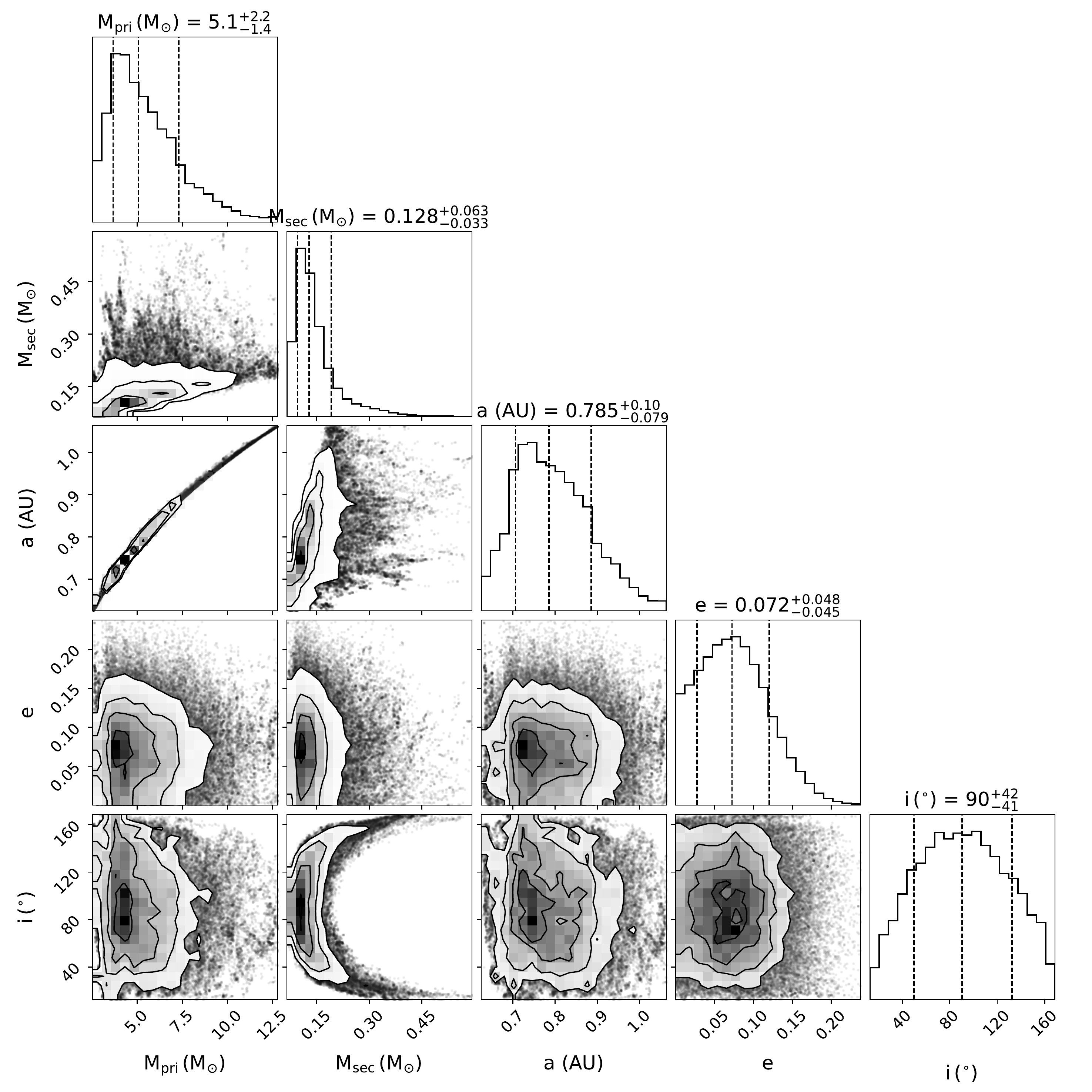}
\includegraphics[width=0.49\textwidth]{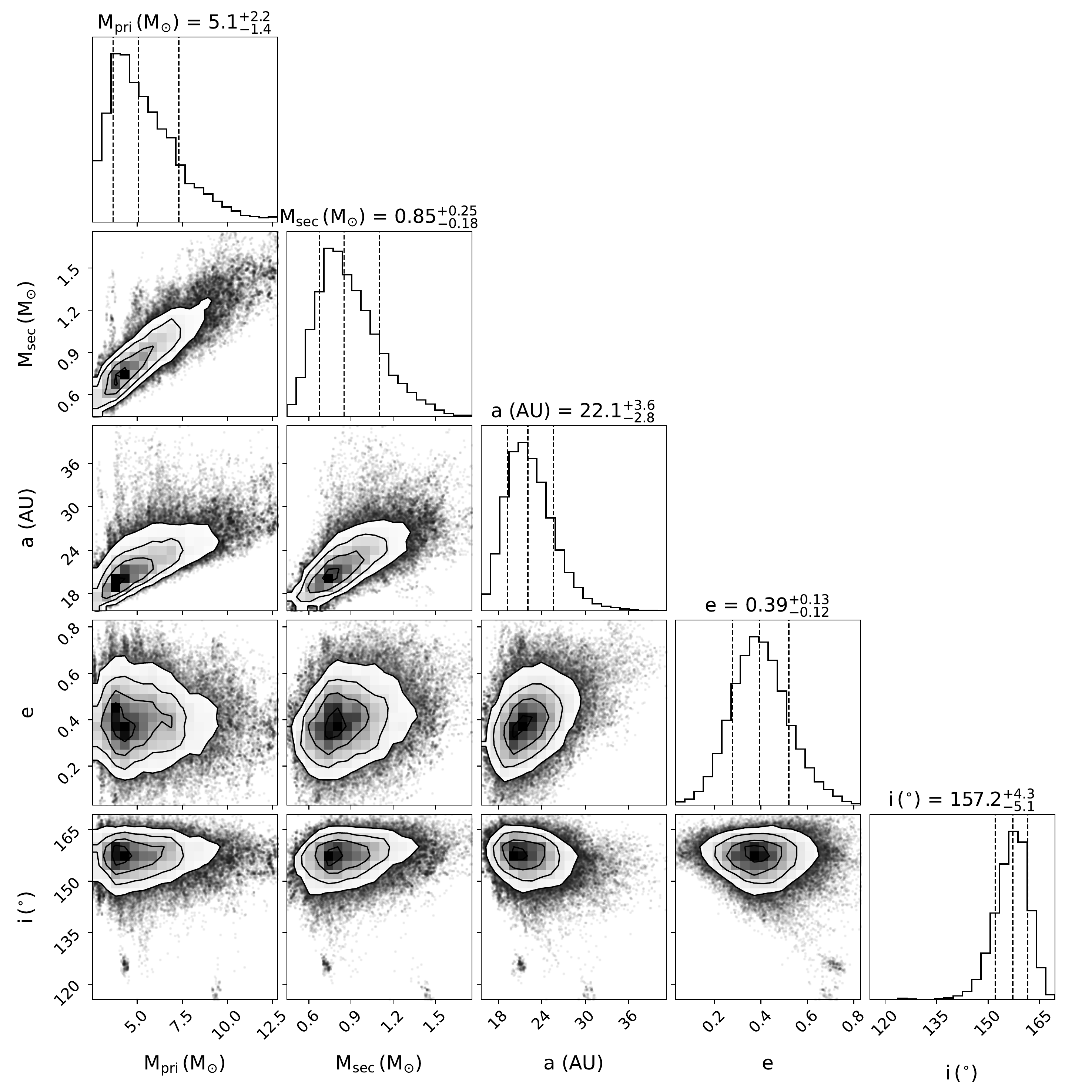}
\caption{\label{Fig:Corner_HD218356triple} Corner plots of the inner (left) and outer (right) orbit of HD\,218356.}
\end{figure*}

\FloatBarrier
\clearpage

\section{Two possible fits for HD\,201657}\label{AppHD201657}

\begin{figure*}[h]
\centering
\includegraphics[width=0.87\textwidth]{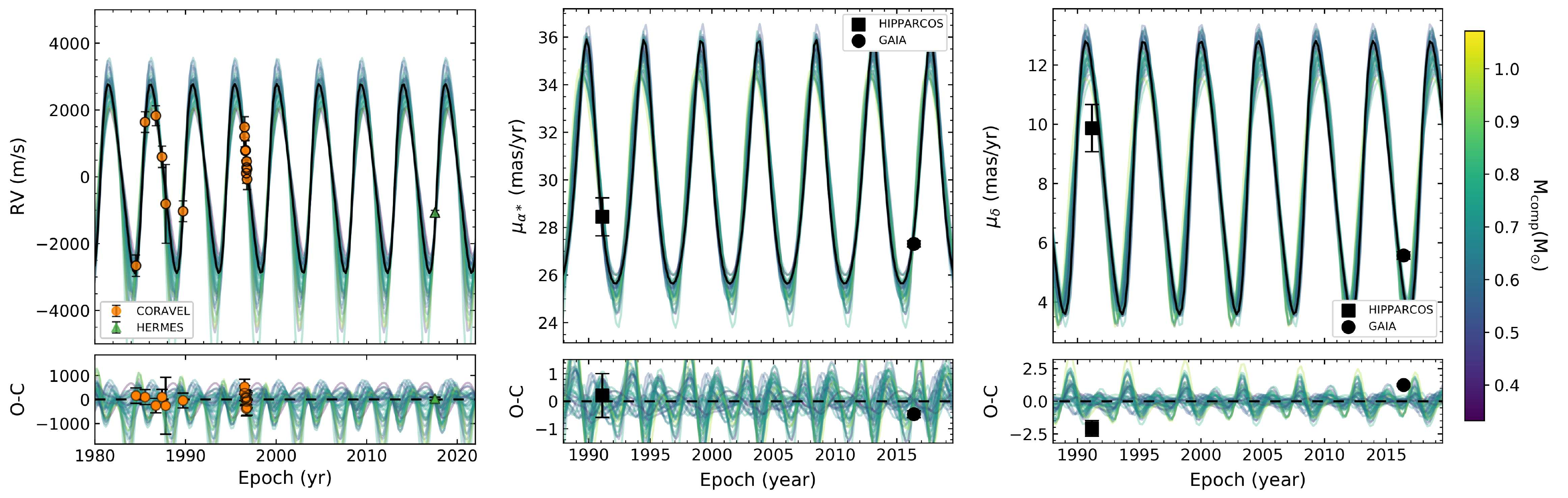}
\caption{\label{Fig:HD201657-short} Possible orbit for HD\,201657 with a smaller eccentricity. Compatible with the orbit published by \cite{Jorissen19}.}
\end{figure*}
\begin{figure*}[h]
\centering
\includegraphics[width=0.87\textwidth]{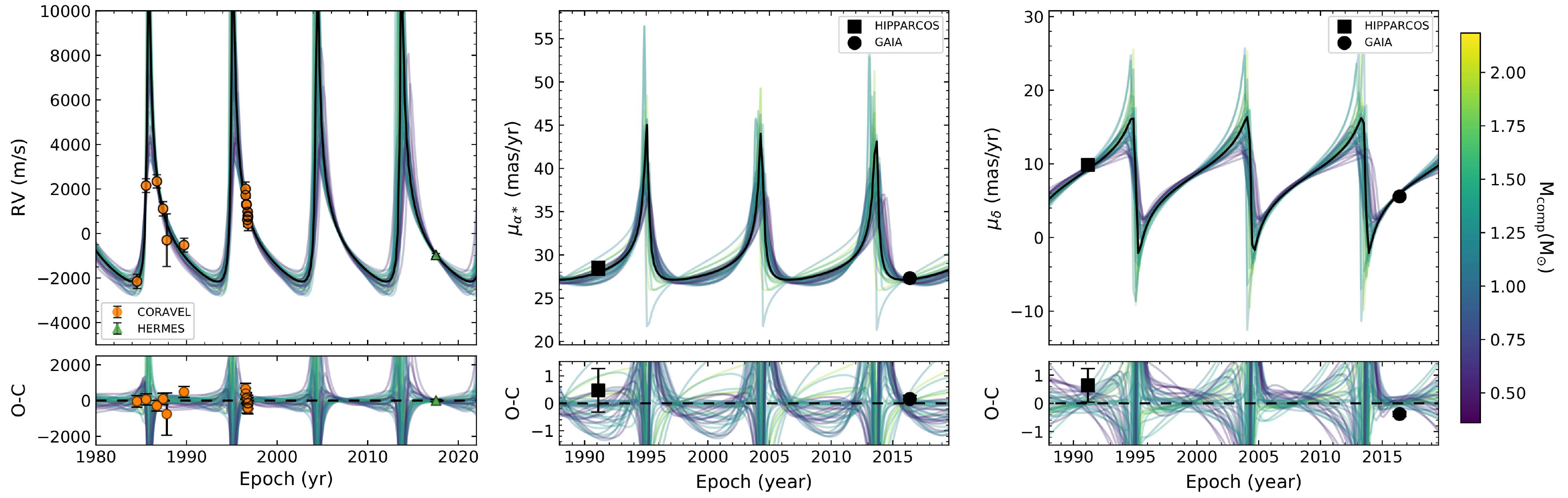}
\caption{\label{Fig:HD201657-long} Possible orbit for HD\,201657 with a larger eccentricity. Twice the period published by \cite{Jorissen19}.}
\end{figure*}
\begin{figure*}[h]
\centering
\includegraphics[width=\textwidth]{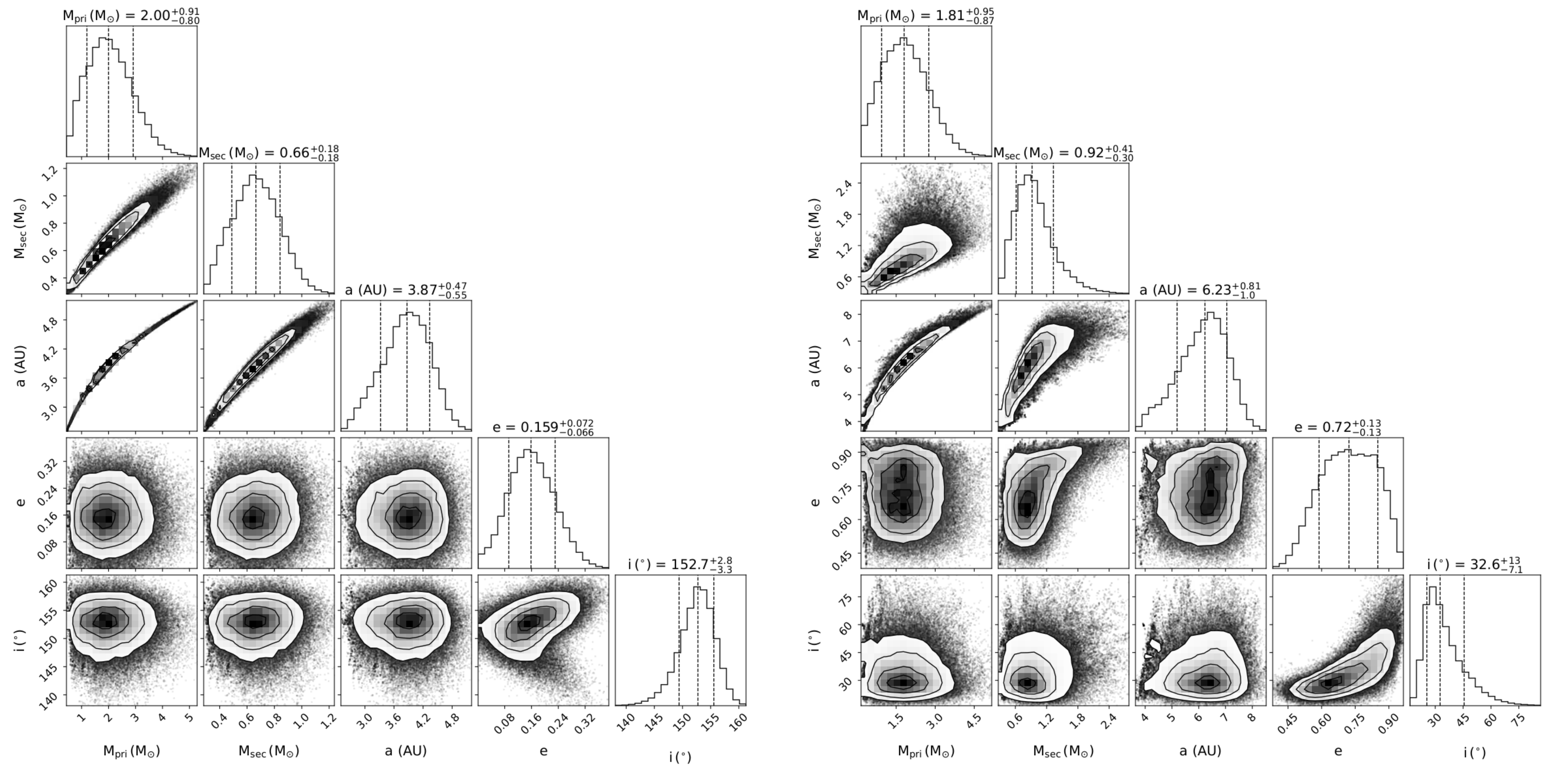}
\caption{\label{Fig:HD201657_corners} Corner plots associated with the fits for HD\,201657 shown in figure \ref{Fig:HD201657-short} (left) and \ref{Fig:HD201657-long} (right).}
\end{figure*}

\end{appendix}
\end{document}